\documentclass[
dsingle,
]{Dissertate}
\usepackage{subfig}
\usepackage{graphicx}
\usepackage{dcolumn}
\usepackage{xcolor}   
\usepackage{bm}        
\usepackage{amssymb}   
\usepackage{physics}
\usepackage{tikz}
\usepackage{xcolor}
\usepackage{float}
\usepackage{dsfont}
\usepackage{mathtools}
 \usepackage{amsthm}
 \usepackage[english]{babel}
\usepackage[normalem]{ulem}
\usepackage{cancel}
\usepackage{url}
\usepackage{bbm}
\usepackage{caption}

\usepackage[capitalize]{cleveref}
\crefname{section}{Sec.}{Secs.}  
\crefname{appendix}{App.}{Apps.}  
\newcommand{\stkout}[1]{\ifmmode\text{\sout{\ensuremath{#1}}}\else\sout{#1}\fi}

\makeatletter
\DeclareRobustCommand{\element}[1]{\@element#1\@nil}
\def\@element#1#2\@nil{%
  #1%
  \if\relax#2\relax\else\MakeLowercase{#2}\fi}
\makeatother
\let\oldaddcontentsline\addcontentsline

\newcommand{\starttocentries}{\let\addcontentsline\oldaddcontentsline}

\newcommand{\nocontentsline}[3]{}
\newcommand{\tocless}[2]
{\bgroup\let\addcontentsline=\nocontentsline#1{#2}\egroup}

\newcommand*{\TPT}{{}^{3}\mathrm{P}_{2}}
\newcommand*{\SSZ}{{}^{1}\mathrm{S}_{0}}
\newcommand{\rme}[3]{\langle#1\vert\vert #2 \vert} 
\makeatletter
\newtheorem*{rep@theorem}{\rep@title}
\newcommand{\newreptheorem}[2]{%
\newenvironment{rep#1}[1]{%
 \def\rep@title{#2 \ref{##1}}%
 \begin{rep@theorem}}%
 {\end{rep@theorem}}}
\makeatother

\theoremstyle{plain}

\newreptheorem{theorem}{Theorem}

\theoremstyle{remark}

\theoremstyle{definition}

\newreptheorem{definition}{Definition}

\renewcommand{\o}{
	\mathchoice
	{{\scriptstyle\mathcal{O}}}
	{{\scriptstyle\mathcal{O}}}
	{{\scriptscriptstyle\mathcal{O}}}
	{\scalebox{.7}{$\scriptscriptstyle\mathcal{O}$}}
}

\renewcommand{\Re}{\operatorname{Re}}

\begin{document}

\title{Quantum Computation Using Large Spin Qudits}
\author{Sivaprasad Thattupurackal Omanakuttan}
\previousdegrees{	B.Sc., Physics, Mahatma Gandhi University, Kerala, 2015\\M.Sc., Physics, Indian Institute of Technology Madras, 2017 
}

\advisor{Ivan H. Deutsch}
\committeeInternalOne{Milad Marvian}
\committeeInternalTwo{Tameem Albash}

\committeeExternal{Michael J Martin}

\degree{Doctor of Philosophy}
\degreeabbrv{Ph.D.}
\field{Physics}
\degreeyear{2024}
\degreeterm{Spring}
\degreemonth{May}
\department{Physics and Astronomy}
\defensedate{March 20\textsuperscript{th}, 2024}

%

\frontmatter
\setstretch{\dnormalspacing}

\setcounter{chapter}{0}  
\chapter{Introduction}
\label{chap:introduction}
The last decade of the $20$th century saw the marriage of two of its great scientific pillars: quantum mechanics and information science and gave birth to the field of quantum information science (QIS) \cite{PRXQuantum.1.020101}.
At the foundational level of scientific understanding, quantum mechanics stands as the most precise theory. 
It delineates the fundamental workings of the world. 
The advent of information science ushered in a new era, bringing forth computers, digital communication, and other transformative devices that have reshaped our daily lives. 
Quantum Information Science (QIS) emerged from inquisitive, curiosity-driven fundamental research, seeking to unravel the implications of merging quantum mechanics with the principles of information science.

QIS tries to harness the power of quantum systems for information processing and as a field lies at the convergence of quantum optics, atomic
molecular and optical (AMO)  physics, condensed matter physics, computer science,  and several other areas of science and engineering \cite{caves2013quantum}.
This has enabled the broad application and integration of tools, techniques, and concepts specific to quantum information science into various domains within theoretical and experimental physics.
QIS has since shown promise in diverse applications, spanning quantum computation, quantum cryptography, quantum sensing, quantum simulation, quantum networking, and more. 
Among these applications, quantum computation stands out as one of the most anticipated, holding significant advantages over classical computers\cite{shor1994algorithms,shor1999polynomial,farhi2001quantum,lloyd1996universal,biamonte2017quantum,aspuru2005simulated}.

Numerous intriguing problems remain beyond the reach of classical computers, not due to inherent unsolvability, but rather because of the astronomical resources needed to address practical instances of these challenges.
The spectacular promise of quantum computers is to use quantum superposition and quantum entanglement to enable new quantum algorithms that tackle problems that require exorbitant resources for their solution on a classical computer \cite{shor1994algorithms,shor1999polynomial,farhi2001quantum,lloyd1996universal,biamonte2017quantum,aspuru2005simulated}.
For example, one has a class of algorithms based on quantum Fourier transform, and includes remarkable algorithms for solving the factoring and discrete logarithm problems, providing a striking exponential speedup over the best-known classical algorithms \cite{shor1994algorithms,shor1999polynomial,kitaev1995quantum}. 
Another example of algorithms is based on Grover’s algorithm for performing quantum searching \cite{grover1996fast,Bennett_1997}. 
The quantum searching algorithm offers a notable quadratic speedup over the best classical algorithms, presenting a significant advancement. Its importance stems from the widespread utilization of search-based techniques in classical algorithms. In many cases, a straightforward adaptation of the classical algorithm enables the development of a faster quantum algorithm, making quantum search particularly impactful.
The exponentially expanding Hilbert space is important for implementing quantum computation, facilitating the storage and processing of information \cite{blume2002climbing}. 

However, the scalability of quantum computation faces limitations imposed by decoherence, which arises from the influence of the external environment on the quantum system. Consequently, it becomes imperative to explore strategies for scaling the system while mitigating the adverse impact of decoherence. 
Unlocking the full power of quantum computation involves comprehending and devising approaches to overcome the effects of decoherence \cite{knill1998resilient,aharonov1997fault,knill2005quantum,raussendorf2007topological}.
Recent years have witnessed significant theoretical and experimental advancements toward realizing the full potential of quantum computation, even in the presence of decoherence \cite{acharya2022suppressing,ryan2022implementing,krinner2022realizing,Bluvstein_Lukin_2023_QEC_Logical}.

In the standard paradigm of quantum information processing (QIP) one encodes information in qubits, the quantum analog of classical bits, by isolating two well-chosen energy levels of the system such that the computation space grows as $2^N$ for $N$ qubits.
In many platforms, one has access and control over multiple levels per subsystem, which can enhance our ability to do QIP in a variety of ways~\cite{wang2020qudits,Blok2021,Gross2021,puri2020bias,Gottesman2001}. 
In particular, one can encode information in base-$d>2$ using $d$-level qudits~\cite{wang2020qudits} such that the computational space goes as $d^N$. 
With a larger state space per subsystem, qudits offer potential advantages for quantum communication~ \cite{PhysRevLett.90.167906}, quantum algorithms~\cite{luo2014geometry,luo2014universal,li2013geometry,lu2020quantum}, and  topological quantum
systems~\cite{cui2015universal,cui2015universal1,bocharov2015improved}.
Quantum computation with qudits can also reduce circuit complexity and can be advantageous in a variety of noisy intermediate scalable quantum(NISQ)-era applications~\cite{brylinski2002universal,lu2020quantum,luo2014universal,li2013geometry,
zobov2012implementation,weggemans2022solving,PhysRevLett.129.160501}.
Qudits may also provide significant advantages in quantum error correction and fault-tolerant quantum computation \cite{campbell2014,PhysRevA.83.032310,gottesman1998fault,campbell2012,Eliot2016}.

This dissertation delves into the realm of quantum computation utilizing spin qudits as its focal point.
The primary emphasis is on addressing and mitigating the adverse impacts of decoherence by leveraging access to qudits. 
One avenue of exploration involves working towards universal quantum computation, where the utilization of qudits allows for quantum computational supremacy with fewer subsystems.
Another facet of this research investigates the feasibility of encoding a qubit within a qudit  for the purpose of fault-tolerant quantum computation.
By harnessing the properties of this qudit with multiple levels, we can establish logical qubits that possess inherent resistance to the impact of dominant noise channels, paving the way for more robust quantum computation.

In the gate-based approach to quantum computation with qubits, a universal gate set consists of single-qubit gates that generate the group SU$(2)$ and one entangling two-qubit gate, such as CNOT ~\cite{divincenzo1995two}. 
This generalizes simply for qudits. 
The universal gate set consists of the generators of single-qudit gates in SU($d$) and an entangling two-qudit gate \cite{muthukrishnan2000multivalued,zhou2003quantum,brennen2005criteria}. 
 The gates that are necessary for the implementation of the universal gate set have been recently implemented for qudits in superconducting transmon~\cite{Blok2021,goss2022high,fischer2022towards} as well as in trapped ions~\cite{ringbauer2021universal,hrmo2022native} up to dimension $d=7$.  In these experiments, one implements qudit gates using constructive methods, e.g., through a prescribed set of Givens rotations~\cite{brennen2005criteria,li2013decomposition}.

 While there has been substantial progress, much work remains to be done to efficiently implement a high-fidelity universal qudit gate set.  
In this dissertation I propose an alternative approach based
on quantum optimal control which was originally developed in NMR~\cite{vandersypen2005nmr} and for coherent control of chemistry~\cite{rabitz2000whither,shapiro2012quantum}, and has been extensively used in quantum information processing~\cite{koch2022quantum}.
This approach yields high-fidelity gates for qudits in the presence of decoherence and can be made robust to experimental imperfections.
As a concrete example that demonstrates the power of the method, we present here an optimal control scheme to implement universal gates in qudits encoded in the nuclear spin of $^{87}$Sr atoms. 
The nuclear spin is a good memory for use in quantum information processing given its weak coupling to the environment and resilience to other background noise~\cite{barnes2021assembly,PhysRevLett.101.170504,daley2011quantum}. 

 Quantum computers are extremely susceptible to environmental noise and imprecise control, which hinders achieving their full computational capacity.
Fault-tolerant quantum computation (FTQC), provides a solution to perform reliable computation even in the presence of imperfect elementary components \cite{knill1998resilient,aharonov1997fault,knill2005quantum,raussendorf2007topological}. 
The cornerstone of FTQC is the threshold theorem, which states that if the error rate of individual components remains below a constant threshold, then arbitrarily long quantum computation can be performed \cite{aharonov1997fault,knill1998resilient,preskill1998reliable,kitaev1997quantum,Aliferis2008fault}. In addition to the value of noise threshold, a critical aspect of FTQC is the resource overhead, quantifying the number of physical systems required to encode logical information.
Despite the formidable challenges, there has been notable experimental progress in FTQC, bringing us closer to harnessing the full potential of quantum computing \cite{acharya2022suppressing,ryan2022implementing,krinner2022realizing,Bluvstein_Lukin_2023_QEC_Logical}.

The conventional approaches for FTQC are mostly devoted to structureless and uncorrelated noise. 
An instance of this is depolarizing noise, where all local Pauli operators have an equal probability. 
However, such decoherence models often entail stringent threshold requirements and result in significant overheads for FTQC
~\cite{knill2005quantum,raussendorf2007topological,svore2006noise,spedalieri2008latency}.
An alternative strategy involves seeking error-correcting codes tailored to the prevalent noise sources of the particular physical platform.
When possible, these tailored approaches can lead to improved thresholds and reduced resource overhead~\cite{Aliferis2008fault,zzPoulin}.
One well-known case is when one noise channel dominates over all other noises. 
For example, the cases in which dephasing noise dominates over bit-flip noise for the qubit shows improved threshold as shown by  Aliferis and Preskill~\cite{Aliferis2008fault} and can be implemented in bosonic systems \cite{puri2020bias,guillaud2019repetition}.
Another case is the Gottesman-Preskill-Kitaev encoding of a qubit in an infinite dimensional oscillator \cite{Gottesman2001} which corrects for displacement errors in a bosonic mode. 
Additionally, in scenarios where erasure errors dominate over Pauli errors,  tailored error-correcting codes have proven advantageous~\cite{Grassl_erasure_1997_PRA,Wu_Puri_Thompson_2022_Nature_erasure,Sahay_Puri_biased_erasure_PRX_2023}. 
By addressing the specific characteristics of dominant noise sources, these tailored methods offer promising avenues to enhance the performance of FTQC.

A similar but much less explored avenue is to encode a qubit in a spin $>1/2$ system (qudit). 
In this context, the angular momentum operators form the natural set of error operators for such encodings, generalizing the Pauli operator basis for qubits. Earlier studies identified quantum error-correcting encodings, but these constructions were not fault-tolerant~\cite{Gross2021,omanakuttan2023multispin}. 
In this dissertation, I investigate how we can achieve FTQC, specifically for a qubit encoded in a spin qudit. 
This approach may be extended to a wide range of physical systems, including semiconductor qubits \cite{Gross2021, gross2021hardware}, ion traps \cite{ringbauer2021universal, Low2020}, atomic systems \cite{omanakuttan2021quantum, Siva_Qudit_entangler_2023, zache2023fermion}, molecules \cite{castro2021optimal,jain2023ae}, and superconducting systems \cite{ozguler2022numerical, Blok2021}, wherein spin qudits offer the means to encode logical qubits.


\section{Outline}
The remainder of this dissertation is organized as follows. 
\cref{chap:Qudecimal_quantum_control} is based on the publication \cite{omanakuttan2021quantum}. 
In this work we study the ability to implement unitary maps on states of the $I=9/2$ nuclear spin in \textsuperscript{87}Sr, a $d=10$  dimensional (qudecimal) Hilbert space, using quantum optimal control. 
\cref{chap:Qudit_entanglers} is based on the publication \cite{Siva_Qudit_entangler_2023}.
Here, we study the generation of two-qudit entangling quantum logic gates using two techniques in quantum optimal control.  
We take advantage of both continuous, Lie algebraic control and digital, Lie group control.
\cref{chap:Qudit_fault}  is based on the publication \cite{Omanakuttanerrorcorrection}.
Here I construct a fault-tolerant quantum error-correcting protocol based on a qubit encoded in a large spin qudit using a spin-cat code, analogous to the continuous variable cat encoding.
The spin-cat codes we develop substantially reduce the resource requirements for fault-tolerance in that a single atom can encode the logical qubit, with only minimal repetition given the structure of the noise. 
An important innovation is the development of a CNOT gate that preserves the structure of the noise at the logical level.  We do so in a way that also is well-aligned with experimental capabilities.
\cref{chap:Qudit_leakage} is based on the publication \cite{Omanakuttancooling}.
Here I present ideas of converting leakage errors to erasure errors when quantum information is encoded in the nuclear spin in the electronic ground state.
After doing so, erasure can be efficiently corrected by standard error correction protocols. 
This protocol for erasure conversion is compatible with a scheme to cool the atoms while preserving the coherence, generalizing previous work on this problem~\cite{Reichenbach_cooling}. 
Lastly, I summarize all of our work in \cref{chap:summary_and_outlook} and suggest potential avenues of research for future work.

\section{List of Publications}
Below is a chronological list of the papers that I coauthored during my PhD.
Not all
works listed here appear as chapters in this dissertation
\begin{itemize}

\item  \cite{omanakuttan2021quantum} 
S.~Omanakuttan, A.~Mitra, M. J.~Martin, and I. H.~Deutsch. Quantum optimal control of ten-level
nuclear spin qudits in $^{87}$Sr. \href{https://link.aps.org/doi/10.1103/PhysRevA.104.L060401}{\emph{ Phys. Rev. A,} 104, L060401 (2021).}

    \item \cite{PhysRevD.104.123026} H.~Duan, J. D.~Martin and S.~Omanakuttan. Flavor isospin waves in one-dimensional axisymmetric neutrino gases.\href{https://link.aps.org/doi/10.1103/PhysRevD.104.123026}{\emph{Phys. Rev. D,} 104, 123026 (2021).}
  
    \item \cite{omanakuttan2023scrambling} S.~Omanakuttan, K.~Chinni, P. D.~Blocher, and P. M.~ Poggi. Scrambling
and quantum chaos indicators from long-time properties of operator distributions. \href{https://link.aps.org/doi/10.1103/PhysRevA.107.032418}{\emph{ Phys. Rev. A,} 107, 032418 (2023).}
    
    \item \cite{anupampra} A. Mitra, S. Omanakuttan, M. J. Martin, G. W. Biedermann, and I. H.
Deutsch. Neutral-atom entanglement using adiabatic rydberg dressing.\href{https://link.aps.org/doi/10.1103/PhysRevA.107.062609}{\emph{ Phys. Rev. A,} 107, 062609 (2023).}
   
    \item \cite{buchemmavari2023entangling} V. Buchemmavari, S. Omanakuttan, Y.-Y. Jau, and I. Deutsch. Entangling quantum logic gates in neutral atoms via the microwave-driven spin-flip
blockade. \href{https://doi.org/10.48550/arXiv.2212.08799}{\emph{  Phys. Rev. A,} 109, 012615  (2024)}
    \item \cite{blocher2023probing}P. D. Blocher, K. Chinni, S. Omanakuttan, and P. M. Poggi. Probing
scrambling and operator size distributions using random mixed states and local
measurements.\href{https://doi.org/10.48550/arXiv.2305.16992}{  \emph{ Phys. Rev. Research,} 6, 013309 (2024)}
     \item \cite{omanakuttan2023multispin} S. Omanakuttan and J. A. Gross. Multispin Clifford codes for angular momentum
errors in spin systems. \href{https://link.aps.org/doi/10.1103/PhysRevA.108.022424}{\emph{ Phys. Rev. A,} 108, 022424 (2023).}
\item \cite{Omanakuttan2023gkp}  S. Omanakuttan and T. J. Volkoff. Spin-squeezed gottesman-kitaev-preskill
codes for quantum error correction in atomic ensembles. 
\href{https://link.aps.org/doi/10.1103/PhysRevA.108.022428}{\emph{ Phys. Rev. A, } 108, 022428 (2023).}
    \item \cite{Siva_Qudit_entangler_2023}S. Omanakuttan, A. Mitra, E. J. Meier, M. J. Martin, and I. H. Deutsch.
Qudit entanglers using quantum optimal control. \href{https://link.aps.org/doi/10.1103/PRXQuantum.4.040333}{\emph{PRX Quantum,} 4:040333}

    \item \cite{Omanakuttanerrorcorrection} S. Omanakuttan, V. Buchemmavari, J. Gross, I. H. Deutsch, and M. Marvian. Fault-tolerant quantum computation using large spin cat-codes. \emph{arXiv:2401.04271,} 2024.
    \item  \cite{Omanakuttanmetrology}  S. Omanakuttan, J. Gross, and T. J. Volkoff. Quantum error correction
inspired multiparameter quantum metrology. \emph{ in preparation,} 2024.
    \item \cite{Omanakuttancooling}  S. Omanakuttan, V. Buchemmavari, M. J. Martin, and I. H. Deutsch. Converting leakage errors to erasure errors and cooling atoms while preserving coherence in neutral atoms for fault-tolerant quantum computation. \emph{in preparation,} 2024
    \item \cite{Omanakuttanspingkp2}   S. Omanakuttan, T. Thurtell, and B. Q. Baragiola. Bridging the discrete
and continuous variable quantum error correction\emph{ in preparation,} 2024.
\end{itemize}

 \chapter{Quantum Optimal Control of Nuclear Spin Qudecimals in \textsuperscript{87}Sr} 
\label{chap:Qudecimal_quantum_control}

\section{Introduction}
\label{sec:qudecimal_introduction}
Ultracold ensembles of alkaline-earth atoms trapped in optical lattices or arrays of optical tweezers are a powerful platform for quantum information processing (QIP), including atomic clocks and sensors~\cite{Ludlow2015, campbell2017fermi, norcia2019seconds, covey20192000, young2020half}, simulators of many-body physics~\cite{gorshkov2010two, daley2011quantum, mukherjee2011many,banerjee2013atomic,isaev2016spin, kolkowitz2017spin}, and general purpose quantum computers~\cite{Madjarov_Endres_2020_Sr,daley2011quantum,PhysRevLett.98.070501}.  
The ability to optically manipulate coherence in single-atoms via ultranarrow optical resonances on the intercombination lines, together with the ability to create high-fidelity entangling interactions between atoms when they are excited to high-lying Rydberg states~\cite{saffman2010quantum,Saffman_review_2016_Rydberg,Browaeys_2016} provides tools that makes this system highly controllable for such applications.  In addition, fermionic species have nuclear spin. As the ground state is a closed shell, there is no electron angular momentum, and the nuclear spin with its weak magnetic moment is highly isolated from the environment.  Such nuclear spins in alkaline-earth atoms are thus natural carriers of quantum information given their long coherence times and our ability to coherently control them with magnetic and optical  fields.  Nuclear spins are also seen as excellent carriers of quantum information in the solid state as demonstrated in pioneering experiments including in NV-centers~\cite{morishita2020} and dopants in silicon~\cite{soltamov2019,morello2018quantum,Godfrin2017,Leuenberger2003}.  

Using magneto-optical fields,~\cite{Lester2021} recently demonstrated the control of qubits encoded in  two nuclear-spin magnetic sublevels levels in  \textsuperscript{87}Sr. The nuclear spin in this atomic species, however, it is not a two-level system; the spin is $I=9/2$ and there are $d=2I+1=10$ nuclear magnetic sublevels.  Such qudits,  here  ``qudecimals," have potential advantage for QIP.  First and foremost, one can encode a $D=d^{n_d}=2^{n_2}$ dimensional Hilbert space associated with $n_2$ qubits in $n_d=n_2/\log_2 d$ qudits.  While only a logarithmic saving, this is meaningful for the qudecimal ($\log_2 d= 3.32$), especially when trapping and control of each atom is at a premium. This savings extends to algorithmic efficiency, in that the number of elementary two-qudit gates necessary to implement a  general unitary map scales as $O(n_d^2 D^2) = O\left( \frac{n_2^2 D^2}{(\log_2 d)^2} \right)$~\cite{muthukrishnan2000multivalued}.  Moreover, qudit architectures can show increased resilience to noise~\cite{cozzolino2019high} and additional routes to quantum error correction~\cite{gottesman1998fault}. For example, one can protect against dephasing errors by encoding a qubit in a nuclear spin qudit~\cite{Li2017}.  In addition, fault-tolerant operation of a quantum computer may be more favorable based on qudit vs. qubit codes~\cite{PhysRevA.83.032310,campbell2014}.

While QIP with qudits has great potential, there are substantial hurdles.  State preparation and readout are more challenging for systems with $d>2$.  Moreover, quantum logic with qudits is more complex.  Universal quantum logic with qubits can be achieved with a set of logic gates that include the unitary-generators of SU(2) on each qubit, plus one entangling gate between qubits pairwise.  In the case of qudits, in addition to the entangling gate, we require unitary-generators of SU($d$) for each subsystem~\cite{muthukrishnan2000multivalued,zhou2003quantum,brennen2005criteria,luo2014universal}. Unlike qubits, the Lie algebra of such gates are not spanned by the native Hamiltonians, and thus implementation of this generating set is not straightforward.  Different approaches have been studied to implement SU($d$) gates~\cite{Moreno2018,neeley2009emulation,Low2020,sawant2020,moro2019}. One approach is to specify an arbitrary SU($d$) unitary matrix through a sequence of so-called Givens rotations acting between pairs of levels~\cite{O'Leary2006}.   In a landmark experiment, the Innsbruck group employed this construction to experimentally demonstrate universal quantum logic with qudits in a trapped ions ion~\cite{ringbauer2021universal}, with performance similar to qubit quantum processors.

An alternative powerful approach to implementing universal quantum logic is to employ the tools of quantum optimal control.
In this paradigm, one numerically searches for a time-dependent waveform that achieves the desired SU($d$) unitary map when one has access to a Hamiltonian that makes the system universally ``controllable"~\cite{Merkel2009,jurdjevic1972control,goerz2015optimizing, koch2016controlling,Frey2020,brockett1973lie,schirmer2002degrees}.  Optimal control is a powerful and flexible approach that does not require specific pairwise Givens rotations, can be high-fidelity, and can be made robust to imperfections such as inhomogeneities through the tools of robust control~\cite{anderson2015accurate, goerz2015optimizing, glaser2015training, koch2016controlling}. In seminal work, the Jessen group used optimal control to demonstrate high-fidelity control of qudits encoded in the hyperfine spin levels of ground-state cesium~\cite{Paul_experiment_Cs_2007, Smith2013}.  This flexible control has found potential application in studies of quantum simulation~\cite{Poggi2020}. 


\section{Background}
\label{sec:qudecimal_background}

In this chapter we build on this approach to study implementation of SU(10) gates on the nuclear spin of $^{87}$Sr-based on quantum optimal control. A nuclear-spin encoding may have long-term advantages compared to hyperfine states that couple electron and nuclear spins, in its strongly reduced sensitivity to to background magnetic fields and resilience against decoherence driven by photon scattering from optical tweezers or lattices ~\cite{PhysRevLett.98.070501, dorscher2018lattice}.  Weak coupling to the environment, of course, comes with increased challenges of weak coupling to control fields.  We will show, nonetheless, that with reasonable experimental parameters one can implement high-fidelity qudecimal logic, with low decoherence.

We consider open loop-control in a Hilbert space with finite dimension $d$, governed by a Hamiltonian $H[\mathbf{c}(t)] = H_0 + \sum_\lambda c_\lambda(t) H_\lambda$ where $\mathbf{c}(t)=\{c_\lambda(t)\}$ is the set of time-dependent classical control waveforms. The system is said to ``controllable" if the set of Hamiltonians, $\{H_0, H_\lambda\}$, are generators of the Lie algebra SU($d$).  Then $\exists \hspace{0.1cm} \mathbf{c}(t)$ \hspace{0.2cm } such that $U[\mathbf{c},T]=\mathcal{T}\left[\exp\left(-i\int_0^T H[\mathbf{c}(t)]dt\right)\right]=U_{\mathrm{tar} }$ for any target unitary matrix $U_{\mathrm{tar} }=\mathrm{SU(d)}$ in this space.  The minimal time $T$ for which this is possible is known as the ``quantum speed limit" (QSL)~\cite{caneva2009optimal} . 
See \cref{chap:app_qudecimal_qudits} for additional details of the quantum control protocol used here.

\begin{figure}[!ht]
\includegraphics[width=0.45\textwidth]{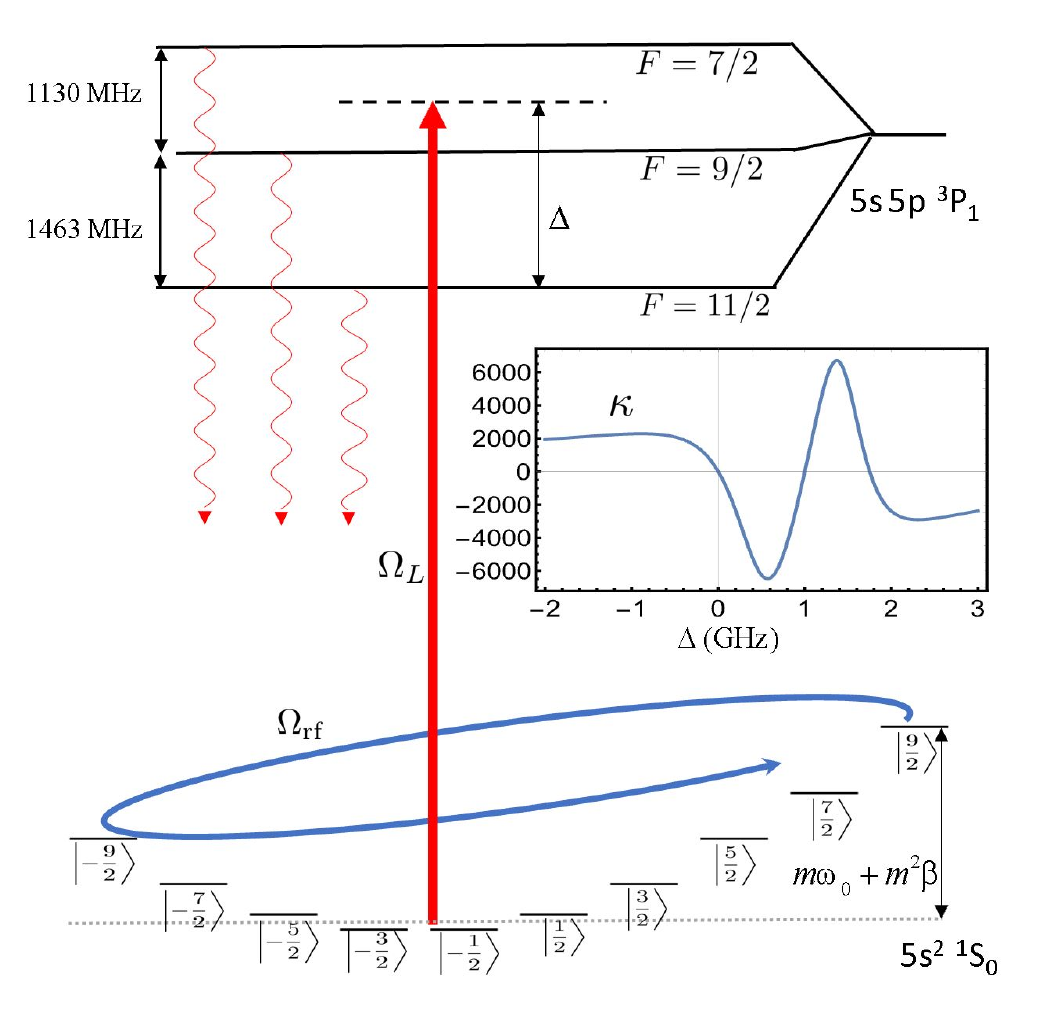}
\centering
\caption{  Schematic for magneto-optical control.  The qudecimal is encoded in the ten magnetic sublevels of the nuclear spin, $\ket{-9/2}\rightarrow \ket{9/2}$, in the $5s^2$ $^1S_0$ ground state. Their levels are shifted by a linear Zeeman effect due to a bias magnetic field and a quadratic tensor AC-Stark effect induced by an off-resonant laser beam, polarized along the quantization axis, and detuned $\Delta$ between the hyperfine levels of the $5s5p$ $^3P_1$ intercombination line.  Control of the qudecimal is then achieved with a phase modulated radio-frequency magnetic field, co-rotating at the bare Larmor precession frequency, whose amplitude causes Rabi rotations at frequency $\Omega_{\mathrm{rf}}$. The figure of merit for the control is the ratio of the AC-Stark shift to the photon scattering, $\kappa$ , shown in the inset (see text).}
\label{fig:level_diagram_strontium}
\end{figure}

One can achieve quantum controllability of the nuclear spin qudecimal  through magneto-optical interactions.  We combine magnetic spin resonance in the presence of an off-resonant laser field as depicted in \cref{fig:level_diagram_strontium}.  The Hamiltonian acting on the nuclear spin in the $5s^2$ $^1S_0$ ground state takes the form $H=H_{\mathrm{mag}} +H_{\mathrm{LS}}$. Here $H_{\mathrm{mag}} = -\boldsymbol{\mu}\cdot \mathbf{B}(t)$ is the magnetic spin-resonance Hamiltonian, with $\boldsymbol{\mu} = g_I \mu_N \mathbf{I}$ the nuclear magnetic dipole vector operator and $\mathbf{B}(t) = B_\parallel \mathbf{e}_z + B_T \Re\left[(\mathbf{e}_x + i \mathbf{e}_y) \mathbf{e}^{-i\left(\omega_\text{rf} t +\phi(t)\right)}\right]$ the magnetic field consisting of a strong bias defining the quantization axis $\mathbf{e}_z$ and a transversely rotating rf-magnetic field with a time dependent phase $\phi(t)$.
Taken alone, the $H_{\mathrm{mag}}$ generates only SU(2) rotations of nuclear spin.  
To achieve full SU($d$) control we add a light-shift Hamiltonian due to the AC-Stark effect, $H_{LS}=-\alpha_{zz}(\omega_L) \left| E_0 \right|^2/4$ where $\alpha_{zz}(\omega_L)$ is the $zz$-component of atomic AC-polarizability tensor operator for a laser field at frequency $\omega_L$ linearly-polarized along the quantization axis, $\mathbf{E}_L(t) = \mathbf{e}_z \Re \left(E_0 e^{-i\omega_L t}\right)$. 
The form of $\alpha_{zz}$ depends on the atomic structure and the detuning of the laser from atomic resonance.  
In particular, when the detuning is not large compared to the hyperfine splitting in the excited state, the polarizability has an irreducible rank-2 tensor component $\alpha_{zz} = \alpha^{(2)} I_z^2$ (there is also a trivial scalar term proportion to the identity)~\cite{deutsch2010quantum}, where $I_x,I_y,I_z$ are the nuclear spin operators along the three Cartesian coordinates. 
This quadratic spin twist together with the linear Larmor precession yields a set of control Hamiltonians $\{I_x, I_y, I_z^2\}$ sufficient to generate the Lie algebra SU($2I+1$) for an arbitrary spin $I$~\cite{Giorda2003}.  
Such control was first demonstrated in the alkali atom cesium, for the hyperfine spin $F=3$ in the electronic ground state, in order to generate nonclassical spin states in the $d=7$ dimensional Hilbert space~\cite{Paul_experiment_Cs_2007}.  

Importantly, the size of tensor polarizability $\alpha^{(2)}$ depends on the ratio of the excited state hyperfine splitting to the laser detuning~\cite{deutsch2010quantum} , achieving its maximum when these are of the same order.  Thus, to achieve high-fidelity control, one must tune sufficiently close to resonance, while avoiding photon scattering that leads to decoherence. Critically, in alkaline-earth atoms, the first excited $^3P_1$ states have long lifetimes and large hyperfine splittings. This leads to a very favorable figure of merit for optimal control, as measured by the ratio of the characteristic tensor light shift to the photon scattering rate  $\gamma_s$,  $\kappa \equiv \alpha^{(2)} \left| E_0 \right|^2 /4 \gamma_s$.  For example, in \textsuperscript{87}Sr, the hyperfine splitting between the $F=7/2$ and  $F=9/2$ levels in the singly-excited $5s5p \; ^3P_1$ state is $\omega_{\mathrm{HF}}/2\pi=1130$ MHz, while the spontaneous emission linewidth is $\Gamma/2\pi = 7.5$ kHz.   For a scattering rate averaged over all magnetic sublevels \cite{deutsch2010quantum}, we find that when we detune  about halfway between these resonances, we obtain the maximum figure of merit $\kappa =6.8 \times 10^3$ (see Fig. \ref{fig:level_diagram_strontium}).  In contrast, $\kappa =18.6$ for $F=3$ hyperfine spin in the cesium ground state when the laser is tuned halfway between the $F=3$ and $F=4$ hyperfine levels in the excited $6P_{1/2}$ D1-resonance.  This small figure of merit limited the fidelity to around $0.85$ for the arbitrary state preparation.  A factor of $364$ increase in the figure of merit for alkaline earths shows the potential power of this approach to yield high-fidelity quantum optimal control of the nuclear spin qudit.

We consider control of the nuclear spin qudecimal with on-resonance rf fields  on resonance with the Zeeman splitting,  $\Delta E_0 = |g_I| \mu_N B_\parallel$, where $g_I \mu_N/h  = -184$~Hz/Gauss in \textsuperscript{87}Sr~\cite{olschewski1972messung}. In the rotating frame, the control Hamiltonian is
\begin{equation}
H(t)=\Omega_{\mathrm{rf}} \left( \cos[c(t) \pi]I_x+\sin[c(t) \pi] I_y\right)+\beta I_z^2,
\label{eq:Control_Hamiltonian}
\end{equation}
where $\Omega_\text{rf} = -g_I \mu_N B_T $ is the  rf-Rabi frequency and $\beta = \alpha^{(2)} \left|E_0\right|^2/4$ is the strength of the tensor light shift (here and to follow $\hbar=1$). Note, for a rotating rf-field, there is no rotating wave approximation, and this Hamiltonian is valid even when $\Omega_\text{rf} \ge \omega_\text{rf}$. Here the  control waveform is solely the rf-phase $c(t) \equiv \phi(t)/\pi$.   It was proven in \cite{Merkel2008} that varying $c(t)$ is sufficient to achieve universal control the system. 

\section{Numerical Results}
\label{sec:qudecimal_numerical_results}
We consider two classes of quantum control tasks, preparation of a target pure state $\ket{\psi_\text{tar}}$ and implementation of a unitary map $U_\text{tar}$.  Optimal control follows by maximizing the relevant fidelity,
\begin{eqnarray}
\mathcal{F}_{\psi}[\bm{c},T]&=&\left|\bra{\psi_{\text{tar}}}U[\bm{c},T]\ket{\psi_0}\right|^2,\\
\mathcal{F}_U[\bm{c},T]&=&\left|\Tr\left(U^{\dagger}_{\text{tar}}U[\bm{c},T]\right)\right|^2/d^2.
\label{eq:fidelity}
\end{eqnarray} 
This is achieved by discretizing the control waveform and then numerically maximizing the fidelity with gradient ascent.   In a series of works, the Rabitz group showed that the fidelity landscape is favorable for this purpose~\cite{
rabitz2004quantum,Hsieh2008}.  We choose here a piecewise constant parameterization (as in ~\cite{Merkel2008}) and write the control function as a vector $\mathbf{c} = \{c(t_j) | j=1,\dots,n\}$ where $t=j\Delta t$ and $n=T/\Delta t$, parameterizing waveforms that are constant over the duration $\Delta t$.  A minimal choice of $n$ depends on the number of parameters necessary for the control task; for state-maps $n_\text{min} = 2d-2$ and for arbitrary SU($d$) maps $n_\text{min} = d^2-1$.  In practice, we choose $n$ to be a larger than $n_\text{min}$ which improves the fidelity landscape when $T$ is close the the QSL. To numerically optimize $\mathcal{F}$ we use a variation of the well-known GRAPE algorithm~\cite{khaneja2005optimal}. 
See \cref{chap:app_qudecimal_qudits} for further details on the choice of parameterization and optimization.
 
\begin{figure}
\centering
\includegraphics[width=\textwidth]{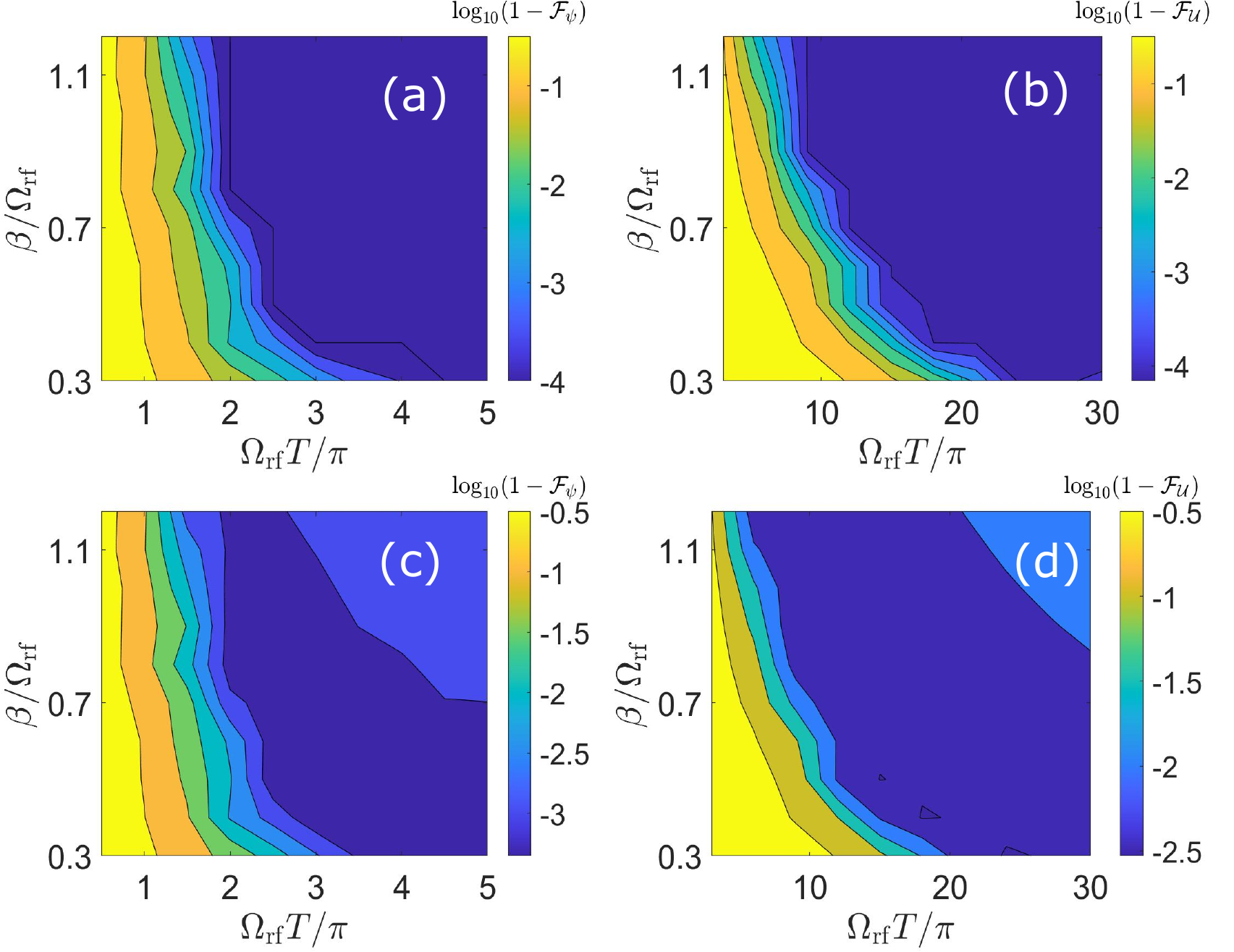}
\caption{Fidelity of objectives found by optimal control as a function of the strength of AC-stark shift, $\beta$, and the total time $T$, in units of the rf-Rabi frequency $\Omega_\text{rf}$. Predictions based on closed-unitary evolution for state-maps (a) and  SU(10) unitary-maps (b) averaged over $120$ Haar-random target states and $10$ Haar-random target SU(10) matrices, respectively.  The control waveforms are piecewise constant, over times $\delta t = T/n$.  For state maps we choose $n=120$ time steps; the unitary maps we take $n=500$. The bottom layer gives the similar figures in the presence of decoherence using the master equation, Eq. (5): state fidelity(c), Eq. (6): and process fidelity (d).}
\label{fig:decohrence_quantum_control}
\end{figure}

For a fixed value of $\Omega_\text{rf}$, the optimal choice of $\beta$ and total time $T$ are found empirically. Figures \ref{fig:decohrence_quantum_control}a(b) show the infidelity, $1-\mathcal{F}$, for state preparation (unitary maps), when averaged over 20 Haar random target vectors (10 random unitary maps).   
As expected, when $T \rightarrow \infty$ the infidelity is essentially zero, when the number of steps $n > n_{\mathrm{min}}$. 
The QSL is highly dependent on the value of $\beta$.
As expected, the optimal choice is $\beta \approx \Omega_\text{rf}$ as this provides the optimal mixing between Larmor precession and one-axis twisting.  The characteristics of state preparation and unitary maps are  similar in nature. The major difference between these two cases is that unitary mapping requires more time for the simple reason that unitary mapping has $d^2-1$ parameters compared to the $2d-2$ for the state preparation.  The quantum speed limit at $\beta=\Omega_{\mathrm{rf}}$ is $T_* \approx 1.5 \pi/\Omega_{\mathrm{rf}}$ for state preparation and $T_* \approx 8 \pi/\Omega_{\mathrm{rf}}$ for SU(10) unitary maps.
 
In principle, one can achieve arbitrarily high fidelity with increasing $T$.  
In practice $T$ is limited by the coherence time of the system.  Here, the coherence time is fundamentally limited by decoherence arising from photon scattering and optical pumping due to the off-resonant light-shift laser.  
We model the effects of decoherence in the state preparation protocols using the Lindblad Master equation \cite{deutsch2010quantum},

\begin{eqnarray}
\frac{d\rho[\bm{c}, t]}{dt} &=&-i \comm{H_\text{eff}[\bm{c}]}{\rho[\bm{c},t]}+ \Gamma\sum_{i}W_q\rho[\bm{c},t] W_{q}^{\dagger}  \nonumber \\ 
&\equiv& \mathcal{L}[\bm{c}]\left[ \rho[\bm{c},t]\right].
\label{eq: evolution of the density matrix}
\end{eqnarray}
where the jump operators for optical pumping between magnetic sublevels describing absorption followed by emission of a $q$-polarized photon are $W_q$,
\begin{equation}
W_q=\sum_{F'}\frac{\Omega/2}{\Delta_{FF'}+i\Gamma/2}(\bm{e}_q^{*}.\bm{D}_{FF'})(\vec{\epsilon}_L.\bm{D}_{FF'}^{\dagger}).
\end{equation}
Here $\bm{D}_{FF'}^{\dagger}$ are the dimensionless dipole raising operators from ground state manifold $F=I$ to the excited state manifold $F'$, as defined in \cite{deutsch2010quantum}. $H_\text{eff}[\mathbf{c}] = H[\mathbf{c}] -i\Gamma \sum_q W_q^\dag W_q$/2 is the non-Hermitian control Hamiltonian, Eq. (\ref{eq: evolution of the density matrix}), now including absorption of the laser light.  

 For gates, we define a $d^2 \times d^2$ superoperator matrix acting on the density matrix.  For the open quantum system, the superoperator describing the evolution of an arbitrary input state is the Completely Positive (CP)-map, $\mathcal{E}[\bm{c},T]=\mathcal{T}\left(\exp\{\int_0^T \mathcal{L}[\bm{c}(t')]\}dt'\right)$, where $\mathcal{L}$ is the Lindbladian superoperator of the master equation,  defined implicitly in Eq. (\ref{eq: evolution of the density matrix}).


 
  We compared the output in the open quantum system dynamics given the  ideal control solution $\mathbf{c}$ found in closed-system optimization. The fidelities for state preparation and full SU(10) maps are, respectively,
\begin{eqnarray}
\mathcal{F}_\psi[\bm{c},T] &=& \Tr{\rho_{\psi_\text{tar}} \rho[\bm{c},T]}, \\
\mathcal{F}_U[\bm{c},T] &=& \left|\Tr{\mathcal{E}_{U_\text{tar}}^{\dagger} \mathcal{E}[\bm{c},T]}\right|/d^2.
\label{eq:effective Fidelity}
\end{eqnarray}
Here $\rho_{\psi_\text{tar}}=\ket{\psi_\text{tar}}\bra{\psi_\text{tar}}$ is the target state and $\rho[\bm{c},T]$ is the solution to the master equation.  $\mathcal{E}_{U_\text{tar}} = U_\text{tar}^{*}\otimes U_\text{tar}$ is the CP-map corresponding to the target unitary gate and $\mathcal{E}[\bm{c},T]$ is the CP-map with decoherence.   Eq. (\ref{eq:effective Fidelity}) is the ``process fidelity," a key quantity of interest in determining the thresholds for fault-tolerant quantum computation \cite{schulte2011}.

Numerical results are given in \cref{fig:decohrence_quantum_control} for both state preparation and unitary mapping.
In contrast to closed-system control,  \cref{fig:decohrence_quantum_control}c and \cref{fig:decohrence_quantum_control}d show that there is an island  where the  infidelity is smallest. 
This reflects the trade off between coherent control and decoherence.
There is an optimal total time of evolution $T$ than larger than the QSL but not too large when compared to the optical pumping time.  
In addition, the optimal choice of  $\beta$ is now smaller than we found for the closed quantum system, as increased tensor-light shift is accompanied by increased photon scattering. Including decoherence, for the case of state preparation, averaged over $20$ random states, we find the fidelity $\langle \mathcal{F}_\psi \rangle \approx 0.9997$.  
Here the island of high fidelity is large, occurring for  $\beta< 1.2$. For the case of unitary mapping the island of lowest infidelity  occurs for  $\beta<1.2$ where the fidelity $\langle \mathcal{F}_U \rangle \approx 0.9970$ which is averaged over $10$ Haar random unitaries.  We emphasize that these qudecimal maps act on a 10-dimension Hilbert space.  Thus a fair comparison of the effective fidelity acting on qubits is $\langle \mathcal{F} \rangle_\text{qubit}= \langle \mathcal{F} \rangle_\text{qudecimal}^{0.3}$, since, in principle, one can encode more than 3 qubits in a qudecimal

Coherence is also limited when there are inhomogenieties arising from uncertainties in the Hamiltonian parameters such as the laser intensity and detuning.  When the decoherence time is longer than than the inhomogeneous dephasing time, one can mitigate this with the numerical tools of robust control~\cite{PhysRevLett.82.2417, PhysRevA.58.2733, anderson2015accurate}.  We consider here an uncertainty in the tensor light shift arising from the thermal velocity of the atoms.  To perform robust control, we replace the control Hamiltonian by $H[\bm{c}] \rightarrow H'[\bm{c},\epsilon]=H[\bm{c}]+\epsilon I_z^2 $, where $\epsilon$ is the variation in $\beta$ around the fiducial value, and define a new objective function as the average fidelity, $\langle \mathcal{F}[\mathbf{c},T] \rangle =\int d\epsilon \; p(\epsilon)\mathcal{F}[\mathbf{c},T, \epsilon]$.  While in principle one can design inhomogeneous control with detailed knowledge of the probability distribution $p(\epsilon)$, in practice, when the standard deviation of the distribution $\delta$ is sufficiently narrow, it is sufficient to simultaneously optimize at two points\cite{anderson2015accurate},  and choose the objective function as 
\begin{equation}
\langle \mathcal{F}[\mathbf{c},T] \rangle = (\mathcal{F}[\mathbf{c},T, \epsilon=+\delta]+\mathcal{F}[\mathbf{c},T, \epsilon=-\delta])/2.
\end{equation}

\begin{figure}[!ht]

\includegraphics[width=\textwidth]{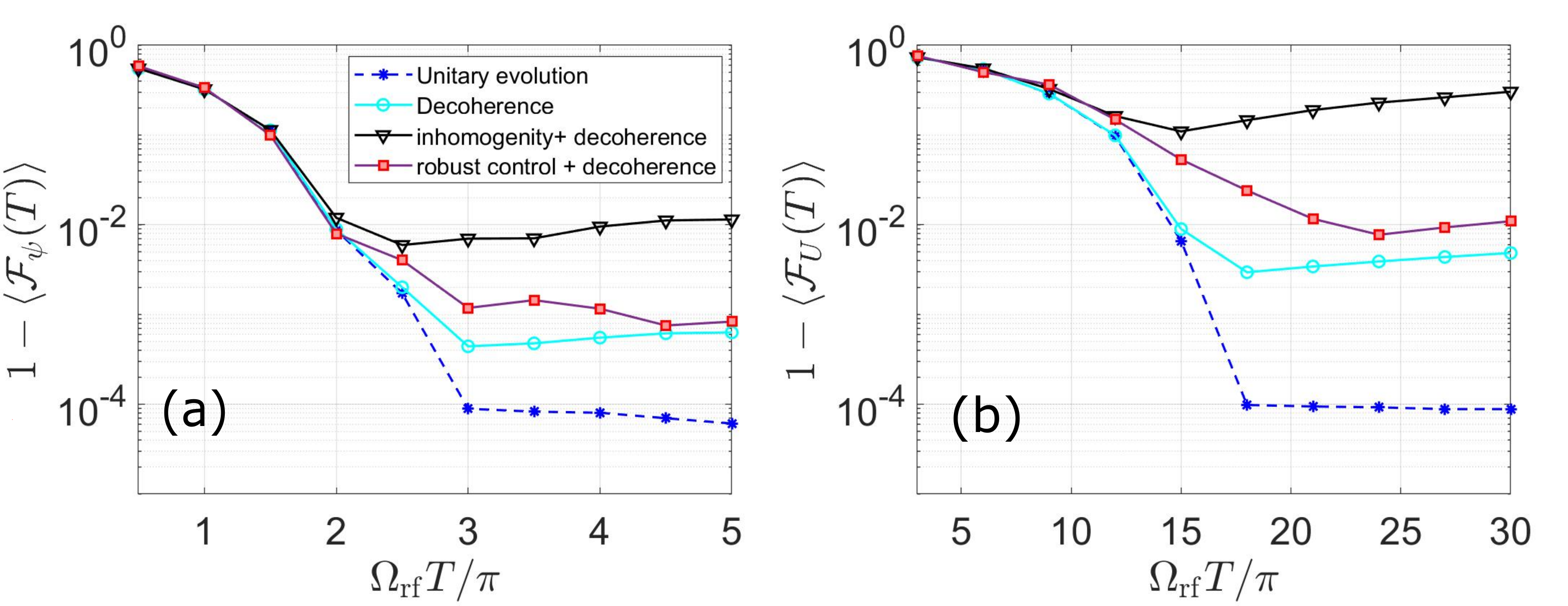}

\caption{ Comparison of infidelity with and without decoherence and robust control to counteract dephasing due to inhomogeneities at the level of $.5\%$ of $\beta$ and $\beta=0.4\Omega_{\mathrm{rf}}$.  (a) state preparation (averaged over $20$ Haar-random target states),  (b) SU(10) mapping (averaged over $10$ Haar-random unitary matrices). Robust control can largely remove dephasing and achieve almost same the infidelity seen due solely to decoherence.}
\label{fig:inhomogenity} 
\end{figure}

The numerical results of robust control are shown in Fig.~\ref{fig:inhomogenity} for $\beta=0.4\Omega_{\mathrm{rf}}$ and an error of  $\delta = .005 \beta$. We see that robust control outperforms the bare waveforms, even in presence of decoherence, but one does not reach the fidelity without any inhomogeneity due to optical pumping occurring over the extended time of the control pulses. For the parameters chosen here, we find that for state preparation one could achieve a fidelity of $\langle \mathcal{F}_\psi \rangle \approx 0.9992$ in a time $T={4.5}\pi/\Omega_\text{rf}$, and for unitary mapping one achieved a fidelity  $\langle \mathcal{F}_U \rangle \approx 0.9923$  in a time $T={24}\pi/\Omega_\text{rf}$. 
Other practical considerations such as the bandwidth needed for rapidly varying waveform may limit the speed of operation (see \cref{chap:app_qudecimal_qudits}).

\section{Conclusion and Summary}
\label{sec:qudecimal_conclusion_and_summary}
In this chapter, we have shown that in the presence of fundamental decoherence and small inhomogeneities, quantum optimal control allows for the realization of high-fidelity arbitrary state maps and SU(10) qudecimal gates acting on nuclear spin in the ground state of \textsuperscript{87}Sr. 
While we proposed one protocol that leverages the strong tensor light shift induced by a laser tuned near the $^3P_1$ hyperfine manifold, the richness of magneto-optical controls in\textsuperscript{87}Sr provides multiple possible approaches, \textit{e.g.}, by employing the tensor light shift when tuned near the $^3P_0$ clock state.  Quantum optimal control of nuclear spins should find a variety of applications in QIP, including metrological enhancement with qudits~\cite{Noris2012}, quantum simulation~\cite{Poggi2020,Blok2021}, and universal quantum computation~\cite{daley2011quantum}. 
For the latter additional components are necessary. One must enable readout of all 10 magnetic sublevels though appropriate shelving and fluorescence protocols~\cite{boyd2007nuclear}. 
Most importantly, we must study the implementation of entangling gates consistent with qudit logic. 
Advances in Rydberg-state control for alkaline earth atoms show great promise in this direction~\cite{Madjarov_Endres_2020_Sr}. 
Finally, while we have studied here two extremes of the control tasks, state preparation and SU(10) maps, optimal control allows for arbitrary partial isometries to encode a $d'<10$ qudit in the qudecimal.  
For example one can encode a qubit in the logical states $\ket{0}=\ket{M_I=9/2}$, $\ket{1}=\ket{M_I=-9/2}$ and potentially protect it from dephasing noise, analogous to a cat-code~\cite{Li2017} or other encodings of a qubit in a large spin that leverages the available interactions and dominant error channels~\cite{Gross2021}.  The flexibility of arbitrary control provides avenues to explore the best approach to encoding and error mitigation.

 \chapter{Qudit entanglers using quantum optimal control}
\label{chap:Qudit_entanglers}

\section{Introduction} 
\label{sec:introduction}

In the gate-based approach to quantum computation with qubits, a universal gate set consists of single-qubit gates that generate the group SU$(2)$ and one entangling two-qubit gate, such as CNOT ~\cite{divincenzo1995two}. 
This generalizes simply for qudits. 
The universal gate-set consists of the generators of single-qudit gates in SU($d$) and an entangling two-qudit gate \cite{muthukrishnan2000multivalued,zhou2003quantum,brennen2005criteria}. 
 Unlike qubits, where native Hamiltonians can be used to naturally implement the desired gate set, qudits require more complex protocols. 
 The gates that are necessary for the implementation of the universal gate set have been recently implemented for qudits in superconducting transmon~\cite{Blok2021,goss2022high,fischer2022towards} as well as in trapped ions~\cite{ringbauer2021universal,hrmo2022native} up to dimension $d=7$.  In these experiments, one implements qudit gates using constructive methods through a prescribed set of Givens rotations~\cite{brennen2005criteria,li2013decomposition}.

Following the ideas from the previous chapter, here we study an alternative approach based
on quantum optimal control for the  implementation of entangling gates between two qudits. 
We study qudit entangling gates for any $k \leq d$ within the $d$-dimensional Hilbert space of each subsystem.
As a concrete example that demonstrates the power of the method, we present here an optimal control scheme to implement entangling gates in qudits encoded in the nuclear spin of $^{87}$Sr atoms.  
The ground state of the $^{87}$Sr is also studied in a recent paper as a possible candidate for qudit encoding with entangling interaction enabled by the Rydberg blockade \cite{zache2023fermion}. 
Also, the recent significant achievements of quantum information processing using the Rydberg blockade~\cite{Levine_Pichler_gate,bluvstein2022quantum,Saffman_Nature_2022}  make this an ideal platform for exploring quantum computation.  Using a combination of a tunable radio-frequency magnetic field and interactions that arise when atoms are excited to high-lying Rydberg states, the atomic qudit is fully controllable.
We find that one can use quantum optimal control to implement  high-fidelity entangling qudit gates even in the presence of decoherence arising from the finite Rydberg-state lifetime.


\section{Controllability} 
\label{sec:controllability}
A complete universal gate set for qudits requires one entangling gate.
A standard choice is the CPhase gate, which is the generalization of CZ gate for qubits, defined 
\begin{equation}
\mathrm{CPhase}\ket{j}\ket{k}=\omega^{jk}\ket{j}\ket{k},
    \label{eq:CPhase_gate}
\end{equation}
where $\omega=\exp(2\pi i/d)$ , the $d$-th primitive root of identity for a subsystem of dimension $d$ and $0\leq j,k\leq d-1$. We can see that for $d=2$ we recover the CZ gate. 
This gate is locally equivalent to the qudit-analog of the CNOT gate, known as CSUM gate,
\begin{equation}
C_{\mathrm{SUM}}\ket{i}\ket{j}=\ket{i}\ket{i\oplus j (\text{mod } d)}
    \label{eq:CSUM_gate}
\end{equation}
by the Hadamard gate for qudits,  $H_d\ket{j}=\frac{1}{\sqrt{d}}\sum_i\omega^{ij}\ket{i}$. 
Previous works have studied how to implement these gates through a well-defined sequence of  maps generated by one-qudit and two-qudit Hamiltonians~\cite{brennen2005criteria,muthukrishnan2000multivalued,vlasov2002noncommutative,brylinski2002mathematics}. We study here the use of numerical optimization and the theory of optimal control.

\subsection{Lie algebraic approach }
\begin{figure*}
	\includegraphics[width=0.98\textwidth]{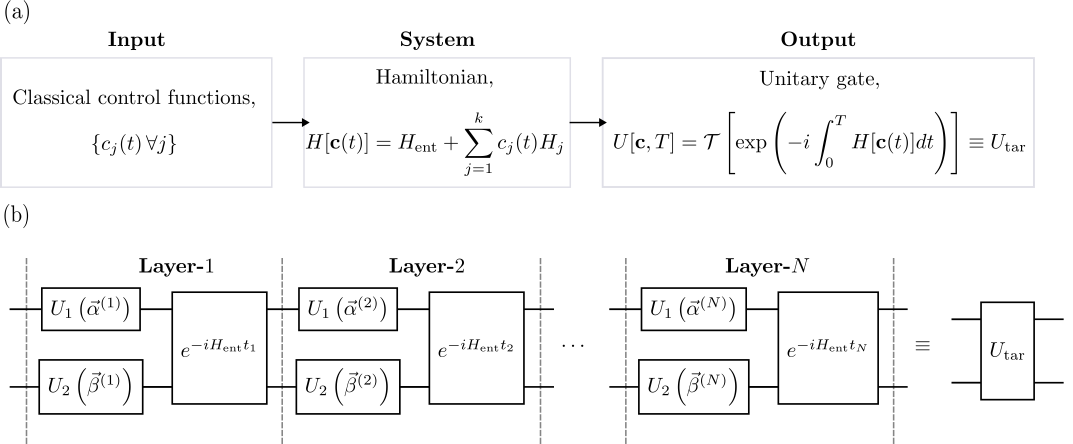}
\caption{ \textbf{Comparison of Lie algebra versus Lie group approach for quantum control.} (a) Schematic of the continuous-time Lie algebraic approach for quantum control. 
The physical systems are governed by the time-dependent Hamiltonian, $H[\mathbf{c}(t)] = H_{\mathrm{ent}} + \sum_{j=1}^{k} c_j(t) H_j$, here with a time-dependent entangling Hamiltonian, $H_{\mathrm{ent}}$. The time-dependent waveforms $\{c_j(t)\}$ are found through numerical optimization, and this defines the target unitary map of interest through the solution to the time-dependent Schr\"{o}dinger equation. 
(b) Schematic for a digital, Lie group approach to quantum control of entangling two-qudit gates. The target unitary is achieved through a discrete series of layers consisting of unitary maps from a given family. One layer of the scheme consists of single-qudit gates on each subsystem and an entangling interaction between them, applied for a given time $t_j$. Through numerical optimization, one finds the parameters of the local $SU($d$)$-gates and the entangling time $t_j$ in each layer.}
	\label{fig:approach_all}
\end{figure*}

In the Lie algebraic approach to quantum control which we also studied in the last chapter, we consider a Hamiltonian of the form $H[\mathbf{c}(t)] = H_0 + \sum_{j=1}^{k} c_j(t) H_j$, where $\mathbf{c}(t)=\{c_j(t)\}$ is the set of time-dependent classical control waveforms, and $H_0$ is called the drift Hamiltonian. The system is said to be ``controllable" if the set of Hamiltonians, $\{H_0, H_1,H_2,\hdots, H_k\}$, are generators of the desired Lie algebra, e.g., $\mathfrak{su}{(d)}$. Then $\exists \hspace{0.1cm} \mathbf{c}(t)$ \hspace{0.2cm } such that $U[\mathbf{c},T]=\mathcal{T}\left[\exp\left(-i\int_0^T H[\mathbf{c}(t)]dt\right)\right]=U_{\mathrm{tar} }$ for any target unitary in desired Lie Group, e.g., $U_{\mathrm{tar}} \in \mathrm{SU}(d)$.
In addition, we require $T \ge T_*$, where $T_*$ is known as the ``quantum speed limit time," which sets the minimal time needed for the system to be fully controllable.


We consider here open-loop control determined by a well-defined Hamiltonian of the general form,
\begin{equation}
H(t)=H^{(1)}(t)+ H^{(2)}(t)+H_{\mathrm{ent}},
    \label{eq:entangling_qudit_Hamiltonian_1}
\end{equation} 
where $H^{(i)}(t)$ are time-dependent Hamiltonians acting on the individual subsystems, and $H_{\mathrm{ent}}$ is  the interaction that entangles them.  Here we include the time dependence in the Hamiltonian that acts on the individual system as these will be generally easier to implement experimentally. In this formulation, $H_{\mathrm{ent}}=H_0$, is the drift Hamiltonian.
However, one could in principle include time dependence in the entangling Hamiltonian as well and this may achieve faster gates.

\subsection{Lie group approach}
\label{sec:lie_group_approach}
 In the digital, Lie group approach to quantum control, we consider a  family of unitary maps in the desired group  that are easily implementable, $U(\lambda_j)$, where $\{\lambda_j\}$ are the parameters that specify the unitary matrices at our disposal. The relevant Lie group of interest here is $\mathrm{SU}(d^2)$, the group of two-qudit unitary matrices in $d^2$ dimensions, where the overall phase is removed. The system is controllable if  $\forall U_{\mathrm{tar}} \in \mathrm{SU}(d^2)$, $\exists \{\lambda_i\}$ such that  $\prod_{j=1}^{k}U(\lambda_j)=U_{\mathrm{tar} }$. Similar to the Lie algebraic quantum control approach, the goal is to find  $\{\lambda_j\}$ through numeric optimization, e.g., via gradient-based methods.

For the case of two-qudit gates, a controllable Lie group structure is given as,
\begin{equation}
    U_{\lambda_j}=U_{\mathrm{ent}}*(U_1\otimes U_2),
    \label{eq:Lie group_approach}
\end{equation}
where $U_{1,2}\in \mathrm{SU}(d) $ and  $U_{\mathrm{ent}}=\exp(-i H_{\mathrm{ent}}t) \notin \mathrm{SU}(d)\otimes \mathrm{SU}(d)$.
Thus, we can achieve the target gate to the desired fidelity by intertwining a sequence of local $\mathrm{SU}(d)$ gates and the available entangling interaction  in alternating layers of single qudit gates and entangling gates, as shown in Fig.~\ref{fig:approach_all}(b).
This approach is similar to the construction based on Givens rotation~\cite{ringbauer2021universal}.
Here, the possibility of accessing arbitrary local $\mathrm{SU}(d)$ gates makes this protocol very powerful. A schematic comparison of both these approaches is shown in Fig.~\ref{fig:approach_all}. 

\subsection{Physical Platform: Rydberg atoms }
To make these ideas concrete, we consider the implementation of entangling gates in neutral atoms using the strong van~der~Waals interactions between atoms in high-lying Rydberg states.
We use the Rydberg dressing paradigm in which one adiabatically superposes the Rydberg state into the ground states to introduce interactions between dressed ground states ~\cite{johnson2010interactions, keating2015robust, jau2016entangling, zeiher2016many, zeiher2017coherent, borish2020transverse}. 
Rydberg dressing has been studied with multiple applications including  the  dynamics of interacting spin models ~\cite{zeiher2016many, zeiher2017coherent, borish2020transverse} as well as to prepare metrologically-useful states ~\cite{Kaubruegger_Zoller_metrology}.
Entanglement between neutral atoms via Rydberg dressing has been theoretically proposed for creating qubit entangling gates ~\cite{keating2015robust, mitra_martin_gate,anupampra} and experimentally implemented ~\cite{jau2016entangling,martin2021molmer,schine2022long}. 
 The dressing approach has a potential advantage in that it exhibits reduced sensitivity to some noise sources~\cite{keating2015robust,mitra_martin_gate,schine2022long}. 
 For the specific protocol based on optimal control, the utilization of Rydberg dressing confines our operations to the qudit subspace, as one can work with the dressed basis \cite{anupampra,mitra_martin_gate}.
 This restriction effectively reduces the dimension of the Hilbert space for optimization from $(2d)^2$ to $d^2$ for a $d$ dimensional qudit. 
 This dimension reduction significantly accelerates the numerical optimization of the pulses required for quantum control.

We study here encoding a qudit in the spin of $^{87}$Sr. 
 To implement entangling two-qudit control, we will make use of the excitation to the $5sns\hspace{0.1cm}^3 S_1$ Rydberg series from one of the metastable $5s5p\hspace{0.1cm}^3 P_J$ first excited states in the triplet series.  For optimal control based on the combination of rf-driven Larmor precession and Rydberg dressing one can compare different choices of metastable states.  One natural choice is the ${^3P_0}$ clock state, whose spin is essentially solely nuclear, and thus robust in the presence of magnetic field noise.  By contrast, the ${^3P_2}$ state involves electronic angular momentum with a large magnetic dipole moment and commensurate sensitivity to noise, including possible tensor light shifts induced by the trapping laser. However, within the specific approach addressed in this study, access to a large magnetic dipole moment enables faster gate operations compared to the Rydberg lifetime. For the ${^3P_2}$ states, the strength of the rf-Larmor precession frequency is closer to that of the available Rydberg dressing interaction. In this regime, the quantum speed limit (i.e., the minimum time required to implement gates) is more favorable compared to the situation that the rf-interaction is much weaker than the Rydberg interaction, as would be the case for the ${^3P_0}$ states. This regime, characterized by similar strengths of competing Hamiltonians, is known to be optimal for achieving the quantum speed limit \cite{omanakuttan2021quantum,buchemmavari2023entangling}. 
We consider here coherently transferring qudits from the $^1{S}_0$ ground state to the $F=9/2$ state hyperfine states of the  $^3{P}_2$ manifold, which provides for faster and more flexible control \cite{trautmann20221}, putting technical noise aside.

To achieve the entangling interaction, we consider Rydberg dressing, generalizing the mechanism discussed in~\cite{jau2016entangling, keating2015robust,  mitra_martin_gate}. 
The AC Stark shift (light shift) associated with a dressed state when a laser is tuned near a Rydberg resonance is modified for two atoms because of the Rydberg blockade. 
The deficit between the two-atom light shift and twice the one-atom light shift determines the entangling  energy~\cite{keating2015robust}. 
For the case of qudits, the same physics holds, but now with a multilevel structure and a spectrum of entangling energies. 
When the spectrum is nonlinear, the system is controllable.

\begin{figure*}
    \includegraphics[width=\textwidth]{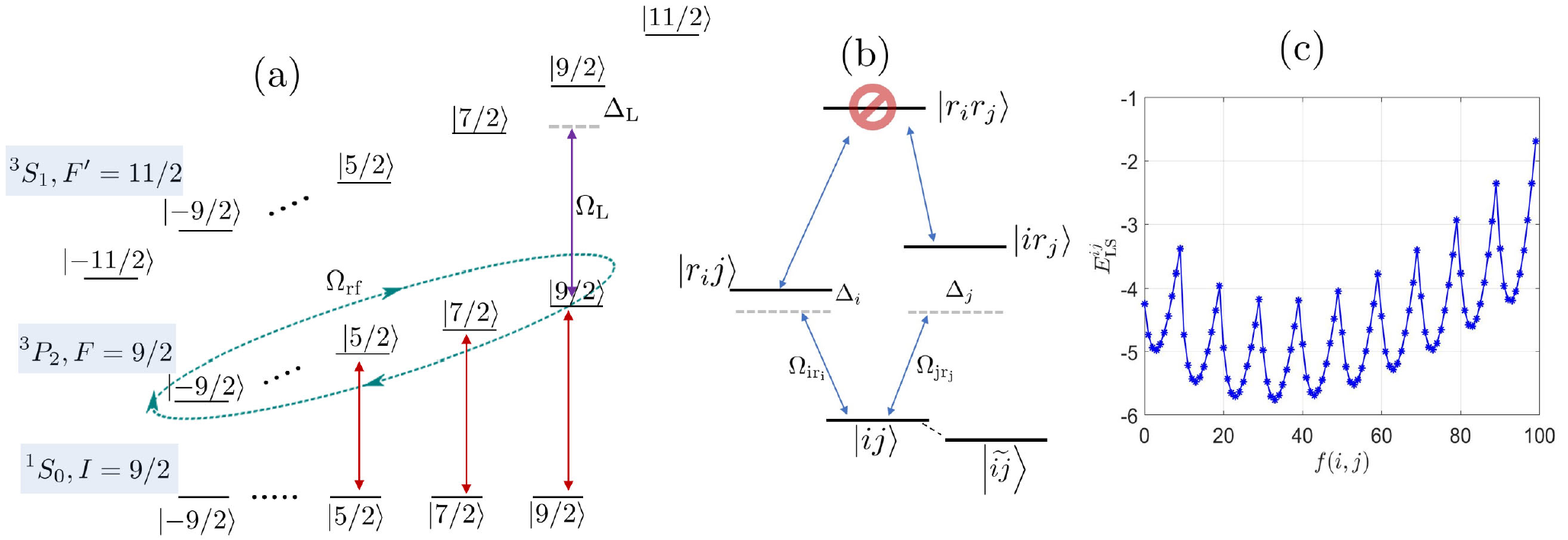}
\caption{\textbf{Schematic for designing two-qudit entangling interactions in $^{87}$Sr neutral atoms.} (a) A $k\le d$-dimensional qudit is encoded in memory in the nuclear spin with $d=10$ magnetic sublevels in the electronic ground state $(5s^2) \;^1S_0$. When the gate is to be performed, the $k$ levels (here $k=3$) are transferred coherently to the  metastable clock states $ (5s5p)\; ^3P_2, F=11/2$ in the presence of a bias magnetic field.  The system becomes controllable  by adiabatically dressing the $^3P_2$ with Rydberg character through the application of a near-resonant laser with Rabi frequency $\Omega_{\mathrm{L}}$ and detuning $\Delta_{\mathrm{L}}$ with respect to the hyperfine manifold $(5sns)\; ^3S_1, F'=9/2$ in the Rydberg series. 
Control is then achieved through the application of a phase-modulated rf-field with Rabi rate $\Omega_{\mathrm{rf}}$ which acts on the dressed states to generate a nonlinear Larmor precession. The entanglement arises due to the Rydberg blockade. 
The coupling of the state of two qudits for a perfect blockade as depicted in (b), where $i$ is a state from the first qudit and $j$ is from the second qudit, excited by two Rabi frequencies and detunings determined by the Clebsch-Gordan coefficients and Zeeman shifts. The state $\ket{ij} \to  \ket{\widetilde{ij}}$ is the dressed state given in Eq. \eqref{eq:dressed_states_qudits}.
The spectrum of eigenvalues of the entangling Hamiltonian Eq.\eqref{eq:entangling_Hamiltonian} is given in (c) as a function of $i$ and $j$ where the function chosen is $f(i,j)=10i+j; 0\leq i,j<10$. The spectrum indicates 10 parabolas, where each parabola corresponds to the effect of a single state in the first atom sees due to all the states in the second atom. This nonlinear spectrum arises through a combination of the tensor AC Stark shift and the Rydberg blockade, making the system controllable, allowing us to implement any symmetric two qudit gate in this system of interest. }
	\label{fig:set_up_figure}
\end{figure*}

 Fig.~\ref{fig:set_up_figure} depicts the basic scheme. Those levels of the qudit that we chose to participate in the gate are  excited from the ground  ${^1S_0}$ to the first excited ${^3P_2}$ state. 
 The Rydberg states in $^{87}$Sr have well-resolved hyperfine splitting. We consider UV dressing laser near the resonance between the ${^3P_2}$, $F=9/2$ hyperfine manifold and the ${^3S_1}$, $F'=11/2$ Rydberg hyperfine states. 
 In the presence of a bias magnetic field, due to the difference in the g-factors, the two manifolds will be differently Zeeman shifted. The different magnetic sublevels that define the qudit will thus be differently detuned to the Rydberg magnetic sublevels. Due to this and the Clebsch-Gordan coefficients associated with the different transitions, each sublevel will be differently dressed (equivalently, there is a tensor light shift). When two atoms are dressed, the effect of the Rydberg blockade modifies the spectrum as discussed above. 
 
An example of two sublevels (one from each atom) is shown in Fig.~\ref{fig:set_up_figure}(b). 
Diagonializing this atom-laser Hamiltonian under the approximation of a perfect Rydberg blockade yields the representation
\begin{equation}
    H_{\mathrm{ent}}=\sum_{ij} E^{ ij}\ket{\widetilde{ij}}\bra{\widetilde{ij}},
    \label{eq:entangling_Hamiltonian}
\end{equation}
where the tilde indicates dressed states,
\begin{equation}
\ket{\widetilde{ij}}=C_{ij}\ket{ij}+C_{r_ij}\ket{r_ij}+C_{ir_j}\ket{ir_j},
    \label{eq:dressed_states_qudits}
\end{equation}
and $E^{ij}$ are the light shifts originating from these interactions.
The spectrum of the entangling Hamiltonian shown in Fig.~\ref{fig:set_up_figure}(c) gives us insight into the controllability of the system. 
In the chosen order, the spectrum reveals the structure of $10$ quadratic potentials arising from a combination of the tensor light shift and Rydberg blockade. 
This inturn creates a Hamiltonian that has support on spherical tensor operators with rank $K\leq 2$ and makes the Hamiltonian controllable; further details are discussed in Appendix~(\ref{sec:controllability_of_the_Hamiltonian}).

  The time-dependent Hamiltonian necessary for the Lie algebraic control can be chosen as phase-modulated Larmor precession, $H_{\mathrm{mag}} = -\boldsymbol{\mu}\cdot \mathbf{B}(t)$, with $\boldsymbol{\mu} = g_F \mu_B \mathbf{F}$ the magnetic dipole vector operator, and where $\mathbf{B}(t) = B_\parallel \mathbf{e}_z + B_T \Re\left[(\mathbf{e}_x + i \mathbf{e}_y) \mathbf{e}^{-i\left(\omega_\text{rf} t +\phi(t)\right)}\right]$. 
Defining the auxilary subspace, $a$, for the levels in hyperfine manifold $\{5s 5p\hspace{0.1 cm}^3P_2,\hspace{0.1 cm} F=9/2\}$ and the subspace, $r$, for the levels $\{5s ns \hspace{0.1 cm}^3S_1,\hspace{0.1 cm} F'=11/2\}$ in the Rydberg hyperfine manifold, we have  $g_F(r)/g_F(a)\approx 2$. 
Thus defining the Zeeman shift $\omega_0= g_F(a)B_\parallel $, the Larmor precession frequency $\Omega_{\mathrm{rf}}= g_F(a)B_T $, and choosing rf drive on resonance in the $a$-manifold, $\omega_{\mathrm{rf}}=\omega_0$, in the co-rotating frame at $\omega_0$, the Hamiltonian is
\begin{equation}
\begin{aligned}
H_{\mathrm{mag}}^{(a)}(t)&=\Omega_{\mathrm{rf}} \left[\cos \phi(t) F_x^{a} +\sin \phi(t) F_y^{a}\right],\\
H_{\mathrm{mag}}^{(r)}(t)&=2\Omega_{\mathrm{rf}} \left[\cos \phi(t) F_x^{r} +\sin \phi(t) F_y^{r}\right]+\omega_{0}F_z^{r},
\end{aligned}
    \label{eq:Hamiltonian_frame}
\end{equation}
where $F_i^{a}, F_i^{r}$ are the spin angular momentum operators in the respective subspaces along axis $i \in \{x, y, z\}$.



As the $H_\mathrm{mag}$ acts on the laser-dressed states defined in Eq.~(\ref{eq:dressed_states_qudits}), which are superpositions of $a$ and $r$ states that have different $g$-factors, one needs to find the action of the magnetic interaction in the dressed basis. Due to the nonlinearity, the action of the rf-magnetic driving on the dressed states is no longer simple Larmor precession. Considering a global rf-magnetic interaction, the $H_\mathrm{mag}$ acts on both qudits as
\begin{small}
    \begin{equation}
\begin{aligned}
    \left(H_\mathrm{mag}(t)\otimes \mathds{1}+\mathds{1}\otimes H_\mathrm{mag}(t)]\right)\ket{\widetilde{ij}}&=C_{ij}\left[H_{\mathrm{mag}}^{(a)}(t)\otimes H_{\mathrm{mag}}^{(a)}(t)\right] \ket{ij}\\
    &+C_{r_ij}\left[H_{\mathrm{mag}}^{(r)}(t)\otimes H_{\mathrm{mag}}^{(a)}(t)\right] \ket{r_ij}\\
    &+C_{ir_j}\left[H_{\mathrm{mag}}^{(a)}(t)\otimes H_{\mathrm{mag}}^{(r)}(t)\right] \ket{ir_j}.
\end{aligned}
\end{equation}
\end{small}

Thus in the dressed basis, the Hamiltonian is $H(t)=\widetilde{H}\left[\phi(t)\right]+H_{\mathrm{ent}}$,
where the action of the magnetic field in the dressed basis is given by the Hamiltonian,
\begin{equation}
\begin{aligned}
&
\widetilde{H}\left[\phi(t)\right]
=\sum_{i,j,k,l}\bra{\widetilde{ij}}H_\mathrm{mag}(t)\otimes \mathds{1}+\mathds{1}\otimes H_\mathrm{mag}(t)]\ket{\widetilde{kl}} \ketbra{\widetilde{ij}}{\widetilde{kl}}.
    \label{eq:dressed_rotation}
\end{aligned}
\end{equation}
By modulating the phase $\phi(t)$ one can generate any target unitary gate.


\section{Numerical Methods} 
\label{sec:Numerical Methods}
We consider encoding a $k$-dimensional qudit in the $d=10$ dimensional Hilbert space associated with $10$ magnetic sublevels of the nuclear spin of $^{87}$Sr. To implement gates based on optimal control for $k<10$, we use techniques based on the structure of  partial isometries. A partial isometry of dimension $k \le d$ in a physical system of dimension $d$ is defined as,
\begin{equation}
V_\mathrm{{tar}}=\sum_{i=1}^{k}\ket{f_i}\bra{e_i}
\label{eq:Partial_isometry_definition}
\end{equation}
where $\{\ket{e_i}\},\{\ket{f_i}\}$ are two orthonormal bases for the qudit. The unitary of maps of interest then has the form, 
\begin{eqnarray}
U_{\mathrm{tar}}=V_{\mathrm{tar}}+V_{\perp},
\end{eqnarray}
where $V_{\perp}$ acts on the orthogonal subspace, with dimension $d-k$. To find the  control waveform, one then optimizes the fidelity between the target isometry and the isometry generated using quantum control~\cite{pedersen2007fidelity}
\begin{equation}   \mathcal{F}_V[{c},T]=\left|\Tr\left(V^{\dagger}_{\text{tar}}V[{c},T]\right)\right|^2/k^2. 
\end{equation}


\subsection{Numerical results for Lie algebraic approach}

As discussed in Sec.~\ref{sec:controllability}c, one can implement an arbitrary entangling gate through a combination of Rydberg dressing and phase-modulated Larmor precession driven by rf-fields. 
Because our control Hamiltonian is symmetric with respect to the exchange of the qudits, we consider here symmetric gates, with global control. We seek, through numerical optimization, the  time-dependent rf-phase, $\phi(t)$. 
To achieve this we employ the well-known GRAPE algorithm~\cite{khaneja2005optimal}. 
To implement GRAPE, we discretize the control waveform, $\phi(t)$, and  numerically maximize the fidelity by gradient ascent.  We choose here a piecewise constant parameterization (as in ~\cite{anderson2013unitary}) and write the control waveform as a vector $\mathbf{c} = \{\phi(t_j)/\pi  \;|\; j=1,\dots,n\}$ where $t=j\Delta t$ and $n=T/\Delta t$. The waveform is thus a series of square rf-pulses with constant amplitude and phase over the duration $\Delta t$. 

The minimum number of elements in the control vector $\mathbf{c}$ is determined by the number of parameters needed to specify the target isometry. 
A $K$-dimensional partial isometry is defined by the $K$ columns in a $D \times D$-dimensional unitary matrix.
Hence, to find the number of free parameters for a $K$-dimensional isometry one can count the number of parameters needed to specify $K$ orthonormal vectors uniquely in a  $D$-dimensional vector space. This is given by 
\begin{equation}
    \begin{aligned}
        n_{\mathrm{min}}(K,D)=&\sum_{j=1}^{K}2(D-j)-1+K-1\\
        =&2\left[KD-\frac{K(K+1)}{2}\right]+K-1\\
        =&2KD -K^2-1,
    \end{aligned}
    \label{eq:number_of_free_parameters}
\end{equation}
where in the first line, we subtracted one from the parameter count in since the overall phase of the isometry is neglected. Eq. \eqref{eq:number_of_free_parameters} recovers  well-known limits. When  $K=1$ and $D=d$,  $n_{\mathrm{min}}=2d-2$, which is the number of free parameters needed to specify a pure state in a $d$-dimensional Hilbert space. 
When $K=D=d$, $n_{\mathrm{min}}=d^2-1$, which is the number of free parameters needed to specify a special unitary map in $d$-dimensions.

In the Lie algebraic protocol for designing entangling gates, the  control Hamiltonian, as well as the target unitary matrices, are symmetric under the exchange of qudits.
In this case, one can work in the symmetric subspace for two qudits. Using the hook length formula~\cite{frame1954hook}, the dimension of the symmetric subspace of the total vector space and isometry is, 
\begin{equation}
D=\frac{d(d+1)}{2}, K=\frac{k(k+1)}{2}.
\label{eq:effective_parameters_unitary}
\end{equation}
Thus, using Eq.~(\ref{eq:number_of_free_parameters}), we find the  number of free parameters required for the two-qubit entangling unitary given in Table~\ref{tab:Lie_algebra_parameters}. 

\begin{table}
\centering
 \begin{tabular}{ |c|c| } 
 \hline
 $k$ & $n_{\mathrm{min}}(K,D)$ \\
\hline 
 2 & 320 \\
 \hline
3& 623  \\
 \hline
5 &  1424\\
\hline
7& 2295\\
\hline
\end{tabular}
    \caption{The minimum number of parameters required for  encoding a partial isometry of dimension $k$ in the $d=10$ dimensional Hilbert space according to Eq.~\eqref{eq:number_of_free_parameters} for the prime dimensions $k\leq 10$ with $K$ and $D$ given by Eq.~\eqref{eq:effective_parameters_unitary}}
    \label{tab:Lie_algebra_parameters}
    
\end{table}

\begin{figure*}
\centering
\includegraphics[width=0.98\textwidth]{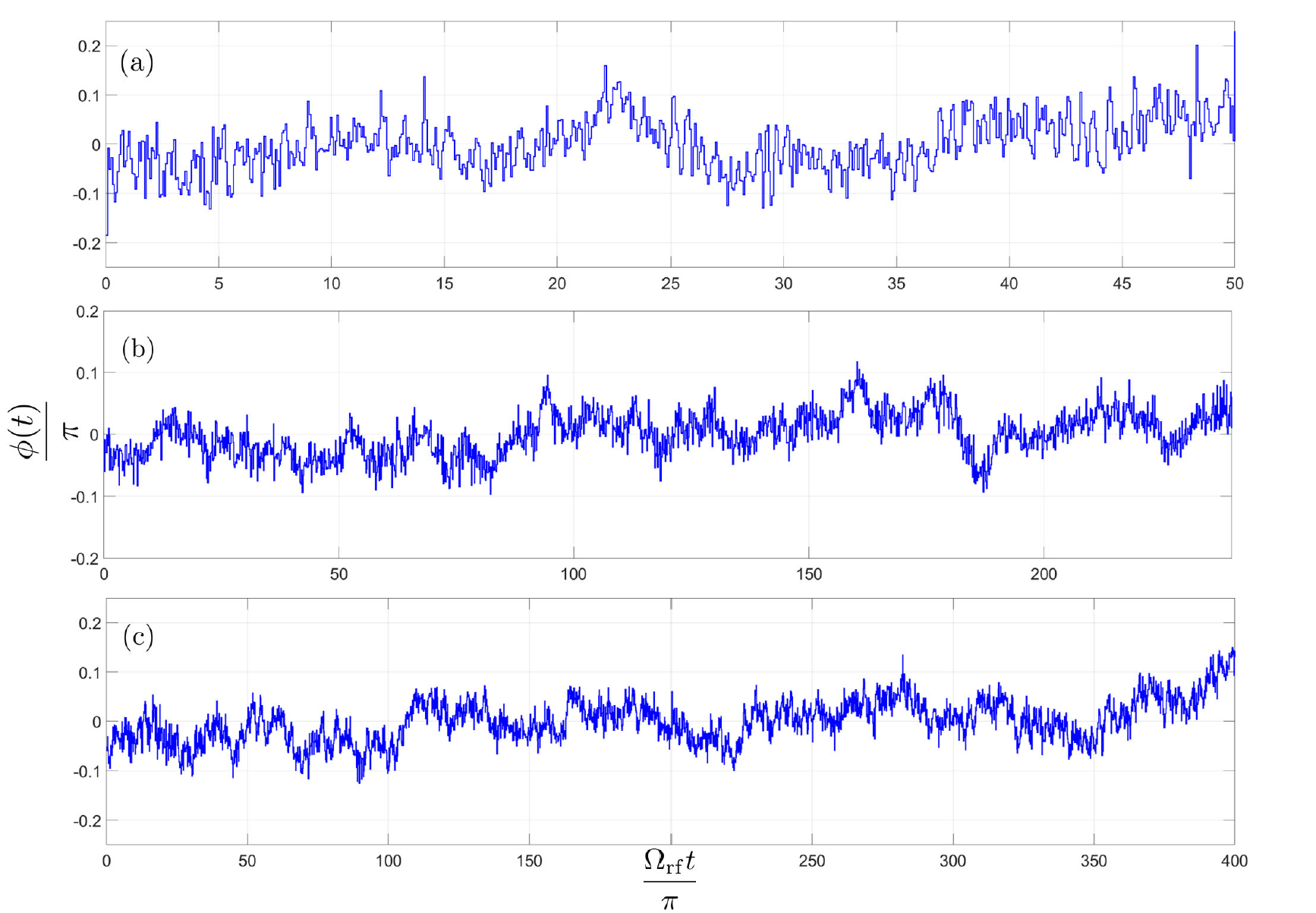}
    \caption{\textbf{Waveforms of the CPhase gate.} 
    Quantum control is achieved by modulating the phase of an rf-field as a function of time, $\phi(t)$. We parameterize this by a piecewise constant waveform. The figure shows proof-of-principle examples of $\phi(t)$ that generate the CPhase gate, optimized using the GRAPE algorithm for different qudit dimensions. (a) The case of the $d=3$ for a total time of $T=50 \pi/\Omega_{\mathrm{rf}}$ with $700$ piecewise constant steps. 
    (b) The case of the $d=5$ for a total time of $T=240 \pi/\Omega_{\mathrm{rf}}$ with $1600$  piecewise constant steps. (c)  The case of the $d=7$ for a total time of $T=400 \pi/\Omega_{\mathrm{rf}}$ with $2500$ piecewise constant steps. 
    For all of these calculations, the rf-field is on resonance with the Zeeman splitting $\omega_{\mathrm{rf}}=\omega_0$ and we choose the rf-Larmor frequency $\Omega_{\mathrm{rf}}=\omega_{\mathrm{rf}}$. Control is achieved by Rydberg dressing with laser Rabi frequency $\Omega_{\mathrm{L}}=6 \Omega_{\mathrm{rf}}$.}
    \label{fig:Qudits_simulations}
\end{figure*}

Proof-of-principle numerical examples of  waveforms that generate the CPhase gate are given in Fig.~\ref{fig:Qudits_simulations}. 
The figure gives the $\phi(t)$ as a piecewise constant function of time, obtained using the GRAPE algorithm. We consider prime-dimensional qudits, the cases of most interest in quantum algorithms. Fig.~\ref{fig:Qudits_simulations}(a) shows the case of the $k=3$, a qutrit encoded in $d=10$. The total time is $T=50 \pi/\Omega_{\mathrm{rf}}$, which is divided into $700$ intervals for the quantum control. 
Fig.~\ref{fig:Qudits_simulations}(b) shows an example waveform for the case of $k=5$. Here, the total time is $T=240 \pi/\Omega_{\mathrm{rf}}$,  divided into $1600$ intervals. 
Similarly, Fig.~\ref{fig:Qudits_simulations}(c) shows the case of $k=7$ in our $d=10$ level system. 
The total time is $T=400 \pi/\Omega_{\mathrm{rf}}$, divided into 2500 time intervals. This controllable Hamiltonian can also be used to generate other two-qudit gates. The qudit generalization of the M$\o$lmer-S$\o$rensen gate, as is given in the Appendix~\ref{sec:Molmer_sorenson_gate}.

The waveforms found here are a proof-of-principle set of square pulses and are not intended to be taken as the best choice for experimental implementation. In practice, one can design and optimize for much smoother waveforms using well-known techniques by imposing additional constraints on bandwidth and slew rate. Alternatively, one can optimize in the Fourier domain or in any other complete basis of functions using the techniques of gradient optimization of analytic controls (GOAT) \cite{machnes2015gradient}.

 \subsection{Numerical results for Lie group approach}
In the Lie group control protocol discussed in Sec.\ref{sec:controllability}c  we parameterize the target unitary map as 
\begin{equation}
    \begin{aligned}
     U_{\mathrm{tar}}&=\prod_{j}U_{\lambda_j},\\
     &=\prod_{j} e^{-i H_{\mathrm{ent}}t_j}\,
    U_1(\vec{\alpha}^{(j)})\otimes U_2(\vec{\beta}^{(j)}).
\end{aligned}
\label{eq:Lie_group_approach_1}
\end{equation}
The control parameters $\{\lambda_i\}$ consist of the set of times $\{t_i\}$ and the $2(d^2-1)$ parameters $\vec{\alpha}^{(j)}$, $\vec{\beta}^{(j)}$, which  specify each of the local $\mathrm{SU}(d)$ unitary maps.  We can parameterize these according to
\begin{equation}
    \begin{aligned}
     U_i(\vec{\alpha}^{(j)})=\exp(-i\sum_{i=1}^{d^2-1}\alpha_i^{(j)}\Lambda_i),
    \end{aligned}
\end{equation}
where $\Lambda$ is the generalized Gell-Mann matrices that span the Lie algebra $\mathfrak{su}(d)$. The matrices can be categorized as,
\begin{equation}
    \begin{aligned}
    \text{symmetric: }&\Lambda_{jk}^{x}=\ketbra{j}{k}+\ketbra{k}{j},\\
    \text{anti-symmetric: }&\Lambda_{jk}^{y}=-i\ketbra{j}{k}+i\ketbra{k}{j},\\
    \text{diagonal: }&\Lambda_{l}^z=\sum_{j=1}^{l}\ketbra{j}{j}-l\ketbra{l+1}{l+1}.
    \end{aligned}
\end{equation}
The task of the numerical optimization, thus, is to find the set of times of the entangling interaction $\{t_j\}$, and the expansion coefficients of the Gell-Mann matrices $\{\alpha_i^{(j)}\}$ and  $\{\beta_i^{(j)}\}$. We denote this whole set of parameter as $\{\lambda_j\}=\{t_j,\vec{\alpha}^{(j)},\vec{\beta}^{(j)}\}$.

We define one layer of the control as consisting of a pair of local SU$(d)$ gates followed by the entangling Hamiltonian for a time $t_j$. 
The total number of free parameters for a CPhase gate is $d^2(d^2+1)/2$, as follows from Eq.~(\ref{eq:effective_parameters_unitary}) for a symmetric gate in $\mathrm{SU}(d^2)$.
Thus, the minimum number of layers  required to obtain the CPhase gate is given by
\begin{equation}
\begin{aligned}
N_{\text{min}}\left(2(d^2-1)+1\right)=&\frac{d^2(d^2+1)}{2}\\
N_{\text{min}}=&\frac{d^2(d^2+1)}{2(2d^2-1)}.
\end{aligned}
\label{eq:minimum_number_Lie_group}
\end{equation}
  The numerical results for the minimum number of layers needed in the system are given in Table~\ref{tab:Lie_group_asymmetric} for the cases of $d=3,5,$ and $d=7$. In practice, we find that one needs more than this minimum number of layers to implement the target unitary gate with high fidelity. This improves the optimization landscape for gradient ascent~\cite{larocca2018quantum}.

 For our case under study, we choose the same entangling Hamiltonian as we used in the Lie algebraic approach given in Eq. (\ref{eq:entangling_Hamiltonian}). However, unlike that approach, we interleave the entangling interaction with local single-qudit SU($d$) gates. Implementation of this requires another layer of optimization. As we do not have access to native Hamiltonians proportional to the Gell-Mann matrices, to implement local qudit gates we can employ local SU($d$) optimal control~\cite{omanakuttan2021quantum}. From a practical perspective, this might be implemented directly in the $^3 P_2$ manifold, either through a combination of tensor-light shift and rf-driven Larmor precession similar to~\cite{omanakuttan2021quantum}, or alternatively through a combination of microwave-driven Rabi oscillations between different hyperfine levels in $^3 P_2$ and rf-driven Larmor procession as in~\cite{anderson2013unitary}. In either case, optimal control can be used to find the relevant experimental waveform that generates the desired local $\mathrm{SU}(d)$ gates.

\begin{table}
\centering
 \begin{tabular}{ |c|c|c|c| } 
 \hline
 $d$ & $N_{\mathrm{min}}$ & $N_{\text{local}}$  &$N_{\text{global}}$\\
\hline 
 3 & 3 &6 & 7\\
 \hline
5 & 7 & 10 & 12\\
 \hline
7 & 13 & 14 &15\\
\hline
\end{tabular}
    \caption{The number of layers of primitive gates in the Lie group approach required to achieve the CPhase gate. The theoretical minimum is $N_{\rm{min}}$ according to Eq.~\eqref{eq:minimum_number_Lie_group}. If we allow locally addressable single qudit gates, the number of layers required is $N_{\rm{local}}$. If we have only global control but allow for a sign change in the entangling Hamiltonian, the number of layers required is $N_{\rm{global}}$}
    \label{tab:Lie_group_asymmetric}
    
\end{table}

In this analysis, we included locally addressable control on each qudit. Though the CPhase gate is symmetric under exchange, we find that this symmetry breaking is necessary for effective optimization of this parameterization, similar to that seen in~\cite{PhysRevResearch.3.023092}.
An alternative protocol is to employ symmetric global control of the local unitaries, $\vec{\alpha}^{(j)}=\vec{\beta}^{(j)}$, but to reverse the sign of the entangling Hamiltonian $H_{\mathrm{ent}}\to - H_{\mathrm{ent}}$ in alternating layers. This allows for effective optimization, and the corresponding result is given in Table~(\ref{tab:Lie_group_asymmetric}).

\subsection{Decoherence}
In a closed quantum system, quantum optimal control employing either the Lie algebraic or the Lie group approaches can be used in principle to implement any qudit entangling gate to any desired fidelity. In our numerical optimization, we took the target infidelity to be $10^{-3}$. In the  absence of decoherence, we could achieve that target in a reasonable time for $d \le 5$. For $d=7$, more time is required. However,  the fundamentally achievable fidelity is limited by decoherence associated with the particular physical platform. For the system at hand, decoherence occurs due to the finite lifetime of the Rydberg states, which predominantly leads to leakage and loss outside the computational basis. In that case, we can model the gate as generated by a non-Hermitian effective Hamiltonian, $H_\mathrm{eff}[c(t)]$, where the Hermitian part is the control Hamiltonian and the anti-Hermitian represents decay out of the Rydberg states. The fidelity of interest is given by
\begin{equation}
\mathcal{F}_V[\bm{c},T]=\left|\Tr\left(V^{\dagger}_{\text{tar}}V_{\mathrm{eff}}[\bm{c},T]\right)\right|^2/d^2,
\label{eq:fidelity_1}
\end{equation} 
where  $V_{\mathrm{eff}}[\mathbf{c},T]=\mathcal{T}\left[\exp\left(-i\int_0^T H_{\mathrm{eff}}[\mathbf{c}(t)]dt\right)\right]$. Here  the decay amplitude from a dressed state is $\gamma_{\mathrm{decay}}^{ij}=|C_{r_ij}|^2\Gamma_{r_i}+|C_{ir_j}|^2\Gamma_{r_j}$, which in turn gives the effective Hamiltonian as 
\begin{equation}
     H_{\mathrm{ent}}^{\mathrm{eff}}=\sum_{ij} \left(E_{\mathrm{LS}2}^{ij}-i \gamma_{\mathrm{decay}}^{ij}/2\right)\ket{\widetilde{ij}}\bra{\widetilde{ij}}.
     \label{eq:effective_entangling_interaction}
\end{equation}

\begin{figure}
\centering
	\includegraphics[width=0.8\textwidth]{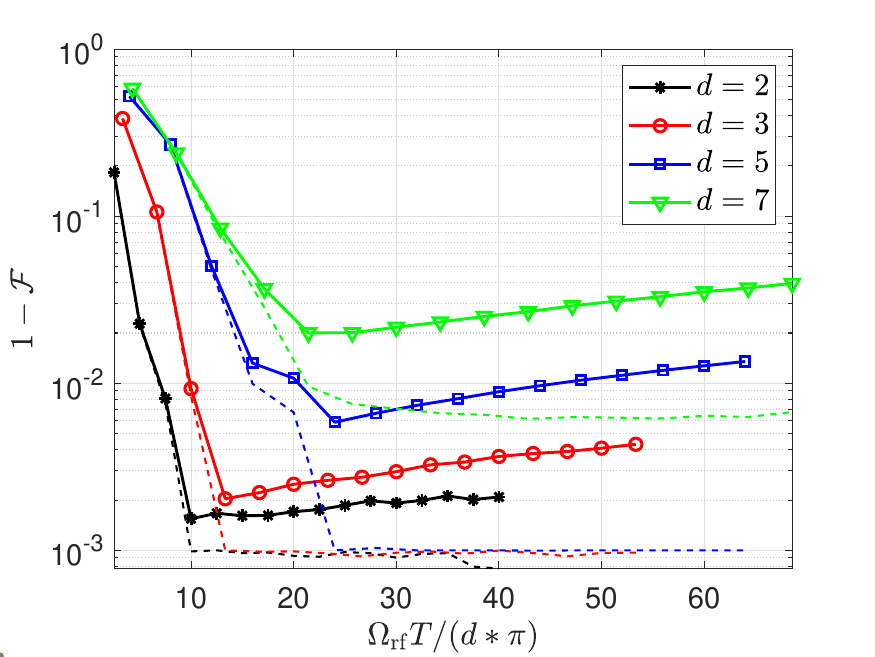}
	\caption{\textbf{Infidelity as a function of time.} 
	Simulated infidelity with and without decoherence as a function of control time divided by the dimension $d$ for CPhase gate with different prime dimensions with $d\leq 10$, as found using  Lie algebraic quantum control and the GRAPE algorithm. Decoherence  due to Rydberg decay outside the computational basis is included through an imaginary part of the Hamiltonian. We take the Rydberg lifetime to be $140 \mu$s and choose the rf-Larmor frequency to be $\Omega_{\mathrm{rf}}/2\pi=10$ MHz. In the absence of decoherence (dashed lines), for a time greater than the ``quantum speed limit" (the time required to obtain ideal fidelity) we achieve a minimal error (infidelity) of $10^{-3}$ due to our threshold in the numerics for $d\le 5$. This speed-limit time increases as we increase the qudit dimension, which in turn results in an increased decay in maximum fidelity.	For the CPhase gate, we obtain a fidelity of  $0.9985$, $0.9980$, $0.9942$, and $0.9800$ for $d=2, d=3, d=5, \text{ and } d=7$ respectively. For all of these calculations, we have taken the dressing laser Rabi frequency to be $\Omega_{\mathrm{L}}=6 \Omega_{\mathrm{rf}}$ and  the lifetime of the Rydberg states to be $140\mu$s.}
	\label{fig:Fidelity figure 1}
\end{figure}

With this model for decoherence in hand, the numerical results for the Lie algebraic approach are  given in Fig.~\ref{fig:Fidelity figure 1}, which shows the infidelity as a function of time for a CPhase gate for different dimension isometries. We focus here on the case of the prime dimensional qudits. In contrast to closed-system control, in the presence of decoherence, infidelity decreases at first and then increases. This is due to the fact there is an optimal time of evolution, larger than the quantum speed limit, but not too large when compared to the coherence time of the system. As expected, one needs more time as the qudit dimension increases, which in turn results in an increase in the minimum infidelity one could achieve in each of these cases as shown in Fig.~\ref{fig:Fidelity figure 1}. 
We obtain a maximum fidelity of 0.9985, 0.9980, 0.9942, and 0.9800 for $d=2$,  $d=3$, $d=5$, and $d=7$ respectively for the CPhase gate.
Note, the values of fidelity for different dimensional qudits should be considered in the context of a particular application. 
For example, the threshold for fault tolerance for qudits, in general, is larger for larger $d$~\cite{anwar2014fast,watson2015fast}. For the particular scheme considered in~\cite{anwar2014fast}, the threshold for $d=2$, $d=3$, $d=5$, and $d=7$ are close to $0.008$, $0.012$, $0.0135$, and $0.015$ respectively. Hence, the proof-of-principle fidelity obtained here is promising and can be further optimized.

In the Lie group approach, we can use the effective Hamiltonian to describe the evolution when the Rydberg dressing is employed. In this case, we have,
\begin{equation}
    \begin{aligned}
     U_{\mathrm{tar}}^{\mathrm{eff}}&=\prod_{j}U_{\lambda_j},\\
     &=\prod_{j} e^{-i H_{\mathrm{ent}}^{\mathrm{eff}}t_j} U_1(\vec{\alpha}^{(j)})\otimes U_2(\vec{\beta}^{(j)}).\\
\end{aligned}
\label{eq:Lie_group_approach_eff}
\end{equation}
 We neglect here any decoherence associated with the local SU($d$) gates. Thus the fidelity including the decoherence effects is given as,
\begin{equation}
\mathcal{F}_{\mathrm{eff}}=\left|\Tr\left(U^{\dagger}_{\text{tar}}U_{\mathrm{tar}}^{\mathrm{eff}}\right)\right|^2/d^2,
\label{eq:Lie_group_fidelity}
\end{equation}
\begin{figure}
\centering
	\includegraphics[width=0.8\textwidth]{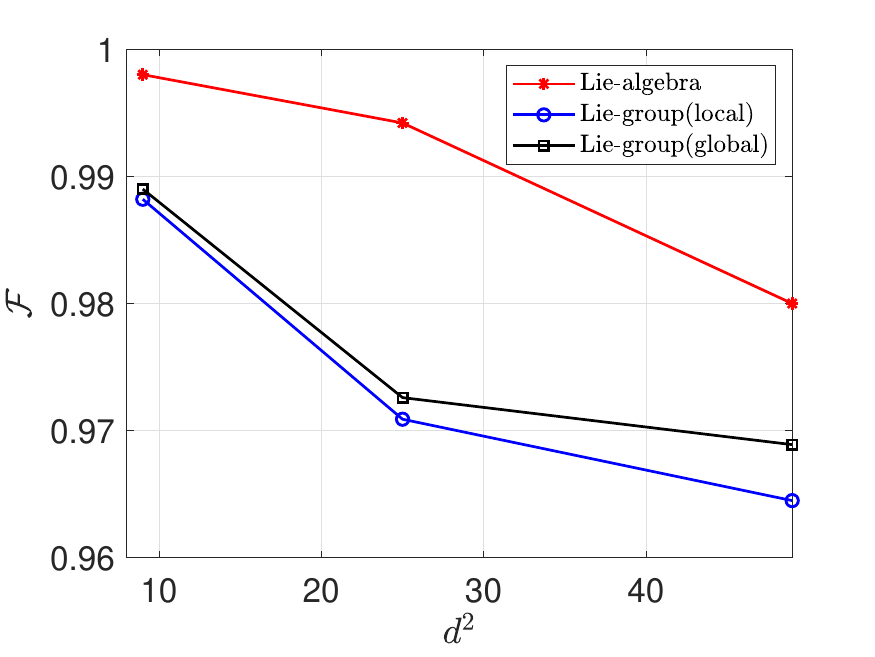}
	\caption{A comparison of the optimized fidelity, $\mathcal{F}$ of the CPhase gate achieved for the Lie algebraic and Lie group approaches (including both local single-qudit control and only global control)  is plotted as a function of the total Hilbert space dimension $d^2$, for the qudits of dimension $d=3,5,$ and $d=7$. For all of these simulations, we have taken the parameters given in Fig.~{\ref{fig:Fidelity figure 1}}.}
	\label{fig:comparison_approach}
\end{figure}

\begin{figure}
\centering
	\includegraphics[width=0.8\textwidth]{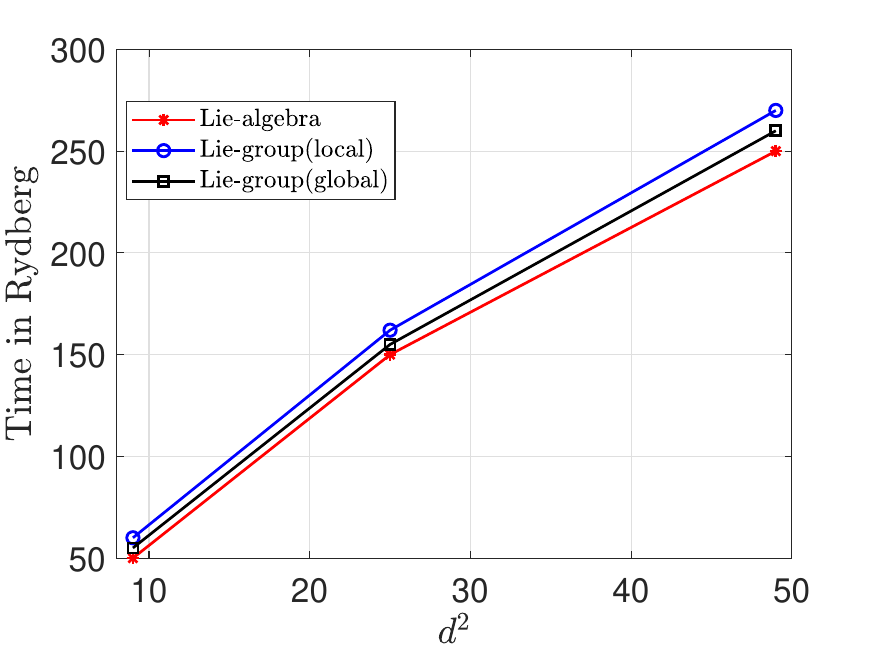}
	\caption{A comparison of the minimum time spent in the Rydberg state to implement the  CPhase gate achieved for the Lie algebraic and Lie group approaches (including both local single-qudit control and only global control)  is plotted as a function of the total Hilbert space dimension $d^2$, for the qudits of dimension $d=3,5,$ and $d=7$.
 For all of these simulations, we have taken the parameters given in Fig.~{\ref{fig:Fidelity figure 1}}.
 Thus the time required for the Lie algebraic control is smaller than the Lie group control which in turn contributes to the fidelity.}
	\label{fig:comparison_approach_time}
\end{figure}

A comparison of the fidelities achieved based on the Lie algebraic and Lie group approaches is given in Fig.~\ref{fig:comparison_approach} for $d=3,5,$ and $d=7$. 
The results suggest that the Lie algebraic protocol slightly outperforms the Lie group protocol in the presence of decoherence.  
This difference in the performance can be attributed to the time spent in the Rydberg state for these two approaches, as shown in Fig.~\ref{fig:comparison_approach_time}. Fundamentally, we can understand this from the fact that the Lie algebraic approach has more control parameters as compared to the Lie group protocol. 
Thus, based on the Magnus expansion~\cite{merkel2009quantum,jurdjevic1972control,brockett1973lie}, the nested commutators which are at the heart of controllability become easier to achieve.  Both approaches yield high fidelities in large dimensional qudits.
Nevertheless, the Lie group approach may be preferable when considering the complexity necessary for experimental control.
 The difference in the behavior of Lie-group(local) to Lie-group(global) is due to the fact that for the global approach we allow $H_{\mathrm{ent}}\to - H_{\mathrm{ent}}$ in alternating layers. 

 In general, a key experimental consideration for the successful implementation of open-loop quantum control is the effect of uncertainties in Hamiltonian parameters. These can be mitigated to some degree using the tools of robust quantum control \cite{anderson2015accurate, goerz2015optimizing, glaser2015training, koch2016controlling}.  Such techniques are generalizations of spin-echo type composite pulses which can be useful when there is sufficient coherence time.  With a detailed  understanding of the dominant inhomogeneities, robust optimal control can be used to implement suitable composite waveforms for  qudit entanglers on any platform.

The specific experimental foundation of this proposal is well-motivated by existing literature, particularly the work of the Jessen group~\cite{anderson2013unitary}.  
One particular issue discussed above is the trap-induced differential light shifts between the ground state and excited state $\TPT$ manifold \cite{PhysRevResearch.5.013219}.
It will be necessary to mitigate motional dephasing arising from vector- and tensor-shifts, which induce an $m_F$-dependence on polarizability, thus inducing possible motional dephasing between $m_F$ levels.
The easiest way around this problem is to operate with a linearly-polarized optical trap, with polarization vector aligned at the ``magic angle'' \cite{PhysRevX.8.041054} and corresponding magic wavelength \cite{doi:10.1126/science.1148259} for the $\SSZ \rightarrow \TPT$ transition.
This allows intra-state coherence within the $\TPT$ $F=9/2$ (and other $F$-levels) manifold, and inter-state (\textit{i.e.}, optical qubit) coherence between the $\SSZ$ and $\TPT$ $F=9/2$. We can also mitigate motional effects via high-fidelity ground-state cooling \cite{kaufman2012cooling, thompson2013coherence, lester2014raman}.

\section{Conclusion and Outlook} 
\label{sec:conclusions_and_future_work}

Quantum computation with qudits has potential advantages when compared with architectures employing qubits. Implementing gates for qudit-based quantum computation is fundamentally more challenging, as the generators for these gates are not native Hamiltonians on physical platforms. One way to overcome this challenge is to use the tools of quantum optimal control, whereby we combine native Hamiltonians with time-dependent waveforms that drive the system in order to implement a universal gate set with high fidelity. 

In this chapter, we introduced two classes of numerical methods of quantum optimal control for implementing the qudit entangling gates, an essential component of the universal gate set. The first approach is based on continuous-time driving given a controllable Hamiltonian with tunable parameters and uses the Lie algebraic structure of the control problem. The second approach is more ``digital," using the Lie group structure to design a family of unitary maps that can be applied in sequence to achieve any nontrivial entangling gate of interest. 

As a specific example, we studied encoding a qudit in the nuclear spin of $^{87}$Sr, a species of atoms that is particularly important in quantum information processing. The nuclear spin can accommodate a qudit of dimension $d\le 10$. We have previously studied protocols for implementing single-qudit gates in SU($d$). To implement entangling gates we studied how we make two atoms interact using the well-known Rydberg blockade mechanism, and in particular, we studied Rydberg dressing schemes. Using this we are able to generate any two-qudit entangling gate, both using the Lie algebraic and Lie group based approaches.

We also studied how the fundamental effects of decoherence introduced by the finite lifetime of the Rydberg states reduce the gate fidelity. To model this we used a nonHermitian Hamiltonian and found that even when including decoherence, one could achieve high fidelity for these qudit entanglers. Given the flexibility of arbitrary control, we can seek the best  approach to encoding qudits and mitigating errors. 

Finally, while we have studied a particular case study in the context of neutral-atom quantum computing, the general methods we have developed here can be applied in other platforms,  including trap ions transmon qudits, and nanomagnets \cite{petiziol2021counteracting,chiesa2021embedded}, which also have natural encoding and control Hamiltonians.

 \chapter{Fault-tolerant quantum computation using large spin cat-codes}
\label{chap:Qudit_fault}

\section{Introduction}
{\label{sec:introduction_qudit_fault}}

In this chapter, we  develop more efficient error-corrected quantum processors by taking advantage of the larger Hilbert spaces that can be controlled in individual subsystems for a given physical platform.
While many platforms offer access to multiple levels, the focus is often on isolating two well-defined levels for qubit-based computations. However, a more advantageous approach emerges when we exploit these multiple levels to create qubits naturally resilient to dominant noise channels \cite{Gottesman2001,gross2021hardware,omanakuttan2023multispin,Omanakuttan2023gkp}.  
In this chapter we will consider encoding a qubit in a spin-$J$ system, corresponding to a qudit with $d=2J+1$ levels \cite{omanakuttan2021quantum,Siva_Qudit_entangler_2023,zache2023fermion}. 
By harnessing the properties of this qudit with multiple levels, we can establish logical qubits that possess inherent resistance to the impact of dominant noise channels, paving the way for more robust quantum computation.


Other works in this direction have previously explored the concept of encoding a qubit in a large spin~\cite{Gross2021, gross2021hardware, omanakuttan2023multispin}.
In this context, the angular momentum operators form the natural set of error operators for such encodings, generalizing the Pauli operator basis for qubits. Earlier studies identified quantum error-correcting encodings, but these constructions were not fault-tolerant~\cite{Gross2021,omanakuttan2023multispin}.  Here, our main objective is to investigate how we can achieve Fault-Tolerant Quantum Computation (FTQC), specifically for a qubit encoded in a large spin. 
This approach may be extended to a wide range of physical systems, including semiconductor qubits \cite{Gross2021, gross2021hardware}, ion traps \cite{ringbauer2021universal, Low2020}, atomic systems \cite{omanakuttan2021quantum, Siva_Qudit_entangler_2023, zache2023fermion}, molecules \cite{castro2021optimal,jain2023ae}, and superconducting systems \cite{ozguler2022numerical, Blok2021}, wherein spin qudits offer the means to encode logical qubits.

We direct our attention to a specific encoding we call the ``spin-cat encoding." 
This choice is motivated by the cat encodings employed in bosonic continuous variable systems~\cite{puri2020bias,guillaud2019repetition}, used to correct photon loss errors, the dominant errors for the continuous variable systems.  Similarly, spin-cat encoding can rectify the dominant error operators in spin systems, namely, the linear and quadratic angular momentum operators. Physically, these arise from uncontrolled Larmor precession of the spins and optical pumping between magnetic sublevels.
To achieve fault tolerance with spin-cat encoding, we develop two key ingredients.
First, we show how to implement a universal gate set that preserves the limited error space of interest.  An essential element here is the ``rank-preserving CNOT" gate that ensures that one does not convert correctable errors into uncorrectable ones. 
Second, aiming at a more easily implemented scheme, we develop a measurement-free error correction {gadget} for spin systems that require fresh ancilla spins {and data-ancilla operations but no measurements}. As we will show, this scheme effectively utilizes the rank-preserving CNOT gate in conjunction with standard phase flip error correction to address and correct angular momentum errors.

A distinctive aspect of the spin-cat encoding, setting it apart from other spin encodings~\cite{Gross2021,omanakuttan2023multispin,kubischta2023family,kubischta2023not}, is its unique structural composition.
In contrast to these earlier methods, the error subspaces in the spin-cat encoding partition the physical space into two-dimensional subspaces where logical operations act identically.
This gives the structure of a stabilizer code, a feature that plays a pivotal role in enabling fault-tolerant schemes for error correction.

{\section{Generalization of cat code for Qudits/spin systems}}{
\label{sec:spin_cat_codes}}

\begin{figure*}
\centering
 \captionsetup[subfigure]{twoside,margin={3.8cm,3cm}}
    \subfloat[]{\includegraphics[width =0.35\columnwidth]{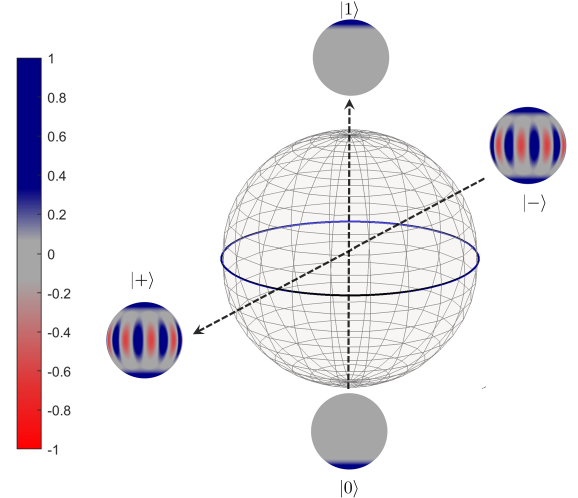} \label{fig:Fig_1_a}}
    \subfloat[]{\includegraphics[width=0.65\columnwidth]{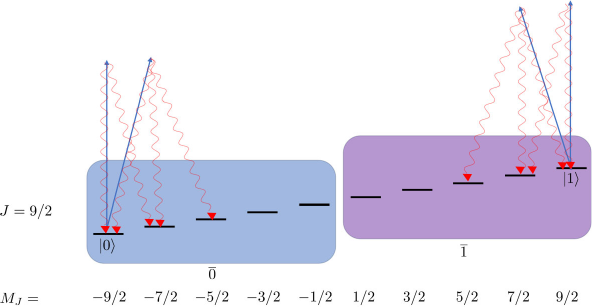}  \label{fig:Fig_1_b}}
    \caption{Qubit encoded in a spin using spin-cat states.
    (a) The Bloch sphere for the qubit encoded in a spin. The two spin-coherent states (stretched states) are the computational basis states, lying on the $Z$-axis and the spin-cat states then lie along the $X$-axis.  The spin Wigner function of the states is shown and its strong negativity indicates that spin-cats are highly nonclassical. (b) The spin-cat encoding of a qubit in spin $J=9/2$, $d=2J+1=10$ levels.
    The correctable errors divide the qudit into two subspaces, $\bar{0}$ and $\bar{1}$, shown as blue and purple boxes, respectively.  One physical error channel is optical pumping, corresponding to the absorption of photons (blur arrows) followed by spontaneous emission (wavy red arrows), which can lead to amplitude damping.}
       \label{fig:bloch_sphere}
\end{figure*}

In this section, we introduce our encoding, present the most prevalent types of noises in spin systems, and look at how they affect an encoded qubit.
We consider quantum information encoded in large spins with angular momentum $J$, a qudit of dimension $d=2J+1$. The space of local errors on a spin system is spanned by the irreducible spherical tensor operators ${T}^{(k)}_q(J)$~\cite{sakurai2014modern,klimov2008generalized,varshalovich1988quantum}  which are polynomials in the spin angular momentum components, $\{{J}_x, {J}_y, {J}_z \}$ of order $k$, {with $2k+1$ components. 
The qudit operator space is spanned by the basis of tensors from $k=1$ to $k=2J$.} 
In most platforms, physical errors are associated with low rank-$k$ tensors for $J \gg 1$.
For example, erroneous Larmor precession caused by noisy magnetic fields are generated by the SU(2) algebra, or rank-1 tensors.  
When controlled by laser light, as in atomic systems, optical pumping arising from photon scattering can lead to rank-2 errors. 
Higher rank errors are rare, as they involve multi-photon processes or higher rank tensor perturbations. 
We thus design codes that can correct any errors in the space spanned by the Kraus operators in the set {of linear and quadratic spin operators} $\{{T}^{(1)}_q(J), {T}^{(2)}_q(J)\}$ \cite{omanakuttan2023multispin}. 
For $J\gg1${, this} is a substantially reduced error space (dimension 8) compared to the total space of all possible errors (dimension $(2J+1)^2-1$).

To design a spin-encoding that can efficiently correct this biased noise structure, we consider the bosonic cat encoding of a qubit~\cite{puri2020bias}. In this encoding, the  qubit states $\ket{0}$ and $\ket{1}$ are chosen to be, 
\begin{equation}
    \ket{C_{\alpha}^{\pm}}\propto \ket{\alpha}\pm \ket{-\alpha},
\end{equation}
where $\ket{\alpha}$ is a coherent state of a single bosonic mode, for e.g., a mode of a microwave cavity as in superconducting systems. 
When the dominant source of noise is photon loss, this encoding exhibits a biased noise channel where increasing the amplitude $\alpha$, exponentially suppresses bit flip errors when compared to phase flip errors. 
It has been shown that by using simple codes such as a repetition code to correct phase flips, one can take advantage of this bias in the noise to achieve significant improvement in the threshold for FTQC \cite{Aliferis2008fault,puri2020bias} for cat qubits. 

 In this work, we pursue a similar approach for finite-dimensional spin systems and consider the spin-cat encoding with,

\begin{equation}
    \begin{aligned}
  \ket{\pm}&\equiv \frac{1}{\sqrt{2}}\left(\ket{J,-J}\pm\ket{J,+J}\right),
    \end{aligned}
 \label{eq:spin_cats}
    \end{equation}
{where now $\ket{0}=\ket{J,-J}$ and $\ket{1}=\ket{J,J}$ are the spin coherent states along the physical quantization ($z$) axis. 
We call this the spin-cat encoding}. 
  Similar to previous works based on continuous variable bosonic cat states \cite{puri2020bias,guillaud2019repetition}, {the spin cat states are defined along the $1$-axis of the qubit Bloch sphere}; see \cref{fig:Fig_1_a}. Note that, unlike the coherent states in the continuous variable setting, the spin coherent states are perfectly orthogonal to each other.

Despite utilizing a similar encoding, there are significant differences between the dominant sources of noise and the easy-to-implement operations in the spin system compared to bosonic cats. Thus, this encoding requires the development of new error-correction procedures that we address in this work.  Central to the continuous variable cat encoding, as explored in~\cite{puri2020bias,guillaud2019repetition}, is the reduction in bit-flip errors. 
The key to this bias is the presence of an energy gap between the excited state manifold and the logical subspace, that scales with $\lvert\alpha\rvert^2$. While this encoding offers significant advantages compared to standard qubit-based encoding, the leakage to these excited states can have detrimental effects on the energy-protected qubits. Dissipative stabilization can be employed to overcome these errors \cite{PhysRevLett.128.110502}.  

In contrast, in spin-cat encoding, we use an alternative approach for fault tolerance.
We consider a primary layer of encoding where we correct for the physically relevant errors and then use a second layer of concatenation to achieve fault-tolerant quantum computation.
We can achieve this because the physically relevant errors are a small subset of all the possible errors for the encoded qubit.  For the spin-cat encoding, these physically relevant errors are composed of spherical tensors of rank-1 and rank-2, as described above. 
The key goal of the first layer of the encoding is to correct for these rank-1 and rank-2 errors.  Our protocol is fault-tolerant because the universal gates and error correction performed in the first layer of encoding do not convert lower-rank spherical tensor operators to higher-rank operators.  We call this ``rank-preserving'' error correction.  It is a generalization of the bias-preserving error correction where the dominant error for the encoded qubit is a single Pauli-error. In the second layer of encoding, the relevant errors are Pauli errors on the logical qubit, which can be corrected by any standard error correction protocol.

\vspace{0.7cm}
{\subsection{Error characterization}
\label{subsec:Error_characterization}}
To categorize the  relevant errors that can be corrected for the spin-cat encoding, it is useful to define the generalized ``kitten states''  as,
\begin{equation}
    \begin{aligned}
    \ket{\pm}_m&=\frac{1}{\sqrt{2}}\left(\ket{0}_m\pm\ket{1}_m\right).
    \end{aligned}
\end{equation}
where,
\begin{equation}
\begin{aligned}
\ket{0}_m&=\ket{J,-J+m}\equiv\ket{-J+m}\\
\ket{1}_m&=\ket{J,J-m}\equiv \ket{J-m}.
\end{aligned}
\label{eq:spin_cat_codes}
\end{equation}
The case $m=0$ is the spin-cat state. 
The total Hilbert space of the spin-cat encoding decomposes to $d/2$ qubit subspace where each of the qubit subspaces is spanned by the kitten states $\ket{\pm}_m$.
Thus we can write,
\begin{equation}
\mathcal{H}_d=\bigoplus_{i=0}^{\frac{d}{2}}\mathcal{H}_2^{(i)},
\label{eq:decompostion_space}
\end{equation}
where each $\mathcal{H}_2$ is a kitten subspace and $\mathcal{H}_d$ is the total Hilbert space of the qudit.
These subspaces are preserved by rotations about the spin quantization $z$-axis and by $\pi$ pulses around axes in the equatorial plane that exchange $\ket{\pm J}$.

We also define the following  projectors onto $\bar{0}$ and $\bar{1}$ subspaces that define correctable errors,
\begin{equation}
\begin{aligned}
    \Pi_{\overline{0}}&= \sum_{k=0}^{\lfloor J-1/2\rfloor}\ketbra{-J+k}{-J+k},\\
    \Pi_{\overline{1}}&= \sum_{k=0}^{\lfloor J-1/2\rfloor}\ketbra{J-k}{J-k}.\\
    \end{aligned}
    \label{eq:projectors}
\end{equation}
See \cref{fig:bloch_sphere} for an illustration. 

The relevant errors on the spin-cat encoding that we aim to correct are a combination of amplitude and phase errors. 
The amplitude errors are defined by the following transformation,
\begin{equation}
\ket{\pm}_m \to \sum_{k=0}^{\lfloor \frac{2J-1}{2}\rfloor} c_k \ket{\pm}_k,
\label{eq:amplitude_error}
\end{equation}
where $c_k$ is an arbitrary complex number. 
The  phase error is given by the transformation,
\begin{equation}
\ket{+}_k \to \ket{-}_k.
\label{eq:phase_error}
\end{equation}
{Physically, these occur as follows.}  First, consider spin rotations,
\begin{equation}
\begin{aligned}
U_{Z}&=\exp(-i\theta J_z),\\
U_{X}&=\exp(-i\theta J_x).
\end{aligned}
\end{equation}
For $\theta \ll 1$ their actions action on the spin-cat states is
\begin{equation}
\begin{aligned}
U_{Z}\ket{\pm}&\approx \left(\mathds{1}-i\theta J_z\right) \ket{\pm} = \ket{\pm}-i\theta J \ket{\mp},\\
U_{X}\ket{\pm}&\approx \left(\mathds{1}-i\theta J_x\right) \ket{\pm} \\
&= \ket{\pm}-i\theta \frac{\sqrt{J}}{\sqrt{2}} \ket{+}_1.
\end{aligned}
\label{eq:error_prob_rotation}
\end{equation}
Thus, the effect of $U_{Z}$ is to introduce a phase error on the spin-cat states whereas $U_{X}$ generates an amplitude error that takes a cat state to a kitten state with $m=1$. 
The ratio of probabilities of amplitude errors to phase errors {due to random rotation errors} goes as $1/J$, and hence approaches zero for large values of $J$.

Next, we consider errors resulting from optical pumping associated with photon scattering. For example, given a laser photon linearly polarized along the quantization axis, followed by the emission of $q=0,\pm1$ helicity photon, the Lindblad jump (Kraus) operators $W_q$ are given by~\cite{Deutsch2000},  
\begin{equation}
    \begin{aligned}
        W_{0}&=\beta T^{(2)}_0,\\
        W_{+1}&=  i \alpha T^{(1)}_{-1}- \beta \sqrt{\frac{3}{4}}T^{(2)}_{-1},\\
          W_{-1}&= i \alpha T^{(1)}_{1}+\beta \sqrt{\frac{3}{4}}T^{(2)}_{1}.
    \end{aligned}
    \label{eq:jump_opt_simplified}
\end{equation}
where $\alpha, \beta$ are real numbers that depend on the atomic structure and the states being excited by a near resonance laser. (See \cref{sec:Ratio_optical_pumping_errors} details.) Optical pumping can include rank-2 tensors as it involves two photons. 
The effect of optical pumping introduces both amplitude errors that change the kitten subspace~\cref{eq:amplitude_error}, and phase errors as given in~\cref{eq:phase_error}.
In contrast to errors that result from rank-1 SU(2) rotation, in optical pumping, it is equally important to correct both amplitude damping and phase errors and ultimately, we must do so fault-tolerantly.

Amplitude errors up to rank $K=\lfloor 2J-1/2 \rfloor$ can be corrected by identifying whether the system is  in 
a specific kitten state with a given $m$ value.
To correct for the phase errors, we concatenate the spin-cat code {in} a repetition code  {with logical states},
\begin{equation}
\begin{aligned}
\ket{+_\mathrm{L}}&=\ket{+}\ket{+} \ket{+},\\
\ket{-_\mathrm{L}}&=\ket{-} \ket{-} \ket{-}.
\label{eq:concat_spin_cat}
\end{aligned}
\end{equation}
While we consider a three-qubit repetition code here and throughout \cref{sec:syndrome_extraction} for simplicity, in \cref{sec:Fault_tolerance_and_threshold} we will look at repetition codes with more than three qubits in order to calculate the threshold for fault-tolerance. One can then perform the corresponding error correction steps similar to the approach taken in the continuous variable encoding~\cite{puri2020bias,guillaud2019repetition}.   We call this the ``logical-level encoding" to differentiate it from the physical-level encoding in \cref{eq:spin_cat_codes}.  

More formally, in \cref{sec:KL_conditions} we show that the  logical-level encodings in \cref{eq:concat_spin_cat} can correct {any single spin} angular momentum errors of the form,
\begin{equation}
\begin{small}
\mathcal{E}_{K}=\left\{
J_x^{l}J_y^{m}J_z^{n}; 0\leq l+m+n\leq K=\lfloor\frac{2J-1}{2} \rfloor\right\}.
\label{eq:error_channel}
\end{small}
\end{equation}
In practice we can restrict our attention to quadratic polynomials.

{\subsection{The irreducible spherical tensor basis}}{\label{subsec:Spherical_tensor_basis}}

The irreducible spherical tensor basis provides a natural basis to characterize the action of the error operators. In the basis of the magnetic sublevels, the normalized tensors are~\cite{varshalovich1988quantum} 

\begin{equation}
    T^{(k)}_q(J) = \sqrt{\frac{2k+1}{2J+1}} \sum\limits_{m,m'=-J}^{J} C_{Jm';kq}^{Jm} \ketbra{J,m}{J,m'},\ 
\end{equation}
where $C_{J,m';kq}^{J,m}= \braket{J,m}{J,m';k,q}$ are the Clebsch-Gordan coefficients.
The spherical tensor operators of rank-$k$ are the solid harmonics consisting of polynomials on the angular momentum operators of order $k$.
To track how errors occur, it is convenient to introduce the following linear combination of the spherical tensor operators, 

\begin{equation}
    \begin{aligned}
        S^{(k)}_q(J)& = \frac{1}{\sqrt{2}}\left[T^{(k)}_q(J) + (-1)^k T^{(k)}_{-q}(J) \right],\\  
A^{(k)}_q(J) &=\frac{1}{\sqrt{2}}\left[T^{(k)}_q(J) - (-1)^k T^{(k)}_{-q}(J) \right],\\
S^{(k)}_0(J) &= T^{(k)}_0(J).
\label{eq:basis}
    \end{aligned}
\end{equation}
for  $0\leq k \leq 2J+1$  and $q>0$.
It is straightforward to check that these operators form another orthonormal basis for a spin-$J$ system, i.e.,

\begin{equation}
\begin{aligned}
   \Tr{\left(S^{(k)}_q\right)^{\dagger} S^{(k')}_{q'}}&=\Tr{\left(A^{(k)}_q \right)^{\dagger}{A}^{(k')}_{q'}}=\delta_{k,k'}\delta_{q,q'}, 
   \\ \Tr{\left(S^{(k)}_q\right)^{\dagger} A^{(k')}_{q'}}&=0,
\end{aligned}
\end{equation}
for $0\leq k,k'\leq 2J+1$, $0\leq q \leq k$, and $0\leq q' \leq k'$.
The action of the operators on the cat and kitten states are given ({for $q>0$}) as,

    \begin{equation}
    \begin{aligned}
        S^{(k)}_q\ket{\pm}_l & =  \sqrt{\frac{2k+1}{2(2J+1)}} \left[(-1)^k C^{J,-J+l-q}_{J,-J+l;k,-q}\ket{\pm}_{l-q} +C^{J,-J+l+q}_{J,-J+l;k,q}\ket{\pm}_{l+q}\right],\\
         A^{(k)}_q\ket{\pm}_l&= \sqrt{\frac{2k+1}{2(2J+1)}} \left[(-1)^k C^{J,-J+l-q}_{J,-J+l;k,-q}\ket{\mp}_{l-q} -C^{J,-J+l+q}_{J,-J+l;k,q}\ket{\mp}_{l+q}\right],\\
         S^{(k)}_0\ket{\pm}_l&=
         \begin{cases}
         \sqrt{ \frac{2k+1}{2J+1}}  C^{J,-J+l}_{J,-J+l;k,0}\ket{\pm}_l, & \text{if }\ k \text{ mod } 2=0 \\
      \sqrt{ \frac{2k+1}{2J+1}}  C^{J,-J+l}_{J,-J+l;k,0}\ket{\mp}_l, & \text{otherwise.}
    \end{cases}
     \end{aligned}
     \label{eq:action_F_LM}
\end{equation}

Note that the states {on the righthand side of the equations} are not normalized, as the operators $S^{(k)}_q,A^{(k)}_q$ are not unitary.
They are the Kraus operators corresponding to the {relevant} errors.

The action of the Kraus operator $S^{(k)}_q$  is the amplitude error given in \cref{eq:amplitude_error}.
The Kraus operator $S^{(k)}_0$ flips the kitten states for $k\;\mathrm{mod}2$=1 which corresponds to the phase error in \cref{eq:phase_error}; the Kraus operators $A^{(k)}_q$ change the value of the kitten state and also flip their sign. 
This corresponds to the action of both amplitude and phase error.  This basis of the Kraus operators tracks whether the error is amplitude, phase, or the product of two.
The correctable single spin errors can be written in terms of the new basis as,
\begin{equation}
    \mathcal{E}_{K}=\left\{S_{q}^{(k)},{A}_q^{(k)} \; \vert \; 0\leq k\leq K ,-k\leq q\leq k\right\},
\label{eq:error_channel_basis}
\end{equation}
where $K=\lfloor\frac{2J-1}{2} \rfloor$.

The logical encoding defined in \cref{eq:concat_spin_cat} introduces a biased logical qubit so that the rate of bit flip errors is exponentially suppressed compared to the phase flip errors as a function of the total value of spin $J$. 
Any uncorrectable amplitude error at the physical level of the spin-cat encoding is transformed into a bit-flip error on the logical qubit. 
In \cref{fig:rotation_error_comparison} we compare the ratio of uncorrectable amplitude error to phase error for rotation error.
It is evident that even for modest values of $J=5/2, 7/2, \text{ and } 9/2$, the bit-flip error rate for the logical qubit is significantly suppressed compared to phase-flip errors.

The proposed encoding can be considered a generalized version of the Shor code,
\begin{equation}
\begin{aligned}
\ket{0}&=\frac{1}{\sqrt{8}}\left(\ket{\uparrow}^{\otimes 2J+1 }+\ket{\downarrow}^{\otimes 2J+1 }\right)^{\otimes 3}\\
\ket{1}&=\frac{1}{\sqrt{8}}\left(\ket{\uparrow}^{\otimes 2J+1 }-\ket{\downarrow}^{\otimes 2J+1 }\right)^{\otimes 3}\\
\end{aligned}
\label{eq:extended_shor_code }
\end{equation} 
For the Shor code \cite{shorcode}, the inner encoding protects against bit-flip errors and the outer encoding protects against phase-flip errors. 
In our case, the inner layer protection originates from the encoding of the qubit in the spin-$J$ qudit, $\ket{\uparrow}^{\otimes 2J+1 } = \ket{J,J}$,  $\ket{\downarrow}^{\otimes 2J+1 } = \ket{J,-J}$.

\vspace{0.7cm}
{\section{Universal gate set and Rank-Preserving CNOT gate }}
{\label{sec:universal_gate_set}}
In this section, we establish a set of universal fault-tolerant operations for spin-cat qubits.  As discussed above, similar to~\cite{Aliferis2008fault,puri2020bias}, our strategy is to first correct for the dominant errors by encoding the biased qubit in a repetition code $\mathcal{C}_1$.
 After performing error correction corresponding to code $\mathcal{C}_1$, we obtain a logical qubit with reduced (but less biased) effective errors. We can then achieve FTQC by employing another level of concatenation using a generic  CSS code $\mathcal{C}_2$, as long as the effective noise strength is below the threshold of the code $\mathcal{C}_2$.

 To construct the universal gate sets, we target the following  physical level gates,
\begin{equation}
    \{\mathcal{P}_{\ket{0}},\mathcal{P}_{\ket{+}},\mathcal{M}_{X},\mathcal{M}_{Z},\mathrm{CNOT},ZZ(\theta),X,Y,Z\}.
    \label{eq:physical_level_gates}
\end{equation}
{We require these spin-cat qubit operations to be ``rank-preserving" so that they do not convert correctable errors into uncorrectable ones. Using this gate set,} one can construct the following {logical} universal gate set {for} $\mathcal{C}_1$, 
\begin{equation}
     \{\mathcal{P}_{\ket{0}_{\mathrm{L}}},\mathcal{P}_{\ket{+}_{\mathrm{L}}},\mathcal{M}_{X_{\mathrm{L}}},\mathcal{M}_{Z_{\mathrm{L}}},\mathrm{CNOT}_{\mathrm{L}}\}\cup \{\mathcal{P}_{\ket{i}_{\mathrm{L}}}, \mathcal{P}_{\ket{T}_{\mathrm{L}}}\}.
      \label{eq:logical_level_gates}
\end{equation}
Here $\mathcal{P}$ denotes state preparation, and $\mathcal{M}$ represents the measurement operators.
  To prepare the magic states $\mathcal{P}_{\ket{i}_{\mathrm{L}}}, \mathcal{P}_{\ket{T}_{\mathrm{L}}}$, we can utilize rank-preserving $ZZ(\theta)$ at the physical level, similar to the bias-preserving case of qubits~\cite{zzPoulin} and cat codes~\cite{puri2020bias}.
\vspace{0.7cm}
{\subsection{Single qubit gates}}  

To ensure fault tolerance, a gate $U$ must not turn correctable errors into uncorrectable errors in a specific level of encoding, i.e., we require that
\begin{equation}
U \mathcal{E}_K U^{\dagger} \in \mathcal{E}_K,
\label{eq:fault_tolerant_condition}
\end{equation}
where $\mathcal{E}_K$ represents the set of correctable errors for the spin-cat encoding as defined in \cref{eq:error_channel_basis}.
Further, to prevent the propagation of correctable errors into uncorrectable ones during subsequent computations, the gates $U$ should act on states for which an error has occurred in the same manner as they act on states within the logical subspace.  Specifically, 
these gates must exhibit identical behavior whether the states are in the cat subspace or the kitten subspace with $m>1$, the subspace corresponding to amplitude damping errors. 

 By building the gates $U$ in the universal gate set using operations solely from the spin-$J$ representations of $\mathrm{SU}(2)$, we can guarantee the condition in \cref{eq:fault_tolerant_condition}. To see this, recall the definition of spherical tensor operators \cite{klimov2008generalized,sakurai2014modern}:
\begin{equation}
    U T_{q}^{(k)} U^{\dagger}= \sum_{-k\leq q'\leq k} D_{q,q'}T_{q'}^{(k)},
    \label{eq:spherical_tensor_operators}
\end{equation}
where $U=e^{-i \theta \hat{\mathbf{n}}.\mathbf{J}}$ is a spin-$J$ $\mathrm{SU}(2)$ rotation operator and 
\begin{equation}
    D_{q,q'}=\bra{k,M=q'} \exp(-i \theta \hat{n}.\mathbf{J}) \ket{k,M=q},
    \label{eq:Wigner_D_matrix}
\end{equation}
are the elements of Wigner $D$-matrices \cite{sakurai2014modern}.
As a result, $\mathrm{SU(2)}$ operators do not change the rank of spherical tensor operators.
Using the above relationships for the basis of errors introduced in \cref{eq:basis}, we get,
\begin{equation}
\begin{aligned}
    U S^{(k)}_q U^{\dagger}&= \sum_{q'}\left( g_{q,q'} S^{(k)}_{q'}+\widetilde{g}_{q,q'}A^{(k)}_{q'}\right)\\
     U A^{(k)}_{q} U^{\dagger}&=\sum_{q'} \left( h_{q,q'} S^{(k)}_{q'}+\widetilde{h}_{q,q'}A^{(k)}_{q'} \right)
\end{aligned}
\end{equation}
where the coefficients $\{g_{q,q'},\widetilde{g}_{q,q'},h_{q,q'},\widetilde{h}_{q,q'}\}$ are given in \cref{sec:coefficients_rotation}.
Therefore, the SU(2) rotations do not change the rank of the error operators and obey the condition given in  \cref{eq:fault_tolerant_condition}.

For the remainder of this chapter, we consider the case of $J$ half-integer (even $d$).  These schemes can be easily refashioned for odd $d$ with minor modifications.
The single-qubit Pauli gates for the qubit encoded in the spin-qudit  can be implemented using the following general $\mathrm{SU}(2)$ operations,
 \begin{equation}
 \label{eq:logical_paulis}
 \begin{aligned}
 X&=\exp(-i\pi J_x),\\
 Y&=\exp(-i\pi J_y),\\
 Z&=\exp(-i\pi J_z).
 \end{aligned}
 \end{equation} 
 These are easily implemented by Larmor precession of the spin.
 
In contrast, and critically, the Hadamard gate $H$ for the spin-cat encoding, defined by
\begin{equation}
    \begin{aligned}
    H\ket{0}&=\ket{+}.\\
    H\ket{1}&=\ket{-},
    \label{eq:Hadamard_gate}
    \end{aligned}
\end{equation}
{\em cannot} be  achieved by $\mathrm{SU}(2)$ operations alone. To see this, note that an SU(2) rotation preserves the projection of the spin onto a rotated axis.  As $\ket{0}$ and $\ket{1}$ are spin coherent states (so-called ``stretched states"), an SU(2) rotation cannot be used to prepare a cat state, which is a superposition of spin coherent states. Therefore, 
\begin{equation}
    H \mathcal{E}_K H^{\dagger}\not \in\mathcal{E}_K.
\end{equation}
The essential feature of our protocol is to circumvent this restriction by using ancilla qubits and rank-preserving CNOT gates to effectively apply a Hadamard gate that preserves the set of correctable errors.\\

{\subsection{Rank-preserving CNOT gate}}
{\label{sec:CNOT_gate}}

In this section, we develop a rank-preserving CNOT gate, the key ingredient to realize the universal gate set, using only $\mathrm{SU}(2)$ operations. For concreteness, we provide a detailed protocol {based on the platform of neutral-atom quantum computing~\cite{Brennen1999, Deutsch2000, Jaksch_gate_2000, Saffman_review_2016_Rydberg,Browaeys_review_2020_Quantum}, which has shown increasing promise for scalable FTQC~\cite{bluvstein2022quantum,Cong_Lukin_QEC_Rydberg_PRX_2022,Singh_Bernien_2022_dual_array,Saffman_Nature_2022,Ebadi_Lukin_2022_MIS_optimization}. 
In particular, we consider $^{87}$Sr atoms which we studied in detail in the previous two chapters, with a spin-qudit encoded in the nuclear spin  $I=9/2$, providing a qudit with $d=10$ levels~{\cite{omanakuttan2021quantum,Siva_Qudit_entangler_2023}}.  

Note, when considering the physical spins of atoms, in standard notation $I$ is the nuclear spin, $J$ is the total angular momentum of the electrons, and $F$ is the total electronic angular momentum plus nuclear spin. Our qudit is encoded in spin $I$ in the electronic ground state with $J=0$ for $^{87}$Sr, so that $F=I=9/2$.  
In this section, the spin angular momentum in which we encode the qudit is $\mathbf{F}$.  In the other sections of this chapter, we use $\mathbf{J}$ to denote a generic spin, without reference to its physical encoding.

We target a CNOT gate for the spin-cat encoding that operates the same for all kitten states. 
As discussed above (see \cref{eq:projectors}), we divide the qudit into ``left" and ``right" subspaces, with projectors onto them $\Pi_{\overline{0}}$ and $\Pi_{\overline{1}}$ respectively.  
The gate is formally given as,
\begin{equation} \label{eq:cnot_def}
    \mathrm{CNOT}= \Pi_{\overline{0}} \otimes \mathds{1}+\Pi_{\overline{1}} \otimes X,
\end{equation}
 where $X=\exp(-i\pi F_x)$.  
That is, we apply a $\pi$-rotation (NOT) to every kitten subspace of the target atom if the control atom is in the $\overline{1}$-subspace (the amplitude damped states of $\ket{1}$ we can correct), and the identity, if the control atom is in the $\overline{0}$-subspace (the amplitude damped states of $\ket{0}$ we can correct).  Clearly, if the amplitude damping takes an atom from the $\overline{0}$ to $\overline{1}$ space, or vice versa, the error cannot be corrected.

The protocol for implementing this gate is presented in \cref{fig:Fig_2_b}. 
We note that this protocol requires individual addressing of the atoms.  In step I of the protocol, the population from the ground state memory is coherently transferred to an auxiliary state where it is more easily controlled.  In $^{87}$Sr, we utilize the auxiliary hyperfine state, $\ket{5\mathrm{s}5\mathrm{p};\; ^3{P}_2; \; F=9/2, M_F}$ with hyperfine quantum numbers $F=9/2, M_F$. This manifold possesses a large magnetic dipole moment and a long lifetime. For the control atom, only the population of $\overline{1}$-subspace is transferred to the auxiliary manifold, whereas for the target atom, the population from both $\overline{1}$-subspace and $\overline{0}$-subspace is transferred.
Both of these are facilitated by an effective $\pi$-pulse between the ground and the auxiliary states, which one can implement using quantum optimal control, as discussed below.

In step II, an effective $\pi$-pulse is applied on the control atom between the auxiliary and the Rydberg state.  In step III, we apply the same $\pi$-pulse on the target atom. Due to the Rydberg blockade, this population exchange only occurs when the control atom is in $\overline{0}$-subspace. If the state of the control atom is in $\overline{1}$-subspace, the population from the auxiliary state of the target atom is blockaded from transferring to the Rydberg state. 

Subsequently in step IV, using a global interaction and quantum optimal control, we {simultaneously} implement a $X=\exp(-i\pi F_x)$ rotation in the auxiliary manifold and an identity operator in the Rydberg manifold {of the target atom}.  The net effect is that if the control atom is in $\overline{1}$-subspace an $X$ gate has been applied to the target atom and if the control atom is in  $\overline{0}$-subspace the identity operator has been applied on the target. We transfer all the states back to the ground state by applying steps III-I in reverse order. The whole procedure implements the desired rank-preserving CNOT gate for the spin-cat encoding in \cref{eq:cnot_def}.

\begin{figure*}
\centering
 \captionsetup[subfigure]{oneside,margin={3.7cm,0cm}}
   \subfloat[][]{\includegraphics[width =0.55\columnwidth]{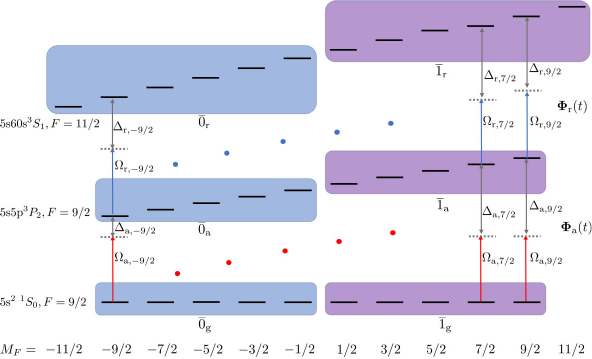} \label{fig:Fig_2_a}}
    \subfloat[]{\includegraphics[width =0.45\columnwidth]{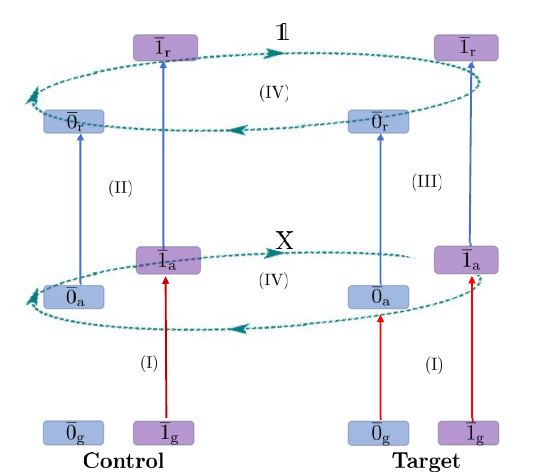}  \label{fig:Fig_2_b}}
\caption{ Protocol for implementing a rank-preserving CNOT-gate in neutral atomic $^{87}$Sr based of optimal control and the Rydberg blockade. The spin-cat qubit is encoded in the nuclear spin, ${I}=F=9/2$, in the electric ground state, $5\mathrm{s}^2 \; ^1{S}_0$.  (a) Detailed level diagram and protocol; (b) High-level schematic. When a gate is to be performed, the qudit is excited from the ground-state memory to the long-lived auxiliary metastable state,   $5\mathrm{s}5\mathrm{p}\;  ^3{P}_2$, ${F}=9/2$. Entangling interactions occur through excitation from the auxiliary state to the Rydberg state, $5\mathrm{s}60\mathrm{s} \; ^3S_1$, ${F}=11/2$. The error-correctable subspaces, $\overline{0}$ and $\overline{1}$, are represented by blue and purple colored boxes respectively, in the ground (g), auxiliary (a), and Rydberg (r) manifolds. The gate is performed in four steps.  Step I: Using quantum optimal control the population from the ground state is transferred to the auxiliary state while preserving coherence between magnetic sublevels. Each two-level resonance, $\ket{\mathrm{a},M_F}\rightarrow\ket{\mathrm{r},M_F}$, has a detuning $\Delta_{\mathrm{a},M_F}$ and Rabi frequency $\Omega_{\mathrm{a},M_F}$.  For the control atom, we only promote the population from the $\bar{1}$-subspace, whereas for the case of the target atom, we promote the population from both the $\overline{0}$ and $\overline{1}$ subspaces to the auxiliary state (see main text for details).  Step II: Using $\pi$-polarized light, local addressing, and quantum control, transfer the population from the auxiliary to Rydberg states only for the control atom. 
Step III: Apply the same pulse to the target atom.  Due to the Rydberg blockade, this population transfer only occurs when the control atom is in $\overline{0}$-subspace; for the  $\overline{1}$-subspace the population is otherwise blockaded.
Step IV: Using global rf-phase-modulated optimal control, we perform the SU(2) rotation $X=\exp(-i\pi F_x)$ in the auxiliary manifold and simultaneously the identity operator in the Rydberg manifold. The result is a CNOT gate -- if the control atom is in $\overline{1}$-subspace we apply an $X$ gate to the target atom if the control atom is in  $\overline{0}$-subspace we implement an identity operator $\mathds{1}$. Finally, we will transfer all the states back to the ground state by reversing steps III-I, thus implementing a rank-preserving CNOT gate for the spin-cat encoding.} 
\label{fig:fig_cnot}
\end{figure*}

\begin{figure*}
 \captionsetup[subfigure]{oneside,margin={2.5cm,0cm}}
    \subfloat[]{\includegraphics[width =0.33\columnwidth]{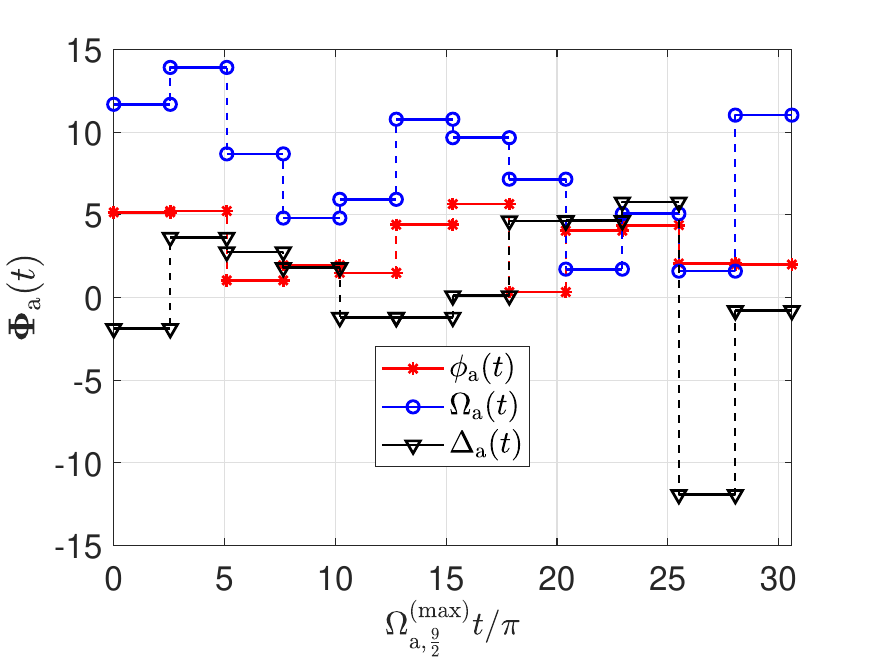} \label{fig:Fig_waveform_a}}
    \subfloat[]{\includegraphics[width=0.33\columnwidth]{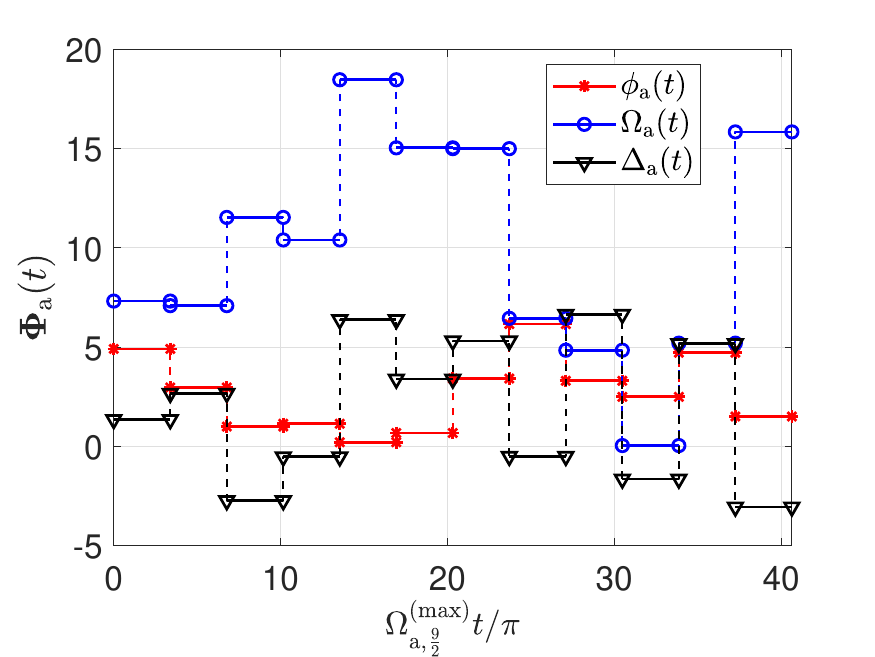}  \label{fig:Fig_waveform_b}}
        \subfloat[]{\includegraphics[width=0.33\columnwidth]{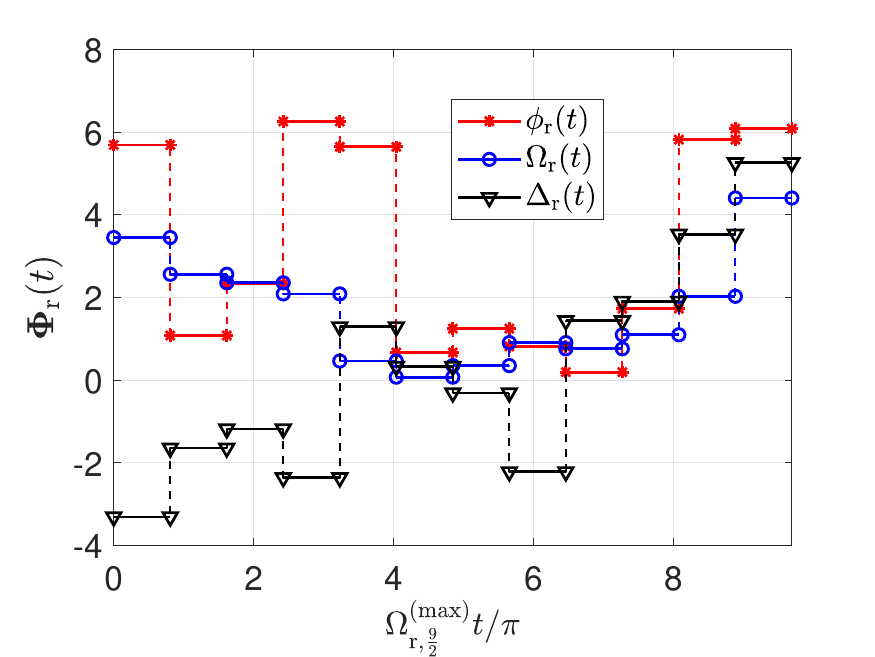}  \label{fig:Fig_waveform_c}}
    \caption{Examples of control waveforms that achieve the transfer of populations between spin manifolds while preserving the coherence between magnetic sublevels.  Based on Hamiltonian~\cref{eq:control_hamiltonian_cnot}, we modulate the lasers' amplitude, detuning, and phase, as piecewise constant functions of time.  Using the GRAPE optimal control we find the target isometries.    (a) The waveform that implements $V_{\mathrm{tar}}^{(\mathrm{C})}$, which transfer population from $\overline{1}_\mathrm{g}$-subspace to $\bar{1}_{\mathrm{a}}$-subspace while the population in the $\bar{0}_\mathrm{g}$-subspace is unchanged.
    (b) The waveform that implements $V_{\mathrm{tar}}^{(\mathrm{T})}$, which transfer population from $\overline{1}_\mathrm{g}$-subspace to $\bar{1}_{\mathrm{a}}$-subspace and $\overline{0}_g$-subspace to $\bar{0}_{\mathrm{a}}$-subspace .
    (c) The waveform that implements $V_{\mathrm{tar}}^{(\mathrm{Ryd})}$ that transfers the population from the auxiliary states to the Rydberg states.
  For all these three cases we divide the time into $12$ equal time steps.}
    \label{fig:transition_waveform}
\end{figure*}

In steps I and II of the rank-preserving CNOT gate, one needs to implement the transfer of population from the ground to the auxiliary manifold and from the auxiliary manifold to the Rydberg manifold, respectively.  This can be achieved by an effective $\pi$-pulse between these respective states and using quantum optimal control.
In both these cases we use the control Rabi Hamiltonian
\begin{equation}
    \begin{aligned}
        H_e(t)=& \sum_{M=-\frac{9}{2}}^{\frac{9}{2}} -\Delta_{e,M}(t)\ketbra{e,M}\\
     &+ \Omega_{e,M}(t)\left[e^{i\phi_{e}(t)}\sigma^+_{e,M}+\mathrm{h.c}\right].
     \label{eq:control_hamiltonian_cnot}
    \end{aligned}
\end{equation}
To simplify the notation we have denoted the two excited metastable manifolds by $e$, where $e=\mathrm{a}$ (auxiliary states) and $e=\mathrm{r}$ (Rydberg states).  Together with the ground-state manifold,
\begin{equation}
\begin{aligned}
    \ket{r,M}&\equiv\ket{5\text{s}60\text{s};\; ^3S_1;\hspace{0.1 cm} F=\frac{11}{2},M_F=M},\\
    \ket{a,M}&\equiv\ket{5\text{s}5\text{p};\; ^3P_2;\hspace{0.1 cm} F=\frac{9}{2},M_F=M},\\
    \ket{g,M}&\equiv\ket{5\text{s}^2;\; ^1S_0,;\hspace{0.1 cm} F=\frac{9}{2},M_F=M},
\end{aligned}    
\end{equation}
and
\begin{equation}
        \sigma^+_{e,M} \equiv \ketbra{e,M}{e',M}.
 \end{equation}
 where $e'=\mathrm{g}$ (for the interaction between the ground and auxiliary states) and $e'=\mathrm{a}$ (for the interaction between auxiliary and Rydberg states).
The control task is achieved by modulation of the amplitude, detuning, and phase of the exciting lasers. 
The time-dependent Rabi frequency and detuning are, 
 \begin{equation}
     \begin{aligned}
         \Omega_{e,M}(t)&=\mathcal{C}_{e,M}\Omega_{e}(t),\\
         \Delta_{e,M}(t)&=\Delta_{e}(t)+\delta_{e,M},
     \end{aligned}
 \end{equation}
 where $\mathcal{C}_{e,M}$ is the ratio of Clebsch-Gordan coefficients,
 \begin{equation}
    \mathcal{C}_{e,M}=\frac{\bra{F,M}\ket{1,0;F,M}}{\bra{F,\frac{9}{2}}\ket{1,0;F,\frac{9}{2}}}. 
 \end{equation}
$\Delta_{e}$ is the detuning, and $\delta_{e,M}$ is the additional detuning due to the relative Zeeman shift. 
 To implement the particular target unitary map interest ($U_{\mathrm{tar}}$) we consider modulation of the amplitude, detuning, and phase of the two lasers that drive the $\ket{g}\rightarrow\ket{a}$ transitions and the $\ket{a} \rightarrow \ket{r}$ transitions.  
 As given in detail in \cref{sec:lie_group_approach} one can use the  GRAPE algorithm to find the optimal control parameters $\Phi=\{\Omega_{e}(t), \Delta_{e}(t),\phi_{e}(t)\}$ that maximizes the fidelity with the target map $U_{\mathrm{tar}}$ 
 \begin{equation}
     \mathcal{F}[\Phi]=\frac{1}{{d}^2}\left| \trace\left\{ U_{\mathrm{tar}}^{\dagger}U[\Phi,T]\right\}\right|^2, 
 \end{equation}
where ${d}$ is the dimension of the qudit and $U[\Phi,T]=\mathcal{T}\left[\exp\left(-i\int_0^T H[\Phi(t)]dt\right)\right]$ is the solution to the time-dependent Schr\"{o}dinger equation. 

We consider partial isometries for our target maps as given in detail in \cref{sec:Numerical Methods}. 
These have fewer constraints than unitary transformations and hence require fewer resources (time, bandwidth etc.).
For the case of the rank-preserving CNOT gate, one needs to implement three target isometries.
Firstly, on the control atom (C) we need to transfer the population from the $\overline{1}$-subspace of the ground manifold to that of the auxiliary manifold while keeping the population in the $\overline{0}$-subspace unchanged.  
The isometry we need to implement is,
\begin{equation}
    \begin{aligned}
        V_{\mathrm{tar}}^{(\mathrm{C})}&=\sum_{M=-\frac{9}{2}}^{-\frac{1}{2}}\ketbra{\mathrm{a},M}{\mathrm{a},M}+\sum_{M=\frac{1}{2}}^{\frac{9}{2}}\ketbra{\mathrm{a},M}{\mathrm{g},M}.
    \end{aligned}
\end{equation}
Secondly, we seek to transfer the entire population from the ground manifold to the auxiliary manifold on the target atom (T).  The isometry is
\begin{equation}
    \begin{aligned}
        V_{\mathrm{tar}}^{(\mathrm{T})}&=\sum_{M=-\frac{9}{2}}^{\frac{9}{2}}\ketbra{\mathrm{a},M}{\mathrm{g},M}.
    \end{aligned}
\end{equation}
Finally, we need to implement an isometry that transfers the population from the auxiliary manifold to the Rydberg manifold,
\begin{equation}
    \begin{aligned}
        V_{\mathrm{tar}}^{(\mathrm{Ryd})}&=\sum_{M=-\frac{9}{2}}^{\frac{9}{2}}\ketbra{\mathrm{r},M}{\mathrm{a},M}.
    \end{aligned}
\end{equation}
All three can be implemented using the Rabi Hamiltonian.

As a proof of principle, we numerically optimize a piece-wise constant waveform based on the well-known GRAPE algorithm for quantum optimal control~ \cite{Merkel2009,merkel2009quantum,jurdjevic1972control,goerz2015optimizing}.  
Example waveforms that implement the target isometries are given in \cref{fig:transition_waveform}. 
The total time required is $4\pi/\Omega_{\mathrm{rf}}$, where $\Omega_{\mathrm{rf}}$ is the rf-Larmor precession rate, chosen to be resonant with the Zeeman splitting in the auxiliary auxiliary manifold.  
To achieve high fidelity control, we have divided the time into $12$ equal time steps.  In practice, other parameterizations could be used to yield smoother waveforms if bandwidth is limited.
 
\begin{figure}[!ht]
    \centering
    \includegraphics[width =1\columnwidth]{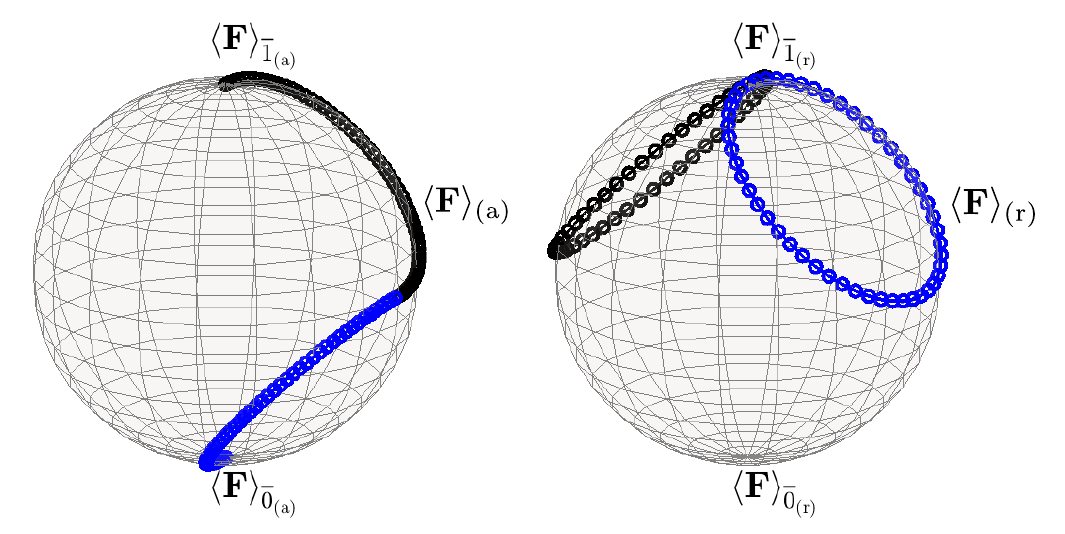} \label{fig:two_parameter_a}
    \caption{Evolution of the spin vector $\langle \mathbf{F} \rangle$ for the auxiliary (a) and Rydberg (r) manifolds resulting from rf-driven Larmor precession with time-varying phases. Optimal control is based on Hamiltonian~\cref{eq:Hamiltonians} for the piece-wise constant phases and total time $T_{\mathrm{tot}}=\sqrt{2}\pi/\Omega_{\mathrm{rf}}$. 
    The blue and black dots correspond to the first and second steps respectively (see text).
    An $X=\exp(-i\pi F_x)$ gate acts on the auxiliary manifold and transfers the population from $\bar{1}_{\mathrm{a}}$ to $\bar{0}_{\mathrm{a}}$ and vice-versa.However, for the   Rydberg manifold, the pulse sequence acts as an identity operator, and the population in the  $\bar{0}_{\mathrm{r}}$ and  $\bar{1}_{\mathrm{r}}$ subspaces remain unaffected.
    \label{fig:independent_evolution_of_excited_and_ground_state}}
\end{figure}

Another important ingredient for the rank-preserving CNOT gate in  \cref{fig:fig_cnot} is that we need to apply an rf-pulse that rotates the auxiliary $^3{P}_2$ state and the Rydberg $^3{S}_1$ state differently. For the case of the rank-preserving CNOT gate, one needs to implement an $X$ gate in the auxiliary manifold and identity in the Rydberg manifold.
This can be achieved because of the different magnetic $g$-factors of the two spin manifolds. 
For our specific choice of Rydberg manifold and auxiliary manifold $g_{\mathrm{r}}/g_{\mathrm{a}} \approx 2$~\cite{urech2023single}. The Hamiltonian describing Larmor precession in each of the excited manifolds, driven by an rf-magnetic field oscillating at frequency $\omega$, in the presence of a basis magnetic field is then
\begin{equation}
\begin{aligned}
H_a&=\Omega_{\mathrm{rf}} \left[\cos(\omega t+\phi) F_x +\sin (\omega t+\phi) F_y\right] +\omega_0 F_z,\\
H_r&=2\Omega_{\mathrm{rf}} \left[\cos(\omega t+\phi) F_x +\sin (\omega t+\phi) F_y\right] +2\omega_0 F_z.
\end{aligned}
\end{equation}
Here $\Omega_{\mathrm{rf}}$ is the Larmor precession frequency and $\omega_0$ is the Zeeman shift induced by the bias B-field in the $^3P_2$ auxiliary manifold.  The spin angular moment operators act in the respective manifolds.  Going to the rotation frame of the rf-oscillation, using the unitary operator $U=\exp(-i\omega t F_z)$, and choosing the rf-frequency to be off-resonant with $\omega=4/3\omega_0$, gives
\begin{equation}
\label{eq:Hamiltonians}
\begin{aligned}
H_a^{\mathrm{rot}}&=\Omega_{\mathrm{rf}}\left[\cos (\phi)F_x +\sin (\phi )F_y\right]-\frac{1}{3}\omega_0 F_z.\\
H_r^{\mathrm{rot}}&=2\Omega_{\mathrm{rf}} \left[\cos (\phi)F_x +\sin (\phi)F_y\right]+\frac{2}{3}\omega_0 F_z.
\end{aligned}
\end{equation}
Because of the finite detuning, the total Larmor precession frequency in the auxiliary and Rydberg manifold is then
\begin{equation} 
\begin{aligned}
   \Omega_{\mathrm{a}}&=\sqrt{\Omega_{\mathrm{rf}}^2+\frac{\omega_0^2}{9}}, \\
    \Omega_{\mathrm{r}}&=\sqrt{4\Omega_{\mathrm{rf}}^2+\frac{4\omega_0^2}{9}}= 2\Omega_{\mathrm{a}}.
\end{aligned}
    \label{eq:effective_Rabi_frequency}
\end{equation}
Since the total Larmor frequency of the auxiliary auxiliary and Rydberg manifolds are different, one can use optimization techniques such composite pulses \cite{levitt1986composite} or quantum optimal control \cite{Merkel2009,goerz2015optimizing} to implement separate unitaries in the auxiliary and Rydberg manifold. 

For example, when $\Omega_{\mathrm{rf}}=\omega_0/3$ using optimal control one achieves an $X$ gate in the auxiliary manifold and the identity in the Rydberg manifold by taking the phase to be a piece-wise constant function time, corresponding to a series of rf-pulses, and a total time, $T_{\mathrm{tot}}=\frac{\sqrt{2}\pi}{\Omega_{\mathrm{rf}}}$.
The resultant dynamics for the auxiliary and Rydberg manifold are given in \cref{fig:independent_evolution_of_excited_and_ground_state}. Since the optimization is purely geometric in nature the same pulse schemes work for any value of the spin as long as the $g$-factors have this ratio.
For further details on the optimization see \cref{sec:Rotating_the_ground_and_excited_manifold}.

The protocol described above can be generalized for other entangling gates.
One can optimize rf-phases in \cref{eq:Hamiltonians} to implement the identity operator in the Rydberg manifold and $R(\theta)=\exp(-i\theta \hat{\bm{n}}.\mathbf{F})$, an SU(2) operator, in the auxiliary manifold. 
Thus one can implement the gate $ZZ(\theta)=\exp{-i\theta Z\otimes Z}$ with any angle $\theta$, up to local $Z$ rotations, for the spin-cat qubits.

 \vspace{0.7cm}
{\subsection{State preparation and Measurement}}{
\label{sec:state_preperation_and_measurement}}
To complete the universal gate set, one needs to implement the state preparation and measurement at the physical level given in \cref{eq:physical_level_gates}. 
$\mathcal{P}_{\ket{0}}$, which is the preparation of the spin coherent state can be achieved with high fidelity using optical pumping  \cite{chow2022high}.
Also, $\mathcal{M}_Z$, which is the measurement in the $\ket{F, M_F}$ basis can be achieved with high fidelity in principle~\cite{barnes2021assembly,ringbauer2021universal}.
However, $\mathcal{P}_{\ket{+}}$ and  $\mathcal{M}_X$ are not straightforward to implement without an SU(2) Hadamard gate.  We describe here new approaches unique to spin-cat encoding and the rank-preserving CNOT gate.
\vspace{0.7cm}
{\subsubsection{Preparation of the spin-cat state}
\label{subsubsec:Preperation_of_the_cat_state}}
We can generate the spin-cat state $\ket{+}$ using multiple approaches. For example, one can use quantum optimal control by considering the controllable Hamiltonian
\begin{equation}
H(t)=\Omega_{\mathrm{rf}} \left( \cos \phi(t) F_x+\sin \phi(t)  F_y\right)+\beta F_z^2.
\label{eq:Control_Hamiltonian_ec}
\end{equation}
This can be implemented in atomic systems using a combination of tensor light shifts and rf rotations~\cite{Paul_experiment_Cs_2007}.  
For the specific case of $^{87}$Sr, we have previously studied how this can be implemented with high fidelity through the tensor light shift imparted on the ground-electronic state nuclear spin~\cite{omanakuttan2021quantum}. Using quantum optimal control protocols one can generate the state $\ket{+}$ from an initial state $\ket{F, M_F=F}$.

The light-shift will also be accompanied by decoherence to photon scattering and optical pumping.
We study this in \cref{sec:Ratio_optical_pumping_errors} to calculate  the fidelity for the state preparation,
\begin{equation}
    \mathcal{F}_{\mathrm{state}}=\bra{+} \rho\ket{+}.
\end{equation}
For the particular choice of $^{87}$Sr, we find the fidelity for quantum optimal control is $\mathcal{F}_{\mathrm{state}}=0.9998$.

\subsubsection{Measurement of $X$}

\begin{figure}
\centering
    \includegraphics[width=\columnwidth]{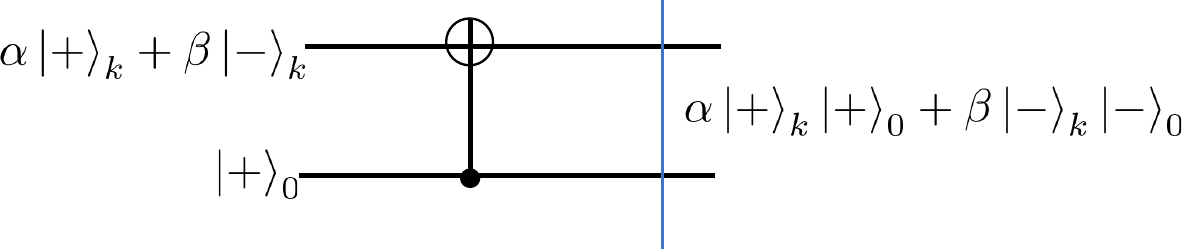}
\caption{Circuit diagram implementing $\mathcal{M}_X$. 
Consider an initial state $\alpha\ket{+}_k+\beta\ket{-}_k$, where $0\leq k \leq \lfloor \frac{2J-1}{2}\rfloor$, 
The action of the CNOT gate  for an ancilla state $\ket{+}_0\equiv\ket{+}$ gives us the state, $\alpha \ket{+}_k \ket{+}+\beta \ket{-}_k\ket{-}$, thus to identify whether the state is in $\ket{+}_k$ or $\ket{-}_k$, we need to measure whether the ancilla is in $\ket{+}_0$ or $\ket{-}_0$.
One can achieve this using a destructive measurement, for more details  (see \cref{eq:X_measurement}). }
\label{fig:X_measurement}
\end{figure}
To measure the $X$ operator ($\mathcal{M}_X$), we need to identify whether the state is in $\ket{+}_k$ or $\ket{-}_k$ for $0\leq k \leq \lfloor \frac{2J-1}{2}\rfloor$. 
We cannot implement the $X$ measurement fault-tolerantly by applying a  Hadamard followed by measuring in the computational basis since Hadamard is not an SU(2) rank-preserving gate. 
{To surmount these challenges, similar to~\cite{puri2019stabilized}, we use an ancilla-assisted measurement protocol, where measurement errors will lead to syndrome errors without disturbing the encoded data.}  Hence, we implement the $X$-measurement by adding an ancilla qubit in the spin-cat state $\ket{+}_0$, applying a CNOT gate, and then destructively measuring the ancilla.
Since the ancilla is measured destructively and discarded, we do not need to implement the $X$-measurement using rank-preserving operators.

The circuit diagram which implements the measurement is shown in \cref{fig:X_measurement}. After the application of the CNOT gate, the joint state of the system is $\alpha \ket{+}_k \ket{+}_0+\beta \ket{-}_k\ket{-}_0$. 
Measuring whether the ancilla is in $\ket{+}_0$ or $\ket{-}_0$ gives the value of $X$ on the data qubit. 
To {measure the ancilla in the} $\ket{\pm}_0$ basis, we use quantum optimal control techniques to implement the required transformation to the $\mathcal{M}_z$ basis using $\mathrm{SU}(d)$ optimal control.  We employ the control Hamiltonian in \cref{eq:Control_Hamiltonian_ec} to implement  the isometry \cite{Siva_Qudit_entangler_2023},
\begin{equation}\label{eq:X_measurement}
V_{\mathrm{targ}}= \ket{F, M_F=F}\bra{+}+\ket{F, M_F=-F}\bra{-}.
\end{equation}
In practice, this operation will be accompanied by decoherence, and the actual map we implement may be written as
\begin{equation}
V= e^{-\int \mathcal{L}(t)dt} V(0),
\end{equation}
where
\begin{equation}
    V(0)=\ket{+}\bra{+}+\ket{-}\bra{-}.
\end{equation}
and $\mathcal{L}(t)$ is the Lindbladian. 
Thus the fidelity for the implementation of the isometry is defined as
\begin{equation}
    \mathcal{F}_{\mathrm{iso}}=\frac{1}{4}\lvert\Tr(V_{\mathrm{targ}}V^{\dagger})\rvert^2.
    \label{eq:accuray_equation}
\end{equation}
  As an example, we consider the effect of photon scattering and optical pumping that accompanies the tensor light shift.  In our simulation, we achieve fidelity of $\mathcal{F}_{\mathrm{iso}}=0.999$ for  $^{ 87}$Sr in the presence of optical pumping described above.

We have now constructed all the required operations at the level of the qubit encoded in the spin, as given in  \cref{eq:physical_level_gates}. We can use these operations {to implement a universal gate set on the spin-cat qubits and} to construct the error correction and logical operations of the $\mathcal{C}_1$  code \cite{Aliferis2008fault}. (See \cref{sec:implementing_the_logical_operator} for the implementation of logical operations in $\mathcal{C}_1$.)
 
Generalizations of rank-preserving gate sets at the physical level can reduce the circuit size for specific applications   \cite{guillaud2019repetition,PRXQuantum.4.030311}.
For example, we can easily generalize our construction of the CNOT gate in \cref{sec:CNOT_gate} to implement a Toffoli gate in spin systems as discussed in~\cref{sec:Toffoli_gate}. 
The scheme is similar to the $\mathrm{CCZ}$ gate implemented in \cite{Levine_Pichler_gate}. 
This utilizes the capability to move neutral atoms in tweezer arrays, arranging the nearest neighbors to interact via the Rydberg blockade, while leaving the next-to-nearest neighbors unaffected. With access to such a gate, similar to the recent development in the bosonic system \cite{guillaud2019repetition}, we can implement the following operations,
\begin{equation}
\{P_{\ket{\pm}},\mathcal{M}_{X},\mathrm{CNOT},\mathrm{Toffoli}\}.
    \label{eq:universal_gate_set_1}
\end{equation}
Such a gate set can be used to construct more efficient fault-tolerant logical-level operations.

\vspace{0.7cm}
{\section{Syndrome Measurement and error recovery}}{
\label{sec:syndrome_extraction}}

The design of error correction gadgets plays a major role in determining the threshold of tolerable noise and also the required overhead of fault-tolerant schemes mainly due to the fact that current fault-tolerant designs require many rounds of error correction to control the spread of errors. The standard method to perform an error recovery is to measure the syndromes to identify the errors and then correct the errors by applying an appropriate unitary operator. This is the approach we take to correct the phase errors. We use a repetition code of size $n$, capable of correcting up to $\lfloor (n-1)/2 \rfloor$ phase errors. In this case, the $(n-1)$ syndrome measurements for phase error correction are
\begin{equation}
\label{eq:syndrome_phase_detection}
\mathcal{S}_{\text{phase}}=\{X_1X_2,X_2X_3,\hdots, X_{n-1}X_n\}.
\end{equation}
These syndrome measurements can be implemented according to the standard circuits in \cref{fig:circuit_phase_error_correction} (for $n=3$)
using the universal operations described in \cref{sec:universal_gate_set}.

When the probability of phase errors is larger than amplitude errors in each spin, increasing the size of the repetition code $n$ can reduce the probability of logical phase errors. However, increasing $n$ will increase the probability of logical amplitude errors due to the increase in the number of the required CNOT gates for the syndrome circuits. Therefore we can choose the optimal $n$ that brings the two types of errors to the same level, determined by the noise threshold required by the outer CSS code $\mathcal{C}_2$.


\begin{figure}
\centering
\includegraphics[width=0.8\columnwidth]{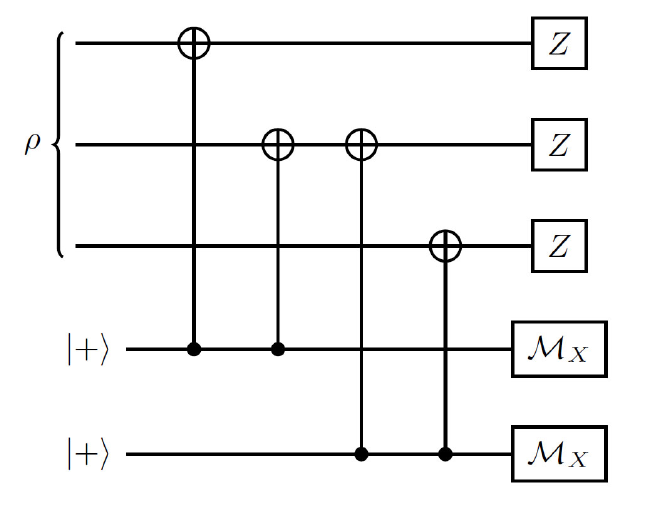}
\caption{Circuit for error correction of a phase error for a qubit encoded in $3$ spins. 
The error correction is achieved by measuring the syndromes $\{X_1X_2,X_2X_3\}$ followed by  $Z=\exp(-i\pi J_z)$ gate(s) according to the syndrome outcomes.
}
\label{fig:circuit_phase_error_correction}
\end{figure}

For the case of amplitude damping, one approach to diagnose the syndrome is to perform nondestructive measurements to identify the amplitude errors, for example, by measuring $J_z^2$.  In practice this can be difficult to implement experimentally. (In this section and below we return to denote a generic spin $J$, without reference to a specific platform.) Instead, we take advantage of the cat encoding and the unique properties of our proposed CNOT gate to coherently apply the recovery map using fresh ancilla without performing any measurement. Our construction is a new example of measurement-free {quantum} error correction (MFQEC) \cite{PhysRevLett.117.130503,PhysRevA.97.012318,li2011recovery,premakumar2020measurement} motivated by the experimental constraints of spin systems.

To describe our proposed error recovery, we first observe that we can ``swap" the state of two qubits encoded in the kitten states. Let 
\begin{equation}
    \begin{aligned}
        \ket{\psi}_k&=\alpha \ket{+}_k+\beta\ket{-}_k,\\
        \ket{\phi}_l&=\gamma \ket{+}_l+\delta\ket{-}_l,\\
    \end{aligned}
\end{equation}
where $\alpha,\beta, \gamma, \text{ and } \delta$ are arbitrary complex amplitudes.
Three applications of our proposed CNOT gate, as shown in Fig.~\ref{fig:Fig_swap_a}, implement the following transformation (see \cref{sec:measurement_free_ec} for a proof):
\begin{equation}
    \begin{aligned}
        \ket{\psi}_k& \otimes \ket{\phi}_l \rightarrow \ket{\phi}_k \otimes \ket{\psi}_l 
    \end{aligned}
\end{equation}

We expect this construction, which implements the swap of kitten states, to find applications beyond error correction, in particular in algorithmic subroutines native to qudits platforms, but in this work, we focus on its application in amplitude correction.


If we replace one of the input states with a fixed cat state, $\ket{+}_0$, then the recovery circuit can be simplified to the circuit in Fig.~\ref{fig:Fig_swap_b}.  
Therefore amplitude errors can be corrected by consuming fresh ancilla qudits in the cat state, $\ket{+}_0$, and applying two CNOT gates. The operation coherently transfers the qubit that is damped at level $k$ back to level $0$, which is our encoded qubit.
In \cref{sec:measurement_free_ec}, we show that the action of this quantum channel, after tracing the extra subsystem, is exactly equivalent to a recovery channel implemented by measuring $J_z^2$ and then applying a unitary correction to transfer the state into the {$k=0$} subspace.

\begin{figure}[!ht]
 \captionsetup[subfigure]{oneside,margin={3.2cm,0cm}}
\subfloat[]{\includegraphics[width =0.45\columnwidth]{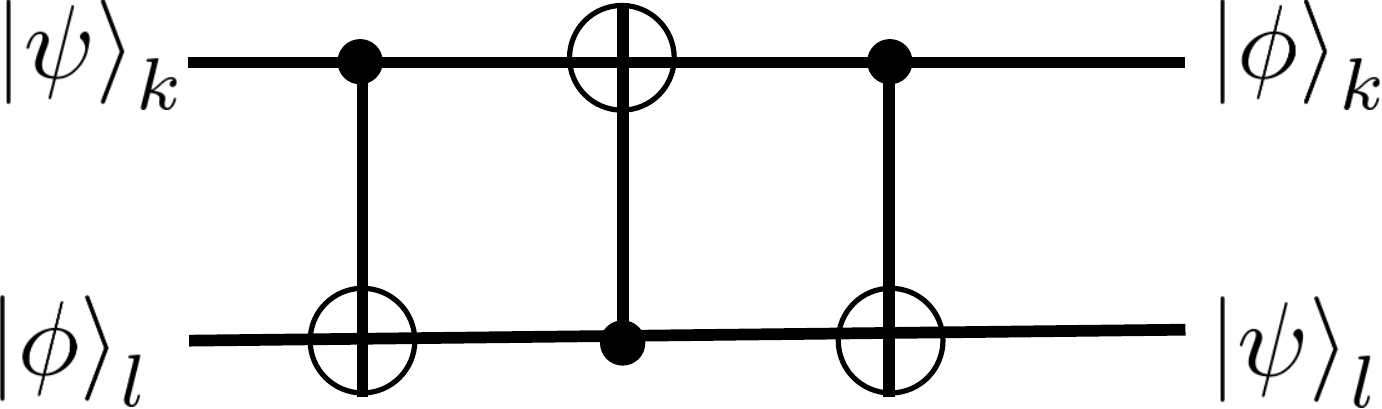} \label{fig:Fig_swap_a}} \hspace{0.5cm}
    \subfloat[]{\includegraphics[width=0.45\columnwidth]{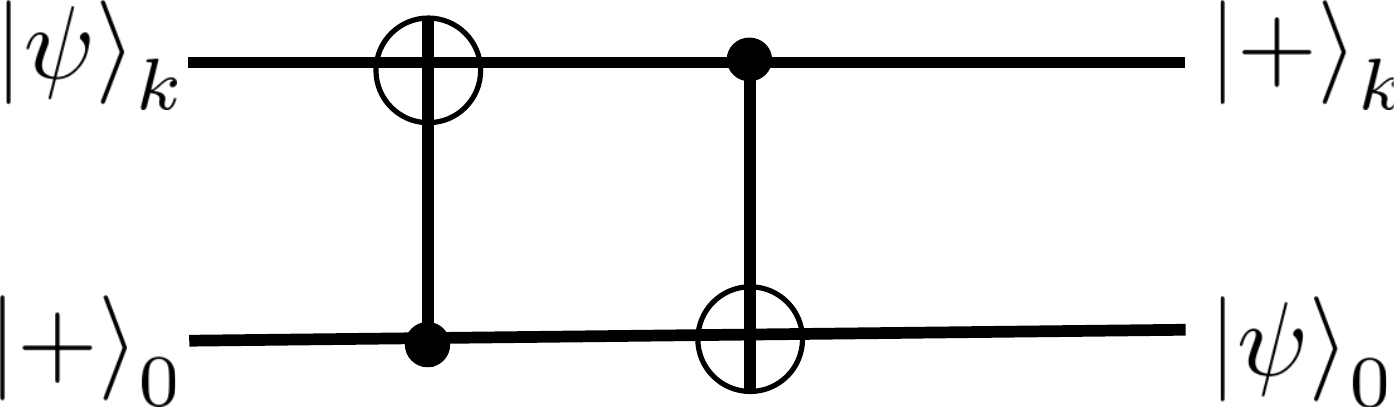}  \label{fig:Fig_swap_b}}
\caption{ (a) General circuit for swapping the state of the two qubits in two different kitten subspaces.
(b) The circuit that swaps the information between the data and ancilla, when the initial state of the ancilla state is $\ket{+}_0$.
}
\label{fig:SWAP_gate_qubit_1}
\end{figure}

\vspace{0.7cm}
{\subsection{Error correction for Optical Pumping}}
To see how phase and amplitude error correction combines to correct any local errors, it is illuminating to describe the procedure for correcting a dominant noise channel in atomic systems, optical pumping. (The details of optical pumping are discussed in \cref{sec:Ratio_optical_pumping_errors}).
In particular, consider the example of absorption of a linear $\pi$-polarized laser photon, followed by the spontaneous emission of a circularly polarized $\sigma_{+}$ photon. 
This process results in mapping   $\ket{J,J}$ to $\ket{J,J-1}$ and also annihilating any amplitude in the state  $\ket{J,-J}$. On the cat states, this transformation can be re-written as,
\begin{equation}
    \begin{aligned}
        \ket{+} &\to \ket{J,J-1}= \frac{\ket{+}_1-\ket{-}_1}{\sqrt{2}},\\
          \ket{-} &\to -\ket{J,J-1}= -\frac{\ket{+}_1-\ket{-}_1}{\sqrt{2}}.
    \end{aligned}
\end{equation}
Consider an arbitrary logical state $\ket{\psi}=\alpha \ket{+}_\mathrm{L}+\beta \ket{-}_\mathrm{L}$. The action of the optical pumping on the first qudit gives
\begin{equation}
    \ket{\psi}\to \frac{\ket{+}_1-\ket{-}_1}{\sqrt{2}}\otimes \left(\alpha\ket{+}_0\ket{+}_0-\beta\ket{-}_0\ket{-}_0\right)\equiv\ket{\phi}.
\end{equation}
Now we can consider the states after the phase and amplitude error correction steps. (As these error correction steps commute with each other,  the order in which we perform  them is irrelevant.)
The phase error correction is specified by the syndromes $X_1X_2$ and $X_2X_3$. If we measure both the syndromes as $+1$, the state $\ket{\phi}$ collapses to,
\begin{equation}
    \ket{\phi} \to \alpha \ket{+}_1\ket{+}_0\ket{+}_0+\beta \ket{-}_1\ket{-}_0\ket{-}_0.
\end{equation}
If the syndrome measurement gives outcome $-1$ and $1$ for the syndrome $X_1X_2$ and $X_2X_3$ the state becomes,
\begin{equation}
    \ket{\phi} \to \alpha \ket{-}_1\ket{+}_0\ket{+}_0+\beta \ket{+}_1\ket{-}_0\ket{-}_0.
\end{equation}
Applying the correction unitary $Z_1$ corresponding to this syndrome yields
\begin{equation}
    \alpha \ket{+}_1\ket{+}_0\ket{+}_0+\beta \ket{-}_1\ket{-}_0\ket{-}_0\equiv\ket{\phi}_{\mathrm{ph}}.
    \label{eq:state_after_phase_correction}
\end{equation}
The same state is achieved after performing the correction for the other two possible syndromes. Thus the state after the phase error correction collapses to the state \cref{eq:state_after_phase_correction}. 

Next, we can apply measurement-free amplitude error correction by consuming three ancilla states $\ket{+}_0$, which gives,
\begin{equation}
\begin{aligned}
    V_s^{\otimes^ 3}\ket{\phi}_{\mathrm{ph}}\ket{+}_0\ket{+}_0\ket{+}_0
    &=\ket{+}_1\ket{+}_0\ket{+}_0 \otimes \ket{\psi}
\end{aligned}    
\end{equation}
Tracing out the first three subsystems yields the initial state $\ket{\psi}$ in the three ancilla subsystems. 
The error correction scheme developed here thus corrects the optical pumping errors.

This quantum error correction gadget is especially well suited to the neutral atom platform due to the ability to move atoms mid-circuit. Firstly, the swap gates are easy to implement as we can move individual ancillas and data atoms into a pairwise configuration to apply the CNOT gates parallelly. Secondly, at the end of the protocol, the ancilla atoms can be used as the new data atoms by simply moving them into the right positions.

{\section{logical CNOT gate and Fault-tolerant threshold}
\label{sec:Fault_tolerance_and_threshold}}

\begin{figure}
    \centering
    \includegraphics[width=\columnwidth]{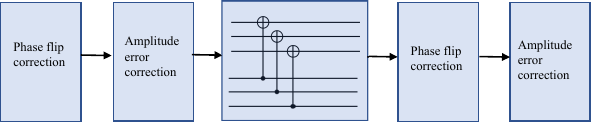}
    \caption{The error corrected {logical CNOT} gadget.
    The logical $\mathrm{CNOT}$ gate is implemented  by applying a physical CNOT gate between each qubit (encoded in the {spin}) of the control and target blocks transversely. 
    Error correction steps are performed before and after the logical CNOT. We apply a total of $r_1$ rounds of phase error correction and $r_2$ rounds of amplitude error correction.}
    \label{fig:fault_tolerant_cx_gadget}
\end{figure}

In this section, we provide lower bounds on the noise level that can be tolerated in our proposed spin-cat code, while still achieving fault-tolerant quantum computation. As discussed in \cref{sec:universal_gate_set}, to achieve fault-tolerance, we need to guarantee that the effective noise strength in our implementation of the logical gadgets of the inner code  $\mathcal{C}_1$, as specified by \cref{eq:logical_level_gates}, is below the noise threshold needed for the outer code $\mathcal{C}_2$ used in concatenation. 

In this concatenated scheme, the main source of error is the logical CNOT gate of $\mathcal{C}_1$, and hence, an upper bound on its failure probability will provide an estimate for the threshold of all $\mathcal{C}_2$ gadgets \cite{puri2020bias,Aliferis2008fault}. The logical CNOT gadget for the code $\mathcal{C}_1$ can be realized using transversal physical CNOT gates between two code blocks, accompanied by error correction procedures to correct phase and amplitude errors, which is illustrated in \cref{fig:fault_tolerant_cx_gadget}. For the sake of generality, we consider each logical CNOT gadget to consist of $r_1$ applications of phase error correction and $r_2$ applications of amplitude error correction. We define $r=r_1+r_2$ and denote the number of the data qudits in each code block by $n$. 
The recovery operation for phase error correction is determined by majority voting of the $r_1$ rounds of syndrome measurement.

We start by estimating the probability of dephasing errors. 
In this case, the analysis is similar to the analysis of biased cat qubits in bosonic systems \cite{puri2020bias}. 
Suppose each physical CNOT gate causes (independent) dephasing errors on the target and control qubits with probability $\epsilon$. 
During the application of {each} phase correction or amplitude correction procedures, every qudit is acted upon by at most two physical CNOTs.  
Hence, after $r_1$ repetitions of phase corrections and $r_2$ repetitions of amplitude corrections 
the probability of dephasing error on each qudit, in both the control and target block, will be at most $2r\epsilon$. 
After the implementation of error correction steps, the next step is to implement the transversal CNOTs between the control and target blocks of data qudits.
This operation can propagate phase errors from the target block to the control block. Therefore, after the action of the transversal CNOT gates,  the probability of dephasing error on each qubit of the target and control blocks is at most $2r\epsilon+\epsilon$ and $4r\epsilon+\epsilon$ respectively.

A logical error would occur if more than $(n+1)/2$ qubits are faulty in either the target or the control code blocks. 
Thus the upper bound on the logical phase error probability in the control and the target blocks can be  given as (keeping only the dominant term),
\begin{equation}
\begin{aligned}
\mathcal{\epsilon}_{\mathrm{target}}^{\mathrm{phase}}&\leq \binom{n}{\frac{n+1}{2}}(2r\epsilon+\epsilon)^{(n+1)/2},\\ \mathcal{\epsilon}_{\mathrm{control}}^{\mathrm{phase}}&\leq \binom{n}{\frac{n+1}{2}}(4r\epsilon+\epsilon)^{(n+1)/2}.
\end{aligned}
\end{equation}

To account for the possible errors in the syndrome measurements in the phase error correction step, we repeat measurements of $(n-1)$ syndromes in the control and the target blocks $r_1$  times and take the majority vote to apply error correction. 
A logical error happens if the syndrome is incorrect for at least $(r_1+1)/2$ rounds of this procedure. 
Each syndrome measurement requires two physical CNOT gates and we also need to account for state preparation and measurement errors used in each syndrome measurement, both of which can be performed with much higher accuracy compared to the rank-preserving CNOT gate. 
Also one needs to account for the dephasing error induced by the amplitude error correction following the phase error correction which has two physical CNOT gates.
Therefore the upper bound on the probability of a dephasing error in each syndrome bit is at most $6\epsilon$. 
As a result, the upper bound on the logical error for the syndrome measurement is given by (only keeping the dominant term):
\begin{equation}
    \mathcal{\epsilon}_{\mathrm{ec}}^{\mathrm{phase}}\leq 2(n-1)\binom{r_1}{\frac{r_1+1}{2}}(6\epsilon)^{\frac{r_1+1}{2}}.
    \label{eq:syndrome_error}
\end{equation}

Next, we establish an upper bound on the probability of logical errors resulting from amplitude errors on the control and target, just before the amplitude error correction step. An amplitude error on an individual qudit occurs when a minimum of $k_{\mathrm{max}} = (2J-1)/2$ jumps has taken place. 
This can be determined by summing the probabilities of $k_{\mathrm{max}}$ jumps, given a total of $s$ CNOT gates and is expressed as $q(s, k_{\mathrm{max}})$ as given in \cref{eq:amplitude_error_ideal} ($s = 2r$).
Following the error correction steps, the subsequent phase involves implementing transversal CNOT gates between the control and target blocks of data qudits.
This operation, however, has the potential to propagate amplitude errors from the control block to the target block. 
Consequently, after the application of transversal CNOT gates, the probability of amplitude errors on each qubit in the target and control blocks is bounded by
\begin{equation}
    \begin{aligned}
        \mathcal{\epsilon}_{\mathrm{target}}^{\mathrm{amp}}&\leq 2nq(s=2r,k_{\mathrm{max}})+nq(s=1,k_{\mathrm{max}}),\\ \mathcal{\epsilon}_{\mathrm{control}}^{\mathrm{amp}}&\leq nq(s=2r,k_{\mathrm{max}})+nq(s=1,k_{\mathrm{max}}).
    \end{aligned}
    \label{eq:blocks_amplitude_error}
\end{equation}

Next, we provide upper bounds on the probability of logical error in the amplitude error correction procedure.
An ideal implementation of the swap protocol described in \cref{sec:syndrome_extraction}
would correct the amplitude errors by putting back the state into the cat manifold, defined as the support of the projector $\Pi_0$, where 
\begin{equation}
\Pi_l=\ket{+}_l\bra{+}_l+\ket{-}_l\bra{-}_l.
\end{equation}
Imperfect amplitude error correction may arise due to factors such as small random rotations during the swapping process intended for error correction, errors caused by optical pumping, or imperfections in ancilla preparation.
For the case of small random rotation errors and optical pumping, the error operators involve at most two amplitude jumps as discussed in \cref{subsec:Error_characterization}.
Similarly, as discussed in \cref{subsubsec:Preperation_of_the_cat_state}, 
optical pumping and random rotation errors can create at most two amplitude jumps during the preparation of the ancilla state. 
Thus the imperfect amplitude error correction can cause at most four amplitude jumps. This phenomenon is {conceptualized} in \cref{fig:swapping_animation}, where
the population in the cat manifold can leak to $\Pi_i$ for $i=\{1,2,3,4\}$ manifolds with probabilities $p_i$.

Errors in the preparation of the ancilla can in principle result in a  superposition of $\ket{+}_k$ states with $k\leq 4$ instead of $\ket{+}_0$. 
However, the amplitude error correction destroys any coherence between the cat and kitten subspaces, resulting in a mixed state in the cat manifold (see \cref{sec:measurement_free_ec}).
Hence, to find an upper bound on the success probability of amplitude correction, we only need to consider the probability of error in preparing  $\ket{+}_k$ states with $k\leq 4$, rather than an arbitrary state in that subspace.

\begin{figure}
    \centering
    \includegraphics[width=0.8\columnwidth]{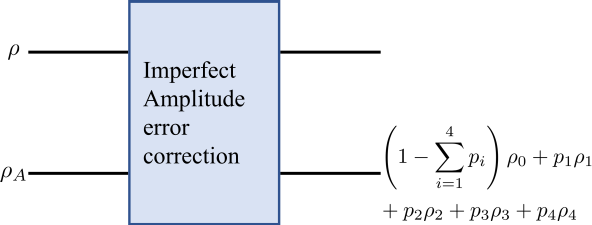}
    \caption{Imperfect amplitude error correction gadget. 
    There are two sources of imperfection one can associate with the amplitude error correction.
    The first one is a rotation error or optical pumping error occurring during the swapping approach to correct amplitude errors.
    The second one is due to imperfect preparation of the ancilla state, where ideally $\rho_A=\ket{+}_0$, however, in a non-ideal setting the ancilla can be in a mixture of $\ket{\pm}_i$ states where $i=\{0,1,2,3,4\}$, due to optical pumping or rotation error during the state preparation.
    For an ideal amplitude error correction, the final state lives in the $  \Pi_0=\ket{+}\bra{+}_0+\ket{-}_0\bra{-}_0$, whereas for a non-ideal setting, there is a small probability to be in other manifold $\Pi_l$. 
    The figure shows when the final state is in the $\Pi_i$ where $i=\{0,1,2,3,4\}$.}
    \label{fig:swapping_animation}
\end{figure}

We denote the failure probability of the amplitude error correction given that the ancilla states is in $\ket{+}_k$  by  $q(s,k_{\mathrm{max}}|k)$ where $s$ is the total number of CNOT gates before the application of error correction, and $k_{\mathrm{max}}$ is the minimum rank of the amplitude errors which create a logical error, i.e.  $k_{\mathrm{max}}=\lfloor (2J+1)/2\rfloor$ in our construction.
This probability can be calculated by adding the probabilities of cascades of single and two jumps that push the population from level $k$ to at least $k_{\mathrm{max}}$ level. 
Assuming the population only leaks to $\Pi_i$ for $i=\{1,2,3,4\}$ the logical error probability after $r_2$ rounds of amplitude error correction can be bounded by 
\begin{equation}
 \begin{aligned}
 \epsilon^{\mathrm{amp}} \leq r_2\left(\sum_{k=0}^4 q(s,k_{\mathrm{max}}|k)p_k\right),
    \end{aligned}
\end{equation}
where $p_0=1-\sum_{i=1}^4p_i$. 
(For a detailed calculation see \cref{sec:upper_bounds_amplitude}.)
As we have $2n$ total qudits, the logical error probability of the amplitude error correction 
blocks for the logical CNOT gate can be bounded by 
  \begin{equation}
 \begin{aligned} \mathcal{\epsilon}_{\mathrm{ec}}^{\mathrm{amp}}\leq 2n\epsilon^{\mathrm{amp}}.
    \end{aligned}
\end{equation}
Note that unlike phase error correction where the measurement is repeated $r_1$ many times and the correction is applied based on a majority vote of syndrome results, amplitude error correction does not involve direct measurement. Therefore repeated applications of amplitude error correction without a phase correction step in between do not provide extra error correction power.

Finally adding up all the probabilities of failures for the various components of the logical CNOT gate, yields an upper bound on its total logical error probability,
\begin{equation}
\begin{aligned}    \epsilon_{\mathrm{logical}}&\leq\mathcal{\epsilon}_{\mathrm{ec}}^{\mathrm{phase}}+\mathcal{\epsilon}_{\mathrm{control}}^{\mathrm{phase}}+\mathcal{\epsilon}_{\mathrm{target}}^{\mathrm{phase}}\\
&+\mathcal{\epsilon}^{\mathrm{amp}}_{\mathrm{ec}}+\mathcal{\epsilon}_{\mathrm{control}}^{\mathrm{amp}}+\mathcal{\epsilon}_{\mathrm{target}}^{\mathrm{amp}}.
\end{aligned}
    \label{eq:error_rate_total}
\end{equation}

To assess the improvement provided by our construction, we provide estimates of $\epsilon$ for various noise parameters that guarantee a logical error $ \epsilon_{\mathrm{logical}}$ below the threshold demanded by the CSS code   $\mathcal{C}_2$. For the CSS code $\mathcal{C}_2$ we use the fault-tolerant construction of \cite{Aliferis_css_code}, with a provable threshold of $\mathcal{\epsilon}_{\mathrm{CSS}}=0.67\times  10^{-3}$.

\begin{figure*}[!ht]
    \centering
     \captionsetup[subfigure]{oneside,margin={3.8cm,0cm}}
     \subfloat[]{\includegraphics[width =0.5\columnwidth]{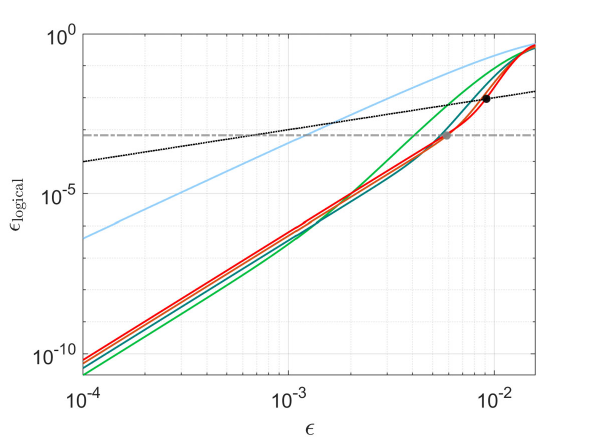} \label{fig:Fig_9_a}}
    \subfloat[]{\includegraphics[width =0.5\columnwidth]{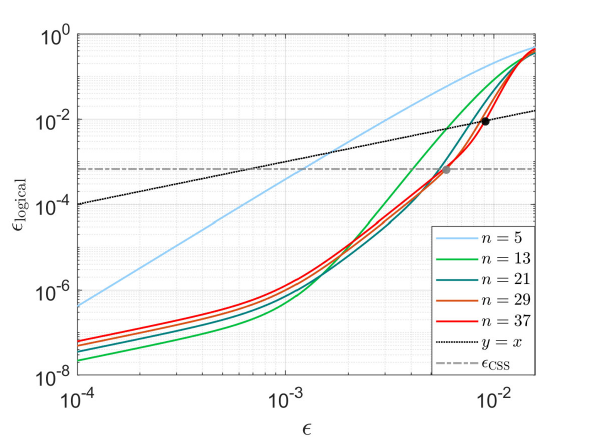}  \label{fig:Fig_9_b}}
    \caption{Logical error as a function of the physical level error (for details of the relation between phase error and amplitude error, see \cref{sec:Ratio_coherent_errors}) for the random rotation error for different value of $n$.
    Also, the threshold one needs to achieve CSS encoding in the second layer of concatenation is given for reference. 
   Figure (a) is for the case of $p_i=0$ for $i\neq 0$ and  figure (b) is for an imperfect ancilla state preparation with   $p_i=10^{-4}$ for $i\neq 0$. 
    We can see whether the swapping error ideal or non-ideal does not affect much except for very low noise and this in turn is because the contribution of the amplitude error is very low for the random rotation error.    
    The black circle shows the intersection of the logical error with $y=x$ line for the optimal case shown here and the gray circle shows the intersection of the $\mathcal{\epsilon}_{\mathrm{CSS}}$ with the logical error for the optimal case.
     The simulation is shown for $r_1=7$ and $r_2=1$.}
    \label{fig:coherent_error}
\end{figure*}

In \cref{fig:coherent_error} we present 
the case of the small rotation error for encoding a qubit in a qudit $J=9/2$ with $r_1=7$, $r_2=1$, and for different choices of $n$. 
The figure on the left assumes no leakage error in the ancilla state preparation, i.e.  $p_i=0$ for $i\neq 0$, and the figure on the right is for a leakage error of   $p_i=10^{-4}$ for $i\neq 0$.
As is evident in the figure, the logical error rates for scenarios with and without leakage error exhibit similar characteristics except for very low noise.
This is expected since for small rotation errors, the probability of amplitude error is exponentially suppressed as a function of $J$ compared to the phase errors, see \cref{fig:rotation_error_comparison} for more details. In particular, we find that for $n=21$, $r_1=7$, and $r_2=1$, the physical error $\mathcal{\epsilon}$ needed to achieve the targeted CSS threshold is less than $0.0054$.

\begin{figure*}[!ht]
    \centering
    \captionsetup[subfigure]{oneside,margin={3.8cm,0cm}}
    \subfloat[]{\includegraphics[width =0.5\columnwidth]{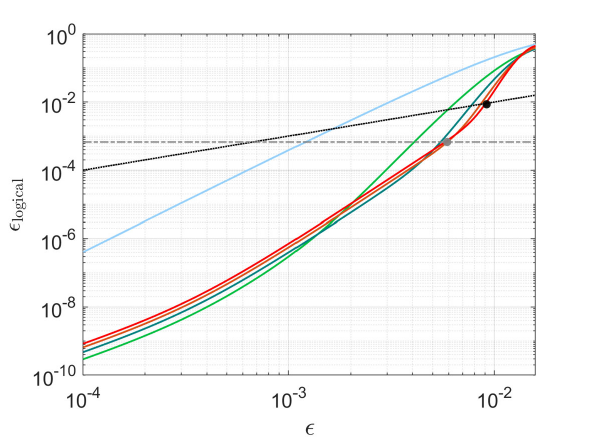} \label{fig:Fig_10_a}}
    \subfloat[]{\includegraphics[width =0.5\columnwidth]{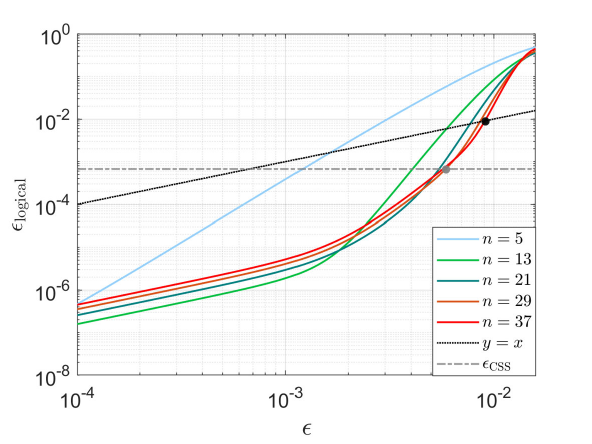}  \label{fig:Fig_10_b}}
       \caption{Logical error as a function of the physical level error (for details to the relation between phase error and amplitude error, see \cref{sec:Ratio_optical_pumping_errors}) for the optical pumping error for different value of $n$.
    The targeted threshold for the CSS code in the second layer of concatenation is given for reference. 
    Figure (a) is for the case of $p_i=0$ for $i\neq 0$ and  figure (b)  is for an imperfect ancilla state preparation with    $p_i=10^{-4}$ for $i\neq 0$. 
    We can see a significant change in the behavior depending on whether the amplitude error correction is ideal or not specifically in the low noise regime. 
    This in turn is due to the fact that for the case of optical pumping, as seen in \cref{sec:Ratio_optical_pumping_errors}, there is a significant contribution to the logical error from the amplitude errors.        
        The black circle shows the threshold value for the optimal value of $n$ and the gray circle shows the intersection of the $\mathcal{\epsilon}_{\mathrm{CSS}}$ with the logical error for the optimal value of $n$.
        The simulation are shown for $r_1=7$ and $r_2=1$.}
    \label{fig:optical_error}
\end{figure*}

Next, in \cref{fig:optical_error} we explore the impact of stronger photon scattering and optical pumping on the encoding of a qubit in a qudit with $J=9/2$ with $r_1=7$,  $r_2=1$, considering various choices of $n$. 
(For the case of $J=9/2$, we get $\alpha=0.0137$ and $\beta=0.2$ in \cref{eq:jump_opt_simplified} for stronger photon scattering and optical pumping.
Details of the noise model and parameters can be found in \cref{sec:Ratio_optical_pumping_errors}.) The left panel is the case with no leakage error $p_i=0$ for $i\neq 0$, while the right panel incorporates a leakage error with   $p_i=10^{-4}$ for $i\neq 0$.

As can be seen in the figure, for the ideal amplitude error correction the behavior of both the rotation error and case when photon scattering and optical pumping are stronger are very similar in nature.
However, when photon scattering and optical pumping are stronger, the imperfect ancilla preparation during amplitude error correction plays a more severe role in the overall logical error of the low noise regime. The competition between the error correction power of the gadget and the extra error due to the increased number of qudits needed to encode a logical qubit leads to identifying a ``sweet spot" that determines the optimal number of qudits needed to encode a logical qubit. In particular, we find that for $n=21$, $r_1=7$, and $r_2=1$, the physical error needed to achieve the targeted CSS threshold is $\mathcal{\epsilon} \leq  0.0053$.

As discussed in detail in \cref{sec:Ratio_coherent_errors}, the primary error source for the considered spin systems is the first-order angular momentum operators, stemming from potential unwanted magnetic fields.
Additionally, there are second-order terms in the angular momentum operators due to optical pumping \cite{deutsch2010quantum,omanakuttan2021quantum}. 
Despite this, the presence of extra levels in the qudit results in a logical error contribution from amplitude errors that is notably lower than that from phase errors. 
Thus the threshold behavior for both these error models only impacts the  low noise regimes.

{\section{Summary and Outlook} \label{sec:discussions_and_future_work}}

To achieve the full power of quantum computing, one needs to execute quantum algorithms on error-corrected logical qubits. However, meeting the demanding requirements for physical qubits and achieving low error rates, essential for error-corrected logical qubits, remains a significant challenge in current quantum implementations~\cite{knill2005quantum,raussendorf2007topological,svore2006noise,spedalieri2008latency}.
Recent advancements in noise-tailored error correction provide a promising avenue for achieving this by substantially alleviating the stringent demands of error-corrected logical qubits~\cite{Aliferis2008fault, zzPoulin, Grassl_erasure_1997_PRA,Wu_Puri_Thompson_2022_Nature_erasure, Sahay_Puri_biased_erasure_PRX_2023}.

In this chapter, we follow this direction and introduce a fault-tolerant quantum computation protocol by encoding a qubit into a spin system, with a spin larger than $J=1/2$.
The general scheme that we introduce in this work is applicable to a wide range of physical spins, including in semiconductors~\cite{Gross2021, gross2021hardware}, atomic ions~\cite{ringbauer2021universal, Low2020}, neutral atoms~\cite{omanakuttan2021quantum, Siva_Qudit_entangler_2023, zache2023fermion}, molecules~\cite{castro2021optimal}, and superconducting systems~\cite{ozguler2022numerical, Blok2021}, where we have spin qudits that can be coherently controlled and entangled.

The specific encoding we consider in this chapter is the spin-cat encoding which draws inspiration from the cat-code encoding for continuous variable bosonic systems~\cite{puri2020bias,Aliferis2008fault}.  For this implementation we develop techniques to perform reliable computation in the presence of dominant noise in spin systems, taking advantage of natively available interactions. One key factor that distinguishes the spin-cat encoding from the other encodings of a qubit in a qudit is that the total Hilbert space of the spin-cat encoding decomposes into a direct sum of qubit subspace.
This induces the structure of a stabilizer code, a feature that plays a pivotal role in enabling fault-tolerant schemes for error correction.

Spherical SU(2) tensor operators provide a basis in which to characterize the error channels and identify the set of correctable errors. The dominant error sources for encoding a qubit in a spin are the rank-1 SU(2) rotations and the rank-2 tensors which can arise, e.g., from optical pumping between magnetic sublevels.  Our codes are constructed with these physical errors in mind. We use the concatenation scheme of~\cite{Aliferis2008fault} to perform fault-tolerant computation. 
In addition to using an inner repetition code that corrects phase errors, we correct for amplitude-damping errors by consuming fresh ancilla spins and performing measurement-free error correction natively for spin systems.

As a concrete application of our proposed scheme, we focus on the encoding of a qubit in the nuclear spin of $^{87}$Sr, characterized by a spin of $9/2$. In this scenario, we systematically build a universal gate set for fault-tolerant quantum computing, leveraging the available interaction mechanisms.
A pivotal element in the formulation of the physical-level gate is the rank-preserving CNOT gate. 
We elaborate on the implementation details of this gate, by taking advantage of the metastable states available in $^{87}$Sr and the well-known Rydberg blockade. In addition to the swap gadget that helps us correct amplitude errors, this CNOT gate is used in the construction of a universal gate set.

We also studied the threshold for fault-tolerant error correction and found that it is much higher than found in standard protocols of error correction with physical qubits, and it is similar to the threshold observed in bosonic cat-codes \cite{puri2020bias}.
As a result, our approach demonstrates a significant reduction in the required overhead and exhibits higher fault tolerance thresholds compared to conventional qubit-based techniques.

Our work represents another example of designing resource-efficient fault-tolerant schemes by taking advantage of the native noise characteristics of a given hardware. In contrast to the earliest work in quantum error correction where models were constructed for hypothetical qubits and generic noise models, efforts are being made to develop error correcting codes that are symbiotic with the control methods and noise structures of physical quantum systems \cite{Gross2021,omanakuttan2023multispin,puri2020bias,puri2019stabilized,Cong_Lukin_QEC_Rydberg_PRX_2022,gross2021hardware}. A related direction of research is to engineer qubit encodings with favorable noise properties~\cite{puri2017engineering,Wu_Puri_Thompson_2022_Nature_erasure}. This has been made possible because of the substantial experimental advances in quantum computing~\cite{acharya2022suppressing,ryan2022implementing,krinner2022realizing}.

In a similar vein, the structure of our protocol works well with spin systems and their control methods, regardless of the platform in which they are implemented. It is particularly well-suited for the neutral atom platform, where significant experimental advances have been achieved recently~\cite{Bluvstein_Lukin_2023_QEC_Logical,bluvstein2022quantum,Saffman_Nature_2022}. We have previously explored the use of quantum optimal control of spin-$9/2$ nuclei in $^{87}$Sr atoms for arbitray single qudit gates~\cite{omanakuttan2021quantum} and two-qudit entangling gates~\cite{Siva_Qudit_entangler_2023}, where this protocol would be a natural fit. The unique capabilities of neutral atom platforms, such as reconfigurable connectivity and the ability to implement hundreds of parallel entangling gates~\cite{Bluvstein_Lukin_2023_QEC_Logical} would assist in the implementation of the fault-tolerant protocol we proposed here.

This work opens many directions for future research. One can extend the current protocol for the rank-preserving CNOT gate in neutral atoms to other, more experimental-friendly protocols.  Specifically, one can explore using the geometric phase approach \cite{Levine_Pichler_gate} or Rydberg dressing-based approaches \cite{mitra_martin_gate, schine2022long, martin2021molmer,anupampra}, typically used for entangling gates in qubits, to realize the rank-preserving CNOT gate.  
While we focus on errors caused by random rotations and optical pumping in this paper. Another very important source of errors we didn't consider is leakage out of computational subspace, especially in the form of atom loss in neutral atom platforms. The conventional approach to circumvent these errors is to use leakage reduction units \cite{Suchara_2015_leakage}. We plan to address this by converting these errors into erasure errors, which are easier to deal with \cite{Wu_Puri_Thompson_2022_Nature_erasure}, in the next chapter.
 \chapter{QND Cooling and leakage detection in neutral atoms}
\label{chap:Qudit_leakage}
\section{Introduction}

A dominant source of imperfection in the neutral atom platform is the weak potential that traps atoms.  
Because of this atomic motion will heat as atoms are transported, such as in architectures like~\cite{Bluvstein_Lukin_2023_QEC_Logical,bluvstein2022quantum}, and often atoms will get lost, be it through transport, collisions with background gas, or during gates. 
The heating of the atomic motion will generally degrade the performance of the system.
In trapped atomic ions, sympathetic cooling in shared vibrational motion with a distinct refrigerant atomic species is used to recool atoms after transport\cite{sympathetic_cooling_Wineland,sympathetic_cooling_Wineland_Monroe}. 
Such a direct mechanism is not possible for neutral atoms.  Moreover, atomic loss can lead to catastrophic ``leakage errors" if not appropriately managed. 
We study here a protocol to simultaneously tackle these issues.
Our goals are two-fold.
We seek to perform a quantum nondemolition (QND) measurement of the presence or absence of an atom that does not disturb the quantum information encoded therein. 
In doing so, leakage out of the computational subspace is converted to erasure which can substantially improve fault-tolerant thresholds~\cite{Wu_Puri_Thompson_2022_Nature_erasure}, when compared to traditional leakage reduction units~\cite{Suchara_2015_leakage}.
In addition, this measurement should not heat the atom, and more favorably, simultaneously cool atomic motion. 
This would greatly enhance quantum operation with neutral atoms.

To achieve these goals we revisit a protocol for cooling atoms without decohering quantum information in nuclear spins of 
alkaline earth atoms~\cite{Reichenbach_cooling}.  
Laser light used to cool and detect atoms directly couples only to the electrons, and indirectly to the nuclei, only through the hyperfine interaction.  
In the ground $\mathrm{^1S_0}$ state, there is no hyperfine interaction, thus highly isolating the quantum information encoded therein from the environment. 
By scattering photons from atoms in a way that avoids hyperfine coupling, we can both laser cool atoms and perform QND measurements to detect lost atoms without decoherence. 
Whereas previous work considered decoupling the electron angular momentum from the nuclear spin through the use of a large magnetic field, we consider here a more flexible approach based on large AC-Stark shifts, building on the work of~\cite{Shi_cooling_PRA_2023}. 
Our approach allows us to retain coherence across all magnetic sublevels in the nuclear spin with $I\ge 1/2$, thus compatible with new protocols that employ multiple levels for qudit gates~\cite{Siva_Qudit_entangler_2023} and error correction~\cite{omanakuttan2023multispin}.

The remainder of this chapter is organized as follows.
After establishing the necessary background in \cref{sec:background}, in \cref{sec:qnd_leakage_detection}, we propose and analyze the scheme that converts leakage errors to erasure errors for alkaline-earth atoms through Rayleigh scattering of photons.
In \cref{sec:qnd_cooling}, we augment the protocol to include resolved sideband cooling of the atoms while preserving nuclear spin coherence.
This scheme generalizes and explains in detail the proposed schemes in \cite{Reichenbach_cooling, Shi_cooling_PRA_2023}.
We conclude and explore possible future directions in \cref{sec:conclusions_and_future_work_qnd}.

\section{Background}
\label{sec:background}

\begin{figure}[!ht]
    \centering
    \includegraphics[width=0.9\columnwidth]{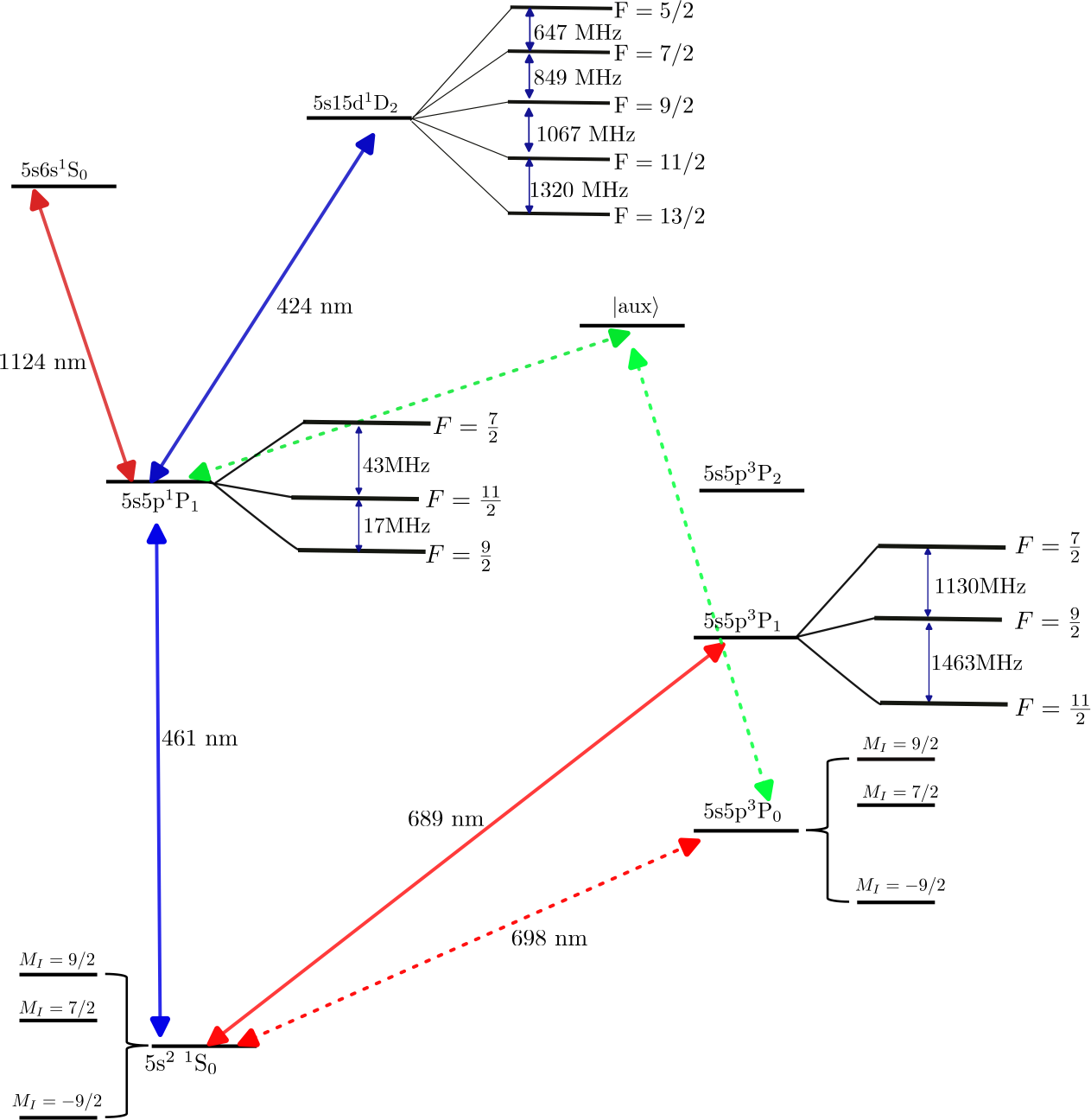}
    \caption{We encode quantum information in the  ground state of $^{87}$Sr, the singlet (5s$^2$ $^1\mathrm{S}_0$).
    One can encode any qudit with dimension $2\leq d\leq 10$ in the ground state.
In the schemes explored in this work, we leverage the rich structure of the excited states of $^{87}$Sr, as detailed in works such as \cite{katori2002spectroscopy,martin2013quantum}. These excited states can exist in either a spin-singlet or triplet configuration.
For the QND leakage detection scheme, we couple the ground state to the excited singlet-state 5s5p $^1\mathrm{P}_1$, which has a very small linewidth and to avoid the hyperfine coupling we work in a far-off resonance and cancel the residual tensor light shift by coupling the ground state to the excited triplet-state 5s5p $^3\mathrm{P}_1$.
For QND cooling we first transfer the state from  the ground state to the excited metastable state 5s5p $^3\mathrm{P}_0$ and then transfer the population to  the state 5s5p $^1\mathrm{P}_1$ using an intermediate state $\ket{\mathrm{aux}}$. 
To overcome the hyperfine coupling we use 
AC Stark shift generated by coupling the excited singlet state to $\mathrm{5s6s ^1S_0}$ and $\mathrm{5s15d ^1D_2}$.
}
    \label{fig:basic_outline}
\end{figure}

Our focus here is on imaging and sideband cooling without decohering the nuclear spin by avoiding the hyperfine coupling using the unique aspects of the alkaline-earth atoms.
In the schemes explored in this work, we leverage the structure of the excited states of $^{87}$Sr as given in \cref{fig:basic_outline}, used throughout this dissertation, and detailed in works such as \cite{katori2002spectroscopy,martin2013quantum}. 
For the QND leakage detection scheme, we couple the ground state to the excited singlet-state 5s5p $^1\mathrm{P}_1$, and to avoid the hyperfine coupling we work in a far-off resonance.
Moreover one can cancel the residual tensor light shift on the ground state that results from the off-resonance excitation by coupling it to the excited triplet-state 5s5p $^3\mathrm{P}_1$.
For QND cooling, we will first transfer the state from the ground state to the excited metastable state 5s5p $^3\mathrm{P}_0$ and then transfer the population to the state 5s5p $^1\mathrm{P}_1$ using an intermediate state $\ket{\mathrm{aux}}$. 
To overcome the hyperfine coupling in the 5s5p $^1\mathrm{P}_1$, we use 
AC Stark shift generated by coupling this state to higher excited states, $\mathrm{5s6s ^1S_0}$ and $\mathrm{5s15d ^1D_2}$.


\subsection{Decoherence free photon scattering}
\label{subsec:decoherence_free_photon_scattering}
\begin{figure}
    \centering
    \includegraphics[width=0.8\columnwidth]{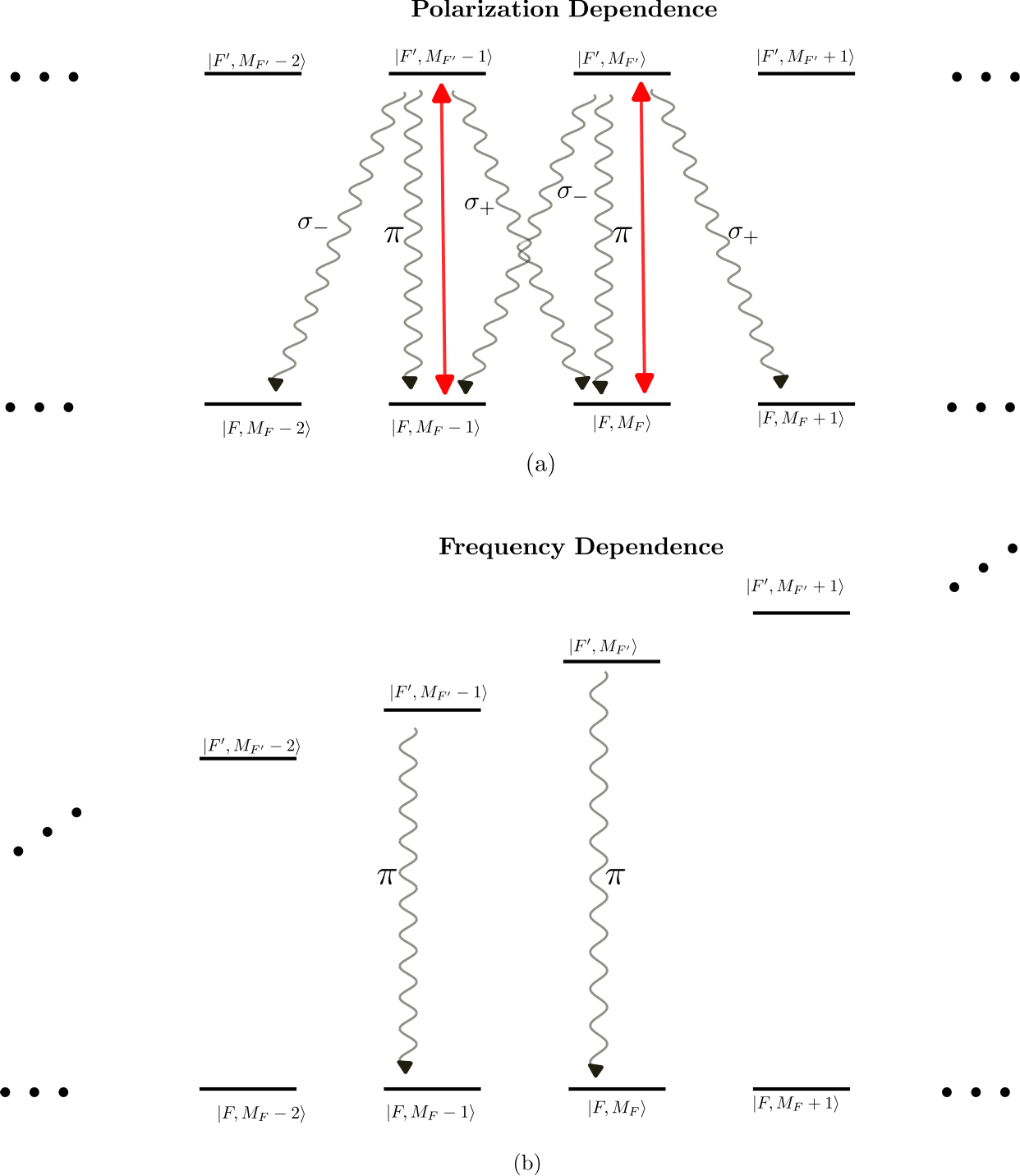}
    \caption{The figure illustrates the concept of photon scattering from an excited state where we encode quantum information in the ground state characterized by magnetic sublevels $M_F$ and total angular momentum $F$. In (a), we demonstrate how the polarization degree of freedom contains information about the specific magnetic sublevel, leading to decoherence. The scattered light can be polarized along $\pi, \sigma_{+}$, or $\sigma_{-}$. 
    Given that the electronic angular momentum accessible to the polarization degree of freedom is $J'=1$, the scattered light polarized along $\pi, \sigma_{+}, \sigma_{-}$ corresponds to electronic angular momentum $ M_F,M_F-1,M_F+1$, respectively. 
    To overcome polarization dependence for decoherence-free scattering, a single polarization degree with equal strength for all $M_F$ sublevels is required, essentially a scalar (constant) term.
In (b), we demonstrate how the frequency of the scattered light contains information about the magnetic sublevel. 
The frequency degrees provide information when the light scattered from each magnetic sublevel has a distinct color. The frequency dependence can arise from the presence of interaction detuning each magnetic sublevel differently.
Thus to achieve decoherence-free scattering, it is necessary to ensure that the different magnetic sublevels are detuned much less compared to the linewidth.
}
    \label{fig:general_theory}
\end{figure}

The requirements for decoherence-free photon scattering for alkaline-earth atoms, which is crucial for achieving QND leakage detection and cooling, can be understood as follows.
Consider the ground state of an atom characterized by total angular momentum $F$, with the quantum information encoded in the magnetic sublevels $M_F$.
Subsequently, a laser interaction is employed to excite the population to an excited state characterized by total angular momenta $F'$ (where multiple angular momenta may exist in the excited state) and magnetic sublevels $M_{F'}$.
As the excited state possesses a finite lifetime, the population will decay from the excited state to the ground state. 
The light emitted during this process can carry information about the magnetic sublevel in which we have encoded the quantum information, resulting in the transfer of population or the loss of coherence. 
The scattered light exhibits two pertinent degrees of freedom: polarization and frequency, as illustrated in \cref{fig:general_theory}.

The polarization degree will have information about the state and leads to optical pumping as given in \cref{fig:general_theory} (a). 
The scattered light can be  $\pi,\sigma_{+}$ or $\sigma_{-}$ polarized relative to the quantization axis. 
Since the electronic angular momentum, which is the degree of freedom to which the polarization degree of freedom has access, $J'=1$, the emission of $\pi,\sigma_{+},\sigma_{-}$ indicates the angular momentum changes to  $ M_F,M_F-1,M_F+1$ respectively.
Thus to overcome the polarization dependence for decoherence-free scattering,  one can only have a single polarization degree with equal strength for all the $M_F$ sublevels.




\begin{figure}
    \centering    
    \includegraphics[width=0.6\columnwidth]{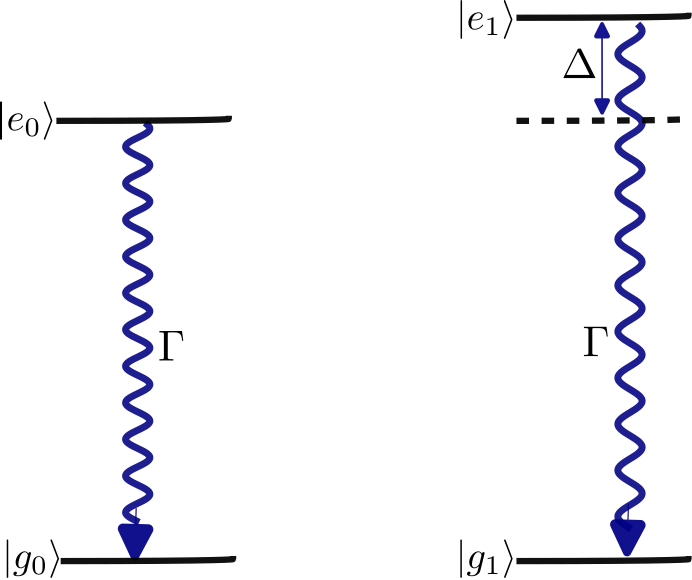}
    \caption{The figure gives the setting of two two-level systems separated by a small detuning $\Delta$ compared to the linewidth $\Gamma$.
}
    \label{fig:two_levels_a}
\end{figure}

The frequency of the emitted photon will have information about the state when the light emitted in spontaneous emission from different magnetic sublevels have different colors as given in \cref{fig:general_theory} (b).
One can consider the frequency dependence originating from the presence of interaction which shifts each of the magnetic sublevels separately such that the color of the scattered light from each of the magnetic sublevels is distinguishable compared to the natural linewidth of the excited state.

To understand this consider the case of two two-level systems separated with a small detuning $\Delta$ compared to the decay rate $\Gamma$.
The key is to analyze the loss of coherence when the states decay from the excited states to the ground state. 
The setting of the problem is given in \cref{fig:two_levels_a} and we have the Hamiltonian of interest given as,
\begin{equation}
    H=-\Delta \ketbra{e_1}{e_1}
\end{equation}
which accounts for the detuning difference between the two excited states.
The jump operators which take into account of the transfer of population from the excited to the ground state is given as,
\begin{equation}
L=\sqrt{\Gamma}\left(\ket{g}_0\bra{e}_0+\ket{g}_1\bra{e}_1\right),
\label{eq:jump_two_levels_a}
\end{equation}
and the  evolution of the system is given by the Lindblad master equation,
\begin{equation}
    \frac{d\rho}{dt}=-i\comm{H}{\rho}+L\rho L^{\dagger}-\frac{1}{2}\left(L^{\dagger} L\rho+\rho L^{\dagger}  L \right).
    \label{eq:Lindblad_master_equation}
\end{equation}
To identify the loss of coherence, consider, 
\begin{equation}
    \begin{aligned}
        \rho_{0,1}^{(g)}&\equiv\bra{g_0} \rho \ket{g_1},\\
        \rho_{0,1}^{(e)}&\equiv\bra{e_0} \rho \ket{e_1}.
    \end{aligned}
\end{equation}
Thus we get,
\begin{equation}
\frac{d\rho_{0,1}^{(e) }}{dt}=-(i\Delta+\Gamma) \rho_{0,1}^{(e)},
\end{equation}
which in turn gives,
\begin{equation}
\rho_{0,1}^{(e)}(t)=\rho_{0,1}^{(e)}(0)e^{-(i\Delta+\Gamma)t}.
\end{equation}
Similarly, one can find that
\begin{equation}
\begin{aligned}
   \frac{d\rho_{0,1}^{(g) }}{dt}&=\Gamma\rho_{0,1}^{(e)}, \\
   \frac{d\rho_{0,1}^{(g) }}{dt}&=\Gamma\rho_{0,1}^{(e)}(0)e^{-(i\Delta+\Gamma)t}. 
\end{aligned}
\end{equation}
Solving the above differential equation,
\begin{equation}
\rho_{0,1}^{(g)}(t)=\frac{\Gamma}{\Gamma-i\Delta}\left(1-e^{-(i\Delta-\Gamma)t}\right)\rho_{0,1}^{(e)}(0),
\end{equation}
which in limit of $\Gamma \gg \Delta$ is,
\begin{equation}
    \rho_{0,1}^{(g)}(t) \approx \left(1-e^{-(i\Delta-\Gamma)t}\right)\rho_{0,1}^{(e)}(0).
\end{equation}
To understand the effect consider an initial state,
\begin{equation}
    \ket{\psi}_0=\sqrt{\frac{1}{2}}\left(\ket{e_0}+\ket{e_1}\right),
\end{equation}
and consider the evolution for a total time $\Gamma T=10$ and for an ideal transfer of population one expects to get the state,
\begin{equation}
    \ket{\psi}_f=\sqrt{\frac{1}{2}}\left(\ket{g_0}+\ket{g_1}\right),
    \label{eq:initial_state_app_1}
\end{equation}
and we can calculate the fidelity under the evolution by the \cref{eq:Lindblad_master_equation} given as,
\begin{equation}
    \mathcal{F}=\bra{\psi}_f\rho(t)\ket{\psi}_f.
    \label{eq:fidelity_two_levels_a}
\end{equation}
\begin{figure}
    \centering    
    \includegraphics[width=0.6\columnwidth]{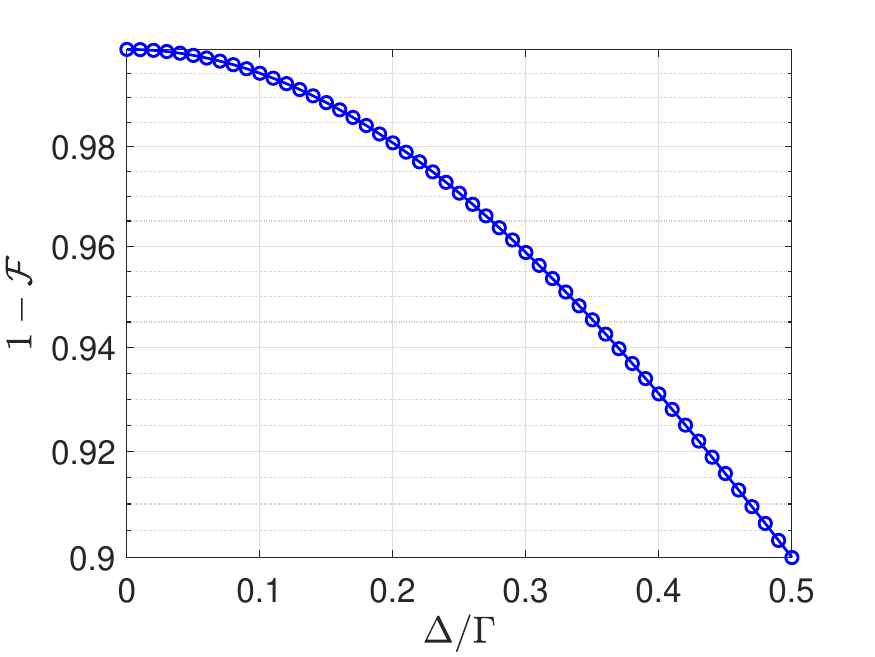}
    \caption{The figure gives the infidelity of the final state as a function of $\Delta/\Gamma$ for the setting given in \cref{fig:two_levels_a} and  \cref{eq:fidelity_two_levels_a}. 
    As we go to a regime in which $\Delta/\Gamma\to 0$, the infidelity goes to $0$, and hence the coherence is preserved in the decay.
    Thus the ratio of the $\Delta/\Gamma$ is the key parameter that determines the loss of coherence for a setting in which the excited states have different detuning.
}
    \label{fig:two_levels_fidelity_a}
\end{figure}
In \cref{fig:two_levels_fidelity_a}, we give the infidelity as a function of $\Delta/\Gamma$ and it is evident that we have good fidelity as  $\Delta/\Gamma \to 0$.
Thus to overcome the frequency dependence for decoherence-free interaction of atoms and photons, if there is no negligible population in the excited state, one needs to ensure that the different magnetic sublevels are shifted much less than the linewidth.

\section{QND leakage detection}
\label{sec:qnd_leakage_detection}

\begin{figure}
    \centering    \includegraphics[width=0.7\columnwidth]{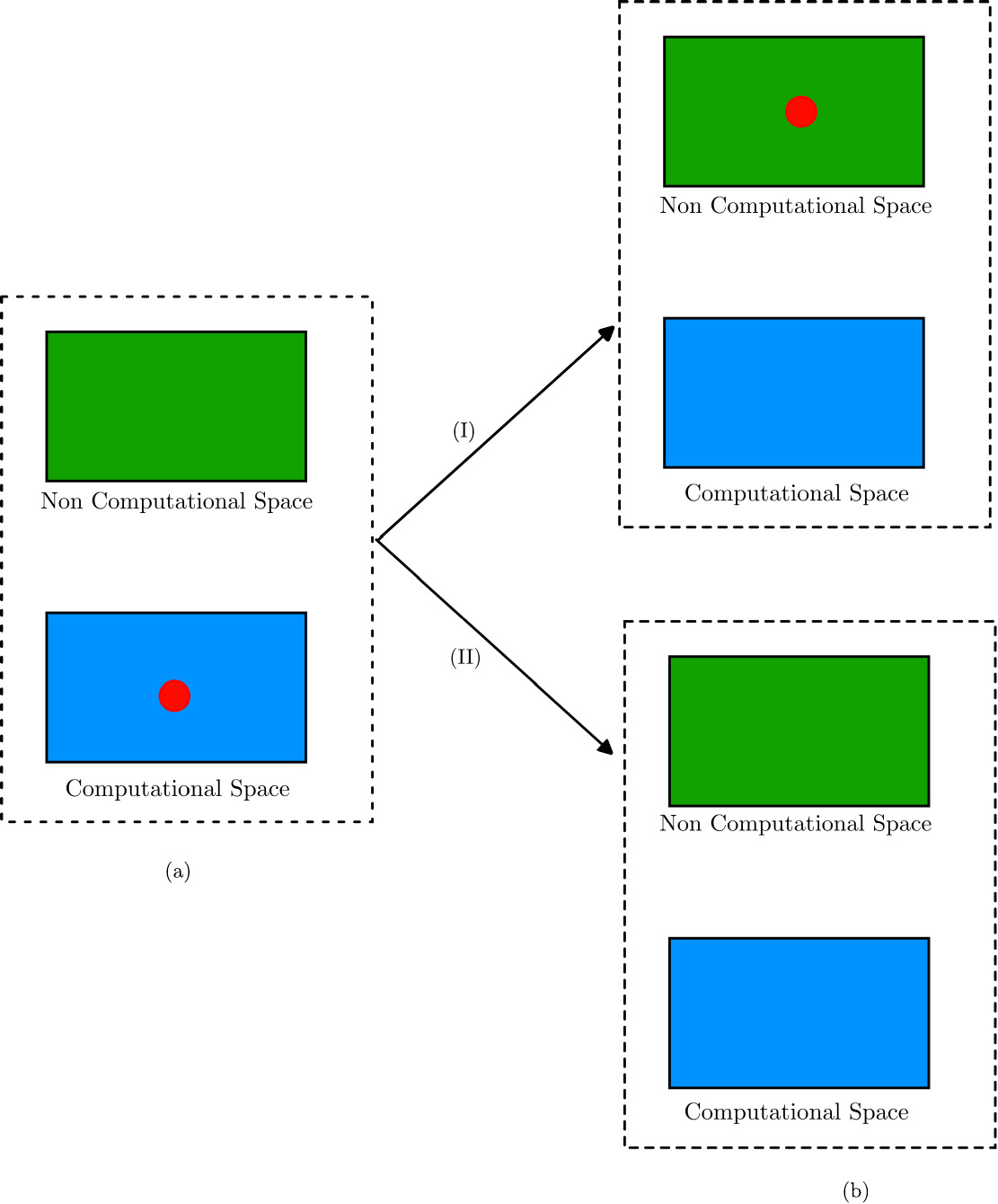}
    \caption{The figure gives the basic idea of leakage error in quantum computation. 
    The leakage error is referred to generically as the errors that take the quantum information outside the computational space of interest.
    In (a), we show a system where we encode our quantum information.
    There are two ways in which leakage errors could occur in neutral atoms as shown in (b).
    (I) the quantum information can sometimes be stuck in the non-computational space like Rydberg states or other metastable states during the computation.
    (II) The quantum information can be completely lost from the atoms, this could in turn be due to the atom losing out of the trap.      
}
\label{fig:leakage_error}
\end{figure}
Leakage errors are errors that take probability amplitude outside the computational subspace as shown schematically in \cref{fig:leakage_error}. 
Population can leak or transfer to metastable states, ionization of atoms from Rydberg states, atom loss, etc.
The traditional way to deal with these errors is so-called ``leakage reduction units''(LRU) \cite{Aliferis_Terhal_leakage_2007,Suchara_2015_leakage} which can be very costly. 
Recently, methods have been developed for alkaline earth atomic systems that convert leakage errors into erasure errors~\cite{Wu_Puri_Thompson_2022_Nature_erasure,Ma_Thompson_Yb_erasure_gate,Scholl_Endres_2023_erasure_simulation} (erasure errors are errors whose locations are known, and these errors can be corrected more easily) during the entangling gate.
These methods only work in specific, well-designed protocols for leakage detection, but are not completely general. 
For other errors like atom loss, which becomes a very important error source for long-depth fault-tolerant measurements with multiple rounds of atom rearrangements, we still need to rely on LRU's which are not resource-friendly. 
In this chapter, we develop schemes that convert leakage errors originating from any possible sources such as entangling interaction or atom loss to erasure errors.

\begin{figure}
    \centering    \includegraphics[width=0.9\columnwidth]{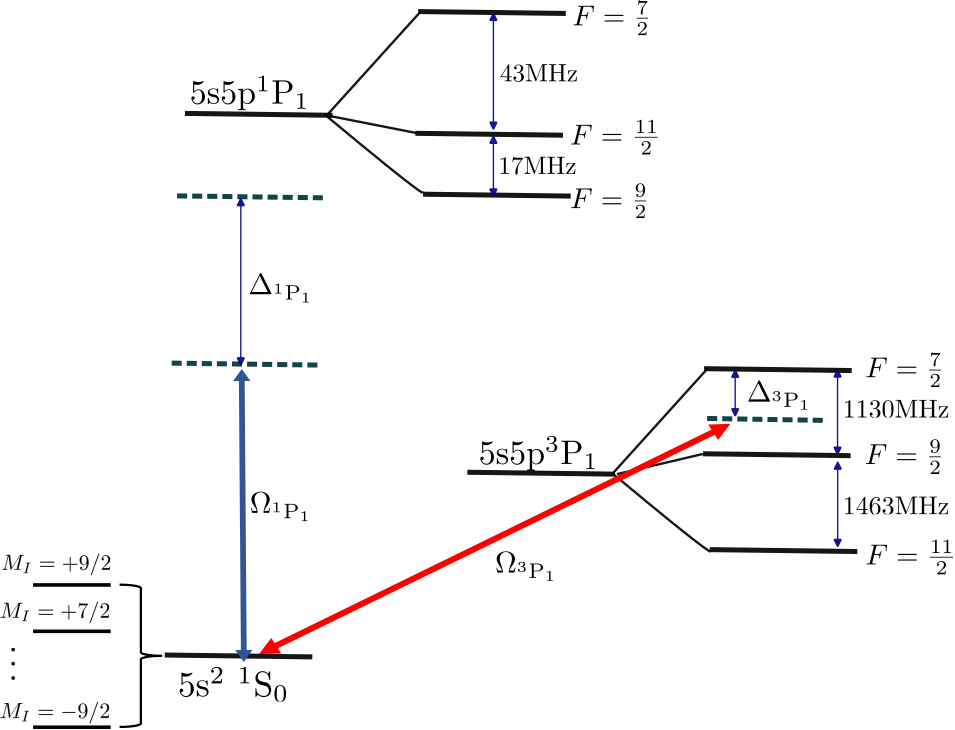}
    \caption{The figure illustrates the setup for detecting the loss of information in the state encoded in the ground state of Sr. We utilize far-detuned light from the singlet P state ($\mathrm{5s5p^1P_1}$). To counteract the tensor-light shift interaction from the singlet state, we employ a tensor-light shift interaction by coupling the ground state to the $\mathrm{5s5p^3P_1}$ state. For details on this tensor-light shift interaction with minimal decoherence, refer to \cite{omanakuttan2021quantum}.
A crucial aspect of the scheme is the small hyperfine splitting in the state $\mathrm{5s5p^1P_1}$, allowing us to identify operational regimes where schemes can be devised such that the scattered light from this state contains no information about the state encoded in the ground state $\mathrm{5s^2\text{ }^1S_0}$.
}
    \label{fig:basic_outline_qnd_leakage}
\end{figure}

Figure~(\ref{fig:basic_outline_qnd_leakage}) gives the setting for detecting the loss of information for the state encoded in the ground state of  $^{87}$Sr without destroying the coherence. 
Consider  a state in the computational subspace, 
\begin{equation}
\ket{\psi}=\sum_{M_I=-\frac{9}{2}}^{\frac{9}{2}}\alpha_{M_I}\ket{\mathrm{5s^2 \text{ }^1S_0}, \text{ }M_I},
\end{equation}
where  $\sum_{M_I} \lvert \alpha_{M_I} \rvert^2=1$. 
The goal of the QND leakage detection scheme is to measure whether the atom is in this computational subspace or not without destroying the quantum information in the state.

The key idea is to scatter photons without inducing optical pumping either flips the spins or has any information about the magnetic sublevels.
First consider the coupling between the $\mathrm{5s^2\text{ }^1S_0}$ and $\mathrm{5s5p \text{ } ^1P_1}$ with  a $\pi$-polarized light. 
One needs to understand whether there is any regime in which the scattered photons contain negligible ``which-way'' information about the nuclear spin sublevels. 
For this one can look at the Lindblad Master equation for a general state $\rho$ of the nuclear spin in the ground state ($\mathrm{5s^2\text{ }^1S_0}$) given as \cite{deutsch2010quantum},
\begin{equation}
\begin{aligned}
\frac{d\rho}{dt}&=-i\left[H_{\mathrm{LS}},\rho\right]+ \Gamma_{\mathrm{^1P_1}}\sum_q W_q^{\mathrm{^1P_1}}\rho \left(W_q^{\mathrm{^1P_1}}\right)^{\dagger}-\frac{1}{2} \{\left(W_q^{\mathrm{^1P_1}}\right)^{\dagger}W_q^{\mathrm{^1P_1}},\rho\},
\end{aligned}
\end{equation}
where
\begin{equation}
H_{\mathrm{LS}}=\Omega_\mathrm{^1P_1}^2\sum_{F'}\frac{1}{4\Delta_{FF'}(\mathrm{^1P_1})} C^{(2)}_{J',F',F} F_z^2 ,
\end{equation}
where, $H_{\mathrm{LS}}$ is the tensor light shift  Hamiltonian.
The coefficient $C^{(2)}$ characterizes the rank-2 irreducible polarizability~\cite{deutsch2010quantum}.
Optical pumping is described by the jump operators \cite{deutsch2010quantum},
\begin{equation}
\begin{aligned}
W_q^{\mathrm{^1P_1}}&=\sum_{F'}\frac{\Omega_{\mathrm{^1P_1}}/2}{\Delta_{FF'}({\mathrm{^1P_1}})+i\Gamma_{\mathrm{^1P_1}}/2} \left[ C^{(0)}_{J'FF'}\left(\bm{e}_q^{*}.\vec{\epsilon}_L\right) \mathds{1}+iC^{(1)}_{J'FF'}\left(\bm{e}_q^{*}\cross\vec{\epsilon}_L\right).\bm{F} \right.\\
&\left.+C^{(2)}_{J'FF'}\left(\frac{\left(\bm{e}_q^{*}.\bm{F}\right)\left(\vec{\epsilon}_L.\bm{F}\right)+\mathrm{h.c}}{2}\right)-\frac{1}{3}\lvert \bm{e}_q^{*}.\vec{\epsilon}_L\rvert^2\bm{F}^2\right],
 \end{aligned}
\end{equation}
where $\Gamma_{\mathrm{^1P_1}}$ is the characteristic linewidth of the excited state, $\vec{\epsilon}_L$ is the polarization of the laser, and $q={-1,0,1}$  represent the polarization of the scattered light.
Working in a far-off resonance regime where the detuning is much bigger than the Hyperfine splitting ($\Delta_{\mathrm{^1P_1}} \gg \delta_{F'}$) and defining $\Delta_{FF'}^{}(\mathrm{^1P_1})=\Delta_{\mathrm{^1P_1}}+\delta_{F'}(\mathrm{^1P_1})$, to lowest order,
\begin{equation}
    H_{\mathrm{LS}} \approx \frac{\Omega_\mathrm{^1P_1}^2}{4 \Delta_\mathrm{^1P_1}} \left(\mathds{1}-\frac{1}{\Delta_\mathrm{^1P_1}}\left[\beta^{(2)}F_z^2\right]\right), 
\end{equation}
where we have used the fact that for the coupling between the electronic angular momentum's $J=0$ and $J'=1$:
\begin{equation}
\begin{aligned}
\sum_{F'} C^{(2)}_{J'FF'}=0,\\
\end{aligned}
\label{eq:sum_of_coefficients}
\end{equation}
and for convenience of notation we have defined,
\begin{equation}
\begin{aligned}
\beta^{(i)}=\sum_{F'}C^{(i)}_{J',F',F} \delta_{F'}(\mathrm{^1P_1}).
\end{aligned}
\end{equation}
For a $\pi$ polarized light keeping the  jump operators  up to $\mathcal{O}(1/\Delta^2)$ we get,
\begin{equation}
\begin{aligned}
W_0&\approx \frac{\Omega_{\mathrm{^1P_1}}}{2\Delta_{\mathrm{^1P_1}}}
\mathds{1}+\frac{\Omega_{\mathrm{^1P_1}}}{2\Delta_{\mathrm{^1P_1}}^2}\left(\beta^{(0)}+i\gamma^{(0)} \right)\mathds{1}+\frac{\Omega_{\mathrm{^1P_1}}}{2\Delta_{\mathrm{^1P_1}}^2}\left(\beta^{(2)}+i\gamma^{(2)} \right)F_z^2,\\
W_{+}&\approx \frac{\Omega_{\mathrm{^1P_1}}}{2\Delta_{\mathrm{^1P_1}}^2} \left(i\left(\beta^{(1)}+i\gamma^{(1)} \right)F_{-}+ \left(\beta^{(2)}+i\gamma^{(2)} \right)\left[\frac{F_zF_{-}+F_{-}F_z}{2}\right]\right),\\
W_{-}&\approx \frac{\Omega_{\mathrm{^1P_1}}}{2\Delta_{\mathrm{^1P_1}}^2} \left(i\left(\beta^{(1)}+i\gamma^{(1)} \right)F_{+}+\left(\beta^{(2)}+i\gamma^{(2)} \right)\left[\frac{F_zF_{+}+F_{+}F_z}{2}\right]\right),
\end{aligned}
\end{equation}
where,
\begin{equation}
\gamma^{(i)}=\frac{\Gamma}{2}\sum_{F'} C^{(i)}_{J',F',F}.
\label{eq:gamma_equation}
\end{equation} 
We highlight a few facts. 
The dominant effect is the scalar term in $W_0$, which corresponds to Rayleigh scattering, and which does not involve any couplings to the magnetic sublevels.  
The remain correction terms (vector and tensor) are due to residual hyperfine coupling but fall off rapidly for large detuning.
Unlike for alkali elements \cite{deutsch2010quantum}, $W_{\pm}$  goes as $1/\Delta_{\mathrm{^1P_1}}^2$  for the alkaline-earth elements in the far-off resonance regime. 
This arises from the fact that for alkaline-earth elements there is no coupling of the electronic and nuclear degrees of freedom in the ground state and thus the dependence come only from hyperfine interaction in the excited states.
Retaining terms up to $\mathcal{O}(1/\Delta_{\mathrm{^1P_1}}^3)$ we find
\begin{equation}
\begin{aligned}
    \frac{d\rho}{dt}&=- i\frac{\Omega_\mathrm{^1P_1}^2}{4\Delta_\mathrm{^1P_1}^2}\left( \beta^{(2)}+\frac{\Gamma_\mathrm{^1P_1}}{\Delta_\mathrm{^1P_1}}\gamma^{(2)}\right)\left(F_z^2\rho-\rho F_z^2\right)+\mathcal{O}\left(\frac{1}{\Delta_{\mathrm{^1P_1}}^4}\right),\\
    &=i\frac{\Omega_\mathrm{^1P_1}^2}{\Delta_\mathrm{^1P_1}^2} \beta^{(2)} \left(F_z^2\rho-\rho F_z^2\right)+\mathcal{O}\left(\frac{1}{\Delta_{\mathrm{^1P_1}}^4}\right),
\end{aligned}    
\end{equation}
where we have used the fact that $\gamma^{(2)}=0$.
Thus the only contribution up to  $\mathcal{O}(1/\Delta_{\mathrm{^1P_1}}^3)$ arises from the coherent light shift evolution which goes like $1/\Delta_{\mathrm{^1P_1}}^2$.
Thus in the limit of far-off resonance,  optical pumping is absent and hence the polarization degree of freedom does not have any information about the magnetic sublevels.

To complete the protocol we must cancel the residual light shift in the ground state without decohering the nuclear spin.
To achieve this we can employ a second laser field and use the large tensor light-shift term when coupling to the $\mathrm{5s5p \text{ }^3P_1}$
manifold, as studied in detail in \cite{omanakuttan2021quantum}. 
We find the appropriate  $\Omega_{\mathrm{^3P_1}}$ that cancels the residual light shift by  detuning about halfway between the hyperfine splitting of $F'=7/2$ and $F'=9/2$ ( $\Delta_{\mathrm{^3P_1}}=635$MHz).
This is the optimal choice as for this detuing the ratio of tensor light to decoherece is minimum \cite{omanakuttan2021quantum}.

\begin{figure}[!ht]
    \centering    \includegraphics[width=0.6\columnwidth]{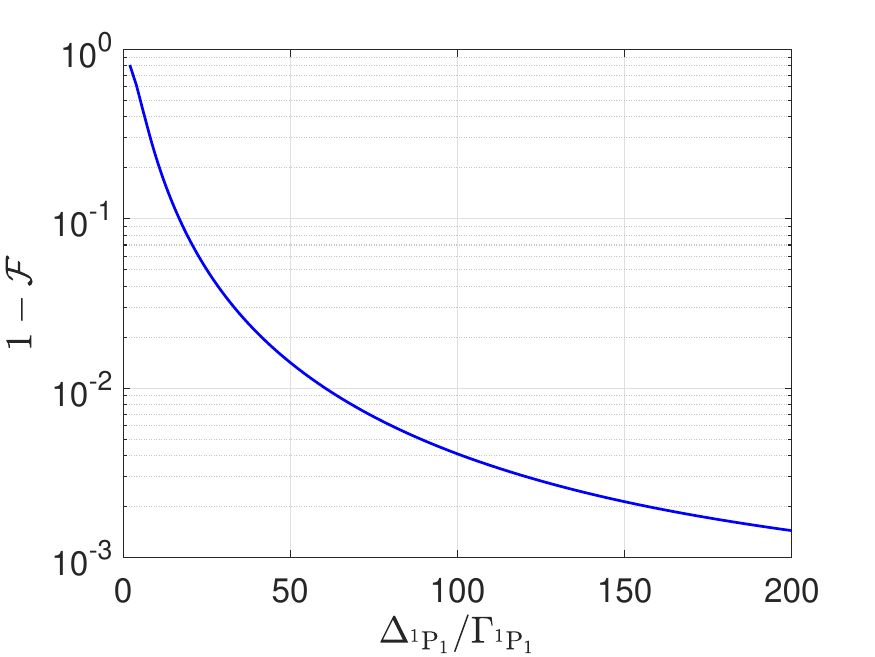}
    \caption{The figure shows the simulation of infidelity as a function of detuning from the singlet state for a time required for scattering $100$ photons for the setting given in \cref{fig:basic_outline_qnd_leakage}.
Lower infidelity indicates  a more effective  QND scheme for leakage detection.
Moving further away from resonance enhances the scheme's effectiveness, approaching an ideal scenario for QND leakage detection.
}
    \label{fig:qnd_leakage}
\end{figure}

To understand the performance of the scheme in \cref{fig:basic_outline_qnd_leakage}, consider the following  initial state, an equal superposition of all magnetic sublevels in the $\mathrm{^1S_0}$ ground state 
\begin{equation}
\ket{\psi}=\frac{1}{\sqrt{10}}\sum_{M_I=-\frac{9}{2}}^{\frac{9}{2}} \ket{\mathrm{5s^2} \text{ } \mathrm{^1S_0}, M_I}.
\label{eq:initial_state_leakage}
\end{equation}
The system evolves according to the master equation,
\begin{equation}
\begin{aligned}
    \frac{d\rho}{dt}&=\Gamma_{\mathrm{^1P_1}}\sum_q W_q^{\mathrm{^1P_1}}\rho \left(W_q^{\mathrm{^1P_1}}\right)^{\dagger}-\frac{1}{2} \{\left(W_q^{\mathrm{^1P_1}}\right)^{\dagger}W_q^{\mathrm{^1P_1}},\rho\}\\
    &+\Gamma_{\mathrm{^3P_1}}\sum_q W_q^{\mathrm{^3P_1}}\rho \left(W_q^{\mathrm{^3P_1}}\right)^{\dagger}-\frac{1}{2} \{\left(W_q^{\mathrm{^3P_1}}\right)^{\dagger}W_q^{\mathrm{^3P_1}},\rho\},
\end{aligned}
    \end{equation}
where we have canceled the tensor light shifts and  $W_q^{\mathrm{^1P_1/^3P_1}}$ are the jump operators for the singlet and triplet $\mathrm{P_1}$ states respectively.
The fidelity of the final state is given as,
\begin{equation}
\mathcal{F}=\bra{\psi}\rho\ket{\psi}.
\label{eq:fidelity_leakage}
\end{equation}
In \cref{fig:qnd_leakage}, we studied the fidelity of the state in \cref{eq:initial_state_leakage} after the time required for Rayleigh scattering $100$ photons, off resonantly, from the $\mathrm{^1P_1}$ state.
From the numerics, one can infer that as we increase$\Delta_{\mathrm{^1P_1}}$  one can recover the ideal fidelity, and for sufficiently large detunings give us a QND leakage detection scheme.

As the QND leakage detection scheme introduced in this work depends on the population decaying from the excited state, the scheme could heat up the atom.
However, one can use the separation of the nuclear and electronic degrees of freedom in the singlet-P state to use ideas of resolved sideband to cool the atoms without decohering the quantum information stored \cite{Reichenbach_cooling,Shi_cooling_PRA_2023}.
In the next section, we study a scheme in which one can cool the information stored in the ground state for any qudit with $d\leq 10$.

\section{QND Cooling}
\label{sec:qnd_cooling}

For quantum information processing with neutral atoms, the unavoidable influence of heating of atomic vibrational necessitates the re-cooling of atoms for arbitrary long quantum computations.
However, standard laser cooling methods \cite{saffman2010quantum,Saffman_review_2016_Rydberg} will inadvertently destroy the quantum information stored in the atomic internal state making them unsuitable for this purpose. 
In this section, we propose a scheme to cool the vibrational motion of the atom without decohering the information encoded in the nuclear spin of the ground state of $^{87}$Sr.

For qudits encoded in the nuclear-spin states, one can achieve resolved-sideband cooling using the special properties of these atoms. 
To preserve the quantum state encoded in a nuclear spin during laser cooling, it is crucial to transfer spin coherences in both excitation and spontaneous decay processes. 
Optical fields interact with atoms through the electric dipole, directly coupling to electrons and influencing nuclear spin states indirectly through the hyperfine interaction. 
A fundamental requirement is thus to excite states with minimal hyperfine coupling and/or to decouple electrons from nuclear spin.
This approach was first introduced in \cite{Reichenbach_cooling}, where one can separate the nuclear and electronic degrees of freedom in the excited singlet state using a large magnetic field. 
However, the required field are very large, making this approach less practical.
In recent work \cite{Shi_cooling_PRA_2023}, a novel approach was introduced to overcome the need of large magnetic field by using laser interaction to create a large AC-Stark shift that dominates over the coupling of  the electronic and nuclear degrees of freedom in the excited states.

The key idea for QND resolved sideband cooling  is carrying out a cooling cycle without decohering information encoded in the nuclear spin.
To achieve this one need to  ensure that there is no information in the spontaneously emitted photon that can give “which way information” about the initial nuclear spin magnetic sublevel $\ket{M_I}$.
Similar to \cite{Shi_cooling_PRA_2023}, we use AC-stark shift to remove the “which way information” of the nuclear spin states in the scattered light. 
In a generalization to the previous work, the scheme work for nuclear spin qudits. 
This capability of cooling quantum information while preserving coherence can further enhance the prospects of qudit based quantum information using $^{87}$Sr where we encode any qudit with $2\leq d\leq 10$.

\subsection{Resolved-Sideband Cooling in $^{87}$ Sr}
\begin{figure} 
    \centering
    \includegraphics[width=\columnwidth]{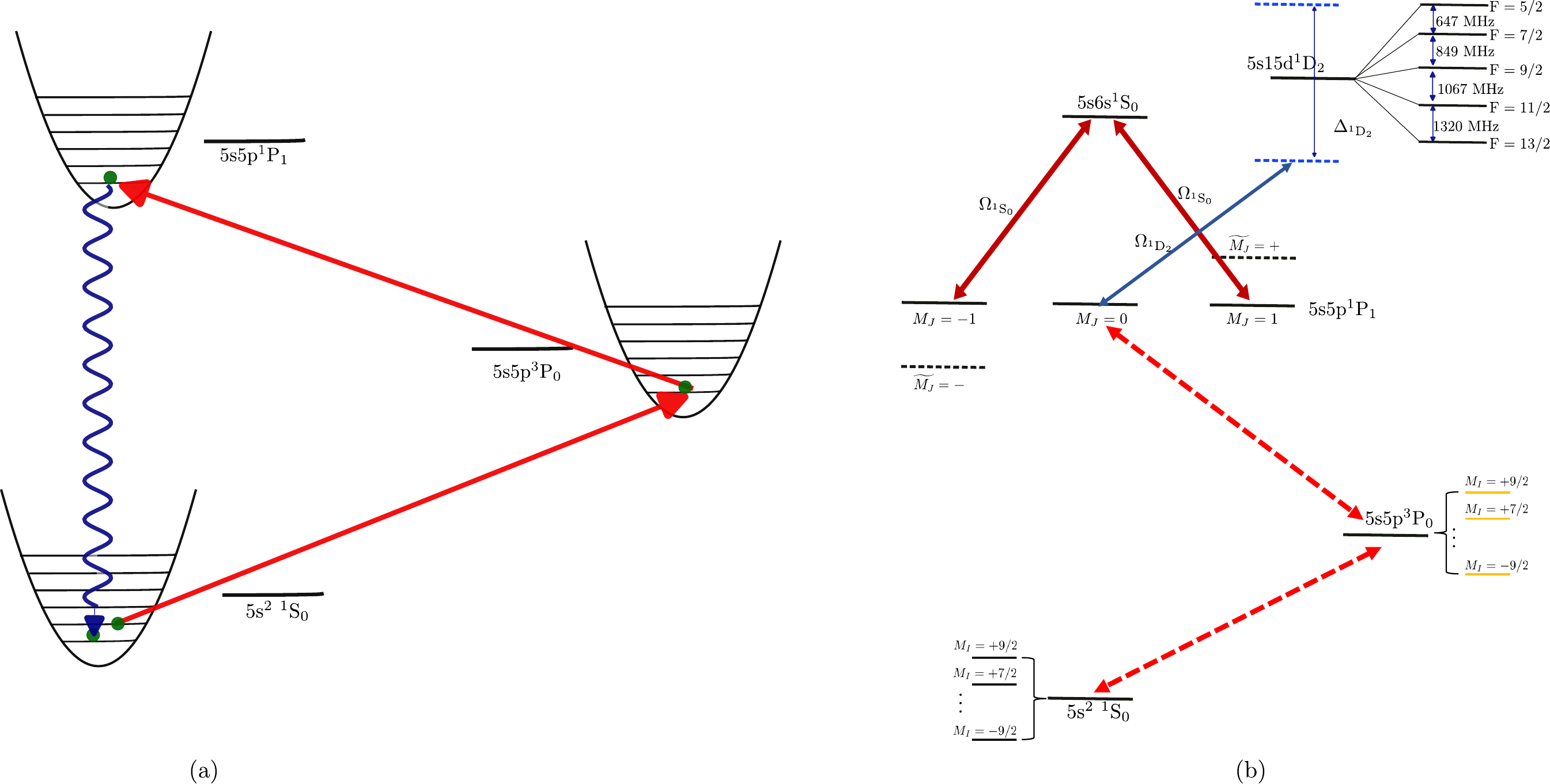}
    \caption{ The figure gives the basic idea of resolved sideband cooling employed for alkaline earth atoms and the specific setting considered in this work.
    (a) shows the resolved sideband cooling, first one excites the population in the ground state using a red sideband transition which lowers the vibrational quantum quantum number. 
    In the next step we transfer the population to a state with a very large linewidth such that the population decays back to the ground state and in this one cyle the vibrational quantum number is reduced by one unit (refer to the text for additional details).
    (b) shows the key ingredient which allows us to overcome the hyperfine splitting interaction in the excited state $\mathrm{5s5p ^1P_1}$. 
    Unlike the previous work \cite{Shi_cooling_PRA_2023}, we use AC stark shift to isolate the $M_J=0$ state in this state, to achieve this we couple the $\mathrm{5s5p ^1P_1}$ to the excited singlet state $\mathrm{5s6s ^1S_0}$ using a light polarized  along the $x$-axis.
    Further to avoid the frequency dependence on the scattered light from the excited state $\mathrm{5s5p ^1P_1}$ we couple the state to the excited D state $\mathrm{5s15d^1D_2}$ with a $\pi$ polarized light far-off resonance. 
     Further details of the results of this coupling are provided in the main text.}
    \label{fig:cooling_setup}
\end{figure}
The resolved-sideband cooling for $^{87}$ Sr follows three steps, as in   previous works \cite{Reichenbach_cooling,Shi_cooling_PRA_2023}, shown in \cref{fig:basic_outline}. 
In the first step using a $\pi$-pulse we coherently excite 
\begin{equation}
    \begin{aligned}
        &\ket{\mathrm{5s^2\text{ } ^1S_0}, M_I}\otimes \ket{n} \\
        \to& \ket{\mathrm{5s5p\text{} ^3P_0},M_F=M_I}\otimes \ket{n-1},
    \end{aligned}
\end{equation}
on the first red sideband, where $n$ is vibrational quantum number. 
In the next step, using a two-photon transition we coherently transfer
\begin{equation}
    \begin{aligned}
        &\ket{\mathrm{5s5p\text{} ^3P_0},M_F=M_I} \otimes \ket{n-1} \\ \to &\ket{\mathrm{5s5p\text{} ^1P_1},M_J=0,M_F=M_I} \otimes \ket{n-1}.
    \end{aligned}
\end{equation}
Assuming a sufficiently tight trap in the Lamb-Dicke regime, in the last step, the short lifetime  of the $\mathrm{5s5p\text{ } ^1P_1}$ leads to the rapid decay
\begin{equation}
\begin{aligned}
    &\ket{\mathrm{5s5p\text{} ^1P_1},M_J=0,M_F=M_I} \otimes \ket{n-1}\\
    \to &\ket{\mathrm{5s^2\text{ } ^1S_0}, M_I}\otimes \ket{n-1},
\end{aligned}    
\end{equation}
thus returning the population back to the ground state but with a decreased vibrational quantum number.
The schematic of the cooling cycle is given in \cref{fig:basic_outline}

In order to achieve perfect resolved sideband cooling in alkaline earth atoms, there are two crucial aspects that require attention: polarization and frequency dependence in the cooling. 
In the following two subsections, we introduce methods to overcome these which arise from the hyperfine interaction present in the excited singlet state $\mathrm{5s5p ^1P_1}$.

\subsection{Overcoming the Polarization dependence}
The polarization dependence of the scattered light and its effect on the nuclear spin state arises from the mixing of the nuclear and electronic spin degrees of freedom in the excited singlet state $\mathrm{5s5p ^1P_1}$.
As the  electronic angular momentum in this state is $J=1$, this occurs via the  hyperfine interaction,
\begin{equation}
H_{\mathrm{hf}}=A (\hat{I}.\hat{J})+Q \frac{3 (\hat{I}.\hat{J})^2+3/2 \hat{I}.\hat{J}-I(I+1)J(J+1)}{2I J(2I-1)(2J-1) },
\label{eq:Hyperfine_Hamiltonian}
\end{equation}
 with dipolar coupling $A/h=-3.4~\mathrm{MHz}$ and the quadrupolar coupling $Q/h=39~\mathrm{MHz}$.
The good quantum number in the state $\mathrm{5s5p^{1}P_1}$ is $F=J+I$ and $M_F$. 
Hence when we transfer the states $\mathrm{5s5p^{3}P_0}$ to $\mathrm{5s5p^{1}P_1}$ with a two-photon $\pi$-polarized light, a single $M_I$ state in the ground state couples to multiple $M_I$ values with different total $M_J$ value in the excited state.
This presence of different $M_J$ values allows for the spontaneously emitted photon to have different polarizations which in turn leads to optical pumping and nuclear spin decoherence.
The key is then to ensure that when we transfer the population from $\mathrm{5s5p^{3}P_0}$ to $\mathrm{5s5p^{1}P_1}$ with a two-photon $\pi$ polarized light, only a single $M_J$ state is allowed.

To achieve this we use a similar scheme to  \cite{Shi_cooling_PRA_2023} and introduce a strong AC-Stark shift to decouple the electronic and nuclear degrees of freedom.
However in \cite{Shi_cooling_PRA_2023}, a $\pi$-polarized light is used and for this case, one can only decouple the electronic and nuclear degrees of freedom for a qubit which is encoded in $\ket{0}=\ket{M_I=-9/2}$ and $\ket{1}=\ket{M_I=-7/2}$.
Here we generalize this scheme to any encoding of the nuclear spin states, which can be used for qudit-based quantum computing or building error-correcting codes as discussed in previous chapters.

To achieve this we consider a resonance coupling between states labelled $a=\mathrm{5s5p^{1}P_1}$ and $b=\mathrm{5s6s^1S_0}$. 
For a  light polarized along $x$-direction, and the interaction Hamiltonian is
\begin{equation}
    \begin{aligned}
        H_{\mathrm{LS}}=&\frac{\Omega_{\mathrm{^1S_0}}}{2\sqrt{2}}\left(\ket{a,M_J=-1}\bra{b,M_J=0}\right.\\
        &\left.-\ket{a,M_J=1}\bra{b,M_J=0}+\mathrm{h.c}\right).
    \end{aligned}
    \label{eq:light_shift_Hamiltonian}
\end{equation}
This leads to an Autler-Townes splitting,  the states $M_J=\pm 1$ are light shifted by $\pm \Omega_{\mathrm{^1S_0}}/2\sqrt{2}$.
For sufficiently large values of Rabi-frequency $\Omega_{\mathrm{^1S_0}}$ the different $M_J$ states are separated as shown in \cref{fig:cooling_setup}b and in this regime, one can solely access the $M_J=0$ state without transferring population to $M_J=\pm 1$.

\begin{figure}
    \centering
    \includegraphics[width=\columnwidth]{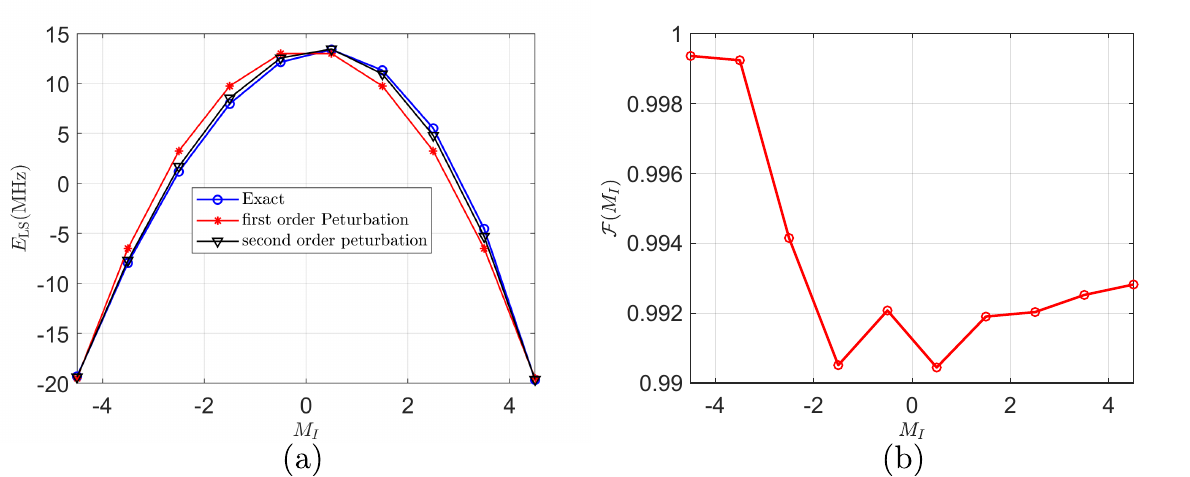}
    \caption{The figure shows the details of the impact of using the interaction between $\mathrm{5s5p^1P_1}$ and $\mathrm{5s6s^1S_0}$.
    This interaction shifts the $M_J=0$ of the $^1P_1$ and the results are shown for  $\Omega_{\mathrm{P}}=1000$ MHz.
    (a) shows the eigenvalues of the eigenstates $M_J=0, M_I$, which are the states of interest, and compares the results with the perturbation theory analysis.
    (b) gives the overlap of the eigenvectors with the states $\ket{M_J=0,M_I}$, the near $1$ overlap indicates that the $M_J=0$ state is isolated via the coupling between $\mathrm{5s5p^1P_1}$ and $\mathrm{5s6s^1S_0}$.
    We need to ensure that there is no information in the spontaneously emitted photon that can give “which way information” about the nuclear spin state.
    Thus, even though we decouple the nuclear spin from the electron with the large AC Stark shift, we still need to cancel the residual detuning to have a QND cooling scheme.}
    \label{fig:cooling_2}
\end{figure}

Consider the regime of large $\Omega_{\mathrm{^1S_0}}$.
Then the hyperfine interaction acts as a perturbation with the unperturbed ground state given as, 
\begin{equation}
    \{\ket{a,M_J=0,M_I},\ket{\widetilde{M}_J=+,M_I},\ket{\widetilde{M}_J=-,M_I}\}.
\end{equation}
where we have the dressed states,
\begin{equation}
\begin{aligned}
    &\ket{\widetilde{M}_J=+}=\frac{1}{\sqrt{2}}\left(\ket{a,M_J=1}-\ket{b,M_J=0}\right),\\
     &\ket{\widetilde{M}_J=-}=\frac{1}{\sqrt{2}}\left(\ket{a,M_J=-1}-\ket{b,M_J=0}\right).
\end{aligned}    
\end{equation}
We now include the hyperfine interaction. 
Focusing on the states $\ket{a,M_J=0,M_I}$, one can use perturbation theory to find out the energy shift on these states to first and second order as,
    \begin{equation}
 \begin{aligned}
 \delta E_{M_I}^{(1)} &=Q'\left[3\left(I(I+1)-M_I^2\right)+I(I+1)J(J+1)\right],\\
 \delta E_{M_I}^{(2)}&=-\frac{2\sqrt{2}}{\Omega_{\mathrm{^1S_0}}} \sum_{M_{I'}} \lvert \bra{M_{I'},M_J=1} H_{\mathrm{hf}}\ket{M_I,M_J=0}\rvert^2 \\
 &+\frac{2\sqrt{2}}{\Omega_{\mathrm{^1S_0}}} \sum_{M_{I'}} \lvert \bra{M_{I'},M_J=-1} H_{\mathrm{hf}}\ket{M_I,M_J=0}\rvert^2,
    \end{aligned}
\end{equation}
where,
\begin{equation}
    Q'=\frac{Q}{2I J(2I-1)(2J-1)}.
\end{equation} 
The first-order shift $\delta E^{(1)}$ arises from the quadrupolar hyperfine terms and leads to a quadratic light shift.  The second-order perturbation, $\delta E^{(2)}$ depends on the strength of the hyperfine coupling compared to the Autler-Townes splitting and can be made negligible with sufficient laser power.
In \cref{fig:cooling_2}, we show the energy shift and overlap of the exact eigenvectors with the perturbative approximation;  we choose here $\Omega_{\mathrm{^1S_0}}=1$ GHz, achievable with an experimentally reasonable intensity.
In \cref{fig:cooling_2}(a), we compare the energy shift to the one obtained from first-order and second-order perturbation theory for state $\ket{M_J=0,M_I}$.
The perturbation theory explains the energy shift and is dominated by quadratic shift.
In \cref{fig:cooling_2}(b), we show an analysis of the eigenvectors and plotted the fidelity,
\begin{equation}
    \mathcal{F}(M_I)=\lvert \bra{n(\Omega_{\mathrm{S},M_I})}\ket{M_J=0,M_I}\rvert^2,
\end{equation}
where $\ket{n(\Omega_{\mathrm{S},M_I})}$ is the exact eigenstates for a Rabi-frequency $\Omega_{\mathrm{^1S_0}}$ and detuning $\Delta_{\mathrm{^1S_0}}$ for the Hamiltonian given in $H=H_{\mathrm{hf}}+H_{\mathrm{LS}}$.
Thus, for on-resonance driving with  $\Omega_{\mathrm{^1S_0}}=1$GHz, the eigenvectors are well approximated by the product state, with a little admixture of $M_J =\pm1$, $\ket{M_J=0,M_I}$.
Hence, the good quantum numbers are $M_J,M_I$ rather than $M_F$ and the polarization degree of freedom of the scattered light does not have any information about the nuclear spin state $\ket{M_I}$.

However, the large energy shift gives a frequency dependence of the spontaneously emitted photon giving a  “which way information” about the nuclear spin state.
 Thus, even though we decouple the nuclear spin from the electron with the large AC Stark shift, we still need to cancel the residual hyperfine interaction. 
 In the next section, we use an additional tensor light shift to overcome this effect.

\vspace{0.7cm}
\subsection{Overcoming the Frequency dependence}
To overcome the residual energy shift, we apply an off-resonant tensor light shift by coupling the  $\mathrm{5s5p^1P_1}$ to the excited state  $\mathrm{5s15d^1D_2}\equiv c$. 
The light shift, with $\pi$ polarized light for state $\ket{M_J=0,M_I}$, is given as,
\begin{equation}
    V_{\mathrm{LS}}^{(ac)}= \sum_{F',M_{F'}}\frac{\Omega_{ac}^2}{4\Delta_{F'}({\mathrm{^1D_2}})}\left\lvert\langle{c,F',M_{F'}}\lvert d_z\ket{a,M_J=0,M_I}\right\rvert^2,
    \label{eq:light_shift_Hamiltonian_d}
\end{equation}
where $\Delta_{F'}({\mathrm{^1D_2}})=\Delta_{\mathrm{^1D_2}}-\left[E_{F'}(c)-E_{M_J=0}(a)\right]=\Delta_{\mathrm{^1D_2}}+\delta_{F'}({\mathrm{^1D_2}})$ and $\Omega_{ac}$ is the Rabi frequency between the $a$ and $c$.

Thus the light-shift interaction involves the coupling of states in $a$ where the good quantum numbers are the uncoupled basis $\ket{M_J,M_I}$ and for $c$ where the good quantum number is the coupled basis, $\ket{M_F}$.
However to find an electric dipole matrix element we need to work in either one of these basis for which one can use either use the  decomposition,
\begin{equation}
    \ket{M_J,M_I}=\sum_F \braket{F,M_F}{I,M_I;J,M_J}\ket{F,M_F},
    \label{eq:expression_basis_uncoupled_to_coupled}
\end{equation}
or 
\begin{equation}
    \ket{F',M_{F'}}=\sum_{I,J}\braket{M_{I},M_{J}}{F,M_{F'}}\ket{M_{I},M_{J}}.
    \label{eq:expression_basis_coupled_to_uncoupled}
\end{equation}
Using \cref{eq:expression_basis_uncoupled_to_coupled}, the light shift interaction coupling $a \to c$ for a $\pi$ polarized light is given as,
\begin{equation}
    \begin{aligned}
V_{\mathrm{LS}}^{(ac)} \propto &\sum_{F',M_{F'}}\frac{\Omega_{ac}^2}{4\Delta_{F'}({\mathrm{^1D_2}})}\left.\lvert \sum_{F,M_F} \bra{c,F',M_{F'}}d_z \ket{a,F,M_F}\bra{F,M_F}\ket{I,M_I;J,M_J=0}\right.\rvert^2,
    \end{aligned}
\end{equation}
A detailed analysis of the above light shift is given in \cref{sec:matrix_element}.
Working in a regime closely detuned to $c,F'=13/2$, one obtains,
\begin{equation}
     V_{\mathrm{LS}}^{(ac)}(M_I)\approx V_0^{ac}\left\lvert\langle F'=\frac{13}{2},M_{F'}=M_I \rvert2,0; \frac{9}{2},M_I \rangle \right \rvert^2,
\end{equation}
where $V_0^{ac}=\frac{\Omega_{ac}^2}{4}\lvert\langle 2,0  \rvert  1,0;1,0\rangle \rvert^2$.
By empirically fitting this as function of $M_I$ we find,
\begin{equation}
\begin{aligned}
    &\left\lvert\langle F'=\frac{13}{2},M_{F'}M_I \rvert2,0; \frac{9}{2},M_I \rangle \right\rvert^2\\
    &\approx 0.3-0.017M_I^2+2.3 \cross10^{-4}M_I^4.
\end{aligned}
        \label{eq:empirical_fitting}
\end{equation}
The quartic behavior is not familiar for a light shift (usually at most quadratic in nature). 
Here it arises from how the nucleus is coupling to the electron through $J=2$ (which is quadrupolar rather than dipolar). 
The dominant quadratic term can be used to cancel the energy light arising from the hyperfine perturbation in the state $a$, which also has a quadratic term from the perturbation theory analysis.

\begin{figure}
    \centering
    \includegraphics[width=\columnwidth]{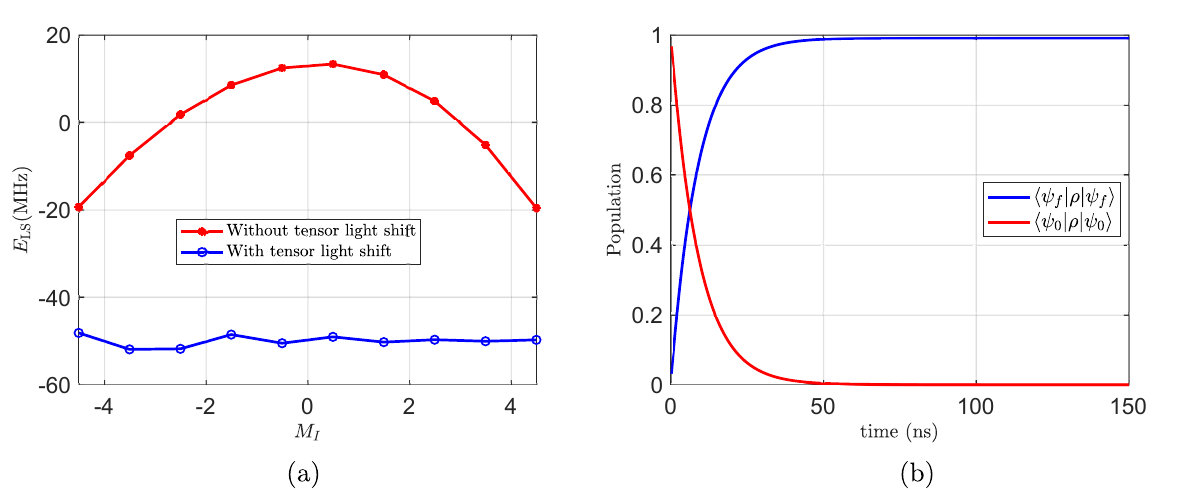}
    \caption{The figure gives the complete analysis of the scheme for QND cooling given in \cref{fig:cooling_setup}.
    (a) gives the energy of the states in $M_J=0$, in the presence and absence of the off resonant light interaction, the off resonant light interaction balances the light shift generated by the hyperfine interaction. 
    Thus the “which way information” about the nuclear spin state is gone and one can cool while preserving coherence.
    In (b) to further illustrate the success of the QND cooling scheme we consider the fidelity of the final states in the cooling scheme.
    The  high overlap of the actual state to the ideal state indicates that the success of the cooling scheme for the parameter regime considered in this work. }
    \label{fig:light_shift_spectrum}
\end{figure}

To further understand the quartic behavior in the tensor light shift, one can expand the $\ket{\mathrm{5s5p^1P_1}, M_J=0}$ in the coupled basis.
Working close to resonance for $F'=13/2$  and using the fact that the  dipole allowed interaction only allows  $F=F'\pm 1$ the only matrix element we need to consider is,
    \begin{equation}
    \begin{aligned}
        &\bra{c,F'=13/2,M_I}d_z\ket{a,1,0;9/2,m_I}        =\bra{c,F'=13/2,M_I}d_z\ket{a,F=11/2,M_I}  \\
        &\hspace{7.3cm}\braket{F=11/2,M_I}{1,0;9/2,M_I},\\
        &= \rme{c,F'=13/2}{d_z}{a,F=11/2}\braket{F'=13/2,M_I}{1,0;F=11/2,M_I}\braket{F=11/2,M_I}{1,0;9/2,M_I},\\
        &=\mathcal{O}^{J',F}_{J,F}\rme{J'=2}{d_z}{J=1}\braket{F'=13/2,M_I}{1,0;F=11/2,M_I}\braket{F=11/2,M_I}{1,0;9/2,M_I},\\
        &=\rme{J'=2}{d_z}{J=1}\braket{F'=13/2,M_I}{1,0;F=11/2,M_I}\braket{F=11/2,M_I}{1,0;9/2,M_I},
    \end{aligned}
    \label{eq:knowing_the_quartic_dependence_2}
\end{equation}
where $\mathcal{O}^{J',F'}_{J,F}$ is the relative oscillator strength defined as,

    \begin{equation}
   \mathcal{O}^{J',F'}_{J,F}=(-1)^{F'+1+F+I}\sqrt{(2J'+1)(2F+1)}\begin{Bmatrix}
       F' & I & J'\\
       J & 1 &F
   \end{Bmatrix}, 
\end{equation}

\noindent and for the case of $J'=2,J=1,F'=13/2,F=11/2$, we get $\mathcal{O}^{J',F'}_{J,F}=1$.
Using the following  property of Clebsch-Gordan coefficients \cite{ME_Rose_angular_momentum},
\begin{equation}
    \braket{j+1,m}{1,0;j,m}=\sqrt{\frac{(j+1)^2-m^2}{(2j+1)(j+1)}},
\end{equation}
 gives,
\begin{equation}
    \begin{aligned}
        &\braket{F'=13/2,M_I}{1,0;F=11/2,M_I}=\frac{1}{2}\sqrt{\frac{169-4M_I^2}{78}}\\
        &\braket{F=11/2,M_I}{1,0;I=9/2,M_I}=\frac{1}{2}\sqrt{\frac{121-4M_I^2}{55}}.
    \end{aligned}
\end{equation}
Thus we get the contribution of the tensor light-shift interaction for the case of close to resonance to $F'=13/2$  is,
\begin{equation}
\begin{aligned}
   V_{\mathrm{LS}}^{ac}= &\lvert \bra{c,F',M_{F'}} d_z \ket{a,J,M_J;I,M_I} \rvert^2\\
    &=V_0^{ac}\left(0.298-0.0169M_I^2+0.000233M_I^4\right).
    \end{aligned}
\end{equation}
This is the approximately the same expression we got from the empirical fitting in \cref{eq:empirical_fitting}.

 Figure~(\ref{fig:light_shift_spectrum}) (a)  shows the shifts of the magnetic sublevels in $M_J=0$ manifold of the $\mathrm{^1P_1}$ state, in the presence and absence of the off resonant light interaction. 
 The additional light shift effectively cancels the residual quadrupolar hyperfine shift which highly supress the  “which way information” about the nuclear spin state in spontaneous emission.
One can understand this using a simple  two-levels system  in \cref{subsec:decoherence_free_photon_scattering}
 
 To further understand this we consider an initial state,
 \begin{equation}
\ket{\psi}_{0}
     =\frac{1}{\sqrt{10}} \sum_{M_I=-\frac{9}{2}}^{\frac{9}{2}} \ket{\mathrm{5s5p ^1P_1},M_J=0,M_I},
     \label{eq:initial_state_cooling}
 \end{equation}
and  the final state to be,
\begin{equation}
    \ket{\psi}_{f}
     =\frac{1}{\sqrt{10}} \sum_{M_I=-\frac{9}{2}}^{\frac{9}{2}} \ket{\mathrm{5s^2\text{ } ^1S_0},M_J=0,M_I}.
\end{equation}
Evolving under the master equation with the Lindbladian jump operator
    \begin{equation}
    L=\sqrt{\Gamma} \sum_{i-\frac{9}{2}}^{\frac{9}{2}}\ket{a_0,M_J=0,M_I=i}\bra{a,M_J=0,M_I=i}
\end{equation}
where $a_0=\mathrm{5s^2\text{ } ^1S_0}$ and $a=\mathrm{5s5p ^1P_1}$.
The  time evolution  for $\Omega_{\mathrm{^1S_0}}=1$ GHz, $\Omega_{\mathrm{^1D_2}}=106$ MHz, and $\Delta_{\mathrm{^1D_2}}=4350$ MHz is given in \cref{fig:light_shift_spectrum}(b).
We have optimized the $\Omega_{\mathrm{^1D_2}}$  and $\Delta_{\mathrm{^1D_2}}$ such that the frequency dependence  of the spontaneously emitted photons is lowest for these values.
From the figure one can see that the excited state decays to the ground state and for the choice of parameter considered the coherence is completely preserved during the decay of the information to the ground state, thus giving us a high-fidelity QND cooling scheme.

\section{Conclusion and Outlook} 
\label{sec:conclusions_and_future_work_qnd}

In this chapter, we devised new schemes to overcome the effect of atom loss and heating which can lead to large errors in neutral atom quantum computing.
These schemes well align with the current hardware development, however, these advances could significantly improve the prospect of neutral atoms for fault-tolerant quantum computation.

First, we consider  a scheme to overcome the effect of loss of information out of the computational subspace in neutral atom quantum computation.
This could be due to atoms lost from the trap or during the gate implementations. 
These errors, in general known as leakage errors, are detrimental for fault-tolerant quantum computation as they are not Pauli errors and thus require separate error correction protocols which could significantly increase the requirements for fault tolerance.
In this chapter, we develop schemes for converting these leakage errors to erasure errors, which can be efficiently corrected by standard error correction protocols.  
We consider this in the context of alkaline earth elements, in particular, we consider the case in which the information encoded in the large nuclear spin ($I=9/2$) in the ground state of $^{87}$Sr.
To detect whether the information stored in the ground state is lost, we studied the regimes of operation in which one can scatter photons from the singlet-$P$ state without decohering the nuclear state in the ground state.
This regime of operation arises from the unique properties of the alkaline earth atoms.

Another critical roadblock for quantum computation with neutral atoms as compared with ions is the lack of a cooling scheme while preserving the coherence, this in turn prevents us from doing arbitrarily long quantum computation. 
In this work, we develop a scheme to cool the atoms while preserving the coherence without a very large magnetic field. The scheme works for alkaline earth atoms where we store the quantum information in the nuclear spin. 
For these atoms, one can use metastable states with narrow linewidth to employ the techniques of resolved sideband cooling. 
Through a combination of AC-Stark shifts we can decouple the electronic and nuclear degrees of freedom and avoid the ``which way information’’ about the nuclear spin state in the spontaneously emitted photons, thereby allowing us to sideband-cool the atom  while preserving  the nuclear spin coherence.

\chapter{Summary and Outlook} 
\label{chap:summary_and_outlook}

Quantum computation holds the promise of surpassing classical computers in performance. 
Despite significant experimental progress, the sensitivity of quantum systems to decoherence and experimental imperfections hinders the realization of practical advantages in quantum computing.
A promising avenue in current research is quantum co-design, a paradigm that capitalizes on the unique capabilities of physical systems rather than relying on hypothetical qubits and generic noise models. 
This work explores the concept of co-design in the context of quantum computation using spin qudits in neutral atoms.
The initial focus involves evaluating the feasibility of qudit-based quantum computation, deviating from the conventional qubit-based paradigm.
Specifically, we explore the practicality of encoding quantum information in the nuclear spin of the ground state of $^{87}$Sr atoms, which offers a Hilbert space of dimension $d=10$.
Currently, the bottle-neck for qudit-based quantum computation is a reliable method to implement high-fidelity  gates.
Leveraging the distinctive features of this physical platform, we develop protocols for high-fidelity universal gate set tailored for qudits in presence of realistic experimental conditions, including both single qudit SU(d) gates and two-qudit entangling gates.
The availability of high-fidelity universal gate set unlocks the power of qudits for multiple paths of quantum information processing  including quantum communication, quantum algorithms, and fault-tolerant quantum computation.   


Similarly in  the field of quantum error correction (QEC), the idea of co-design has attracted significant interest recently where one develops error-correcting codes that harmonize with the control methods and noise structures inherent in physical quantum systems \cite{Gross2021,omanakuttan2023multispin,puri2020bias,puri2019stabilized,Cong_Lukin_QEC_Rydberg_PRX_2022}. 
An important component of such co-design is engineering qubit encodings with favorable noise properties \cite{puri2017engineering}, facilitated by substantial experimental advances in quantum computing \cite{acharya2022suppressing,ryan2022implementing,krinner2022realizing}.
This dissertation contributes to this direction by developing QEC protocols specifically designed for spin qudits and their control methods, capitalizing on their unique capabilities.
This approach proves particularly effective in the neutral atom platform, where recent experimental advancements have been notable \cite{Bluvstein_Lukin_2023_QEC_Logical,bluvstein2022quantum,Saffman_Nature_2022}. 
This native QEC protocol aligns seamlessly with the neutral atom platform's capabilities including re-configurable connectivity and the ability to implement hundreds of parallel entangling gates \cite{Bluvstein_Lukin_2023_QEC_Logical}, enhance the feasibility of implementing the developed QEC protocol presented in this work.

Here we give a brief summary of the main results of this dissertation.
In \cref{chap:Qudecimal_quantum_control} and \cref{chap:Qudit_entanglers}, we implemented universal gates for qudit-based quantum computation using quantum optimal control. The ground state of $^{87}$Sr allows encoding qudits up to dimension $d\leq 10$.
In  \cref{chap:Qudecimal_quantum_control}, we employed quantum optimal control to implement high-fidelity single qudit gates, leveraging the unique atomic structure of alkaline earth atoms. 
Including the effects of decoherence and inhomogeneity, quantum optimal control sequences for $d=10$ were identified, enabling the implementation of single qudit gates with fidelity exceeding $0.99$.
In \cref{chap:Qudit_entanglers} we developed quantum optimal techniques for implementing qudit entanglers. 
Notably, we achieved a CPhase gate with fidelity values of $0.9985$, $0.9980$, $0.9942$, and $0.9800$ for qudit dimensions $d=2$, $d=3$, $d=5$, and $d=7$, respectively.
In comparison to a specific scheme proposed in~\cite{anwar2014fast}, our fault-tolerant threshold for qudit dimensions $d=2$, $d=3$, $d=5$, and $d=7$ is approximately $0.008$, $0.012$, $0.0135$, and $0.015$, respectively.
These results demonstrate a promising proof-of-principle fidelity, which can be further optimized.

In \cref{chap:Qudit_fault}, we investigated a novel scheme for encoding a qubit in a spin-qudit system. 
In this context, the angular momentum operators that generates $\mathrm{SU}(2)$ rotations form the natural set of error operators for such encodings, generalizing the Pauli operator basis for qubits.
Our work introduces a spin-cat encoding, designed to correct dominant errors and establish a fully fault-tolerant scheme for qubit encoding within a spin.
A distinctive feature of the spin-cat encoding is its unique structural composition. 
Unlike earlier methods, the error subspaces in the spin-cat encoding partition the physical space into two-dimensional subspaces where logical operations exhibit identical behavior. This structural characteristic takes the form of a stabilizer code, a key feature facilitating fault-tolerant schemes for error correction.
Another novelty of our work is the introduction of a measurement-free quantum error correction procedure specifically designed to take advantage of the capabilities of neutral atom computing and more generally other platforms with large spins.

In \cref{chap:Qudit_leakage}, we developed schemes for identifying loss of quantum information encoded in the ground state of $^{87}$Sr without damaging the coherence of the encoded state.
In doing so, leakage out of the computational subspace is converted to erasure which can substantially improve fault-tolerant thresholds~\cite{Wu_Puri_Thompson_2022_Nature_erasure}, when compared to traditional leakage reduction units~\cite{Suchara_2015_leakage}.
This protocol is compatible with  a scheme to cool the atoms while preserving the coherence. 
The protocol works for alkaline earth atoms where we store the quantum information in the nuclear spin. 
For these atoms, one can use metastable states with narrow linewidth to employ the techniques of resolved sideband cooling. 
To avoid mixing electronic and nuclear degrees of freedom in the excited state, which we need for cooling, the AC-Stark shift induced near resonances to auxiliary excited states.   
This decouples the electronic and nuclear degrees of freedom and avoids the ``which way information’’ about the nuclear spin state and one can cool while preserving coherence.

The work presented in this dissertation opens new directions of exploration with potentially far-reaching implications. 
The increasing focus on qudits over qubits in quantum computation raises a fundamental question: Is there an optimal choice of dimension $d$ that stands as the most effective for quantum computation? 
Addressing this question is pivotal for advancing our understanding of optimal resources in quantum computation.
Furthermore, the availability of a universal gate set for qudits opens the door to designing entirely new quantum algorithms tailored specifically for qudits.
This shift enables us to explore uncharted territory rather than merely extending existing quantum algorithms for qubits.
The novel encoding of a qubit in a qudit introduced in this dissertation opens additional paths of exploration. 
Similar to continuous-variable cat encoding~\cite{puri2020bias}, the proposed gate set enables the use of other codes, including the topological codes. 
Another direct extension involves developing gate sets for computation by encoding a qudit instead of a qubit, into the large spin.
These extensions deepen our understanding and pave way for innovative applications in quantum information processing.


\appendix 
\chapter{Bandwidth limited Qudecimal Quantum Optimal 
Control}
\label{chap:app_qudecimal_qudits}

In this appendix, we detail the methods we employ for quantum optimal control in more experimentally friendly scenario.
Here we study  bandwidth limited quantum optimal control.

We consider open loop-control to create arbitrary unitary evolution, a problem which is studied extensively in the literature \cite{jurdjevic1972control,brockett1973lie,schirmer2002degrees, goerz2015optimizing, glaser2015training}. In general consider a Hamiltonian in a $d$-dimension Hilbert space of the form, 
\begin{equation}
H(t)=H_0+\sum_{\lambda=1}^{K}c_\lambda(t)H_{\lambda}.
\label{eqn: control Hamiltonian}
\end{equation}
 The system is controllable if we can generate any $U_0\in SU(d)$ using a set of controls $c_\lambda(t)$,  This means that in a finite time $T$, the Hamiltonian evolution given by the Schr{\"o}dinger equation $\dot{U}=-iH(t)U$, maps the identity operator to any arbitrary unitary operator $U_0$ in the group with arbitrary precision. A necessary and sufficient condition for the controllability is the set of Hamiltonians $\{H_0,H_1,H_2,..., H_K\}$ generate the Lie algebra $\mathrm{su}\left(d\right)$.  
 
 In this appendix we consider the control of a nuclear spin $\mathbf{I}$ with dimension $d=2I+1=10$ using a combination of radio-frequency driven Larmor precession and a tensor AC-Stark shift according to the Hamiltonian,
\begin{equation}
H(t)=\Omega_{\mathrm{rf}} \left( \cos[c(t) \pi]I_x+\sin[c(t) \pi] I_y\right)+\beta I_z^2.
\label{eq:Control_Hamiltonian_qudec}
\end{equation}
It was proved in \cite{Merkel2008} that by manipulating the phase $c(t)$ the above system is controllable.

We consider two classes of quantum control tasks: preparation of a target pure state $\ket{\psi_\text{tar}}$ and implementation of a target unitary map $U_\text{tar}$ on an arbitrary input state.  We implement these tasks using quantum optimal control. The goal is to find the waveform $c(t)$ which optimizes the objective function.  As a first step we discretize the control waveform as a piecewise constant function over $n$ equal intervals in the time $T$, $\mathbf{c}=\{c_i = c(t_i) | i=1,\dots N\}$.  Optimal control for state preparation and unitary maps follows by maximizing the relevant fidelity, 
\begin{eqnarray}
\mathcal{F}_{\psi}[\bm{c},T]&=&\left|\bra{\psi_{\text{tar}}}U[\bm{c},T]\ket{\psi_0}\right|^2,\\
\mathcal{F}_U[\bm{c},T]&=&\left|\Tr\left(U^{\dagger}_{\text{tar}}U[\bm{c},T]\right)\right|^2/d^2.
\label{eq:fidelity_qudec}
\end{eqnarray} 
Here $U[\bm{c},T] =\prod_{i=1}^n e^{-iH(c_i)T/n}$.  To find $\mathbf{c}$, we use the well-known gradient based optimization method GRAPE \cite{khaneja2005optimal}. Robust optimization follows when $T$ is sufficiently large compared to the minimal value $T_*$ set by the quantum speed limit~\cite{caneva2009optimal} and $n$ is sufficiently large compared with the minimal number of parameters necessary to specific the control task.  For a $d$-dimensional Hilbert space, $n_{\min} = 2d-2$ for state preparation and $n_{\min} =d^2-1$ for unitary maps.

 
While in principle we can find simple control waveforms with $n$ close to $n_{\min}$, in practice, the resulting discontinuous waveforms may not be exactly realizable in an experimental implementation. To find waveforms that are more experimentally feasible we constrain the maximum jump allowed between $c_i \text{ and } c_{i+1}$ to create a smoother waveform, as was shown in~\cite{Frey2020}.  Another important ingredient is the choice of the initial seed $\mathbf{c}$ to the GRAPE algorithm. A waveform that yields high-fidelity is not unique, and by choosing smoother initial seed, the optimal solution will be smoother as well.    Here we choose the initial condition where $c_i=0 \hspace{0.1cm}\forall i$.  This is sufficiently small so that the time for computer optimization is reasonable, by sufficiently large that we obtain experimentally feasible waveforms, with a maximum of  $c_{i+1}-c_i \leq 0.4 $.

While the quantum control technique described above creates relatively smooth waveforms, there still exist discontinuities which can result in a large slew rate and bandwidth that is outside the range of the physical control. To see how this constraint affects the fidelity, we take a simple model to pass the phase waveform through a low-pass filter,
\begin{equation}
     c(t)=\phi(t)/\pi=\Omega_c\int_{0}^{t} c_{ideal}(\xi)\exp\left[-\Omega_c(t-\xi)\right]d\xi\\
    \label{eq:LIS}
\end{equation}
where $c_{ideal}(\xi)$ is the ideal waveform value one would attain as the output of the GRAPE algorithm in a perfect piecewise approach. The waveforms depend on the choice of the corner frequency, $\Omega_c$, which is related to the bandwidth of the controller.  Examples of filtered waveforms obtained using $\Omega_c = 20 \Omega_{\rm{rf}} $ are given in Fig.~(\ref{fig:control_waveform_convolution}).  The resulting waveforms are continuous functions of time and band-limited. 

\begin{figure}
\includegraphics[width=1\textwidth]{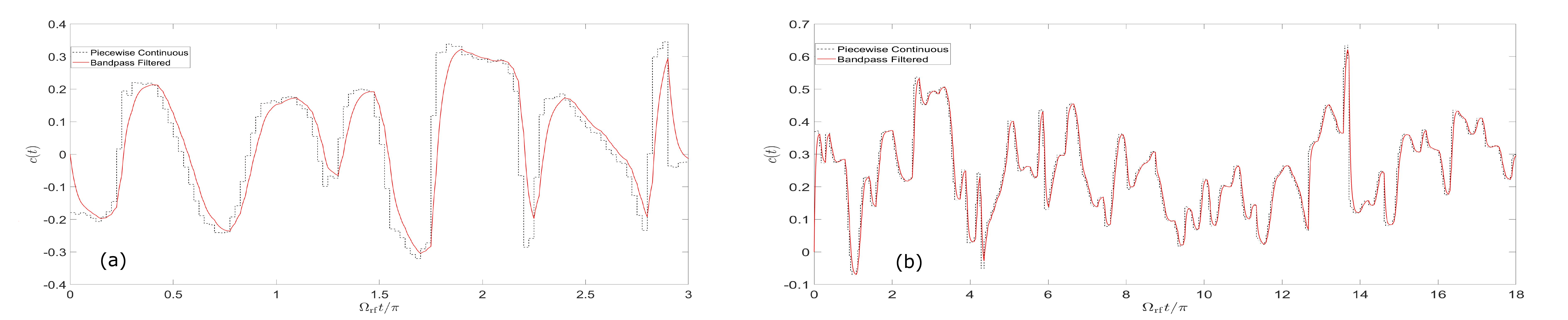}
\caption{Control waveforms for a piecewise constant parameterization, with a limited slew rate (dotted black line) and the waveforms created after the low-pass filter (solid red line ) for the state preparation (a) and unitary mapping (b) with $\Omega_c=10\Omega_{\mathrm{rf}}$.  }
\label{fig:control_waveform_convolution}
\end{figure}

The analysis of the control seed after the low-pass filter shows that there is high fidelity operation can be obtain for $\Omega_c \sim 100 \Omega_{\rm{rf}}$, e.g., $\Omega_{\rm{rf}} = 100$ Hz, $\Omega_{c} = 1$ kHz. The decoherence analysis for the continuous waveforms for the state preparation and  unitary mapping (Eq.(6) and Eq.(7) in the main text)  for $\beta=0.4\Omega_{\mathrm{rf}}$ is given in Fig.~(\ref{fig:control_waveform_convolution_fidleity})

\begin{figure}
\includegraphics[width=\textwidth]{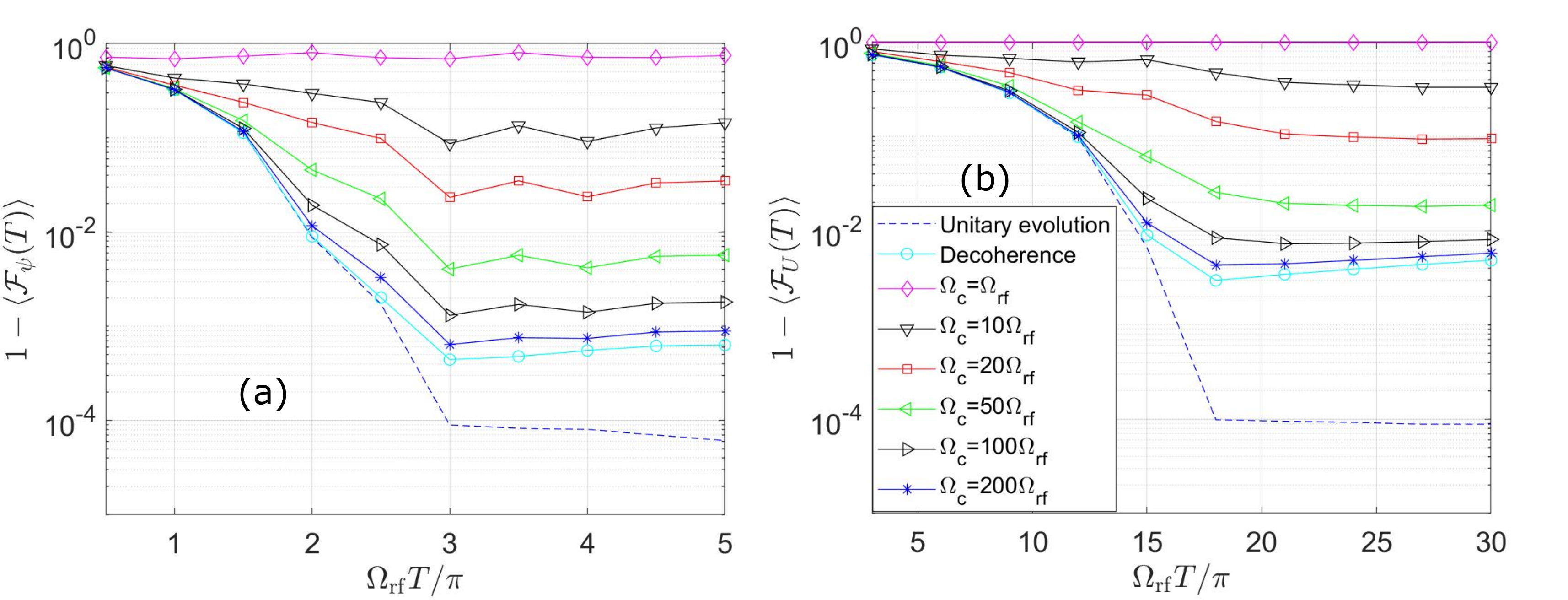}
\caption{The fidelity observed for state preparation (a) and unitary mapping (b) for $\beta=0.4\Omega_{\mathrm{rf}}$ under the full decoherence analysis for different value of the corner frequencey $\Omega_c$.  }
\label{fig:control_waveform_convolution_fidleity}
\end{figure}

\chapter{Quantum optimal control for qudit entanglers}
\label{chap:app_entangling_qudits}

In this appendix, we go into some more details of the implementation of the entangling gates for qudits using quantum optimal control. 
\section{Hyperfine structure of Rydberg states and Clebsch-Gordan coefficients}

As described in the Sec.~\ref{sec:controllability}c, to create entanglement we promote the population from the ground state $^{1}S_0$ to the first excited  $^3P_2$ state, with the hyperfine quantum number  $F=9/2$, and then consider a UV laser to excite the atoms to the $^{3}S_1$ Rydberg series to implement the interaction between atoms with adiabatic dressing (see Fig.~\ref{fig:set_up_figure}). The Rabi frequency characterizing the coupling of the different $m_F$ levels in the $^3P_2$ hyperfine manifold to the $^{3}S_1$ Rydberg states will be different due to the  Clebsch-Gordon Coefficients for these transitions. 
Let $\Omega_{\mathrm{L}}$ be the Rabi frequency on the $\ket{0_a}\to \ket{0_r}$ ($m_F=-9/2$ transition). 
The Rabi frequency experienced by the other levels is then
\begin{small}
\begin{equation}
      \Omega_{r_i}=\frac{\bra{F,m_F=-9/2+i}\ket{1,0; F', m_F=-9/2+i}}{\bra{F,m_F=-9/2}\ket{1,0; F', m_F=-9/2}}\Omega_{\mathrm{L}}, 
    \label{eq:Omega_eff}
\end{equation}
\end{small}
where we have chosen $F=9/2$ and $F'=11/2$, and  a $\pi$-polarized light. In Fig.~\ref{fig:Omega_eff} the Rabi frequencies of the different levels are given as a function of $m_F$, whose parabolic shape describes the tensor light shift, thus giving a natural nonlinearity which arises solely due to well-defined hyperfine structure of $^{87}$Sr.

\begin{figure}
\centering
\includegraphics[width=0.45\textwidth]{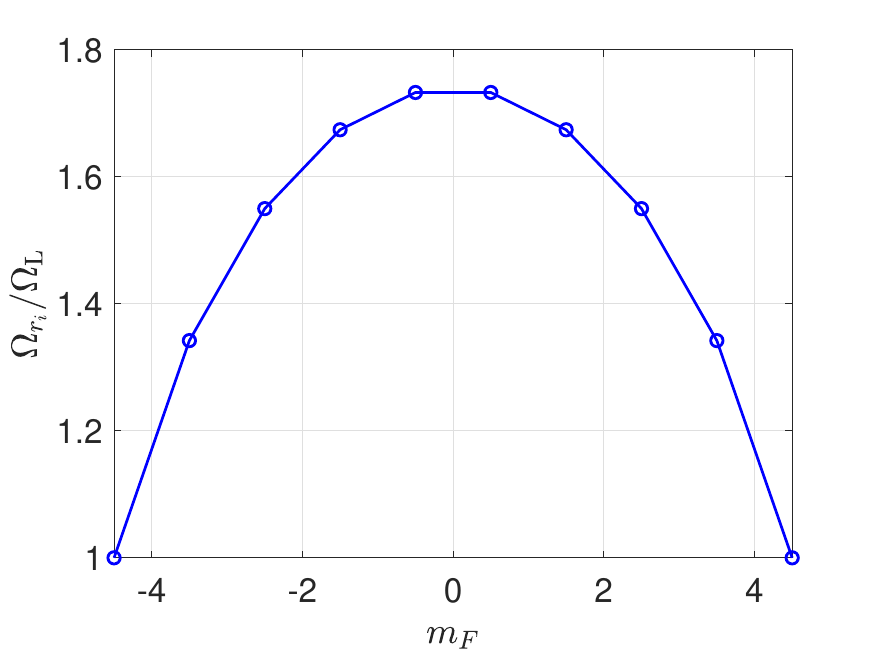}
\caption{Relative Rabi frequency, $\Omega_{r_i}/\Omega_{\mathrm{L}}$, plotted as a function of $m_F$ for $\pi$ polarized light for the $(5s5p)^3P_2F=9/2 \to (5sns) ^3S_1 F'=11/2$ transition to the Rydberg state.
The quadratic function arises due to the tensor polarizability. }
\label{fig:Omega_eff}
\end{figure}


\begin{figure}
\centering
		\includegraphics[width=0.8\textwidth]{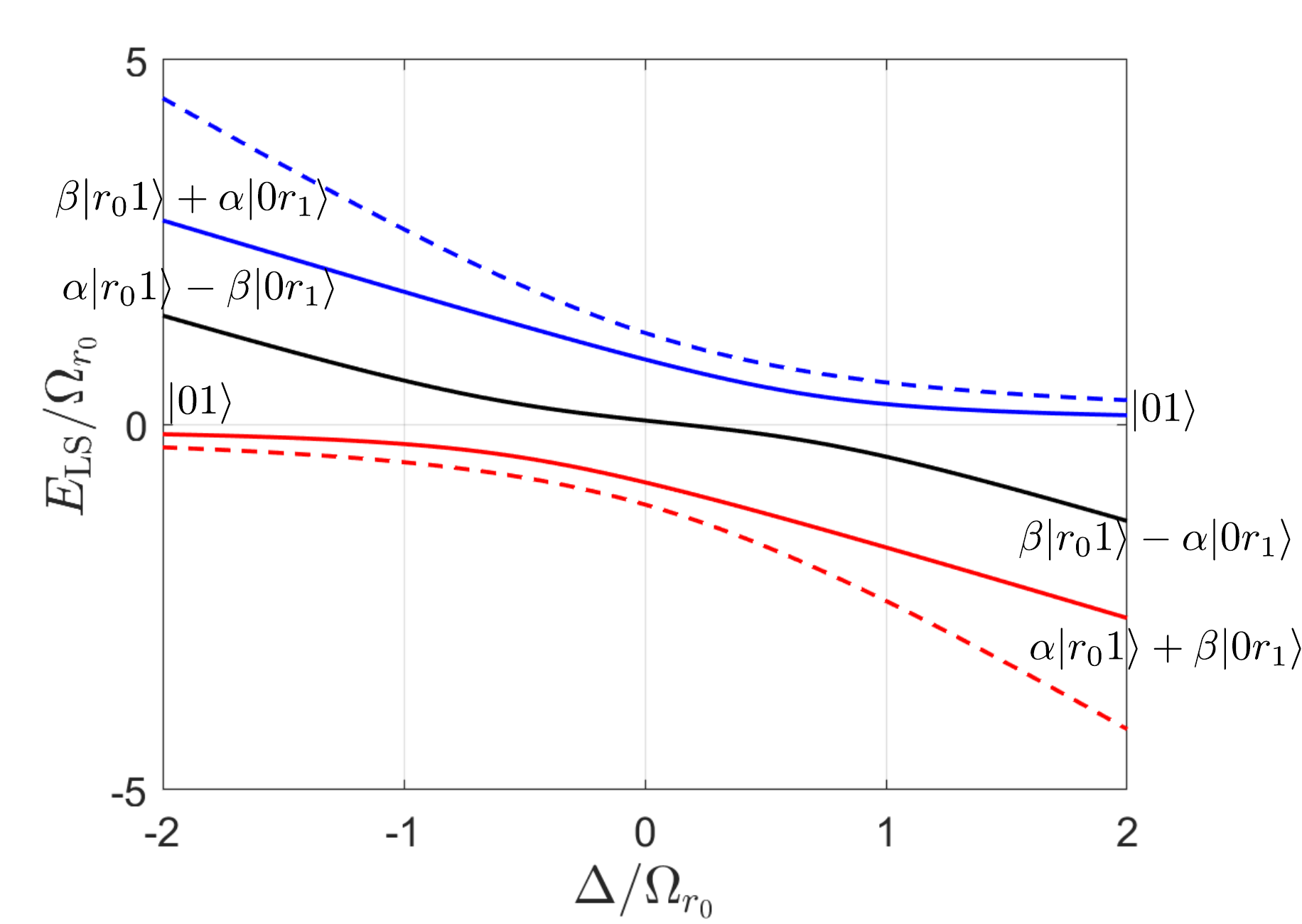}
	\caption{Autler-Townes splitting of the three dressed states as a function of detuning for the Hamiltonian in  Eq.  (\ref{eq:two_atom_Hamiltonian}), where $i=0, \;j=1$, such that $\ket{0} \equiv \ket{^3P_2, \; m_F=9/2}$ and $\ket{1} \equiv \ket{^3P_2, \; m_F=7/2}$. 
    Here $\alpha=\sqrt{7/16}$ and $\beta=\sqrt{9/16}$.
	The dashed line shows the AC Stark shift (light shift) in the absence of a perfect Rydberg blockade. The blue curve adiabatically connects to the clock states for large blue detuning and the red curve for large red detuning. The black curve is a dressed superposition that does not adiabatically connect to the clock states. The dashed lines show the light shifts in the absence of van der Waals interactions between the atoms. The difference between the solid line and the dashed line is the entangling power of the Hamiltonian $H_{2}^{12}$ defined in Eq.(\ref{eq:two_atom_Hamiltonian}).}
	
	\label{fig:eig_figure}
\end{figure}


Consider the Rydberg dressing scheme in Fig.~\ref{fig:set_up_figure}. 
In the perfect blockade regime, the two-atom Hamiltonian coupling of two magnetic sublevels labeled $i$ and $j$ is described by a three-level system, governed by the Hamiltonian,
\begin{equation}
\begin{aligned}
    H_2^{ij}=&-\Delta_i\ket{r_i j}\bra{r_i j}+\frac{\Omega_{r_i}}{2}\left(\ket{r_ij}\bra{ij}+\ket{ij}\bra{r_ij}\right)\\
    & -\Delta_j\ket{ir_j}\bra{ir_j}+\frac{\Omega_{r_j}}{2}\left(\ket{ir_j}\bra{ij}+\ket{ij}\bra{ir_j}\right),
        \label{eq:two_atom_Hamiltonian}
    \end{aligned}
\end{equation}
where $\Delta_i$ determines the detunings due to the differential Zeeman shit. Fig.~\ref{fig:eig_figure} shows the resulting AC Stark shifts on the three dressed states after diagonalizing this Hamiltonian. The dressed ground state is shown in red; the other two dressed states represent Autler-Townes splitting. In the absence of the van der Waals interaction the AC Stark shift (light shift) is the sum of the light shifts of each atom independently (dashed line in Fig.~\ref{fig:eig_figure}. The difference between these is the entangling energy. 

One can understand the entangling power of the Hamiltonian by studying the properties of the dressed energy levels as a function of detuning. Figure~\ref{fig:eig_figure} shows the particular case of $i=0$, $j=1$ for the Hamiltonian in Eq.~(\ref{eq:two_atom_Hamiltonian}), where  $\ket{0} \equiv \ket{m_F=9/2}$ and $\ket{1} \equiv \ket{m_F=7/2}$.
On the red side of detuning and for large detuning, as we start with the bare state and we adiabatically sweep through resonance, the state maps to the superposition of the two Rydberg states. 
Note, this is not an equal superposition as seen in \cite{mitra_martin_gate} due to the fact that the states $\ket{0}$ and $\ket{1}$ couple with different Rydberg Rabi frequency and detuning to the Rydberg states.

\section{Controllability}
\label{sec:controllability_of_the_Hamiltonian}
The quantum system is said to be controllable if, given a time-dependent Hamiltonian $H[\mathbf{c}(t)]$, there exist a time-dependent set of waveforms $\mathbf{c}(t)$, such that the one can generate an arbitrary unitary map. Here we consider those two-qudit unitary maps generated by an entangling Hamiltonian that is symmetric under the exchange of the qudits and thus does not require local addressing. To show that a Hamiltonian is controllable, we use the operator basis of irreducible spherical tensors on spin $j$  defined as \cite{sakurai2014modern,klimov2008generalized},
\begin{equation}
T^{(k)}_q=\sqrt{\frac{2k+1}{2j+1}}\sum_m \braket{j,k+q}{k,1;j,m}\ketbra{j,m+q}{j,m}.
    \label{eq:spherical_tensor}
\end{equation}
These satisfy the fundamental commutation relations,
\begin{equation}
    \begin{aligned}
    \comm{j_z}{T^{(k)}_q}=&q T^{(k)}_q,\\
    \comm{j_{\pm}}{T^{(k)}_q}=&\sqrt{k(k+1)-q(q\pm 1)}T^{(k)}_{q\pm 1}.
    \label{eq:spherical_tensor_commutators}
    \end{aligned}
\end{equation}
The set of operators $T^{(k)}_q$ form a complete orthonormal operator basis. Merkel \emph{et al.}~\cite{merkel2009quantum} showed that given a generating set of Hamiltonians $\{h_i\}$, if 
\begin{equation}
    \Tr{h_i,T^{k}_q}\neq 0
\end{equation}
for $k>2$, the system is fully controllable. That is, the set generates the whole Lie algebra of interest, which thus allows us to implement an arbitrary unitary map on the spin of the system using quantum control. 

We consider two-qudit systems, where the relevant Lie Group is $\mathrm{SU}(d^2)$; here $d^2=100$. We expand the entangling Hamiltonian in the operator basis of spherical tensors with $j=99/2$, spanning the space of dimension $D=2j+1=100$. Fig.~\ref{fig:controllability} shows operator decomposition of the entangling Hamiltonian $H_{\mathrm{ent}}$ in different orders of spherical tensors. One can see in this figure that there are contributions from higher rank tensors,  making the system controllable.
From \cite{Merkel2008}, it is known that if we have a Hamiltonian that has contribution from a spherical tensor with $K\leq 2$, one can combine that with simple $\mathrm{SU}(2)$ rotations to obtain a controllable Hamiltonian by modulating the phase of the $\mathrm{SU}(2)$ rotations. 

\begin{figure}
\centering
\includegraphics[width=0.48\textwidth]{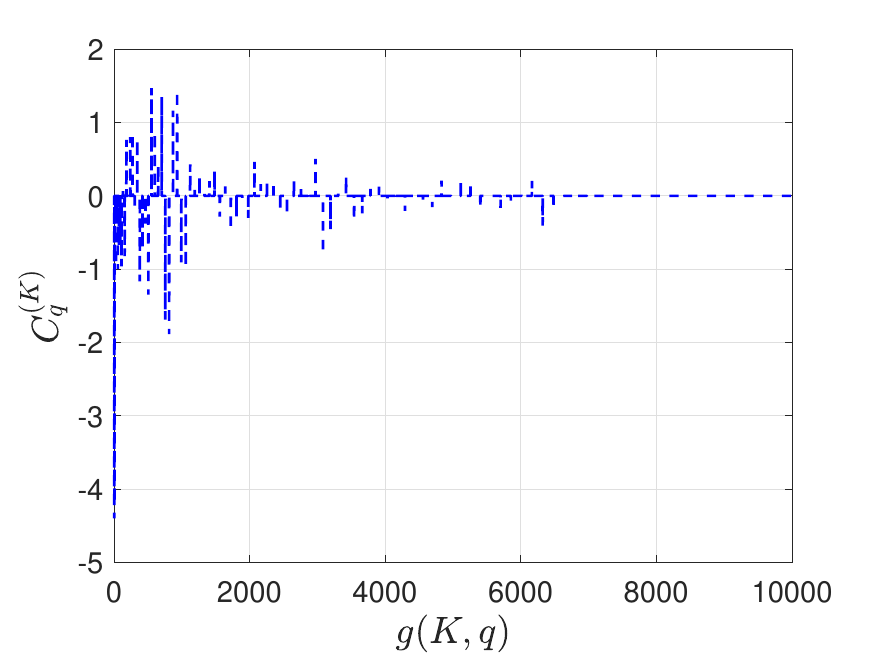}
\caption{The decomposition the entangling Hamiltonian $H_{\mathrm{ent}}$, Eq.~(\ref{eq:entangling_Hamiltonian}) in different orders of spherical tensors, $T^{(K)}_q$, for $j=99/2$,  an operator basis of dimension $D=2j+1=100$, spanning the two-qudit space for $d=10$. The expansion coefficients are given by $C^{(K)}_q= \left| {\rm Tr}(H_{\rm{ent}}T^{(K)\dag}_q)\right|^2$. We have ordered the expansion coefficients according to $g(K,q)=(k+1)^2-1+q$, where $0\leq k\leq j$, and $-k\leq q\leq k$. 
The existence of contributions of at least a single higher-rank tensor ($K\leq 2$) makes the system controllable when combined with time-dependent rf-fields that act locally on the atoms.}
\label{fig:controllability}
\end{figure}

\section{Creating other symmetric qudit entanglers for the Lie algebraic approach}
\label{sec:Molmer_sorenson_gate}
\begin{figure*}

\includegraphics[width=0.98\textwidth]{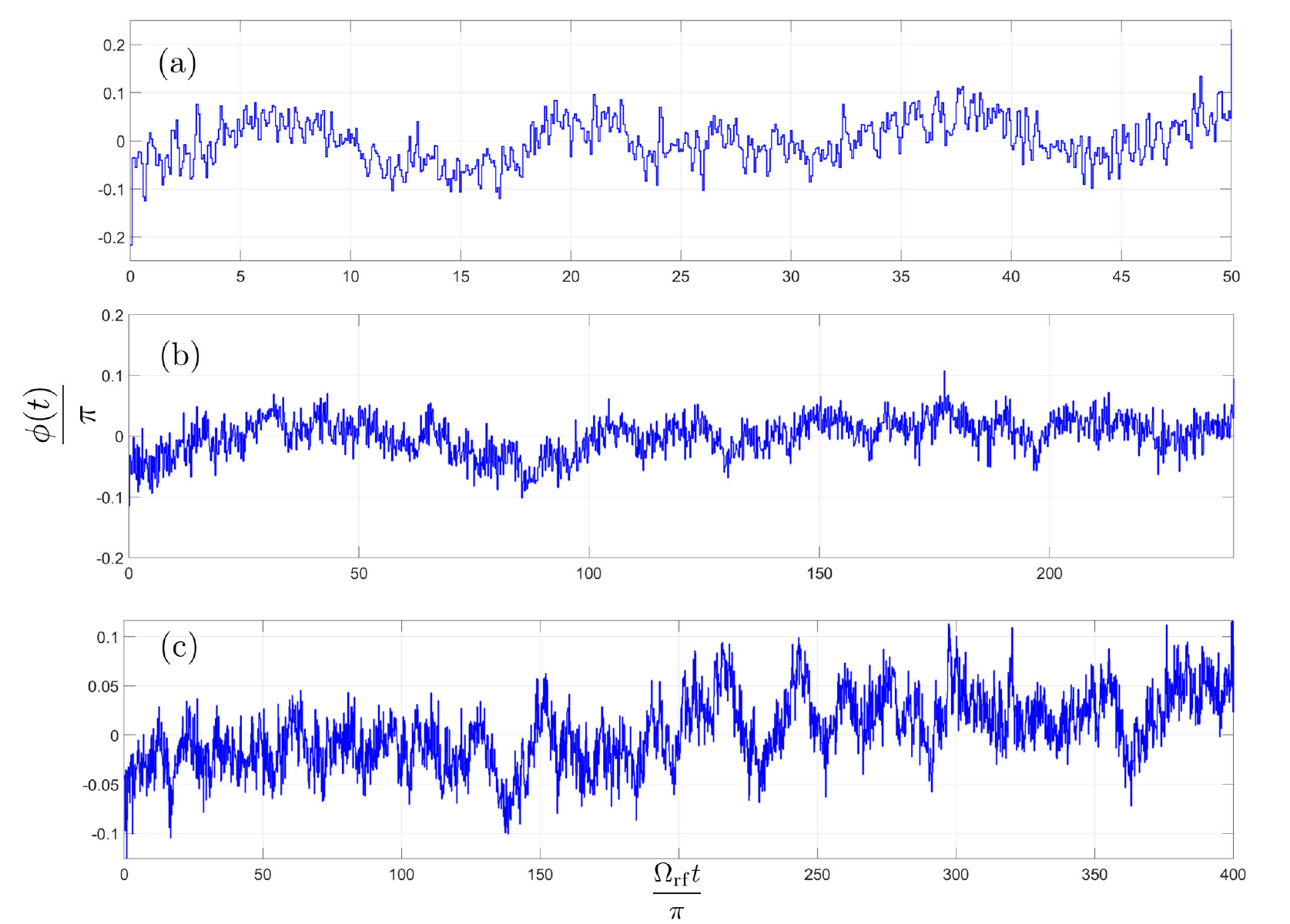}

    
    \caption{The figure gives the $\phi(t)$ that generates the M$\o$lmer-S$\o$renson gate as a function of time for $ \theta=\pi/2$ using the piecewise constant quantum control approach for the Hamiltonian given in Eq.\eqref{eq:entangling_Hamiltonian}. 
    In (a) the case of the $d=3$ for a total time of $\Omega_{\mathrm{rf}}T=50 \pi$ with $700$ piecewise constant steps. 
    In (b) the case of the $d=5$for a total time of $\Omega_{\mathrm{rf}}T=240 \pi$ with $1600$ piecewise constant steps.
    And in (c)  the case of the $d=7$for a total time of $\Omega_{\mathrm{rf}}T=240 \pi$ with $2500$ piecewise constant steps. 
    For all of these calculations we have taken $\Omega_{\mathrm{L}}=6 \Omega_{\mathrm{rf}} $.}
    \label{fig:Qudits_simulations_ms_gate}
\end{figure*}
Since the Hamiltonian described in Eq.\eqref{eq:entangling_qudit_Hamiltonian_1} can be used to create any symmetric two-qudit Hamiltonian, we can also generate the M$\o$lmer-S$\o$renson gate for qudits defined as, 
\begin{equation}
U_{\mathrm{MS}}(\theta)=\exp(-i\theta\frac{J_z^2}{2}).
\end{equation}
where the total angular momentum operator for the two qudits is
\begin{equation}
 J_z =\mathds{1}\otimes j_z + j_z \otimes \mathds{1}. 
\end{equation}

We employ the same procedure for optimal control as we discussed in the main text in designing the waveforms to implement the CPhase gate. Numerical examples of the waveforms that create the M$\o$lmer-S$\o$renson gate for $\theta=\pi/2$ are given in Fig.~\ref{fig:Qudits_simulations_ms_gate}. 
The figure shows $\phi(t)$, the piecewise constant of the control waveform, obtained using the GRAPE algorithm. Fig.~\ref{fig:Qudits_simulations}(a) shows the case of the $k=3$ the qutrit encoded in $d=10$. The total time is $T=50 \pi/\Omega_{\mathrm{rf}}$ and we divide the time into $700$ time steps for the quantum control. In Fig.~\ref{fig:Qudits_simulations}(b) we plot an example waveform for the case of the $d=5$ into our $10$ level system. We have a total time of $T=240 \pi/\Omega_{\mathrm{rf}}$ and we divide the time into $1600$ time steps for the quantum control. 
In Fig.~\ref{fig:Qudits_simulations}(c) we plot an example for the case of the $d=7$ into our $10$ level system. 
We have a total time of $T=400 \pi/\Omega_{\mathrm{rf}}$ and we divide the time into $2500$ time steps for the quantum control.

\chapter{ Fault tolerant quantum computation for a qubit encoded in qudit}
\label{chap:app_fault_qudits}

In this appendix we go into some more details of the qubit encoded in a spin qudit using spin-cat states.

\section{Small Rotation errors}
\label{sec:Ratio_coherent_errors}
\begin{figure*}[t]
\captionsetup[subfigure]{oneside,margin={3.7cm,0cm}}
    \centering
    \subfloat[]{\includegraphics[width =0.5\columnwidth]{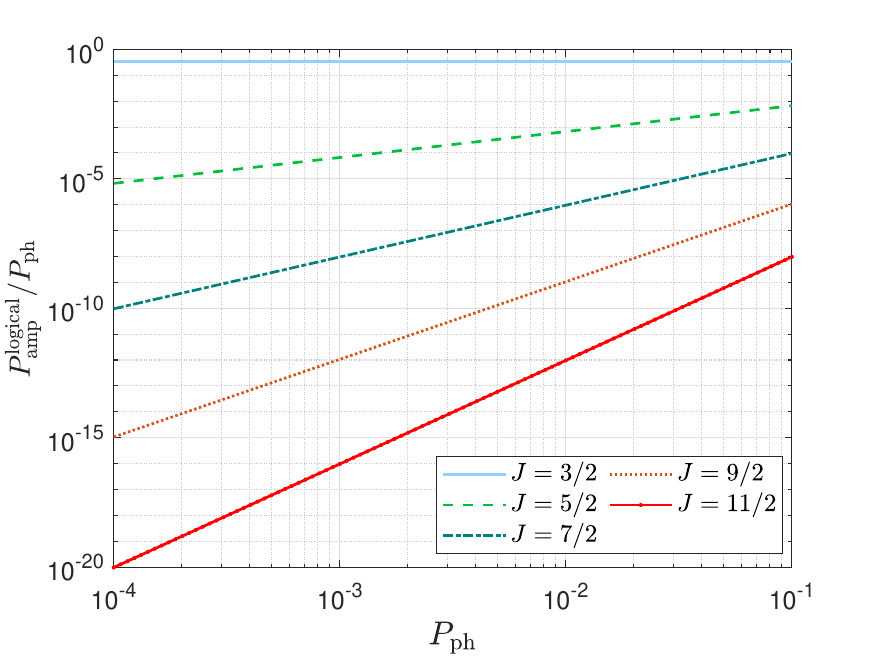} \label{fig:rot_error_a}}\hspace*{-1.9em}
    \subfloat[]{\includegraphics[width =0.5\columnwidth]{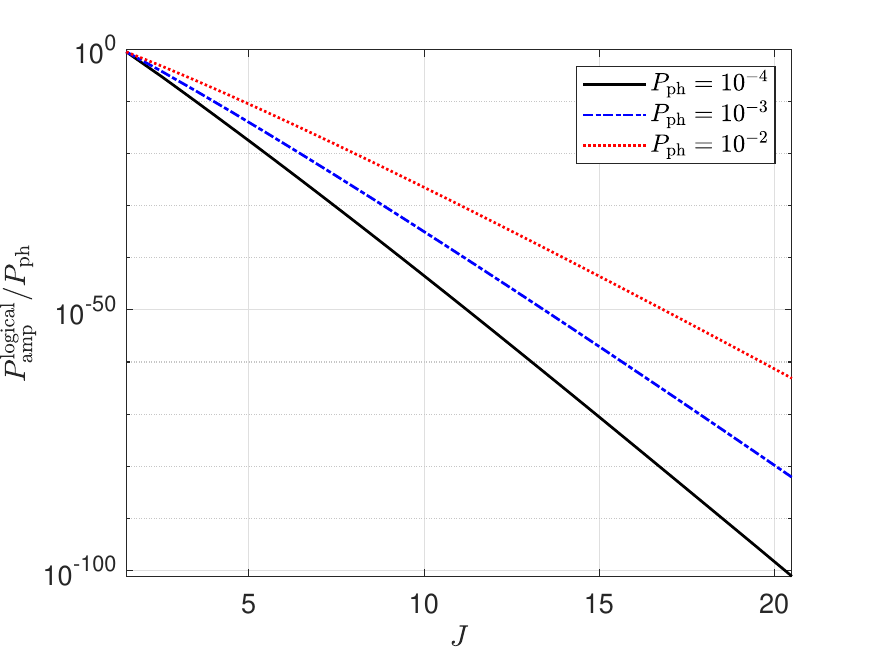}  \label{fig:rot_error_b}}
    \caption{Logical amplitude error probabilities due to rotation errors.
    (a) The ratio of logical amplitude error to phase error is given as a function of phase error.
    The probability of logical errors falls as the overall error rate decreases.
    A logical error occurs when we have $\lfloor (2J-1)/2\rfloor$ amplitude errors and thus as spin  $J$,  increases, the ratio decreases exponentially.
    However, for $J=3/2$, a single amplitude jump creates a logical error and thus the ratio of logical error to phase error is a constant equal to $1/2J$.
    (b)  The ratio of logical error probability due to amplitude errors to phase error for rotation error as a function of spin $J$. We can see that this ratio exhibits an exponential trend, and the logical error becomes negligible for sufficiently large values of $J$. 
  Consequently, there is no need for amplitude error correction in such cases.}
    \label{fig:rotation_error_comparison}
\end{figure*}

A main source of decoherence for a qubit encoded in a spin is small random rotation errors \cite{gross2021hardware,Gross2021}.
As given in \cref{eq:error_prob_rotation}, for the spin-cat encoding the ratio of phase error to amplitude error decreases with spin $J$ as $1/J$.
However, for the spin-cat encoding, we need $\lfloor (2J-1)/2\rfloor$ amplitude errors/jumps for a logical error (logical amplitude error) to occur, such that these errors are not correctable by the amplitude error correction (a logical bit flip error for the encoding in \cref{eq:concat_spin_cat}). 
As such, we look at the probability of such logical amplitude errors in \cref{fig:rotation_error_comparison}.
In \cref{fig:rot_error_a} we show that for a spin $J$, the logical amplitude error decreases with phase error probability, and the decrease shows an exponential behavior with spin $J$.

To further illustrate the exponential suppression of the logical error arising from amplitude errors as a function of spin due to random rotation errors, in  \cref{fig:rot_error_b}, the ratio of logical amplitude error probability to phase error for rotation error is given as a function of spin $J$ for different value of phase error. 
Notably, this ratio exhibits an exponential trend, and for sufficiently large values of $J$, the logical amplitude error becomes negligible.
Consequently, there is no need for amplitude error correction in such cases.

\section{Photon scattering and optical pumping}
\label{sec:Ratio_optical_pumping_errors}
\begin{figure}[!ht]
\centering
\includegraphics[width=0.8\columnwidth]{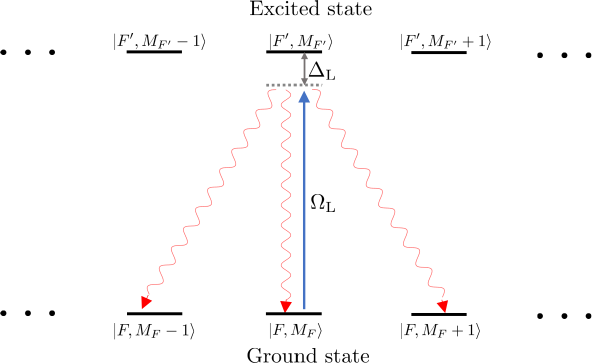}
\caption{The error process corresponding to the photon scattering and optical pumping for encoding a qudit in an atomic spin $\mathbf{F}$. The information is stored in the ground state and is controlled by laser light with Rabi frequency $\Omega_{\mathrm{L}}$ and detuning $\Delta_{\mathrm{L}}$ from an excited state manifold, with spin $\mathbf{F}'$. Absorption of a laser photon (here $\pi$-polarized) is followed by a spontaneous emission given by wavy lines.  The process causes amplitude errors and can collapse a cat-state to a single magnetic sublevel. }
\label{fig:decoherence_origin}
\end{figure}
Another major source of decoherence for the qubit encoded in a spin is the optical pumping arising from photon scattering when the spin are manipulated by laser light.
We consider here optical pumping arising from laser excitation with Rabi frequency $\Omega_{\mathrm{L}}$ and detuning $\Delta_{\mathrm{L}}$  from a dominant resonance. Absorption of a laser with polarization $\vec{\epsilon}_L$ is followed by a spontaneous emission of photon $\mathbf{e}_q$. A schematic of the error process corresponding to the photon scattering and optical pumping for atomic spins is shown in \cref{fig:decoherence_origin} for the case $\vec{\epsilon}_L = \mathbf{e}_0$. 

In this section, the spin angular momentum in which we encode the qudit is $\mathbf{F}$, and  $\mathbf{J}$ is the total angular momentum of the electrons. 
The jump operators for the optical pumping followed by photon scattering are, \cite{deutsch2010quantum}:
\begin{equation}
W_q=\sum_{F'}\frac{\Omega_{\mathrm{L}}/2}{\Delta_{FF'}+i\Gamma/2}(\bm{e}_q^{*}.\bm{D}_{FF'})(\vec{\epsilon}_L.\bm{D}_{FF'}^{\dagger}),
\label{eq:jump_operators_opt}
\end{equation} 
where $\Omega_{\mathrm{L}}$ is the Rabi frequency and $\Delta_{FF'}$ is the detuning between the ground state and excited with total spin $F$ and $F'$ respectively. 
$\Gamma$ is the characteristic linewidth of the excited state, $\vec{\epsilon}_L$ is the polarization of the laser, and $q={-1,0,1}$  represent the polarization of the scattered light.
$\bm{D}_{FF'}$ are the dimensionless raising operators from a ground state with total spin $F$ to an excited state with spin  $F'$ and see \cite{deutsch2010quantum} for a detailed analysis of these operators.

By decomposing the dyadic into irreducible tensors, one can derive a basis independent representation for the jump operators~\cite{deutsch2010quantum},
\begin{equation}
\begin{aligned}
&= \bm{e}_q^{*}.(\bm{D}_{FF'}\bm{D}_{FF'}^{\dagger}).\vec{\epsilon}_L\\
&=C_{J'FF'}^{0}\bm{e}_q^{*}.\vec{\epsilon}_L+iC^{1}_{J'FF'}(\bm{e}_q^{*}\cross\vec{\epsilon}_L).\bm{F}\\
&+C_{J'FF'}^{2}\left[\frac{(\bm{e}_q^*.\bm{F})(\vec{\epsilon}_L.\bm{F})+(\vec{\epsilon}_L.\bm{F})(\bm{e}_q^{*}.\bm{F})}{2}-\frac{1}{3}|\bm{e}_q^{*}.\vec{\epsilon}_L|\bm{F}^2\right]
\label{eq:jump_operators_optical_pumping_basis_independent}
\end{aligned}
\end{equation}
 where $J$ is the electron angular momentum. The above expression involves only angular momentum operators of the form $\bm{F}$ (rank-1) and $\bm{F}^2$ (rank-2), and thus for photon scattering and optical pumping the error operators are linear and quadratic powers of angular momentum operators.
Then the Lindblad master equation gives us: 
\begin{equation}
\begin{aligned}
    \frac{d\rho(t)}{dt} &=-i \left(H_\mathrm{eff}\rho(t)-\rho(t) H_\mathrm{eff}^\dag \right)+ \Gamma\sum_{i}W_q\rho(t) W_{q}^{\dagger}  \nonumber \\ 
&\equiv \mathcal{L}\rho(t).
\end{aligned}
\label{eq:evolution_of_the_density_matrix}
\end{equation}
where $\mathcal{L}$ is the Lindbladian and $H_\text{eff}=H-i\sum_q W_q^{\dagger}W_q/2$.

From the jump operators, one can find the probability of phase errors and amplitude errors by finding the overlap of the jump operators with the basis operators as given in \cref{eq:basis}.


\section{Correctable set of errors}
\label{sec:KL_conditions}
In this section, we find the set of correctable errors for the logical level encoding $\mathcal{C}_1$ in \cref{eq:concat_spin_cat}.
To find the correctable set of errors $\{E_a\}$, one can use the Knill-Laflamme conditions \cite{PhysRevA.55.900}:
\begin{equation}
\bra{\psi_i}E_a^{\dagger} E_b\ket{\psi_j}=C_{ab}\delta_{ij},
\label{eq:Knill-Laflamme conditions}
\end{equation}
where $i,j=\{0,1\}$ represents the codespace of interest.

The local angular momentum errors of interest here are of the form $J_x^lJ_y^mJ_z^n$. 
From the locality assumption of the errors, one can find that for the spin-cat encoding in \cref{eq:concat_spin_cat},
\begin{equation}
\bra{\psi_i}E_a^{\dagger} E_b\ket{\psi_j}=0 \hspace{0.2cm}\forall i\neq j.
\end{equation}
The next condition we need to satisfy for the spin-cat encoding is,
\begin{equation}
\bra{+_\mathrm{L}}E_a^{\dagger}E_b\ket{+_\mathrm{L}}=\bra{-_\mathrm{L}}E_a^{\dagger}E_b\ket{-_\mathrm{L}},
\label{eq:kl_diagonal}
\end{equation}
where the logical states are defined in \cref{eq:concat_spin_cat}.
From the locality assumption of the noise, this condition translates into two cases. 
In the first case the error operators $E_a$ and $E_b$ act on the same physical system, thus for the angular momentum errors the error correction condition in \cref{eq:kl_diagonal} becomes, 
\begin{equation}
    \begin{aligned}
        \bra{+}J_x^lJ_y^mJ_z^n J_x^{l'}J_y^{m'}J_z^{n'} \ket{+}=\bra{-}J_x^lJ_y^mJ_z^n J_x^{l'}J_y^{m'}J_z^{n'} \ket{-}.       
    \end{aligned}
    \label{eq:kl_local_same}
\end{equation}
Using an alternate definition of the spin-cat codes,
\begin{equation}
\begin{aligned}
\ket{\pm }=\frac{\mathds{1}\pm\exp(i\pi J_y)}{\sqrt{2}}\ket{J,-J},
\end{aligned}
\label{eq:basis_states}
\end{equation}
\cref{eq:kl_local_same} transforms into a compact expression:
\begin{equation}
\begin{aligned}
    &\bra{J,J } J_x^lJ_y^mJ_z^n J_x^{l'}J_y^{m'}J_z^{n'}\ket{J,-J}\\
    &=\bra{J,-J } J_x^lJ_y^mJ_z^n J_x^{l'}J_y^{m'}J_z^{n'}\ket{J,J}=0.
\end{aligned}
\label{eq:error_condition_3}
\end{equation}
Plugging the ladder operators,
\begin{equation}
\begin{aligned}
J_{+}&=J_x+i J_y\\
J_{-}&=J_x-i J_y
\end{aligned}
\label{eq:ladder_operators}
\end{equation}
into \cref{eq:error_condition_3}, and using the condition that  one needs at least $2J-1$ operations of $J_{+}$ or $J_{-}$ to make the overlap between the states $\ket{J,J}$ and $\ket{J,-J}$ non-zero, the error correction condition in \cref{eq:kl_local_same} simplifies to,
\begin{equation}
    l+m+n+l'+m'+n' \leq 2J-1.
\end{equation}
Thus we can correct the errors of the form $J_x^lJ_y^mJ_z^n$ if 
\begin{equation}
    l+m+n\leq \lfloor\frac{2J-1}{2}\rfloor.
\end{equation}

The second case for \cref{eq:kl_diagonal} is when the two error operators $E_a$ and $E_b$ act on different physical systems. For the angular momentum errors this simplifies to,
\begin{equation}
    \begin{aligned}
        &\bra{+}J_x^lJ_y^mJ_z^n\ket{+} \bra{+}J_x^{l'}J_y^{m'}J_z^{n'} \ket{+}\\
        &=\bra{-}J_x^lJ_y^mJ_z^n\ket{-} \bra{-}J_x^{l'}J_y^{m'}J_z^{n'} \ket{-}.
    \end{aligned}
\end{equation}

\noindent Again using the \cref{eq:basis_states} and \cref{eq:ladder_operators}, the error correction condition is given as:
\begin{equation}
    \begin{aligned}
        l+m+n &\leq 2J-1,\\
         l'+m'+n' &\leq 2J-1.
    \end{aligned}
\end{equation}

\noindent Hence the  spin-cat encoding can correct all the errors of the form,
\begin{equation}
\begin{aligned}
\mathcal{E}_K=
\left\{
J_x^{l}J_y^{m}J_z^{n}; 0\leq l+m+n\leq K=\lfloor\frac{2J-1}{2} \rfloor\right\}.
\end{aligned}
\end{equation}

\section{Action of the SU(2) operators}
\label{sec:coefficients_rotation}
The Euler angle representation of an $\mathrm{SU}(2)$ operator  $V=\exp(-i \theta \hat{n}.\mathbf{J})$ is,
\begin{equation}
    V(\alpha,\beta,\gamma)=\exp(-i \theta \hat{n}.\mathbf{J})= e^{-i\alpha J_z} e^{-i\beta J_y} e^{-i\gamma J_z}.
\end{equation}
The Wigner $D$ matrix defined in \cref{eq:Wigner_D_matrix} can be expressed in terms of Euler angles as,
\begin{equation}
    \begin{aligned}
        D_{q,q'}(\alpha,\beta,\gamma)&=\bra{k,J_z=q'} \exp(-i \theta \hat{n}.\mathbf{J}) \ket{k,J_z=q}\\
        &=e^{-iq'\alpha} d_{q,q'}(\beta) e^{-iq \gamma}.
    \end{aligned}
\end{equation}
Hence, deriving from the definitions of the spherical tensor operators in \cref{eq:spherical_tensor_operators}, the operators in \cref{eq:basis}, and the inherent properties of the Wigner $d$ matrices,
\begin{equation}
    d_{q, q'}=(-1)^{q-q'}d_{-q,-q'},
\end{equation}
we find the action of an SU(2) rotation acting on the error operator, Eq. (\ref{eq:basis}) is
\begin{equation}
    \begin{aligned}
       & V S^{(k)}_q V^{\dagger}\\
       &= \sum_{q'}\frac{f_{q,q'}(\vec{\theta})}{\sqrt{2}} \left(T^{(k)}_{q'}+(-1)^{q-q'+k}e^{-2i\left(q\alpha+q'\beta\right)}T^{(k)}_{-q'}\right),\\
       &=\sum_{q'} f_{q,q'} S^{(k)}_q+ \frac{\widetilde{f}_{q,q}}{2} \left({F}^{(k)}_q-A^{(k)}_q\right).
    \end{aligned}
\end{equation}
where to lighten the notation we defined,
\begin{equation}
    \widetilde{f}_{q,q}=(-1)^k \left[1-(-1)^{q-q'}e^{-2i\left(q\alpha+q'\beta\right)}\right] f_{q,q'}.
\end{equation}
Thus,
\begin{equation}
    \begin{aligned}
        V S^{(k)}_q V^{\dagger}&=\sum_{q'} g_{q,q'} S^{(k)}_q+ \widetilde{g}_{q,q}A^{(k)}_q
    \end{aligned}
\end{equation}
where we have defined,
\begin{equation}
    \begin{aligned}
         g_{q,q'}&= f_{q,q'}+\frac{\widetilde{f}_{q,q'}}{2},\\
           \widetilde{g}_{q,q'}&=-\frac{\widetilde{f}_{q,q'}}{2}.
    \end{aligned}
\end{equation}
Similarly,
\begin{equation}
    \begin{aligned}
        V A^{(k)}_q V^{\dagger}&=\sum_{q'} h_{q,q'} S^{(k)}_q+ \widetilde{h}_{q,q}A^{(k)}_q
    \end{aligned}
\end{equation}
where again for simplification of notation,
\begin{equation}
    \begin{aligned}
         h_{q,q'}&= \frac{(-1)^k \left[1+(-1)^{q-q'}e^{-2i\left(q\alpha+q'\beta\right)}\right] f_{q,q'}}{2},\\
           \widetilde{h}_{q,q'}&=f_{q,q'}-h_{q,q'}.
    \end{aligned}
\end{equation}
Thus the action of the $\mathrm{SU}(2)$ does not change the rank of the error operators, $ A^{(k)}_q, {F}^{(k)}_q$ and obey the condition given in  \cref{eq:fault_tolerant_condition}.

\section{Rotating the ground and excited manifold differently using optimal control}
\label{sec:Rotating_the_ground_and_excited_manifold}
To implement the rank-preserving CNOT gate in \cref{fig:fig_cnot}, one needs to implement $X=\exp(-i\pi J_x)$ gate on the auxiliary manifold while applying the identity operator on the Rydberg manifold.  
For the specific choice of auxiliary and Rydberg states considered, we have the Hamiltonian in the rotating field as given by \cref{eq:Hamiltonians}.
As we are dealing with $\mathrm{SU}(2)$ representations of the spin $J$, the problem is isomorphic to the simultaneous control of two two-level systems/two qubits with different Rabi frequencies and different detuning. 
The objective would be to apply a Pauli $X$ operation on the first qubit and identity on the second system.
This problem has a Quantum-Speed-Limit(QSL) of $\pi/\Omega_{\mathrm{rf}}$\cite{Qcontrol_QSL_Hegerfeldt}.

Since $\Omega_{\mathrm{r}}=2\Omega_{\mathrm{a}}$, a pulse of length $\pi/\Omega_{\mathrm{a}}$ would cause a full Rabi rotation in the Rydberg manifold and only a half rotation in the auxiliary manifold. 
By choosing the phases of {N such} pulses, $\vec{\phi}$ in \cref{eq:Hamiltonians}, 
 one can use quantum optimal control algorithms to implement the desired transformation.
The minimum number of pulses $N$ required depends on the ratio $\omega_0/\Omega_{\mathrm{rf}}$. 
While a solution with $N=2$ only exists when $\omega_0/\Omega_{\mathrm{rf}}=3$, a solution with $N=3$ is possible if $\omega_0/\Omega_{\mathrm{rf}}<3\sqrt{3}$ for example the case of $\omega_0=5\Omega_{\mathrm{rf}}$ and $T=3\pi/\Omega_{\mathrm{rf}}$ is given in \cref{fig:bloch_3_variable}.
The overall trend is that with an increasing ratio $\omega_0/\Omega_{\mathrm{a}}$, we need a larger $N$. This protocol is similar to \cite{Levine_Pichler_gate}, and takes $N\pi/\Omega_{\mathrm{a}}$, which is longer than the QSL. We can use waveforms with a large number of steps to implement a gate in the minimum time $\pi/\Omega_{\mathrm{rf}}$, as shown in the example below.
\vspace{0.2cm}
\begin{figure}[!ht]
    \centering
   \includegraphics[width =1\columnwidth]{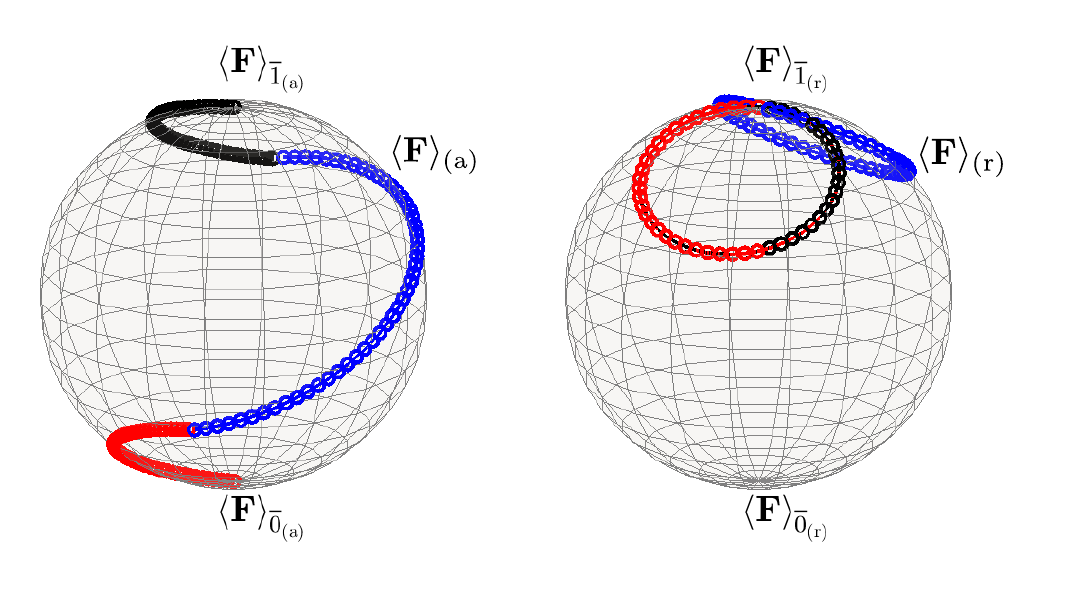} 
    \caption{ Evolutions of the spin vector $\langle \vec{F} \rangle$ for the auxiliary (a) and Rydberg (r) manifolds resulting from rf-driven Larmor precession with time-varying phases in \cref{eq:Hamiltonians} for piecewise constant function with $3$ time steps with a total time  $T_{\mathrm{tot}}=3\pi/\Omega_{\mathrm{rf}}$ and $\omega_0=5 \Omega_{\mathrm{rf}}$. 
    For the specific choice of parameters, an $X$ gate acts on the auxiliary manifold and transfers the population from $\overline{0}_{\mathrm{a}}$-subspace to $\overline{1}_{\mathrm{a}}$-subspace and vice-versa.
However, for the  Rydberg manifold, the pulse sequence acts as an identity operator, and the population in the  $\overline{0}_{\mathrm{r}}$ and  $\overline{1}_{\mathrm{r}}$ subspaces remain unaffected.}
    \label{fig:bloch_3_variable}
\end{figure}

 Using the  Hamiltonians in \cref{eq:Hamiltonians}, one can also optimize the phase $\phi$ to implement a gate $R(\theta)=\exp(-i\theta \hat{\bm{n}}.\mathbf{J})$ in the auxiliary manifold and identity on the Rydberg manifold. 
For example, the pulse scheme for the $R=\exp(i\pi J_z)$ for the auxiliary manifold, which can be used to implement the rank-preserving $\mathrm{CZ}$ gate is given in \cref{fig:zz_pulse}.
The total time is $\Omega_{\mathrm{rf}}T=\pi$ and total time is divided into $10$ equal time steps with  $\omega_0=3\Omega_{\mathrm{rf}}$.

Finally, for $\omega_0\gg\Omega_{\mathrm{rf}}$, a field that is resonant for the auxiliary spin will be far off-resonant for the Rydberg manifold. So we can implement any desired transformation $\mathrm{SU}(2)$ operation in the auxiliary subspace without disturbing the Rydberg manifold populations.
\begin{figure}
   \centering
\includegraphics[width=\columnwidth]{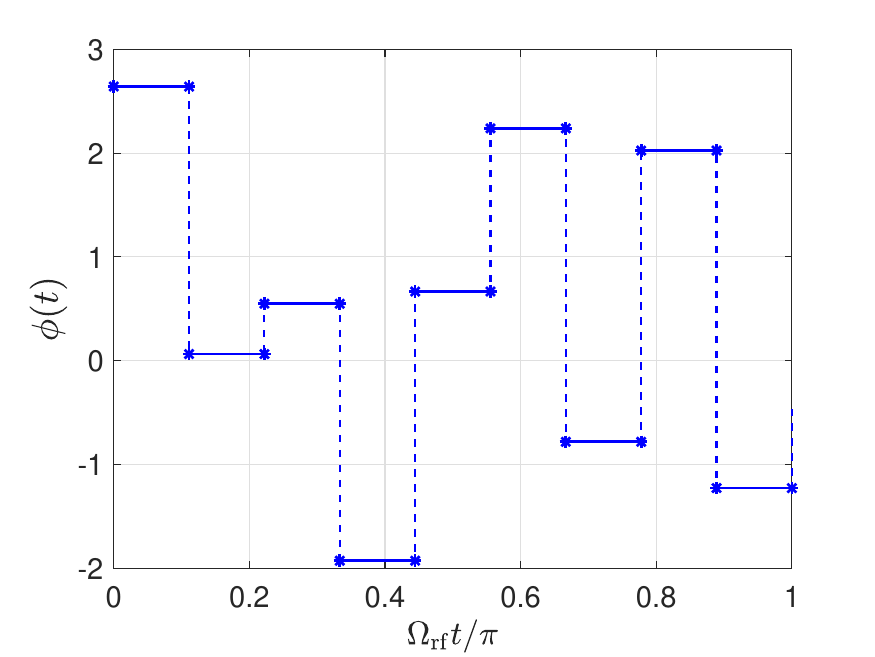}
\caption{ The phase $\phi(t)$ which generates an $R=\exp(i\pi J_z)$ for the auxiliary manifold and an identity in the Rydberg manifold, which can be used to implement the rank-preserving $\mathrm{CZ}$ gate.
The total time is $\Omega_{\mathrm{rf}}T=\pi$, which is divided into $10$ equal time steps with  $\omega_0=3\Omega_{\mathrm{rf}}$ and pulse sequence is found using the quantum optimal control {algorithm} GRAPE.}
\label{fig:zz_pulse}
\end{figure}

 \section{Implementing Hadamard gate from the Physical level gates}
 \label{sec:Hadamard_gate}
 \begin{figure}
    \centering
    \includegraphics[width=0.8\columnwidth]{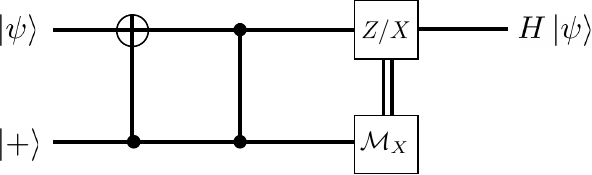}
    \caption{Circuit implementing a fault-tolerant Hadamard gate using the physical level gates for the spin-cat encoding.
    This differs from the standard implementation as we use both CNOT and CZ gate to implement the action of the target unitary of interest. }
    \label{fig:hadamard_gate}
\end{figure}
The  physical level gates for the spin-cat encoding are given as,
 \begin{equation}
    \{\mathcal{M}_z,\mathcal{M}_X,\mathcal{P}_{\ket{+}},\mathcal{P}_{\ket{0}},\mathrm{CNOT},X,Y,Z,ZZ(\theta)\}.  
 \end{equation}
 The Hadamard gate is not in the universal gate set as it does not preserve the rank. Here we show the implementation of the Hadamard gate using the rank-preserving physical level gates and an ancilla qubit.
 The circuit diagram corresponding to a teleportation-based scheme for the Hadamard gate is given in the \cref{fig:hadamard_gate}. 
Consider an initial arbitrary state,
 \begin{equation}
     \ket{\psi}=\alpha \ket{0}_k+\beta\ket{1}_k,
 \end{equation}
 and ancilla state,
 \begin{equation}
     \ket{+}_0=\frac{1}{\sqrt{2}}\left(\ket{0}+\ket{1}\right).
 \end{equation}
 Define $\ket{\phi}=\ket{\psi}\otimes \ket{+}_0$, then
 \begin{equation}
 \begin{aligned}
      &\mathrm{CNOT}\ket{\phi}= \mathrm{CNOT}\ket{\psi}\otimes \ket{+}_0\\
      &=\frac{1}{\sqrt{2}}\left(\alpha \ket{0}_k\ket{0}+\alpha \ket{1}_k\ket{1}+\beta \ket{1}_k\ket{0}+\beta \ket{0}_k\ket{1}\right),
 \end{aligned}    
 \end{equation}
and
 \begin{equation}
 \begin{aligned}
    \mathrm{CZ}\text {  } \mathrm{CNOT}\ket{\phi}&=\left(\alpha \ket{-}_k+\beta\ket{+}_k\right)\ket{+}\\
     &+\left(\alpha \ket{+}_k-\beta\ket{-}_k\right)\ket{-}.
 \end{aligned}     
 \end{equation}
 Thus one can act $Z$ or $X$ gate depending on the measurement of the $X$ operator in the ancilla to get the state,
 \begin{equation}
     H\ket{\psi}=\alpha \ket{+}_k+\beta\ket{+}_k,
 \end{equation}
and implement the action of the Hadamard gate.

\section{Implementing the Logical operator}
\label{sec:implementing_the_logical_operator}

 \begin{figure*}[!ht]
 \captionsetup[subfigure]{oneside,margin={3.7cm,0cm}}
    \centering
    \subfloat[]{\includegraphics[width =0.4\columnwidth]{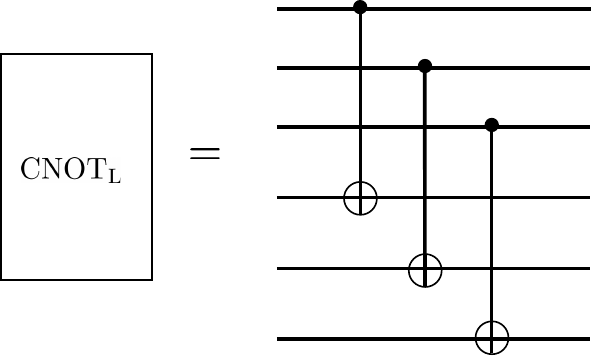} \label{fig:Fig_19_a}}\hspace*{1.9em}
    \subfloat[]{\includegraphics[width =.55\columnwidth]{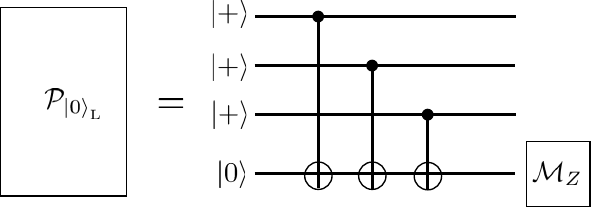}  \label{fig:Fig_19_b}}\\
    \vspace{1.5em}
    \subfloat[]{\includegraphics[width =.45\columnwidth]{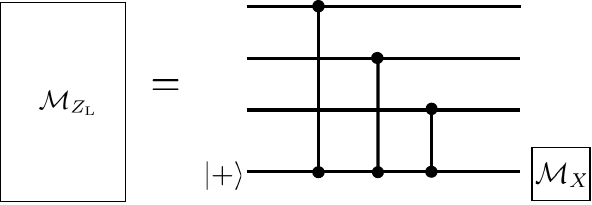} \label{fig:Fig_19_c}}
    \hspace*{1.9em}
    \subfloat[]{\includegraphics[width =0.45\columnwidth]{logical_mz.pdf}  \label{fig:Fig_19_d}}
    \caption{Circuits implementing logical level gates in $\mathcal{C}_1$ using the physical level gates.
    (a)  {Logical $\mathrm{CNOT}$ $\mathrm{CNOT}_{\mathrm{L}}$. To implement $\mathrm{CNOT}_{\mathrm{L}}$, we apply physical $\mathrm{CNOT}$ gates transversally on all qubit pairs}.
(b) Preparation of $\ket{0}_{\mathrm{L}}$ $\mathcal{P}_{\ket{0}_{\mathrm{L}}}$. $\ket{0}_{\mathrm{L}}$ is prepared by initializing the system with the state $\mathcal{P}_{\ket{+}_{\mathrm{L}}}$ and measuring the parity.
To measure the parity we use an ancilla initialized with $\mathcal{P}_{\ket{0}}$ and use physical CNOT gates followed by measuring the $\mathcal{M}_Z$, the final state  is  $\ket{0}_{\mathrm{L}}$ or $\ket{1}_{\mathrm{L}}$ for the measurement outcomes $1$ and $-1$ respectively. 
(c) {The Logical $Z$ measurement $\mathcal{M}_{Z_{\mathrm{L}}}$}. An ancilla state is prepared in $\ket{+}$ and physical CZ gates {with the data qubits are applied} followed by measuring the ancilla in the $X$ basis.
(d) Logical $X$ measurement $\mathcal{M}_{X_\mathrm{L}}$. 
The logical $X$ is measured by applying the physical CNOT gates and then measurement along $X$.}
    \label{fig:logical_gate}
\end{figure*}
In this section, we demonstrate the universal gate set at the logical level with the physical level gates for the spin-cat encoding. 
The rank-preserving physical level gates for the spin-cat encoding are,
\begin{equation}
    \{\mathcal{M}_Z,\mathcal{M}_X,\mathcal{P}_{\ket{+}},\mathcal{P}_{\ket{0}},\mathrm{CNOT},X,Y,Z\}.
\end{equation}
Consider a universal gate set,
\begin{equation}
    \{\mathcal{M}_{Z_\mathrm{L}},\mathcal{M}_{X_\mathrm{L}},\mathcal{P}_{\ket{+}_\mathrm{L}},\mathcal{P}_{\ket{0}_\mathrm{L}},\mathrm{CNOT}_{\mathrm{L}}\},
    \label{eq:logical_level_gate_app}
\end{equation}
where $\mathcal{P}$ refers to preparation and $\mathcal{M}$ denotes measurement.
The  logical preparation of the $\mathcal{P}_{\ket{+}_\mathrm{L}}$ can be done transversally by preparing the $\mathcal{P}_{\ket{+}}$ in the individual systems.
For example in the case of three physical systems, the logical level state preparation is,
\begin{equation}
    \begin{aligned}
         \mathcal{P}_{\ket{+}_\text{L}}= \ket{+}_\text{L}&=\ket{+++}.
              \end{aligned}
\end{equation}
In a similar fashion, the construction of additional logical-level gates follows the approach detailed in \cite{Aliferis2008fault,puri2020bias}.
Comprehensive details for the implementation of all other logical gates are provided in \cref{fig:logical_gate}.
In (a), the $\mathrm{CNOT}_{\mathrm{L}}$ is implemented using the physical CNOT gates.
One can implement the $\mathrm{CNOT}_{\mathrm{L}}$ by transversal application of the CNOT gates.
In (b), the $\mathcal{P}_{\ket{0}_{\mathrm{L}}}$ is prepared by initializing the system with the state $\mathcal{P}_{\ket{+}_{\mathrm{L}}}$ and measuring the parity.
To measure the parity we use an ancilla initialized with $\mathcal{P}_{\ket{0}}$ and use physical CNOT gates followed by measuring the $\mathcal{M}_Z$, the final state  is  $\mathcal{P}_{\ket{0}_{\mathrm{L}}}$ and $\mathcal{P}_{\ket{1}_{\mathrm{L}}}$ for the measurement outcomes $1$ and $-1$ respectively. 
(c) implements the logical measurement of $Z$ with an ancilla state prepared in $\ket{+}$ and physical CZ gates followed by measuring the ancilla in the $X$ basis.
Finally (d) implements the logical measurement by applying the physical CNOT gates and measurement of $X$.

\section{Toffoli gate}
\label{sec:Toffoli_gate}

\begin{figure}
    \centering
    \includegraphics[width=\columnwidth]{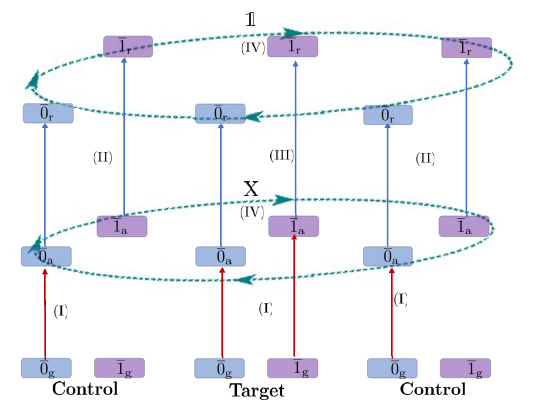}
    \caption{  Protocol for  a rank-preserving Toffoli {gate} for spin-cat encoding using $\mathrm{SU}(2)$ operations. 
 Similar to the rank-preserving CNOT gate \cref{fig:fig_cnot}, we implement the Toffoli gate in the ground state of $^{87}$Sr and the physical setting is the same as given in \cref{fig:Fig_2_a}.
We consider a geometry of atoms such that the nearest neighbors are constrained by the Rydberg blockade, but the next-nearest neighbors are not constrained. 
In step I the population is promoted to the auxiliary manifold in the atoms. 
{In} the control atoms we only promote the population of the $\overline{0}$-subspace whereas for the target atom, the population from both the $\overline{0}$ and $\overline{1}$ subspaces are promoted to the auxiliary state.
In step II, we transfer the population between the auxiliary and the Rydberg manifolds of the control atoms.
In step III, we transfer the population from the auxiliary to the Rydberg manifold of the target atom. However, due to the Rydberg blockade, this population transfer only happens when both the control atoms are in $\overline{0}$-subspaces. 
If even one of the control atoms is in $\overline{1}$-subspace this transition is blockaded. 
Then similar to the rank-preserving CNOT gate, in step IV we implement aa $X=\exp(-i\pi J_x)$ gate in the auxiliary manifold and an identity operator in the Rydberg manifold.   
Finally, we will transfer all the states back to the ground state by acting steps III-I in reverse, thus implementing a rank-preserving Toffoli gate for the spin-cat encoding.}
    \label{fig:fig_toffoli}
\end{figure}
One can generalize the rank-preserving CNOT gate in \cref{fig:fig_cnot} to construct a Toffoli gate, also known as a controlled-controlled NOT gate. 
\cref{fig:fig_toffoli} gives the protocol for creating the rank-preserving Toffoli gate for the spin-cat encoding using only $\mathrm{SU(2)}$ interactions. 
Again, similar to the rank-preserving CNOT gate, the Toffoli gate is implemented in the ground state of $^{87}$Sr.
The key to the scheme is the availability of special geometries for the neutral atoms \cite{Levine_Pichler_gate,bluvstein2022quantum}. 
Here we use a geometry such that {for three linearly arranged atoms,} the nearest neighbors are constrained by the Rydberg blockade, but the next-nearest neighbors are not constrained by it.
{The central atom acts as the target atom while its two neighbors are the control atoms.}

In step I of the Toffoli gate, the population is promoted to the auxiliary state. For the case of the control atoms we only promote the population of the $\overline{0}$-subspace whereas for the target atom, the population from both the $\overline{0}$ and $\overline{1}$ subspaces are promoted to the auxiliary state.
In step II, we use a pulse sequence similar to the \cref{fig:Fig_waveform_c} to transfer the population between the auxiliary and the Rydberg state of the control atoms using \sout{a} $\pi$ polarized light.
In step III we apply the same pulse sequence as in step II to the target atom however, due to the Rydberg blockade, the population transfer between the auxiliary and Rydberg state only happens when both the control atoms are in $\overline{1}$-subspace. 
Then similar to the case of the rank-preserving CNOT gate in step IV, we implement a $X=\exp(-i\pi J_x)$ gate in the auxiliary manifold and an identity operator in the Rydberg manifold.  
Finally, we will transfer all the states back to the ground state by acting steps III-I {in reverse,} thus implementing a rank-preserving Toffoli gate for the spin-cat encoding up to local rotations.

Thus when one of the control atoms is in the $\overline{0}$-subspace, $X$ gate is applied target atom, and when both the control atoms are in the $\overline{1}$-subspace, the target atoms remain unchanged.
This is the Toffoli gate up to a local {$X=\exp(-i\pi J_x)$} rotation on the target atom.

\section{Alternate approaches for cat-state preparation and measurement of $X$ }
One can use alternative approaches than quantum optimal control for cat-state preparation and measurement of $X$. In this section, we detail some of those approaches. 
For example, one can use an adiabatic approach and one-axis twisting to create a spin-cat state.

I) Adiabatic approach. Starting with an initial state $\ket{J,J_z=J}$ and evolving the Hamiltonian 
\begin{equation}
    H(s)=(1-s)J_x-\frac{s}{2J}J_z^2,
\end{equation}
adiabatically {from $s=0$ to $s=1$} guarantees the final state to be close to a cat state $\ket{+}$ \cite{puri2017engineering}. 
This can be implemented in atomic systems using a combination of tensor light shifts and rf rotation~\cite{Paul_experiment_Cs_2007,omanakuttan2021quantum}.

 II) One-axis twisting: Using a time-independent Hamiltonian, $H=\beta J_z^2$, for a certain time $T=\pi/(2\beta)$, one can evolve a spin coherent state along $J_x$ to prepare a high-fidelity cat state.
\begin{equation}
    \ket{+}=\exp(-i\pi J_x)\exp(-i\frac{\pi}{2}J_z^2)\ket{J,J_x=J}.
\end{equation}

Including the effect of decoherence due to photon scattering and optical pumping for  $^{87}$Sr, we find the fidelity for one-axis twisting is $0.9998$ whereas for the adiabatic preparation, one can achieve a fidelity of $0.9889$.

Similarly one can use an alternative approach to measure $X$, in particular, to know if the ancilla state is in $\ket{+}$ or $\ket{-}$. 
We can adiabatically rotate the states using the Hamiltonian
\begin{equation}
    H(s)=-(1-s)J_z^2/(J)+s J_x,
\end{equation}
which implements the following transformations:
\begin{equation}
    \begin{aligned}
        \ket{+}_0&\to \ket{J,J_x=J},\\
        \ket{-}_0 &\to \ket{J,J_x=J-1},
    \end{aligned}
    \label{eq:M_x_transformation_adiabatic}
\end{equation}
and then then measuring  $J_x$.

To evaluate the accuracy of $X$ measurement, we define the target isometry as: 
\begin{equation}    V_{\mathrm{targ}}=\ket{J,J_x=J}\bra{+}+\ket{J,J_x=J-1}\bra{-}.
\end{equation}
The implemented isometry using the adiabatic approach is given as,
\begin{equation}
    V= e^{-\int \mathcal{L}(s)ds} V(0)
\end{equation}
where $\mathcal{L}(s)$ is the Lindbladian including the effects of decoherence and 
\begin{equation}
    V(0)=\ket{+}\bra{+}+\ket{-}\bra{-}.
\end{equation}
Thus the fidelity for the implementation of the isometry is defined as:
\begin{equation}
    \mathcal{F}_{\mathrm{iso}}=\frac{1}{4}\lvert\Tr(V_{\mathrm{targ}}V^{\dagger})\rvert^2.
    \label{eq:accuray_equation_app}
\end{equation}
This approach is similar to the approach taken in bosonic cat qubits \cite{puri2019stabilized}. To measure $J_x$, we first implement the unitary transformation $U=\exp(-i\pi/2 J_y)$ to rotate the basis to $\ket{J,J_z}$ and then perform the readily accessible measurement $\mathcal{M}_Z$ which we can in principle achieve with a fidelity larger than $99\%$ \cite{barnes2021assembly}. 
Including the effects of optical pumping as discussed in \cref{sec:Ratio_optical_pumping_errors}, one can implement this transformation with a fidelity of $\mathcal{F}_{\mathrm{iso}}=0.98$ for the $^{ 87}$Sr nuclear spin qudit.

\section{Error correction without measurement }
\label{sec:measurement_free_ec}

An alternative to syndrome-based quantum error correction is measurement-free quantum error correction (MFQEC) \cite{PhysRevLett.117.130503,PhysRevA.97.012318,li2011recovery,premakumar2020measurement}.
The standard syndrome-based error correction is given by recovery operation:
\begin{equation}
     R(\rho)=\sum_i U_i M_i \rho M_i^{\dagger} U_i^{\dagger},
\end{equation}
where for a general state $\rho$, $M_i$ is the syndrome measurement and $U_i$ is the correction unitary according to the outcome of the syndrome measurement. 

MFQEC is based on the unitary operator $V$, which couples the data and ancilla qubits.
The action of which is given as,
\begin{equation}
    V\ket{\psi} \ket{0}=\sum_i \left(U_i M_i \otimes \mathds{1} \right)\ket{\psi} \ket{i}.
\end{equation}
Defining $\rho=\sum_{kl}\alpha_{kl}\ket{\psi}_k\bra{\psi}_l$, we can find that,
\begin{equation}
    V\rho \otimes \ketbra{0}{0} V^{\dagger}=\sum_{k,l,i,j}\alpha_{kl}U_iM_i\ket{\psi}_k\bra{\psi}_l M_j^{\dagger} U_j^{\dagger} \otimes \ketbra{i}{j}
\end{equation}
 Partial tracing of the ancilla gives,
\begin{equation}
    \rho_{\mathrm{rec}}=\sum_{i}U_i M_i \rho M_i^{\dagger} U_i^{\dagger}.
\end{equation}
Thus the MFQEC is equivalent to syndrome-based error correction and the key for MFQEC is a specific unitary gate between the ancilla and the data.

One can consider a fault-tolerant  MFQEC scheme for the amplitude errors.
The syndrome for the amplitude errors is the eigenvalue of $J_z^2$, which can be extracted by the projective measurement, 
\begin{equation}
    M_k=\ket{+}_k\bra{+}_k+\ket{-}_k\bra{-}_k,
\end{equation}
where $0\leq k\leq (2J-1)/2$.
Recovery unitaries corresponding to  the projective measurement outcomes are
\begin{equation}  
\begin{aligned}    U_k&=\ket{+}_0\bra{+}_k+\ket{+}_k\bra{+}_0+\ket{-}_0\bra{-}_k+\ket{-}_k\bra{-}_0\\
&+\sum_{j\neq k, j\neq 0} \ket{+}_j\bra{+}_j+\ket{-}_j\bra{-}_j,
\end{aligned}
\end{equation}
which takes the state from the subspace,
\begin{equation}
    \{\ket{+}_k, \ket{-}_k\} \to \{\ket{+}_0, \ket{-}_0\}.
\end{equation}
Consider the following unitary operator, using the definitions from \cref{eq:projectors,eq:logical_paulis,eq:cnot_def} the product of three alternating CNOT gates can be written as:
\begin{equation}
\begin{aligned}
     V_s&=\Pi_{\overline{0}}\otimes \Pi_{\overline{0}}+\Pi_{\overline{1}}\otimes \Pi_{\overline{1}}\\
     &+X\Pi_{\overline{0}} \otimes X\Pi_{\overline{1}} +X\Pi_{\overline{1}}\otimes X\Pi_{\overline{0}},
\end{aligned}   
\label{eq:V_s_gate}
\end{equation}
Consider the following states, 
\begin{equation}
    \begin{aligned}
        \ket{\psi}_k&=\alpha \ket{+}_k+\beta\ket{-}_k,\\
        \ket{\phi}_l&=\gamma \ket{+}_l+\delta\ket{-}_l,\\
    \end{aligned}
\end{equation}
where $\alpha,\beta, \gamma, \text{ and } \delta$ are arbitrary complex amplitudes. 
The action of the $V_s$ on the state, $\ket{\xi}=\ket{\psi}_k\otimes \ket{\phi}_l$ gives,
\begin{equation}
    V_s \ket{\xi}=\ket{\phi}_k\otimes \ket{\psi}_l.
    \end{equation}
 Thus $V_s$ gate swaps the information between two kitten or cat states.
 The circuit diagram for the $V_s$  gate for a qubit encoded in the qudit is given in \cref{fig:Fig_swap_a}.

When the second qudit is prepared in  $\ket{+}_0$ state, as shown in \cref{fig:Fig_swap_b}, 
the application of the $V_s$ gate gives
\begin{equation}
   \ket{\phi}= V_s\ket{\psi}\ket{+}_0=\ket{+}_k \otimes \left(\alpha\ket{+}_0+\beta \ket{-}_0\right). 
\end{equation}
The above state can also be written as,
\begin{equation}
    \ket{\phi}=\sum_k U_k M_k \ket{\psi} \ket{+}_k,
\end{equation}
where the notion of data and ancilla qubits are swapped for convenience. 
Thus the unitary operator $V_s$ followed by partial tracing implements the desired recovery operation.
Thus one can correct the amplitude error fault tolerantly using a combination of two rank-preserving CNOT gates and fresh $\ket{+}_0$ state.

For fault-tolerant gadgets, one needs to repeat the phase and amplitude error correction multiple times and one needs to ensure that these two error correction steps commute with each other.
The  phase error correction \cref{fig:circuit_phase_error_correction}  commutes with measurement-free error correction of the amplitude error and the details of the calculation are given in 
\cref{sec:commutativity_of_error_operators}.

\subsection{Upper bounds on the probability of
the logical error in the amplitude error correction}
\label{sec:upper_bounds_amplitude}
In this section, we provide a detailed analysis to find an upper bound on the probability of a logical error in the amplitude error correction used in the error-corrected logical CNOT gadget in \cref{fig:fault_tolerant_cx_gadget}.

First, consider the case where ancilla is prepared perfectly, i.e., we have $\rho_A=\ket{+}_0$ and $p_i=0$ for $i\neq 0$ in \cref{fig:swapping_animation}. In this case, a logical amplitude error occurs after $s$ faulty CNOT gates if they create at least $k_{\mathrm{max}}=\lfloor (2J+1)/2\rfloor$ many jumps, the probability of which we denote by $q(s,k_{\mathrm{max}})$. The number of CNOT gates $s$ is determined by the number of phase error corrections that appear before an amplitude correction, in addition to the two CNOT acting in the amplitude error correction itself. To find the probability $q(s,k_{\mathrm{max}})$, we note that each physical CNOT gate can create one or two jumps with probabilities $p_1$ and $p_2$ respectively, and therefore we need to add the probabilities of cascades of one and two jumps that can create more than $k_{\mathrm{max}}$ jumps. Therefore $q(s,k_{\mathrm{max}})$ can be written as
\begin{equation}
    q(s,k_{\mathrm{max}})=\sum_{i}\lambda_i(s,k_{\mathrm{max}}),
    \label{eq:amplitude_error_ideal}
\end{equation}
where $\lambda_i$ represents the probability of one path such that we have at least $k_{\mathrm{max}}$ jumps. For example, consider the case of $s=4$ and $J=9/2$, then $\lambda_i$ represents all the possible combinations of one and two jumps, such that the total sum of these jumps is at least $5$. 
One such possibility is a combination of $(1,1,1,2)$ where we have one jump occurring at the first three CNOTs and two jumps occurring at the last CNOT.



When the ancilla is imperfect, for example, if it is prepared in $\ket{+}_k$ state rather than $\ket{+}_0$, one needs to find the paths that create $k_{\mathrm{max}}-k$ many jumps. Thus we get, 
\begin{equation}
q(s,k_{\mathrm{max}}|k)=\sum_i \lambda_i(s,k_\mathrm{max}|k),
\end{equation}
where $\lambda_i(s,k_{\mathrm{max}}\lvert k)$ is the probability of a path where we have at least $k_{\mathrm{max}}$ jumps given that we already had $k$ jumps to start with.

We repeat the amplitude error correction $r_2$ many times in one error-corrected logical CNOT gate. Thus the upper bound of the logical amplitude error probability after $r_2$ rounds of error correction in \cref{fig:fault_tolerant_cx_gadget} is,
\begin{equation}
\epsilon^{\mathrm{amp}}=r_2\left(\sum_{k=0}^{4}q(s,k_{\mathrm{max}}|k)p_k\right).
\end{equation}
where $p_k$ is the probability of ancilla starting at $\ket{+}_k$.

\vspace{0.7cm}
\section{Commutativity of the Error correction steps}
\label{sec:commutativity_of_error_operators}

    The error correction for the spin-cat encoding follows two steps. 
The first step is the phase error correction in \cref{fig:circuit_phase_error_correction} and the second step is the measurement-free error correction for correcting amplitude errors given in \cref{fig:SWAP_gate_qubit_1}.
For fault-tolerant gadgets, one needs to repeat these steps multiple times and we need to ensure that these two error correction steps commute with each other such that the errors do not proliferate uncontrollably.
For this, we  need to satisfy,
\begin{equation}
    \mathcal{R}_{\mathrm{amp}} \mathcal{R}_{\mathrm{ph}}\left(\mathcal{E}\left(\rho\right)\right)= \mathcal{R}_{\mathrm{ph}} \mathcal{R}_{\mathrm{amp}}\left(\mathcal{E}\left(\rho\right)\right),
\end{equation}
where $\mathcal{R}_{\mathrm{amp}}, \mathcal{R}_{\mathrm{ph}}$ are the recovery maps corresponding to the amplitude and phase error correction respectively.
The recovery map for the amplitude error can be expressed in terms of the Kraus operators as,
\begin{equation}
    \mathcal{R}_{\mathrm{amp}}(\rho)=\sum_{j,i} M_{j,i}^{\mathrm{amp}} \rho \left(M_{j,i}^{\mathrm{amp}}\right)^{\dagger},
\end{equation}
where,
\begin{equation}
\begin{aligned}
    M^{\mathrm{amp}}_{j,1}&=\left( \bra{j}^{(1)} V_s \ket{+}_0^{(2)}\right)\otimes \mathds{1} \otimes \mathds{1},\\
      M^{\mathrm{amp}}_{j,2}&=\mathds{1} \otimes\left( \bra{j}^{(1)} V_s \ket{+}_0^{(2)}\right)\otimes  \mathds{1},\\
      M^{\mathrm{amp}}_{j,3}&=\mathds{1} \otimes  \mathds{1} \otimes\left( \bra{j}^{(1)} V_s \ket{+}_0^{(2)}\right), \\
\end{aligned} 
\end{equation}
and  $V_s$ is the unitary operator given in \cref{eq:V_s_gate}.
The Kraus operator representation of the phase error correction for spin-cat encoding is,
\begin{equation}
    \mathcal{R}_{\mathrm{ph}}(\rho)=\sum_{i,j} M_{i,j}^{\mathrm{ph}} \rho \left(M_{i,j}^{\mathrm{ph}}\right)^{\dagger}
\end{equation}
where, 
\begin{equation}
\begin{aligned}
    M_{00}^{\mathrm{ph}}&= \sum_{i,j,k}\ket{+}_i\ket{+}_j\ket{+}_k\bra{+}_i\bra{+}_j\bra{+}_k+\ket{-}_i\ket{-}_j\ket{-}_k\bra{-}_i\bra{-}_j\bra{-}_k,\\
    M_{01}^{\mathrm{ph}}&= Z_3\sum_{i,j,k}\ket{+}_i\ket{+}_j\ket{-}_k\bra{+}_i\bra{+}_j\bra{-}_k+\ket{-}_i\ket{-}_j\ket{+}_k\bra{-}_i\bra{-}_j\bra{+}_k,\\
     M_{10}^{\mathrm{ph}}&= Z_1\sum_{i,j,k}\ket{-}_i\ket{+}_j\ket{+}_k\bra{-}_i\bra{+}_j\bra{+}_k+\ket{+}_i\ket{-}_j\ket{-}_k\bra{+}_i\bra{-}_j\bra{-}_k,\\
      M_{11}^{\mathrm{ph}}&= Z_2\sum_{i,j,k}\ket{+}_i\ket{-}_j\ket{+}_k\bra{+}_i\bra{-}_j\bra{+}_k
      +\ket{-}_i\ket{+}_j\ket{-}_k\bra{-}_i\bra{+}_j\bra{-}_k.\\
\end{aligned}    
\end{equation}
To prove the commutativity of the two error correction steps first consider the Kraus operators $M_{00}^{\mathrm{ph}}$ and $M^{\mathrm{amp}}_{j,1}$, we get
\begin{equation}
    \begin{aligned}       
    M_{j,1}^{\mathrm{amp}}M_{00}^{\mathrm{ph}}=\sum_{klm} \bra{j}\ket{+}_k\left(\ket{+}_0^{(b)}\ket{+}_l^{(a)}\ket{+}_m^{(a)}\bra{+}_k^{(a)}\bra{+}_l^{(a)}\bra{+}_m^{(a)}\right)\\+\sum_{klm} \bra{j}\ket{+}_k\left(\ket{-}_m^{(a)}\bra{-}_k^{(a)}\bra{-}_l^{(a)}\bra{-}_m^{(a)}\right),\\
    \end{aligned}
\end{equation}
and,
\begin{equation}
\begin{aligned}
     M_{00}^{\mathrm{ph}}M_{j,1}^{\mathrm{amp}}&= \sum_{k,l,m}\ket{+}_k^{(b)}\ket{+}_l^{(a)}\ket{+}_m^{(a)}\bra{+}_k^{(b)}\bra{+}_l^{(a)}\bra{+}_m^{(a)}\bra{j}^{(a)}\otimes \mathds{1}^{(b)} V_s^{(ab)} \mathds{1}^{(a)}\otimes \ket{+}_0^{(b)}\\     &+\sum_{k,l,m}\ket{-}_k^{(b)}\ket{-}_l^{(a)}\ket{-}_m^{(a)}\bra{-}_k^{(b)}\bra{-}_l^{(a)}\bra{-}_m^{(a)}\bra{j}^{(a)}\otimes \mathds{1}^{(b)} V_s^{(ab)} \mathds{1}^{(a)}\otimes \ket{+}_0^{(b)}.\\ 
\end{aligned}   
\end{equation}
Using the resolution of the identity $ \mathds{1}=\sum_p \ket{+}_p\bra{+}_p+\ket{-}_p\bra{-}_p$ in the above equation yields,
\begin{equation}
    \begin{aligned}     M_{00}^{\mathrm{ph}}M_{j,1}^{\mathrm{amp}}=M_{j,1}^{\mathrm{amp}}M_{00}^{\mathrm{ph}}.
\end{aligned} 
\end{equation}
Thus these two Kraus operators commute with each other. 
Similarly, one can find that,
\begin{equation}
    \begin{aligned}
        \comm{M_{j,2}^{\mathrm{amp}}}{M_{00}^{\mathrm{ph}}}&=0,
         \comm{M_{j,3}^{\mathrm{amp}}}{M_{00}^{\mathrm{ph}}}&=0.\\
    \end{aligned}
\end{equation}
Similar calculations also give,
\begin{equation}
    \begin{aligned}
         \comm{M_{j,2}^{\mathrm{amp}}}{M_{10}^{\mathrm{ph}}}&=0,
         \comm{M_{j,2}^{\mathrm{amp}}}{M_{01}^{\mathrm{ph}}}&=0,\\
         \comm{M_{j,3}^{\mathrm{amp}}}{M_{01}^{\mathrm{ph}}}&=0,
         \comm{M_{j,3}^{\mathrm{amp}}}{M_{11}^{\mathrm{ph}}}&=0,\\
         \comm{M_{j,1}^{\mathrm{amp}}}{M_{10}^{\mathrm{ph}}}&=0,
         \comm{M_{j,1}^{\mathrm{amp}}}{M_{10}^{\mathrm{ph}}}&=0.\\
    \end{aligned}
\end{equation}

\noindent Next, consider the Kraus operators, $M_{10}^{\mathrm{ph}}$ and $M^{\mathrm{amp}}_{j,1}$ we get,
\begin{equation}
    \begin{aligned}      
    M_{j,1}^{\mathrm{amp}}M_{10}^{\mathrm{ph}}&=\sum_{klm} (-1)^k\bra{j}\ket{+}_k\left(\ket{+}_0^{(b)}\ket{+}_l^{(a)}\ket{+}_m^{(a)}\bra{-}_k^{(a)}\bra{+}_l^{(a)}\bra{+}_m^{(a)}
    \right)\\   
    &+\sum_{klm} (-1)^k\bra{j}\ket{+}_k\left(\ket{-}_0^{(b)}\ket{-}_l^{(a)}\ket{-}_m^{(a)}\bra{+}_k^{(a)}\bra{-}_l^{(a)}\bra{-}_m^{(a)}\right),\\    M_{10}^{\mathrm{ph}}M_{j,1}^{\mathrm{amp}}&=\sum_{k,l,m}\bra{j}\ket{+}_k\left(\ket{-}_0^{(b)}\ket{+}_l^{(a)}\ket{+}_m^{(a)}\bra{-}_k^{(a)}\bra{+}_l^{(a)}\bra{+}_m^{(a)}\right)\\
&+\sum_{k,l,m}\bra{j}\ket{+}_k\left(\ket{-}_0^{(b)}\ket{-}_l^{(a)}\ket{-}_m^{(a)}\bra{+}_k^{(a)} \bra{-}_l^{(a)}\bra{-}_m^{(a)}\right),\\
     &\neq M_{j,1}^{\mathrm{amp}}M_{10}^{\mathrm{ph}}.
\end{aligned} 
\end{equation}
Thus these two Kraus operators do not commute with each other, however looking at the full recovery operation, 
\begin{equation}
    \begin{aligned}
       \sum_{j}  M_{10}^{\mathrm{ph}}M_{j,1}^{\mathrm{amp}}  \rho \left(M_{j,1}^{\mathrm{amp}} \right)^{\dagger} \left(M_{10}^{\mathrm{ph}}\right)^{\dagger}
       &=\sum_j \sum_{k,l,m,k',l',m'} \bra{+}_{k'}\ket{j}\bra{j}\ket{+}_kA_{k,l,m} \rho A_{k',l',m'}^{\dagger},\\
       &=\sum_{k,l,m,l',m'} A_{k,l,m} \rho A_{k,l',m'}^{\dagger},
    \end{aligned}
    \label{eq:non_comm_1}
\end{equation}
where we have defined,
\begin{equation}    
\begin{aligned}    A_{k,l,m}&\equiv\ket{-}_0^{(b)}\ket{+}_l^{(a)}\ket{+}_m^{(a)}\bra{-}_k^{(a)}\bra{+}_l^{(a)}\bra{+}_m^{(a)}
&+\ket{-}_0^{(b)}\ket{-}_l^{(a)}\ket{-}_m^{(a)}\bra{+}_k^{(a)} \bra{-}_l^{(a)}\bra{-}_m^{(a)}.
\end{aligned}
\end{equation}
Similarly, we get,
\begin{equation}
\begin{aligned}
       \sum_{j}  M_{j,1}^{\mathrm{amp}}M_{10}^{\mathrm{ph}}  \rho  \left(M_{10}^{\mathrm{ph}}\right)^{\dagger}\left(M_{j,1}^{\mathrm{amp}} \right)^{\dagger}
    &=\sum_j \sum_{k,l,m,k',l',m'} (-1)^{k+k'}\bra{+}_{k'}\ket{j}\bra{j}\ket{+}_kA_{k,l,m} \rho A_{k',l',m'}^{\dagger},\\
    &=\sum_{k,l,m,l',m'} A_{k,l,m} \rho A_{k,l',m'}^{\dagger}.
\end{aligned}
\label{eq:non_comm_2}
\end{equation}
Combining \cref{eq:non_comm_1} and \cref{eq:non_comm_2} gives,
\begin{equation}
\begin{aligned}
      \sum_{j} M_{j,1}^{\mathrm{amp}}M_{10}^{\mathrm{ph}}  \rho  \left(M_{10}^{\mathrm{ph}}\right)^{\dagger}\left(M_{j,1}^{\mathrm{amp}} \right)^{\dagger}
      =\sum_{j}M_{10}^{\mathrm{ph}}M_{j,1}^{\mathrm{amp}}  \rho \left(M_{j,1}^{\mathrm{amp}} \right)^{\dagger} \left(M_{10}^{\mathrm{ph}}\right)^{\dagger}
\end{aligned}  
\end{equation}
Similarly one can find,
\begin{equation}
    \begin{aligned}
        \sum_{j}M_{j,2}^{\mathrm{amp}}M_{11}^{\mathrm{ph}}  \rho  \left(M_{11}^{\mathrm{ph}}\right)^{\dagger}\left(M_{j,2}^{\mathrm{amp}} \right)^{\dagger}
        &=\sum_{j}M_{11}^{\mathrm{ph}}M_{j,2}^{\mathrm{amp}}  \rho \left(M_{j,1}^{\mathrm{amp}} \right)^{\dagger} \left(M_{01}^{\mathrm{ph}}\right)^{\dagger}\\
       \sum_{j} M_{j,3}^{\mathrm{amp}}M_{10}^{\mathrm{ph}}  \rho  \left(M_{01}^{\mathrm{ph}}\right)^{\dagger}\left(M_{j,3}^{\mathrm{amp}} \right)^{\dagger}
       &=\sum_{j}M_{01}^{\mathrm{ph}}M_{j,3}^{\mathrm{amp}}  \rho \left(M_{j,3}^{\mathrm{amp}} \right)^{\dagger} \left(M_{01}^{\mathrm{ph}}\right)^{\dagger}
    \end{aligned}
\end{equation}     
Combining all these we get,
\begin{equation}
     \mathcal{R}_{\mathrm{amp}} \mathcal{R}_{\mathrm{ph}}\left(\mathcal{E}\left(\rho\right)\right)= \mathcal{R}_{\mathrm{ph}} \mathcal{R}_{\mathrm{amp}}\left(\mathcal{E}\left(\rho\right)\right),
\end{equation}
and thus the phase error correction and amplitude error correction commute with each other.

\chapter{QND Leakage detection and Cooling in Alkaline-earth atoms}
\label{chap:app_leakage_qudits}

In this appendix we go into some more details of the quantum non demolition(QND) leakage detection and cooling for alkaline-earth atoms discussed in \cref{chap:Qudit_leakage}.

\section{QND leakage detection in $^{171}$Yb}
\begin{figure}[!ht]
    \centering    \includegraphics[width=0.45\columnwidth]{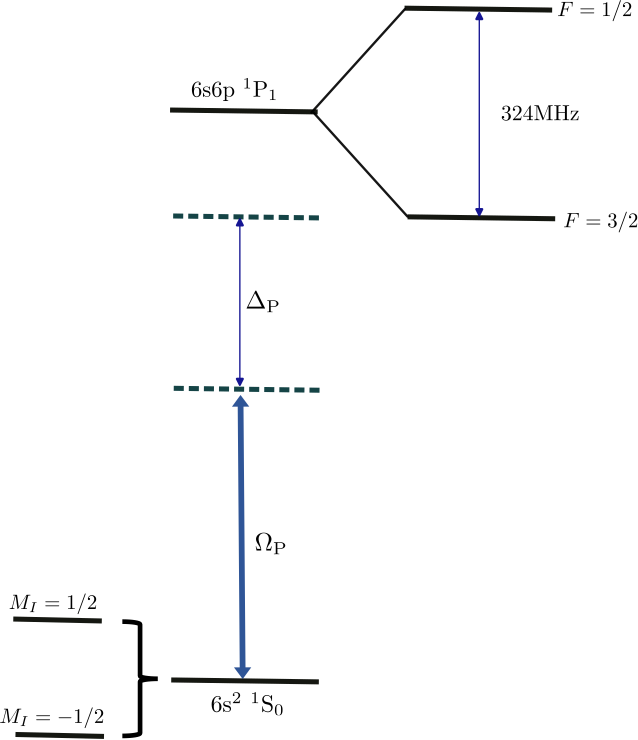}
    \caption{The figure illustrates the setup for detecting the loss of information in the state encoded in the ground state of $^{171}$Yb.
    We utilize far-detuned light from the singlet P state ($\mathrm{6s6p^1P_1}$). 
    For the case of the Yb, there is no tensor light-shift and thus we only need a single laser for detecting the loss of atoms.
    Since the hyperfine is splitting is large compared to the case of Sr, we need to further go off-resonance for a perfect QND leakage detection scheme.
}
    \label{fig:basic_outline_leakage_yb}
\end{figure}

In this section, we outline the extension of the QND leakage detection scheme in $^{171}$Yb.
To understand the working of the leakage detection scheme, one can study the Lindblad Master equation given as,
\begin{equation}
\frac{d\rho}{dt}=-i\comm{H_{\mathrm{LS}}}{\rho}+\sum_q W_q\rho W_q^{\dagger}-\frac{1}{2}\{W_q^{\dagger}W_q,\rho\}.
\end{equation}
The light shift Hamiltonian in the far-off resonance for a $\pi$ polarized light is given as,
\begin{equation}
H_{\mathrm{LS}} \approx \frac{\Omega_\mathrm{P}^2}{4 \Delta_\mathrm{P}} \left(\mathds{1}-\frac{1}{\Delta_\mathrm{P}}\beta^{(0)} \mathds{1}\right).
\end{equation}
The tensor light shift goes to zero as the nuclear spin in the ground state is $1/2$ and there can only be scalar and vector term and for $\pi$ polarized light, the vector term also goes to zero.
The jump operators are given as,
\begin{equation}
\begin{aligned}
W_0&\approx \frac{\Omega_{\mathrm{P}}}{2\Delta} \mathds{1}+\frac{\Omega_{\mathrm{P}}}{2\Delta_{\mathrm{P}}^2}\gamma^{(0)} \mathds{1},\\
W_{+}&\approx \frac{\Omega_{\mathrm{P}}}{2\Delta_{\mathrm{P}}^2} \left(i\gamma^{(1)}F_{-}\right),\\
W_{-}&\approx \frac{\Omega_{\mathrm{P}}}{2\Delta_{\mathrm{P}}^2} \left(i\gamma^{(1)}F_{+}\right),
\end{aligned}
\end{equation}
Thus upto third order in $1/\Delta_{\mathrm{P}}$ we get $d\rho/dt \to 0$.
To understand the performance of the scheme in \cref{fig:basic_outline_qnd_leakage}, consider the following  initial state 
\begin{equation}
\ket{\psi}=\frac{1}{\sqrt{d}}\sum_{i=-\frac{1}{2}}^{\frac{1}{2}} \ket{M_I=i},
\label{eq:initial_state_leakage_yb}
\end{equation}
where $d=2$. 
The fidelity of the final state is given as,
\begin{equation}
\mathcal{F}=\bra{\psi}\rho\ket{\psi}.
\label{eq:fidelity_leakage_yb}
\end{equation}
In \cref{fig:qnd_leakage_yb}, we investigated the fidelity of the state described in \cref{eq:initial_state_leakage_yb} after the time required for detecting $100$ photons. Numerical analysis suggests that increasing $\Delta_{\mathrm{P}}$ improves fidelity, and for sufficiently large detunings, the ideal fidelity can be recovered, establishing a Quantum Non-Demolition (QND) leakage detection scheme.
Compared to the case of $^{87}$Sr for the $^{171}$Yb, we need to go further off resonance for a near ideal QND leakage detection.
\begin{figure}[!ht]
    \centering    \includegraphics[width=0.8\columnwidth]{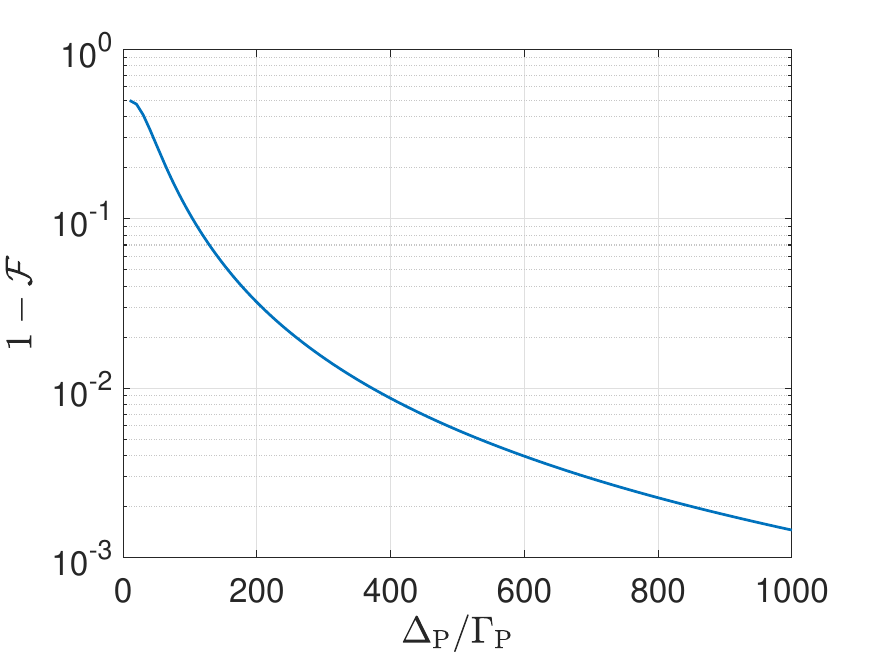}
    \caption{The figure shows the simulation of infidelity as a function of detuning from the singlet state for a time required for scattering $100$ photons for the setting given in \cref{fig:basic_outline_leakage_yb}.
Lower infidelity indicates  a more effective  QND scheme for leakage detection.
Moving further away from resonance enhances the scheme's effectiveness, approaching an ideal scenario for QND leakage detection.
However, compared to the case of $^{87}$Sr, we need to go further off resonance for a near ideal QND leakage detection.
}
    \label{fig:qnd_leakage_yb}
\end{figure}

\section{Perturbation Theory Analysis}
\label{sec:perturbation_theory}
In this section, we detail the perturbation theory calculation for the QND cooling scheme in \cref{sec:qnd_cooling}.
This gives us an understanding of the frequency dependence of the scattered photons.
We have the total Hamiltonian given as,
 \begin{equation}
     H=H_{\mathrm{LS}}+H_{\mathrm{hf}}
 \end{equation}
where the Hyperfine Hamiltonian ($H_{\mathrm{hf}}$) is given in \cref{eq:Hyperfine_Hamiltonian} and the the light shift Hamiltonian ($H_{\mathrm{LS}}$) is given in \cref{eq:light_shift_Hamiltonian}.
For the case when the light shift Hamiltonian is the dominant one, the unperturbed eigenstates are in the product basis $\ket{I,M_I}\otimes \ket{J,M_J}$.
One can find the correction up to the second order in energy as,
\begin{equation}    E_{M_I,M_J}=E_{M_I,M_J}^{(0)}+  \delta E_{M_I,M_J}^{(1)}+ \delta E_{M_I,M_J}^{(2)}
\end{equation}
where $E_{M_I,M_J=0}^{(0)}$ is the unperturbed energy for the $M_J=0$ (the states of interest),
\begin{equation}
    \begin{aligned}
        \delta E_{M_I,M_J=0}^{(1)}&=\bra{M_I,M_J=0} H_{\mathrm{hf}}\ket{M_I,M_J=0}\\
     \delta E_{M_I,M_J=0}^{(2)}&=\sum_{M_{I}' M_{J}'}\frac{\lvert\bra{M_I',M_J'} H_{\mathrm{hf}}\ket{M_I,M_J=0}\rvert^2}{E_{M_I,M_J=0}^{(0)}-E_{M_I',M_J'}^{(0)}}
    \end{aligned}
\end{equation}
We get, $E_{M_I,M_J=0}^{(0)}=0$ and the first order correction is given as,

    \begin{equation}
    \begin{aligned}
          \delta E_{M_I,M_J=0}^{(1)}&=\frac{Q}{2I J(2I-1)(2J-1) }\bra{M_I,M_J=0}3 (\hat{I}.\hat{J})^2-I(I+1)J(J+1)\ket{M_I,M_J=0},
    \end{aligned}
\end{equation}
where we have used the fact that,
\begin{equation}
    \hat{I}.\hat{J}=\frac{1}{2}\left(I_+J_-+I_+J_-\right)+I_zJ_z,
\end{equation}
we get,
  \begin{equation}
    \delta  E_{M_I,M_J=0}^{(1)}=\frac{Q}{2I J(2I-1)(2J-1)}\left[3\left(I(I+1)-M_I^2\right)+I(I+1)J(J+1)\right].
\end{equation}  

Thus from the first-order perturbation theory for the states of interest, the term that depends on the nuclear spin is the $M_I^2$ which creates a quadratic dependence on the nuclear spin of the state.
For $^{87}$Sr we have $Q=39$MHz, which in turn gives the coefficient in front of the quadratic dependence to be $39/72$MHz which is not very small compared to the linewidth $\Gamma=32$MHz.

The second-order perturbation theory for the specific state of interest is given as,
\begin{equation}
    \begin{aligned}
         \delta E_{M_I}^{(2)}&=-\frac{2\sqrt{2}}{\Omega_{\mathrm{^1S_0}}} \sum_{M_{I'}} \lvert \bra{M_{I'},M_J=1} H_{\mathrm{hf}}\ket{M_I,M_J=0}\rvert^2 \\
 &+\frac{2\sqrt{2}}{\Omega_{\mathrm{^1S_0}}} \sum_{M_{I'}} \lvert \bra{M_{I'},M_J=-1} H_{\mathrm{hf}}\ket{M_I,M_J=0}\rvert^2.
    \end{aligned}
\end{equation}

\vspace{0.7cm}
\section{Wigner-Eckart theorem to find the matrix element }
\label{sec:matrix_element}
In this section, we consider the light shift interaction that couples the  $\mathrm{5s5p^1P_1}\equiv a$ to the  $\mathrm{5s15d^1D_2}\equiv c$. 
The light shift Hamiltonian for a $\pi$ polarized light for state the $\ket{M_J=0,M_I}$ with the the uncoupled basis in $c$ is given as,
\begin{equation}
    V_{\mathrm{LS}}^{(ac)}\propto \sum_{F',M_{F'}}\frac{\Omega_{ac}^2}{4\Delta_{F'}}\lvert\langle{c,F',M_{F'}}\lvert d_z\ket{a,M_J=0,M_I}\rvert^2.
    \label{eq:light_shift_Hamiltonian_dd_app}
\end{equation}
where $\Delta_{F'}=\Delta_{\mathrm{D}}-\left[E_{F'}(c)-E_{M_J=0}(a)\right]=\Delta_{\mathrm{D}}+\delta_{F'}$.
To find the matrix element for a particular value of $M_I$, one can use the Wigner-Eckart theorem. 
Notice that,

    \begin{equation}
    \begin{aligned}        \bra{c,F',M_{F'}}d_z\ket{a,J,M_J,I,M_I}&=\bra{F',M_{F'}}\ket{J,M_J,I,M_I}\bra{c,J',M_J;I',M_{I'}}d_z\ket{a,J,M_J,I,M_I}.
    \end{aligned}
\end{equation}
For  a $\pi$ polarized light using the reduced dipole matrix element one obtains,
\begin{equation}    \bra{c,J',M_J;I',M_{I'}}d_z\ket{a,J,M_J,I,M_I}=\langle{c,J'}\lvert\lvert d\rvert \rvert {a,J}\rangle\bra{J',M_{J'}}\ket{J,M_J;1,0}\delta_{M_I,M_{I'}}.
\end{equation}
Thus for $J=1,J'=2$ and $M_{J}=0$, we get,
\begin{equation}
    \begin{aligned}        \bra{c,F',M_{F'}}d_z\ket{a,J,M_J,I,M_I}&\propto\langle{c,J'}\lvert\lvert d\rvert \rvert {a,J}\rangle \langle{F',M_{F'}=M_I}\rvert{2,0;I=\frac{9}{2},M_I}\rangle\bra{2,0}\ket{1,0,1,0},
    \end{aligned}
\end{equation}
which in turn gives,
\begin{equation}
    V_{\mathrm{LS}}^{(ac)}(M_I)=\Omega_{ac}^2  \sum_{F'}\frac{1}{4\Delta_{F'}}\lvert\langle F',M_{F'}=M_I \rvert2,0; \frac{9}{2},M_I \rangle \rvert^2  \lvert\langle 2,0 \rvert \rangle 1,0;1,0\rvert^2 .
    \label{eq:light_shift_Hamiltonian_dd}
\end{equation}
where $\Omega_{ac}^2=E_{\mathrm{L}}^2  \langle{c,J'}\lvert\lvert d\rvert \rvert {a,J}\rangle^2$ and $E_{\mathrm{L}}^2$ is the proportionality constant.
For the case when we detune far away from all the hyperfine states i.e $\Delta_{\mathrm{D}}\gg \delta_{F'}$, we get,
\begin{equation}
    V_{\mathrm{LS}}^{(ac)}(M_I)=V_0^{ac}\left(1-\sum_{F'}\frac{\delta_{F'}}{\Delta_{\mathrm{D}}}\lvert\langle F',M_{F'}=M_I \rvert2,0; \frac{9}{2},M_I \rangle \rvert^2 \right)
\end{equation}
where $V_0^{ac}=\frac{\Omega_{ac}^2}{4}\lvert\langle 2,0  \rvert  1,0;1,0\rangle \rvert^2$.
In the other regime when we work closely detuned to $F'=13/2$, one can obtain,
\begin{equation}
     V_{\mathrm{LS}}^{(ac)}(M_I)\approx V_0^{ac}\lvert\langle F'=\frac{13}{2},M_{F'}=M_I \rvert2,0; \frac{9}{2},M_I \rangle \rvert^2.
\end{equation}
By empirically fitting,
\begin{equation}
    \lvert\langle F'=\frac{13}{2},M_{F'}=M_I \rvert2,0; \frac{9}{2},M_I \rangle \rvert^2\approx 0.3-0.017M_I^2+2.3 \cross10^{-4}M_I^4.
\end{equation}

The quadratic behavior is not familiar for a light shift (usually at most quadratic in nature). 
Here it arises from the way in which the nucleus is coupling to the electron through $J=2$ (which is quadrupolar rather than dipolar). 
This quadratic term indicates that one can cancel the energy light arising from the hyperfine perturbation in the state $a$ which also has a quadratic term from the perturbation theory analysis.

To further understand the light-shift Hamiltonian, one can use the following expansion,

\begin{small}
     \begin{equation}
      \begin{aligned}
    &\left\lvert \bra{F',M_{F'}}d_z\ket{J,M_J,I,M_I} \right\rvert^2=\left\lvert\sum_{F}\bra{F',M_{F'}}d_z\ket{F,M_F}\braket{F,M_F}{J=1,M_J=0,I=\frac{9}{2},M_I}\right\rvert^2,\\
    =&\sum_{F_1,F_2 } \bra{F_1,M_{F_1}}d_z\ket{F',M_{F'}}
    \bra{F',M_{F'}}d_z\ket{F_2,M_{F_2}}
    \braket{F_1,M_{F_1}}{J=1,M_J=0,I=\frac{9}{2},M_I} \\
    &\braket{F_2,M_{F_2}}{J=1,M_J=0,I=\frac{9}{2},M_I},\\
    =&\sum_{F_1,F_2 } \bra{F_1,M_I}d_z\ket{F',M_I}
    \bra{F',M_I}d_z\ket{F_2,M_I}
    \braket{F_1,M_I}{J=1,M_J=0,I=\frac{9}{2},M_I}\\
    &\braket{F_2,M_I}{J=1,M_J=0,I=\frac{9}{2},M_I}.\\
    \end{aligned}
\end{equation}  
\end{small}

Also one can write,
\begin{equation}
    \bra{F_1,M_I}d_z\ket{F',M_I}=\braket{F_1,M_I}{1,0,F',M_I} \langle F_1\lvert\lvert d_z\rvert \rvert F'\rangle,
\end{equation}
thus the light-shift interaction here is not restricted to a single angular momentum ground state $F$.
This is reflected in the fact that the light-shift does not have the familiar scalar-vector-tensor form in terms of the polynomials in $(F_x,F_y,F_z)$.

\setstretch{\dnormalspacing}
\backmatter

\end{document}